\documentclass{article}

\usepackage{ifpdf}
\usepackage{lscape}
\usepackage{longtable}
\usepackage{natbib}
\usepackage{epubtk}
\usepackage{graphicx}

\def\ga{\;\rlap{\lower 2.5pt\hbox{$\sim$}}\raise 1.5pt\hbox{$>$}\;}
\def\la{\;\rlap{\lower 2.5pt\hbox{$\sim$}}\raise 1.5pt\hbox{$<$}\;}
\def\ph{\phantom{0$\pm$0.00}}
\def\pha{\phantom{$\pm$0.00}}
\def\phb{\phantom{$\pm$0.0}}
\def\phc{\phantom{$\pm$0.00}}

\newcommand{\unit}[1]{\,\mathrm{#1}}

\def\P{{\it P }} 

\begin{document}

\title{The Sun in Time: Activity and Environment}

\author{\epubtkAuthorData{Manuel G\"udel}{%
Paul Scherrer Institute, W\"urenlingen and Villigen, \\
CH-5232 Villigen PSI, Switzerland \\
and \\
Max-Planck-Institute for Astronomy, K\"onigstuhl 17, \\
D-69117 Heidelberg, Germany }{%
guedel@astro.phys.ethz.ch}{%
http://www.astro.phys.ethz.ch/staff/guedel/guedel.html}%
}

\date{}
\maketitle

\begin{abstract}
The Sun's magnetic activity has steadily declined during its
main-sequence life. While the solar photospheric luminosity was about
30\% lower 4.6\,Gyr ago when the Sun arrived on the main sequence
compared to present-day levels, its faster rotation generated
enhanced magnetic activity; magnetic heating processes in the
chromosphere, the transition region, and the corona induced
ultraviolet, extreme-ultraviolet, and X-ray emission about 10, 100,
and 1000~times, respectively, the present-day levels, as inferred from
young solar-analog stars. Also, the production rate of accelerated,
high-energy particles was orders of magnitude higher than in
present-day solar flares, and a much stronger wind escaped from the
Sun, permeating the entire solar system. The consequences of the
enhanced radiation and particle fluxes from the young Sun were
potentially severe for the evolution of solar-system planets and
moons. Interactions of high-energy radiation and the solar wind with
upper planetary atmospheres may have led to the escape of important
amounts of atmospheric constituents. The present dry atmosphere of
Venus and the thin atmosphere of Mars may be a product of early
irradiation and heating by solar high-energy radiation. High levels of
magnetic activity are also inferred for the pre-main sequence Sun. At
those stages, interactions of high-energy radiation and particles with
the circumsolar disk in which planets eventually formed were
important. Traces left in meteorites by energetic particles and
anomalous isotopic abundance ratios in meteoritic inclusions may
provide evidence for a highly active pre-main sequence Sun. The
present article reviews these various issues related to the magnetic
activity of the young Sun and the consequent interactions with its
environment. The emphasis is on the phenomenology related to the
production of high-energy photons and particles. Apart from the
activity on the young Sun, systematic trends applicable to the entire
main-sequence life of a solar analog are discussed.
\end{abstract}

\epubtkKeywords{Active stars, Climate, Cool stars, Corona,
		Circumstellar disks, Dynamo, Flares, High-energy
		radiation, Magnetic activity, Magnetic fields,
		Planetary atmospheres, Solar analogs, Solar evolution}

\newpage

\setcounter{tocdepth}{3}
\tableofcontents
\newpage


\section{Introduction}
\label{section:introduction}

The study of the past of our Sun and its solar system has become an
interdisciplinary effort between stellar astronomy, astrophysics of
star and planet formation, astrochemistry, solar physics, geophysics,
planetology, meteoritical science and several further disciplines. The
interest in understanding the past evolution of our star is obvious;
the Sun's radiative energy, the solar wind, and various forms of
transient phenomena (e.g., shock waves, high-energy particle streams
during flares) are key factors in the formation and evolution of the
planets and eventually the biosphere on Earth.

The Sun is, like almost all cool stars, a ``magnetic star'' that
produces magnetic fields through dynamo operation in the
interior. These fields reach the surface where their presence is
noticed in the form of sunspots. However, magnetic activity has much
more far-reaching consequences: solar magnetic fields control
essentially the entire outer solar atmosphere, they heat coronal gas
to millions of degrees, they produce flares whose by-products such as
shock waves and high-energy particles travel through interplanetary
space to eventually interact with planetary atmospheres; the solar
wind is guided by open magnetic fields; this magnetized-wind complex
forms a large bubble, an ``astrosphere'' in interstellar space
containing the entire solar system and protecting it from a high dose
of cosmic rays.

Was the Sun's magnetic activity different in its infancy when planets
and their atmospheres formed, or when it was still surrounded by an
accretion disk? Accumulated direct and indirect evidence indeed points
to a much higher level of magnetic activity in the young Sun, in
particular in its pre-main sequence (PMS) phase and the subsequent
epoch of its settling on the main sequence (MS). Direct evidence
includes meteoritic traces and isotopic anomalies that require much
higher proton fluxes at early epochs at least partly from {\it within}
the solar system (Section~\ref{section:environment} below); indirect
evidence comes from systematic comparisons of the contemporary Sun
with solar analogs of younger age that unequivocally show a strong
trend toward elevated activity at younger ages
(Section~\ref{section:solarflux} below). Interestingly, planetary
atmospheres offer further clues to strongly elevated activity levels:
evidence for a warmer early climate on Mars or the extremely arid
atmosphere of Venus -- a sister planet of the water-rich Earth -- call
for explanations, and such explanations may be found in the elevated
activity of the young Sun (see Section~\ref{section:planets} below). The
study of the early solar activity is the theme of the present review
article.

The main goal of this article is therefore to demonstrate evidence for
a much more active young Sun, and to study the consequences this might
have had for the development of the solar environment, including the
formation and evolution of planets. Our discussion will therefore take
us through the following three major issues:

\begin{itemize}

  \item The young Sun's more rapid rotation induced an internal
  magnetic dynamo that was much more efficient than the present-day
  Sun's. Consequently, stronger surface magnetic fields and/or higher
  surface magnetic filling factors should have induced enhanced
  ``activity'' in all its variations, from larger surface spots to a
  stronger, extended solar wind. If we can observationally probe the
  outer magnetic activity of the Sun, we obtain invaluable diagnostics
  for a deeper theoretical understanding of the internal dynamo.

  \item The solar output largely controls planetary atmospheres and
  their climates. While this is obviously true for the dominant
  optical and infrared emissions of a star like the Sun, the
  irradiation of planetary atmospheres by higher-energy ultraviolet
  and X-ray photons as well as interactions with high-energy particles
  and the solar wind leads to atmospheric alterations that have been
  recognized and numerically simulated only recently. The much higher
  magnetic activity of the young Sun and the resulting higher levels
  of ultraviolet, X-ray, and particle irradiation were therefore of
  prime importance for the early evolution of the planets. \\

  The discovery of extrasolar planets in particular around Sun-like
  stars has also spurred interest in star-planet interactions, e.g.,
  erosion of atmospheres or photochemical reactions, that profit from
  detailed studies in the solar system.

  \item Similarly, at still younger stages of a star's evolution, its
  environment is rich in molecular gas and dust, both in the form of a
  large envelope (in the youngest, protostellar phases of evolution)
  and a circumstellar gas and dust disk (including the T~Tauri phase
  when planets were forming). Star-disk interactions are manifold, and
  their role is fundamental in various respects. The optical and
  ultraviolet radiation heats the disk and therefore primarily
  determines disk structure and the formation and evolution of
  planetary systems.\\
  High-energy emission, in particular X-ray radiation, further heats
  and ionizes parts of the circumstellar disk. Even moderate disk
  ionization will lead to accretion instabilities if weak magnetic
  fields are present. Disk heating by X-rays may also produce extreme
  temperature gradients across the disk that drive complicated
  chemical networks relevant for the later processing of the disk
  material into forming planets and planetary atmospheres.

\end{itemize}

The focus of this review is therefore, on the one hand, on signatures
of magnetic activity across the electromagnetic spectrum, representing
physical processes in the photosphere, the chromosphere, and the
thermal and non-thermal corona of a solar-like star. I will mostly use
young solar-like stars to infer conditions that -- by analogy -- might
have prevailed on the young Sun. On the other hand, I will also
discuss traces that the elevated activity of the young Sun might have
left behind in meteorites and in planetary atmospheres, thus
collecting ``in-situ'' information about the distant past of our own
solar system.

While this article focuses on the conditions on the young Sun and in
the early solar system, it has proven convenient to study the solar
evolution in time systematically from young to old, because a number
of trends become evident that can be calibrated with the
contemporaneous Sun. We thus not only learn about the young Sun, but
we uncover the systematics that made it different from what it is
today. This is the approach I adopt in the present work.

This article will not address issues on the formation and evolution of
the Sun that are related to its internal constitution, with the
exception of cursory reference to the magnetic dynamo that is, of
course, at the origin of all solar magnetic activity. I will
  treat the PMS Sun in separate chapters for three related reasons:
  First, fundamental properties of the PMS Sun were largely different
  from those of the contemporaneous Sun (for example, its spectral
  type, or its photospheric effective temperature). Second, new
  features not present in the modern Sun become dominant key players
  related to activity and environment, among them accretion disks,
  accretion streams, star-disk magnetospheres, outflows, and
  jets. And third, the PMS behavior of the Sun cannot be assessed in
  detail judged from the present-day solar parameters; we can only
  discuss the range of potential evolutionary scenarios now observed
  in a wide sample of PMS stars (e.g., with respect to mass accretion
  rate, disk mass, disk dispersal time, rotation period, etc.).

Numerous review articles are available on subjects related to the
present one. Without intention to be complete, I refer here in
particular to the collection of papers edited by \cite{sonett91} and
\cite{dupree04}, the {\it Cool Stars Workshop} series (the latest
volume edited by \citealt{vanbelle07}), and the {\it Protostars and
 Planets} series (in particular the latest volumes by
\citealt{mannings00} and \citealt{reipurth07}). An early overview of
solar variability (including that of its activity) can be found in
\cite{newkirk80}. \cite{walter91} specifically reviewed knowledge of
the long-term evolution of solar activity as known in the early
nineties, in a similar spirit as the present review; numerous older
references can be found in that work. For summaries of stellar X-ray
and radio emission, I refer to \cite{guedel04} and \cite{guedel02},
respectively. \cite{feigelson99} and \cite{feigelson07} have
summarized PMS aspects of magnetic activity. \cite{glassgold05} have
reviewed the influence of the magnetic activity of the PMS Sun on its
environment, in particular on its circumstellar disk where our planets
were forming. \cite{wood04} has discussed evidence for winds
emanating from solar-like stars, and \cite{goswami00} have summarized
findings related to radionuclides in meteorites and inferences for
the young solar system; the most recent developments in this field
have been reviewed by \cite{wadhwa07}. \cite{kulikov07} and
\cite{lundin07} have provided summaries on interactions between solar
high-energy radiation and particles with planetary atmospheres, in
particular those of Venus, Earth, and Mars.

\newpage


\section{What is a Solar-Like Star?}
\label{section:solar_like_stars}

The present Sun is a G2~V star with a surface effective temperature of
approximately 5780\,K. Stellar evolution theory indicates, however,
that the Sun has shifted in spectral type by several subclasses,
becoming hotter by a few hundred degrees and becoming more luminous
(the bolometric luminosity of the Sun in its zero-age main-sequence
[ZAMS] phase amounted to only about 70\% of the present-day output;
\citealt{siess00}). In understanding the solar past, we must therefore
also consider stars of mid-to-late spectral class G. On the other
hand, alternative evolutionary scenarios have suggested continuous
mass loss from the young Sun at a high rate that would require a
somewhat {\it earlier} spectral classification of the young Sun
\citep{sackmann03}. In any case, magnetic activity in the outer
stellar atmospheres is predominantly controlled by the depth of the
stellar convection zone and stellar rotation, both of which also
evolve during stellar evolution. For our understanding of magnetic
activity, the precise spectral subclass is rather likely to play a
minor role. When discussing ``solar analogs'', I will therefore
concentrate on stars mostly of early-to-mid-G spectral types but will
occasionally also consider general information from outer atmospheres
of somewhat lower-mass stars if available.

The situation is more complex for stars in their PMS stage. The Sun
spent much of its PMS life as a mid-K (K5~IV) star when it moved down
the Hayashi track. But again, the precise spectral subtype matters
even less for magnetic activity in this stage, the more important key
parameters being the age of the star (controlling its total
luminosity, its radius, and the development and therefore the depth of
the convection zone), the presence and dispersal of a circumstellar
disk (controlling mass accretion and, via magnetic fields, the spin of
the star), and the presence and strength of outflows (controlling,
together with accretion, the final evolution of the stellar mass).
A somewhat more generous definition of ``pre-main sequence solar
analogs'' is clearly in order, given that the Sun's history of
rotation, accretion, the circumsolar disk, and the solar mass loss
cannot be precisely assessed. Quite generally, I will take solar-like
stars in the PMS phase to be, from the perspective of ``magnetic
activity'', stars with masses of roughly 0.5\,--\,1.5 solar masses,
covering spectral classes from early G to late K/early M.
 
The expression ``solar twin'' \citep{cayrel89} is occasionally
used. This term should be used solely in the context of a solar analog
with an age close to the Sun's, i.e., of order 4\,--\,6\,Gyr, an age
range in which the internal structure and the rotation period of a
$1\,M_{\odot}$ (and therefore, its activity level) evolve only
insignificantly.  Efforts toward identifying real solar twins have
  been important in the context of putting our Sun into a wider
  stellar context; nearby solar analogs that are essentially
  indistinguishable from the Sun with regard to spectral type,
  effective temperature, gravity, luminosity, age, rotation, and
  magnetic activity \citep{portodemello97} prove that the Sun can be
  robustly used as an anchor to calibrate evolutionary trends -- the
  Sun is not an exception but is representative of its age and mass, a
  conclusion also reached by \cite{gustafsson98} from a rather general
  comparison of the Sun with sun-like stars.\epubtkFootnote{The star
    claimed to be ``the closest ever solar twin''
    \citep{portodemello97}, HR~6060, shows parameters nearly
    indistinguishable from solar values indeed: $L/L_{\odot} = 1.05
    \pm 0.02$; spectral type G2~Va (Sun: G2~V); $B-V = 0.65$ (Sun:
    0.648); $U-B = 0.17$ (Sun: 0.178); $T_\mathrm{eff} = 5789
    \unit{K}$ (Sun: 5777\,K; $\Delta T_\mathrm{eff} = 12 \pm 30
    \unit{K}$); $\log g = 4.49$ (Sun: 4.44; $\Delta\log g = 0.05 \pm
    0.12$); microturbulence velocity $\xi = 1.54 \unit{km s^{-1}}$
      (Sun: $1.52 \unit{km s^{-1}}$; $\Delta\xi = [0.02 \pm 0.04]
	\unit{km s^{-1}}$); element abundances solar within $1\sigma$,
	  in particular $\mathrm{[Fe/H]} = 0.05 \pm 0.06$; Mount
	  Wilson activity index $<S> = 0.174$ (Sun: 0.177 in 1980);
	  rotational velocity $v\sin i < 3.0 \unit{km s^{-1}}$ (Sun: $v
	    = 2 \unit{km s^{-1}}$).}

Table~\ref{table:definitions} gives a list of terms, symbols, and
acronyms used throughout the text.

\begin{table}[htbp]
\begin{center}
\caption{\it Symbols and units used throughout the text}
\label{table:definitions}
\vskip 4mm
    
\begin{tabular}{ll}
\hline\noalign{\smallskip}
Symbol, acronym	       	    & Explanation \\
\noalign{\smallskip}\hline\noalign{\smallskip}
$c$             	    & Speed of light [$\approx 3\times 10^{10} \unit{cm\ s^{-1}}$] \\
$k$             	    & Boltzmann's constant [$\approx 1.38 \times 10^{-16} \unit{erg\ K^{-1}}$] \\
$G$             	    & Gravitational constant [$\approx 6.67 \times 10^{-8} \unit{dyn\ cm^{2}\ g^{-2}}$] \\
$R_*$             	    & Stellar radius [cm] \\
$R_{\odot}$                 & Solar radius [$7 \times 10^{10} \unit{cm}$] \\
$M, M_*$             	    & Stellar or planetary mass [g] \\
$M_{\odot}$                 & Solar mass [$2 \times 10^{33} \unit{g}$] \\
$\dot{M}_\mathrm{w}$        & Mass loss rate (by wind) [$M_{\odot} \unit{yr^{-1}}$] \\
$P$                         & Rotation period [d] \\
$\Omega$                    & Angular rotation frequency \\
$v_\mathrm{rot}$            & Equatorial rotation velocity [$\mathrm{km\ s^{-1}}$] \\
$T$             	    & Temperature, also coronal electron temperature [K] \\
$T_\mathrm{av}$             & Average coronal temperature [K] \\
$T_\mathrm{eff}$            & Effective temperature [K] \\
$T_\mathrm{exo}$            & Exospheric (planetary) temperature [K] \\
$B$             	    & Magnetic field strength [G] \\
$f$             	    & Surface filling factor \\
$L$             	    & Coronal magnetic loop semi-length [cm] \\
$L_\mathrm{X}$              & X-ray luminosity 	[$\mathrm{erg\ s^{-1}}$] \\
$L_\mathrm{bol}$            & Stellar bolometric luminosity [$\mathrm{erg\ s^{-1}}$] \\
$F$             	    & Flux (line- or band-integrated) [$\mathrm{erg\ cm^{-2}\ s^{-1}}$] \\
$Ro$             	    & Rossby number \\
$t, t_6, t_9$               & Stellar age, in Myr, in Gyr \\
$\omega_\mathrm{cycl}$      & Activity cycle frequency \\
ISM                         & Interstellar Medium \\
EM             	            & Emission Measure \\
(ZA)MS             	    & (Zero-Age) Main Sequence \\
PMS                         & Pre-Main Sequence \\
(C/W)TTS                    & (Classical/Weak-line) T~Tauri Star \\
UV                          & Ultraviolet radiation \\
FUV                         & Far-Ultraviolet radiation \\
EUV                         & Extreme Ultraviolet radiation \\
XUV                         & Ultraviolet--to--X-ray radiation \\
IUE                         & {\it International Ultraviolet Explorer} (NASA, ESA, UK)\\
EUVE                        & {\it Extreme Ultraviolet Explorer} (NASA) \\
ROSAT                       & {\it R\"ontgensatellit} (German/UK/NASA X-ray satellite) \\
VLA                         & {\it Very Large Array} radio telescope (USA) \\
MRI                         & Magneto-Rotational Instability \\
CAI                         & Calcium-Aluminum-rich Inclusion \\
(I)FIP                      & (Inverse) First Ionization Potential Effect \\
\noalign{\smallskip}\hline
\end{tabular}
\end{center}
\end{table}

\newpage


\section{The Sun in Time}
\label{section:sunintime}

\subsection{Goals of the ``Sun in Time'' Project}

Solar analogs with different ages and therefore activity levels play
an important role in our understanding of the past magnetic evolution
of the Sun. For stars with masses $\la 1.5\,M_{\odot}$ and ages of at
least a few 100\,Myr, angular momentum loss by a stellar wind brakes
rotation in such a way that it is nearly uniquely determined by the
stellar age \citep{soderblom93}. The only independent variable, the
stellar age, then determines the rotation period and, through the
internal magnetic dynamo, magnetic activity at all levels of the
stellar atmosphere, probably including characteristics of the stellar
wind \citep{wood02, wood05, kulikov06}. The study of a series of
near-solar-mass stars back to ages close to the ZAMS age will
therefore be sufficient to reconstruct the history of our Sun and the
interaction of its magnetic activity with its environment. This is the
goal of the observational ``Sun in Time'' program (see, e.g.,
\citealt{dorren94c}).

Specifically, the ``Sun in Time'' program was established to study the
long-term magnetic evolution of the Sun during the entire MS
lifetime. The primary aims of the program are, (1) to test dynamo
models of the Sun in which rotation is the only significant variable
parameter, and (2) to determine the spectral irradiance of the Sun
over its MS lifetime. Key studies within the ``Sun in Time'' program
comprise the following:

\begin{itemize}

  \item the optical (photospheric) and UV ``Sun in Time'', the latter
  studied with {\it IUE} \citep{messina02, messina03, dorren94}, to
  characterize level and extent of surface magnetic activity;

  \item the FUV ``Sun in Time'' as studied with {\it FUSE}
  \citep{guinan03}, to infer transition-region magnetic activity;

  \item the EUV ``Sun in Time'' as studied with {\it EUVE}
  \citep{ayres97, guedel97b, audard99, ribas05}, to obtain information
  on coronal activity;

  \item the X-ray ``Sun in Time'' as studied by {\it ROSAT} and {\it
  ASCA} \citep{guedel97a, guedel95a, dorren95}, to obtain full
  coverage in coronal temperatures and characterize flares;

  \item the X-ray ``Sun in Time'' as studied spectroscopically with
  {\it XMM-Newton} \citep{telleschi05}, to study the thermal
  stratification and composition of coronae of solar analogs;

  \item spectral evolution of the XUV ``Sun in Time'' \citep{ribas05},
  to find systematics across the UV-to-X-ray range;

  \item the radio ``Sun in Time'' as studied with the VLA
  \citep{guedel94, guedel95c, guedel95a, guedel97a, gaidos00,
  guedel01b}, to measure the production of non-thermal particles in
  magnetic coronae;

  \item magnetic cycles of the ``Sun in Time'' \citep{messina02},
  measured in the optical or in X-rays;

  \item differential rotation of the ``Sun in Time'' \citep{messina03}
  as seen in starspot measurements.

\end{itemize}

A summary of the instruments used for the observations and their
wavelength (or energy) ranges covered is given in
Table~\ref{table:sunintimeobs}. Various project summaries can be found
in \cite{dorren94c}, \cite{guedel98a}, \cite{guinan02a},
\cite{guinan02b}, and \cite{guedel03d}.

\begin{table}[htbp]
\begin{center}
\caption{\it The ``Sun in Time'' Project: Relevant Observations}
\label{table:sunintimeobs}
\vskip 4mm

\begin{tabular}{lrrr}
\hline\noalign{\smallskip}
Instrument                     & Wavelength range &  Energy range  & References$^a$ \\
                               & (~\AA)           &  (keV)         &  \\
\noalign{\smallskip}\hline\noalign{\smallskip}
{\it XMM-Newton} (CCDs)        & 0.83\,--\,83         & 0.15\,--\,15       & 1 \\
{\it XMM-Newton} (gratings)    & 6.0\,--\,38.0        & 0.33\,--\,2.1      & 1 \\
{\it ASCA} (CCDs)              & 1.2\,--\,31          & 0.4\,--\,10	    & 2 \\
{\it ROSAT} (PSPC)             & 5.2\,--\,124         & 0.1\,--\,2.4	    & 2, 3, 4, 5 \\
{\it EUVE} (gratings)          & 80\,--\,760$^b$      & 0.016\,--\,0.15    & 5, 6, 7, 8 \\
{\it FUSE} (gratings)          & 920\,--\,1180        &		    & 9 \\
{\it HST} (gratings)           & 1150\,--\,1730       &		    & 8 \\
{\it IUE} (gratings)           & 1150\,--\,1950       &		    & 10 \\
UBVRI photometry               & 3500\,--\,8300       &		    & 11, 12 \\
{\it VLA} (continuum)          & 3.6\,cm/8.4\,GHz   &		    & 2, 3, 13, 14, 15, 16 \\
\noalign{\smallskip}\hline					     
\end{tabular}

\begin{list}{}{}
\item[$^{\mathrm{a}}$]{References: 
                       1 \cite{telleschi05}; 
		       2 \cite{guedel97a};
		       3 \cite{guedel95a};
		       4 \cite{dorren95};
		       5 \cite{ayres97};
		       6 \cite{guedel97b};
		       7 \cite{audard99};
		       8 \cite{ribas05};
		       9 \cite{guinan03};
		      10 \cite{dorren94};
		      11 \cite{messina02};
		      12 \cite{messina03};
		      13 \cite{guedel94};
		      14 \cite{guedel95c};
		      15 \cite{gaidos00};
		      16 \cite{guedel01b}.
		           }
\item[$^{\mathrm{b}}$]{Wavelengths longer than $\approx$360\,\AA\ are
  subject to strong interstellar absorption}
\end{list}
\end{center}
\end{table}

\subsection{Overview of Stellar Sample}

The program uses a homogeneous sample of single nearby G0-5~V MS stars
(and one G2~IV subgiant) that all have known rotation periods and
well-determined physical properties, including temperatures,
luminosities, metal abundances and ages
(Table~\ref{table:sunintimeobs}). Observations cover spectral ranges
from radio to X-rays (excluding the 360\,--\,920\,\AA\ range, which is a
region of very strong absorption by the interstellar medium). The
principal stellar sample is given in Table~\ref{table:sunintime}; the
samples used for particular studies may vary somewhat, depending on
the availability of appropriate observational data. The parameters
collected in this table are from \cite{dorren94}, \cite{guedel97a},
\cite{guedel98a}, \cite{guedel01b}, \cite{guinan02a}, \cite{ribas05},
and \cite{telleschi05}. Bolometric luminosities have been derived from
absolute visual magnitudes, $M_\mathrm{V}$, by applying standard
bolometric corrections. I briefly summarize the magnetic-activity
properties of a few important ``Young Suns'' that will be discussed
extensively in this paper (further notes can be found in
\citealt{ribas05}).

{\it EK Draconis = HD 129333} is an exemplary near-ZAMS solar analog,
representing the very active young Sun at the time when planetary
atmospheres first formed. EK Dra is a member of the Pleiades Moving
Group at a distance of 34\,pc. \cite{koenig05} estimated an age of only
35\,Myrs, and a mass of $(0.9-1.0) \pm 0.1\,M_{\odot}$. EK Dra's optical
and UV properties have been studied in detail by \cite{dorren94}. With
a rotation period of 2.7\,d, it reveals extreme magnetic activity,
showing a photospheric spot coverage of 6\% (\citealt{dorren94},
larger fractions were given by \citealt{koenig05}), transition-region
line fluxes 20\,--\,100 times larger than the Sun's \citep{dorren94}, a
very hot X-ray emitting corona with a luminosity $\approx$1000~times
the present-day solar X-ray output \citep{guedel97a, telleschi05}, and
an extremely luminous non-thermal radio corona exceeding radio
luminosities of strong solar flares by orders of magnitude
\citep{guedel94, guedel95a}. EK Dra shows rotational modulation from
spots in the optical, and due to inhomogeneous coronal active regions
also in the radio and X-ray regimes \citep{guedel95a}. It further
shows strong evidence for a $\approx$10\,yr activity cycle
\citep{dorren94, dorren95, guedel04, jarvinen05}. \cite{duquennoy91}
reported a low-mass (M-type) companion star, whose mass was found to
be $(0.5 \pm 0.1)\,M_{\odot}$, and \cite{metchev04} discussed evidence
for another $0.2\,M_{\odot}$ companion although the latter has been
questioned by \cite{koenig05}. Note that UV, EUV, X-ray, and radio
emissions are likely to be dominated by the G star, given the low
mass(es) of the companion(s).

{\it 47 Cas~B = HR~581 = HD~12230} is a somewhat mysterious solar
analog in a binary system at a distance of 33.5\,pc. It has not been
revealed in the optical regime given the primary F0~V star's
overwhelming light. But astrometric, X-ray and radio properties
constrain it to be an early-to-mid G-type star, and its likely
membership in the Pleiades Moving Group suggests a near-ZAMS age
\citep{guedel95b, guedel98b}. It shows X-rays and radio emission at an
even more extreme level than EK Dra, perhaps induced by the even
faster rotation; the rotation period has been inferred from periodic
X-ray modulation to be about 1\,d \citep{guedel95b, guedel98b,
  telleschi05}.

{\it $\pi^1$~UMa = HD~72905} and {\it $\chi^1$~Ori = HD~39587:} These
are two members of the Ursa Major stream \citep{king03}, with an
estimated age of 300\,Myr \citep{soderblom93a}. While $\pi^1$~UMa is
considered to be single, $\chi^1$~Ori is orbited by an M-type
companion with an orbital period of about 14\,yrs \citep{han02} and a
mass of 0.15\,$M_{\odot}$ \citep{koenig02}, suggesting that most
observational signatures of magnetic activity are dominated by the
primary.

{\it $\kappa^1$~Cet = HD~20630} is a Hyades-age solar analog. Its age
is not well known but is indirectly inferred from its rotation period
and its X-ray luminosity \citep{ribas05}.

\begin{landscape}

\begin{table}
\begin{center}
\caption{\it The ``Sun in Time'' Sample$^a$}
\label{table:sunintime}

\scriptsize
\begin{tabular}{lrrrlllllllrrll}
\hline\noalign{\smallskip}
Star & HD  & Dist.    & Spectr.& $T_\mathrm{eff}$ & Mass  & Radius	& $M_\mathrm{V}$ & $L_\mathrm{bol}$   & log~$L_\mathrm{X}$ & log~$(L_\mathrm{X}$      & log$L_\mathrm{R}^d$ & $P$           & Age & Age indicator,\\
     & no. & (pc)$^b$ & Type   & (K)        & $(M_{\odot})$ & $(R_{\odot})$ & (m)	     & ($L_{\odot}$)   & (erg/s)$^c$   & $/L_\mathrm{bol})$     & (erg/Hz/s)  & (d) 	  &(Gyr)& Membership \\
\noalign{\smallskip}\hline\noalign{\smallskip}
47 Cas B        & 12230& 33.5  & G~V   & ...	& ...  & ...   & ... & ...     & 30.31 & ...	& 14.91    & 1.0?                 & 0.1  & Pleiades Moving Group   \\	    
EK Dra          &129333& 33.9  & G0~V  & 5870	& 1.06 & 0.91  & 4.96 & 0.90    & 29.93 & -3.61 & 14.18    & 2.75                 & 0.1  &
Pleiades Moving Group   \\	    
$\pi^1$ UMa     & 72905& 14.3  & G1~V  & 5850	& 1.03 & 0.96  & 4.87 & 0.97    & 29.10 & -4.47 & $<$12.67 & 4.68                 & 0.3  & Ursa Major Stream       \\	    
HN Peg          &206860& 18.4  & G0~V  & 5970	& 1.06 & 0.99  & 4.68 & 1.14    & 29.12 & -4.52 & ...	   & 4.86                 & 0.3  & Rotation-Age Relationship$^e$ \\ 
$\chi^1$ Ori    & 39587&  8.7  & G1~V  & 5890	& 1.01 & 1.02  & 4.71 & 1.13    & 28.99 & -4.65 & ...	   & 5.08                 & 0.3  & Ursa Major Stream       \\	    
BE Cet          &  1835& 20.4  & G2~V  & 5740	& 0.99 & 1.02  & 4.83 & 1.02    & 29.13 & -4.46 & ...	   & 7.65                 & 0.6  & Hyades Moving Group     \\	    
$\kappa^1$ Cet  & 20630&  9.2  & G5~V  & 5750	& 1.02 & 0.93  & 5.02 & 0.86    & 28.79 & -4.73 & $<$12.42 & 9.2\phantom{4}       & 0.75 & Rotation-Age Relationship$^f$  \\
$\beta$ Com     &114710&  9.2  & G0~V  & 6000	& 1.10 & 1.10  & 4.45 & 1.41    & 28.21 & -5.52 & $<$12.53 & 12.4\phantom{4}      & 1.6  & Rotation-Age Relationship\\      
15 Sge          &190406& 17.7  & G5~V  & 5850	& 1.01 & 1.10  & 4.56 & 1.29    & 28.06 & -5.64 & ...	   & 13.5\phantom{4}      & 1.9  & Rotation-Age Relationship \\     
Sun             & - &1\,AU & G2~V  & 5777	& 1.00 & 1.00  & 4.83 & 1.00    & 27.30 & -6.29 & ...	   & 25.4\phantom{4}      & 4.6  & Isotopic Dating on Earth \\      
18 Sco          &146233& 14.0  & G2~V  & 5785	& 1.01 & 1.03  & 4.77 & 1.08    & ...	& ...	& ...	   & 23\phantom{.40}      & 4.9  & Isochrones \\		    
$\alpha$ Cen A  &128620&  1.4  & G2~V&5800$^g$&1.10$^h$&1.22$^h$ &4.34& 1.60    & 27.12 & -6.67 & ...	   & $\sim$30\phantom{.40}& 5-6  & Isochrones, Rotation    \\       
$\beta$ Hyi     &  2151&  7.5  & G2~IV & 5774	& 1.10 & 1.92  & 3.43 & 3.70    & 27.18 & -6.41 & ...	   & $\sim$28\phantom{.40}& 6.7  & Isochrones$^i$ \\	            
16 Cyg A        &186408& 21.6  & G1.5~V& 5790	& 1.00 & 1.16  & 4.29 & 1.38    & ...	& ...	& ...	   & $\sim$35\phantom{.40}& 8.5  & Isochrones \\		    
\hline							     
\end{tabular}

\begin{list}{}{}
  \item[$^{\mathrm{a}}$]{Parameters mostly collected from
  \cite{dorren94}, \cite{guedel97a}, \cite{guedel98a},
  \cite{guedel01b}, \cite{guinan02a}, \cite{ribas05}, and
  \cite{telleschi05}.}

  \item[$^{\mathrm{b}}$]{Stellar distances are from the Hipparcos
  Catalogue \citep{perryman97}}

  \item[$^{\mathrm{c}}$]{$L_\mathrm{X}$ refers to the 0.1\,--\,2.4\,keV band as
  measured by ROSAT.}

  \item[$^{\mathrm{d}}$]{For radio observations of further solar
  analogs, see \cite{guedel94} and \cite{guedel01b}.}

  \item[$^{\mathrm{e}}$]{Same rotation period as Ursa Major stream G0V
  members.}

  \item[$^{\mathrm{f}}$]{Possible member of the Hyades Moving Group.} 

  \item[$^{\mathrm{g}}$]{From \cite{chmielewski92}.}

  \item[$^{\mathrm{h}}$]{From \cite{kervella03} based on
  interferometric observations.} 


  \item[$^{\mathrm{i}}$]{Isochrone age from \cite{dravins98};
  $L_\mathrm{X}$ normalized to $1\,R_{\odot}$.}
\end{list}

\end{center}
\end{table}

\end{landscape}

\newpage


\section{The Solar Magnetic Field in Time}
\label{section:magneticfield}

\subsection{The Young Solar Photosphere: Large, Polar Spots}
\label{section:photospheric}

Stellar photospheres provide the crucial interface between the region
of magnetic field generation in the stellar interior and the extended
outer magnetic fields in the corona and in interplanetary
space. Distribution and size of photospheric magnetic features are
thought to reflect the location of the magnetic dynamo in the stellar
interior; the study of photospheres of young, active solar analogs is
therefore important for a closer understanding of the the dynamo
operating under more extreme conditions, but it of course also gives
the boundaries for the extended magnetic-field distribution in the
overlying chromosphere, corona, and the stellar wind.

\subsubsection{Doppler Imaging of Young Solar Analogs}
\label{section:doppler}

The most advanced (indirect) imaging technique to map magnetic
structure in stellar photospheres is Doppler imaging that uses
deformations in spectral-line profiles to map starspots
\citep{vogt83}.

Surface imaging using Doppler reconstruction methods or spot modeling
has been performed for several young, active solar
analogs. Figure~\ref{figure:dopplerimages} shows views of the three
examples discussed below. The colors code for temperature (from
\url{http://www.aip.de/groups/activity/DI/maps/}).

\epubtkImage{}{%
\begin{figure}[htbp]
  \centerline{
    \includegraphics[width=0.329\textwidth]{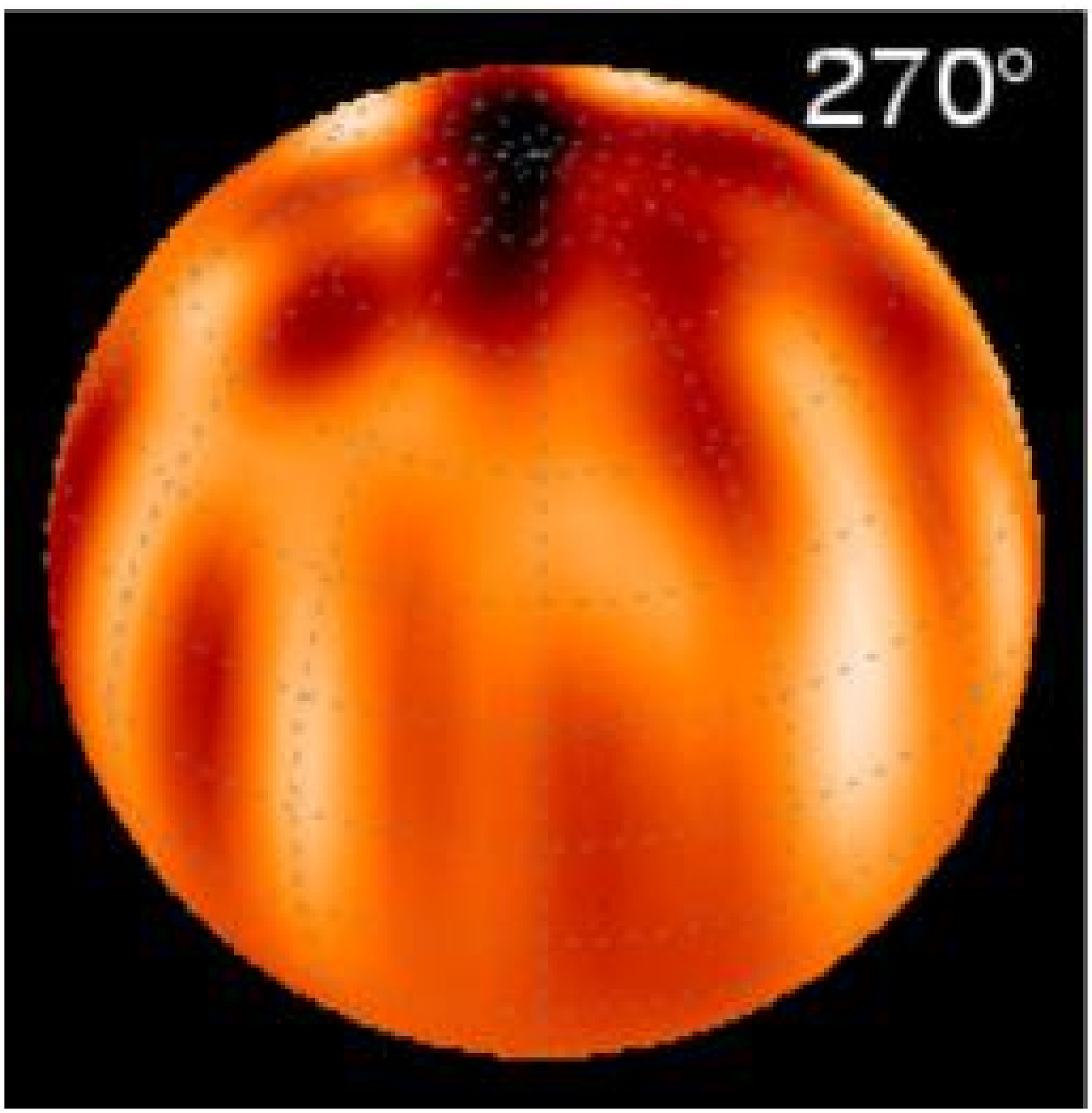}
    \includegraphics[width=0.331\textwidth]{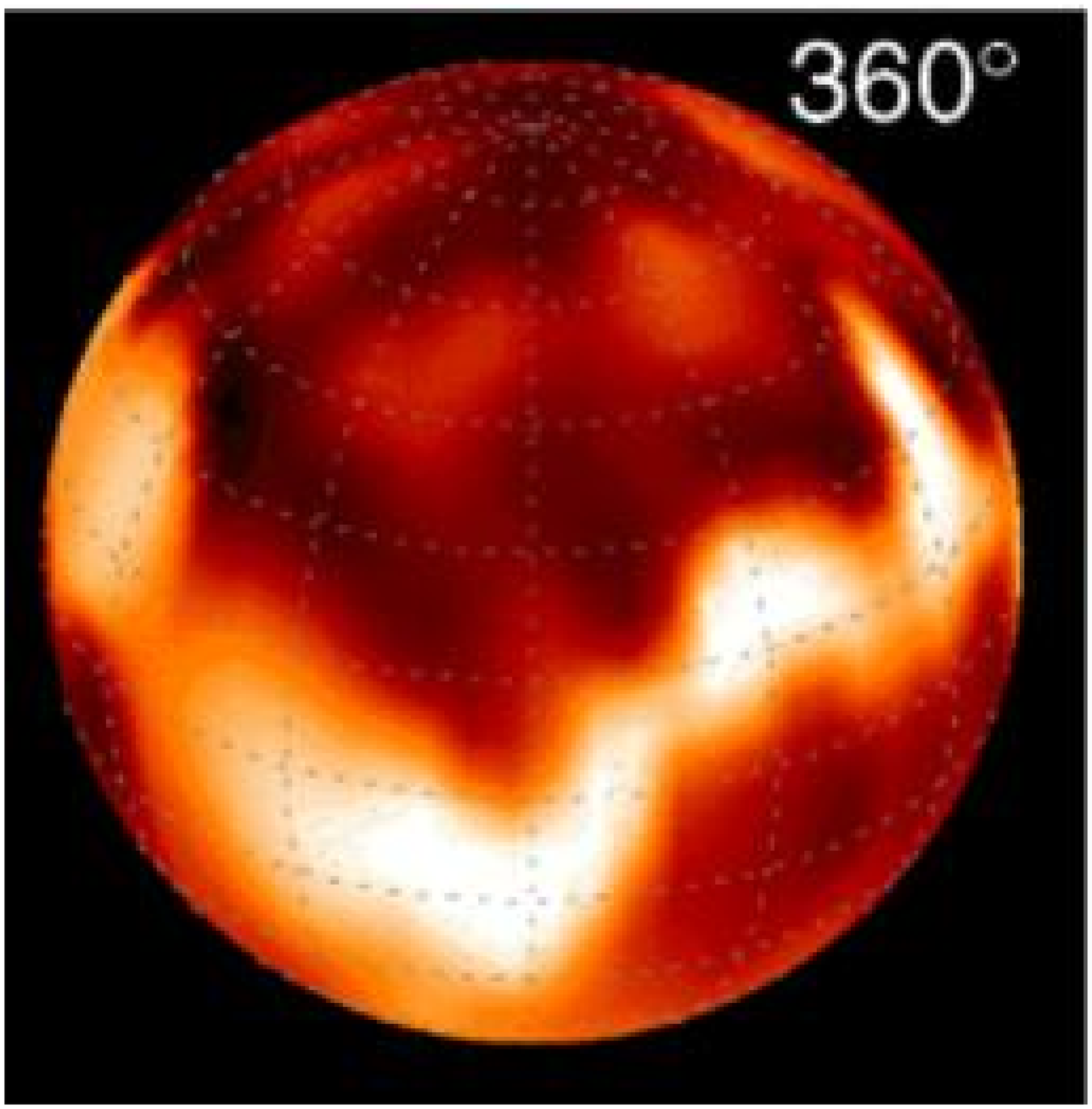}
    \includegraphics[width=0.333\textwidth]{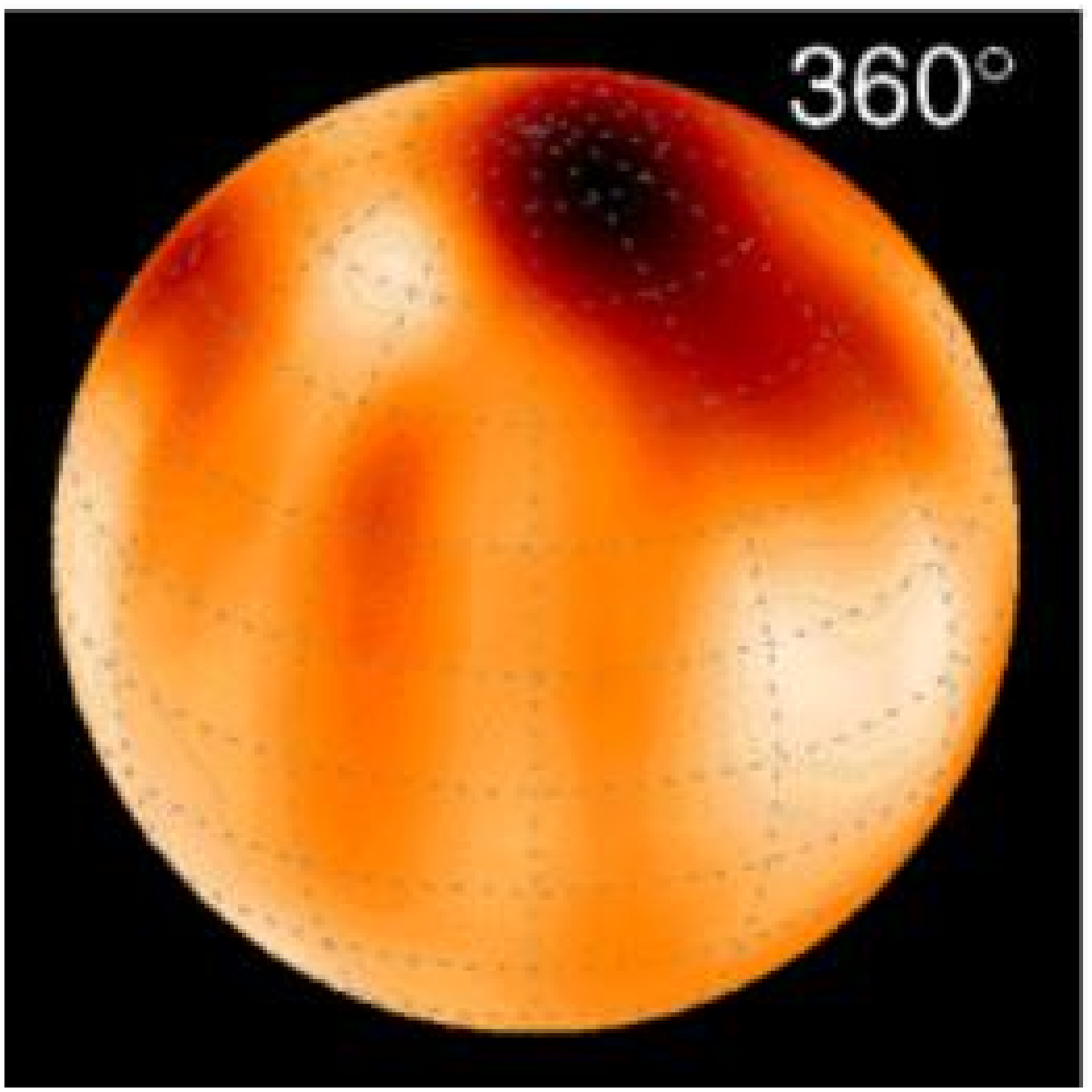}
    }
  \caption{\it Three temperature maps of young, active solar analogs,
  derived from Doppler imaging. From left to right: HD~171488 (P =
  1.34\,d; \citealt{strassmeier03}), HII~314 (P = 1.47\,d;
  \citealt{rice01}), and EK~Dra (P = 2.7\,d; \citealt{strassmeier98})
  (see also \url{http://www.aip.de/groups/activity/DI/maps/}; reprinted with permission).}
\label{figure:dopplerimages}
\end{figure}}

The photosphere of EK~Dra has been extensively mapped using spot
modeling \citep{jarvinen05} and Doppler imaging techniques
\citep{strassmeier98}. Doppler images show spots both close to the
visible pole but also near the equator, the dominant feature being
located at latitudes of 70\,--\,80\,deg \citep{strassmeier98}. This is
supported by radial-velocity variations \citep{koenig05}. The polar
region of EK~Dra is, however, less pronounced than in more active
(binary) stars, perhaps indicating a trend toward lower-latitude spots
for less active stars, as predicted by the \cite{schuessler96} theory
of Coriolis-force driven magnetic flux tubes that converge to the pole
for very rapid rotators (see below).

The Pleiades G dwarf HII~314 (\P = 1.47\,d) has been Doppler-imaged by
\cite{rice01}. Again, high-latitude and polar spots are visible. The
still younger, ``infant Sun'' HD~171488 = V889~Her (\P = 1.34\,d, age
= 30\,Myr), a star in its last stage of evolution toward the ZAMS,
shows very prominent polar spots and various high-latitude dark
features \citep{strassmeier03}. Further Doppler maps have been
presented for the early G-type ultra-fast rotators He~520 (\P =
0.49\,d) and He~699 (\P = 0.61\,d). Again, apart from the prominent
polar spots, there is a definitive band of low-latitude ($l\la
30\mbox{\,--\,}40 \unit{deg}$) spots on these stars as well
\citep{barnes98}.

Although these surface spot distributions vary significantly between
young solar analogs, there is agreement with regard to high-latitude
magnetic activity on all of them \citep{rice01, strassmeier03}.

The most recent Doppler technique uses spectropolarimetric
observations of lines to apply {\it Zeeman-Doppler Imaging} (ZDI),
which results in maps of radial, azimuthal, and meridional magnetic
fields. Successful application to solar analog stars have been
presented by \cite{marsden06} and \cite{catala07}; excellent ZDI maps
have also been derived for the somewhat cooler, young early-K star
AB~Dor \citep{donati03a}, and put into context with coronal X-rays
\citep{jardine03, mcivor03, hussain07}. ZDI images
(Figure~\ref{figure:zdiimages}) have shown non-solar magnetic features
such as high-latitude ``rings'' of azimuthal (toroidal) field
\citep{donati03a, marsden06, catala07}, possibly suggesting that the
responsible magnetic dynamo is located close to the surface (a
``distributed dynamo'') rather than (only) near the tachocline where
azimuthal fields are expected for $\alpha\Omega$-type dynamos. The
field distribution and orientation also serves as an important
diagnostic to study and explain spot lifetimes based on numerical
studies \citep{isik07}.

\epubtkImage{}{%
\begin{figure}[htbp] 
\centerline{\includegraphics[width=0.9\textwidth]{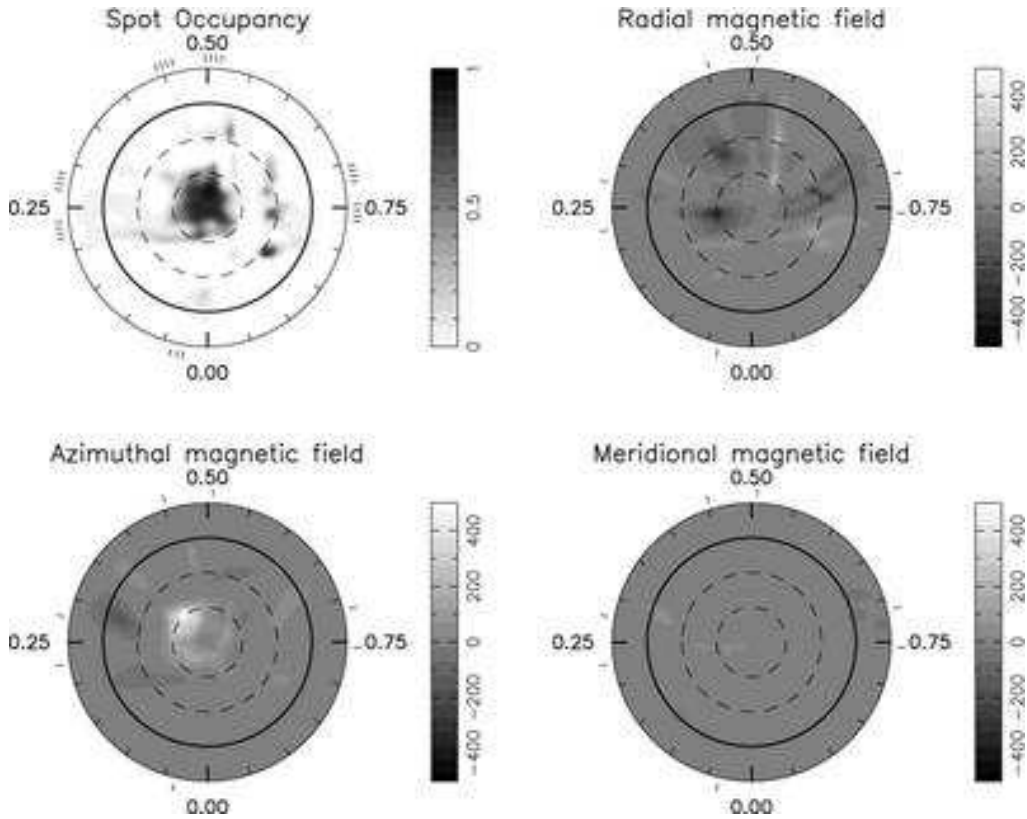}}
\caption{\it  Zeeman Doppler images of the rapidly rotating early G
    dwarf HD~171488, showing, in polar projection, spot occupancy
    (upper left), the radial magnetic field (upper right), the
    azimuthal magnetic field (lower left -- note the ring-like
    high-latitude feature), and the meridional magnetic field (lower
    right) (from \citealt{marsden06}, reprinted with permission of 
    Blackwell Publishing).}
\label{figure:zdiimages}
\end{figure}}

\subsubsection{Polar Spots}
\label{section:polarspots}

Why are there polar spots in magnetically active stars?
\cite{schuessler92} and \cite{schuessler96} suggested that strong
Coriolis forces act on magnetic flux bundles that rise from the dynamo
region at the boundary between the radiative core and the convective
envelope of the star. This force would deflect rising flux to higher
altitudes although, given the size of the radiative core, the maximum
latitude would probably be no more than about 60~degrees. A parameter
study confirms these findings systematically, with flux emergence
latitudes increasing with i) rotation rate, ii) decreasing stellar
mass (i.e., smaller radiative core radii), and iii) decreasing age; a
fraction of the flux tubes will, however, also erupt in
near-equatorial regions \citep{granzer00}. To produce truly polar spot
regions, additional latitudinal transport of flux tubes is still
required. A possibility is an additional pole-ward slip of a segment of
a flux ring in the stellar interior after the eruption of flux at
mid-latitudes in another segment of the same ring
\citep{schuessler96}.

Alternatively, \cite{schrijver01b} explored migration of surface
magnetic fields toward the poles in a model developed by
\cite{schrijver01a}. Here, magnetic bipoles are injected
randomly. These flux concentrations migrate pole-ward in a meridional
flow and are subject to differential rotation. The bipoles can
interact, i.e., fragment, merge, or cancel. The magnetic cycle is
simulated by periodically varying the injection
latitudes. \cite{schrijver01b} simulated a star with a bipole
injection rate 30~times higher than the present-day Sun, corresponding
to a solar analog with a rotation period of 6\,d, i.e., an age of a
few 100\,Myr. The differential rotation profile was assumed to be
identical to the present-day Sun's, and so was the length of the
activity cycle (11\,yrs). In the present-day Sun, the pole-ward
migration of the trailing flux in a bipole cancels with existing
high-latitude flux of opposite polarity relatively rapidly. In the
simulations of the active star, however, the magnetic concentrations
contain more flux, resulting in slower diffusion and a longer lifetime
before cancellation. The result of these simulations is that, first,
there are strong magnetic features accumulating in the polar regions,
and second, nested magnetic rings of opposite polarity form around
the pole (Figure~\ref{figure:schrijver_photos}). These are suggestive
sites of chromospheric and coronal interactions, perhaps leading to
strong coronal heating and flares in these polar regions. As
  described above, ZDI images indeed provide evidence for
  high-latitude ``rings'' of azimuthal (toroidal) field (e.g.,
  \citealt{marsden06}, Figure~\ref{figure:zdiimages}).

Observationally, the picture is more complicated. In contrast to these
simulations and also in contrast to the solar picture, Doppler images
of very active, young stars (Figure~\ref{figure:dopplerimages}) show
intermingling of opposite polarities in longitude also at high
latitudes \citep{mackay04}. Such features can indeed be reproduced if
the latitudes of flux emergence are shifted pole-ward, to 50\,--\,70
degrees, and the meridional flow be made faster \citep{mackay04,
  holzwarth06}. The first modification is of course suggested from the
\cite{schuessler96} theory.

The structure of polar spots (unipolar, multiple bi-polar regions, or
nested rings of different polarity) is indeed also very important for
the large-scale coronal field; unipolar magnetic spots suggest the
presence of more polar open-field lines, therefore concentrating
strong coronal X-ray emission to more equatorial regions, but also
reducing the efficiency of angular momentum removal by the magnetized
wind due to the smaller lever arm compared to equatorial winds
\citep{mcivor03}.

\epubtkImage{}{%
\begin{figure} 
\centerline{\includegraphics[width=0.48\textwidth]{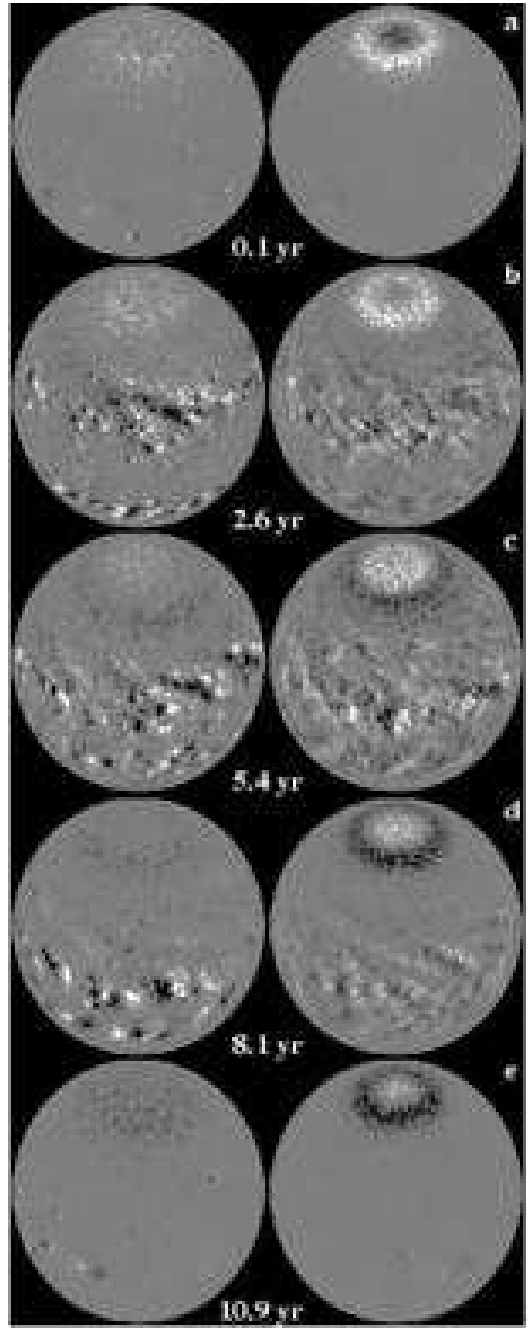}}
\caption{\it Simulations of surface magnetic fields for a star like
  the Sun (left) and an active star with a bipole emergence rate
  30~times higher (right). Various snapshots during the activity cycle
  are shown (from top to bottom). Note the concentration of magnetic
  flux in rings of opposite polarity around the pole of the active
  star \citep[from][reproduced by permission of AAS]{schrijver01b}.}
\label{figure:schrijver_photos}
\end{figure}}

\subsection{Coronal Structure of the Young Sun}
\label{section:coronal_structure}

Stellar magnetic fields are anchored in the photospheres, but they
can unfold into large, interacting, complicated structures in the
solar corona and may reach out into the surrounding ``interplanetary''
space. Mapping the true 3-D structure of outer stellar atmospheres has
therefore been an important goal of stellar coronal physics, but a
challenging one. Apart from the complications in inferring the 3-D
structure of an optically thin gas that is not spatially resolved by
present-day telescopes, the emitting regions (e.g., in X-rays) may not
be identical to what we would like to map as ``magnetic structure'';
further, strong variability in coronal regions on short time scales
may make ``imaging'' challenging. Methods to infer coronal structure
include,

\begin{description}

  \item [1.] reconstruction of magnetic features from light curves
  that are rotationally modulated (at radio or X-ray wavelengths);

  \item [2.] eclipse image reconstruction for eclipsing binaries
  (radio and X-rays);

  \item [3.] theoretical models of magnetic loops, using measurable
  quantities such as emission measures and temperatures as input
  (X-rays); specific models have been developed for flaring magnetic
  loops;

  \item [4.] implications for coronal structure based on electron
  density measurements in X-ray spectra;

  \item [5.] direct imaging using interferometric techniques at radio
  wavelengths.

\end{description}

Most of these methods are applicable to various types of stars, but
results are overall ambiguous. I will not review all aspects of
stellar coronal structure \citep[see][]{guedel04} but concentrate on
results that have been obtained -- at least in part -- specifically
for young solar analogs.

\subsubsection{Magnetic Loop Models and Active Regions}
\label{section:loops}

In the simplest approach, let us assume that the observed X-ray
luminosity $L_\mathrm{X}$ is produced by an ensemble of identical
magnetic coronal loops with characteristic half-length $L$, surface
filling factor $f$, and an apex temperature $T$ used for the entire
loop; then, on using the Rosner--Tucker--Vaiana (RTV,
\citealt{rosner78}) loop scaling law and identifying $L_\mathrm{X} =
\epsilon V$ where $\epsilon$ is the volumetric heating rate (in
$\mathrm{erg\ cm^{-3}\ s^{-1}}$), we obtain

\begin{equation}
  L \approx 6\times 10^{16} \left({R_*\over R_{\odot}}\right)^2
  {f\over L_\mathrm{X}}T^{3.5}\quad \mathrm{[cm]}.
\end{equation}

This relation can only hold if $L$ is smaller than the pressure scale
height. Based on this expression, the luminous, hot plasma component
in magnetically active stars seems to invariably require either very
large, moderate-pressure loops with a large filling factor, or
solar-sized high-pressure compact loops with a very small ($<1\%$)
filling factor \citep{giampapa85, stern86, schrijver89, giampapa96,
  guedel97a, preibisch97, sciortino99}.

\cite{schrijver84} modeled $T$ and the emission measure, EM (=
$n_\mathrm{e}n_\mathrm{H}V$, where $n_\mathrm{e}$ and $n_\mathrm{H}$
are the electron and hydrogen number densities, and $V$ is the volume)
of a sample of coronal sources based on RTV loop models and found
several trends: i) Inactive MS stars such as the Sun are covered to a
large fraction by large-scale, cool (2\,MK) loops of modest size
($0.1\,R_*$). ii) Moderately active dwarfs are dominated by very
compact, high-density, hot ($\approx$20\,MK) loops that require high
heating rates (up to 20~times more than for solar compact active
region loops). iii) The most active stars may additionally form rather
extended loops with heights similar to $R_*$.

Similar studies of loop models varied in their results, although for
magnetically active stars, most authors reported results that require
hot, compact, high-pressure loops (up to several $100 \unit{dyn\ }$
$ \unit{cm^{-2}}$), somewhat reminiscent of flaring loops \citep{stern86,
  giampapa96, maggio97, ventura98, sciortino99}. The hypothesis that
the hottest plasma in magnetically active stellar coronae does not
form in static loops but in flaring active regions will be encountered
again -- see Section~\ref{section:flares}.

If the Sun were entirely covered with active regions, the X-ray
luminosity would amount to only $\approx (2\mbox{\,--\,}3) \times
10^{29} \unit{erg\ s^{-1}}$ \citep{vaiana78, wood94}, short of
$L_\mathrm{X}$ of the most active solar analogs by one order of
magnitude. But the X-ray emission measure distributions of such active
stars show excessive amounts of plasma around 10\,--\,20\,MK
\citep{guedel97a}, which incidentally is the typical range of solar
flare temperatures. This again led to the suggestion that the high-$T$
emission measure is in fact due to the superposition of a multitude of
temporally unresolved flares (see Section~\ref{section:flares}).

\subsubsection{Inferences from Coronal Density Measurements}
\label{section:density}

Comprehensive surveys of stellar coronal electron density
($n_\mathrm{e}$) measurements based on X-ray spectroscopic line-flux
ratios were presented by \cite{ness04} and \cite{testa04}, including a
sample of active solar analogs. These studies concluded that the
surface filling factor (derived from the emission measure, the
measured $n_\mathrm{e}$, and a realistic coronal scale height) of
magnetic loops containing {\it cool} X-ray emitting material increases
from inactive to moderately active stars but then ``saturates'' at
levels of about ten percent. In the most active stars, hot coronal
loops are added, with a sharply increasing filling factor, thus
probably filling the volume left between the cooler coronal magnetic
loops. Observations of rotational modulation in very active solar
analogs suggests, however, that the coronal volume filling remains
significantly below 100\% (see
Section~\ref{section:rotationalmodulation} below).

\subsubsection{Inferences from Rotational Modulation}
\label{section:rotationalmodulation}

Inhomogeneous coronae may reveal signatures in light curves as the
star rotates, although success of this method has been moderate given
that coronal features evolve on time scales shorter than one rotation
(e.g., owing to flares). Two young, near-ZAMS solar analogs have shown
clear signatures of X-ray rotational modulation (EK~Dra,
\citealt{guedel95a}, and 47~Cas~B, \citealt{guedel95b}), pointing to a
filling factor below unity. This is unexpected because such stars are
in the empirical X-ray saturation regime that has often been suggested
to be due to complete filling of the surface with X-ray bright coronal
magnetic loops (see Section~\ref{section:coronal_evolution}
below). EK~Dra also showed evidence for radio rotational
modulation. The depth and length of the modulation
(Figure~\ref{rotmodstars}a) constrains the X-ray coronal height, and
also the electron densities to $n_e > 4 \times 10^{10}
\unit{cm^{-3}}$, in agreement with spectroscopic measurements
\citep{ness04}. This leads to the conclusion that much of the
emitting material is concentrated in large ``active regions''.

\epubtkImage{}{%
\begin{figure}[t!]
\centerline{
\hbox{
\resizebox{0.48\textwidth}{!}{\includegraphics{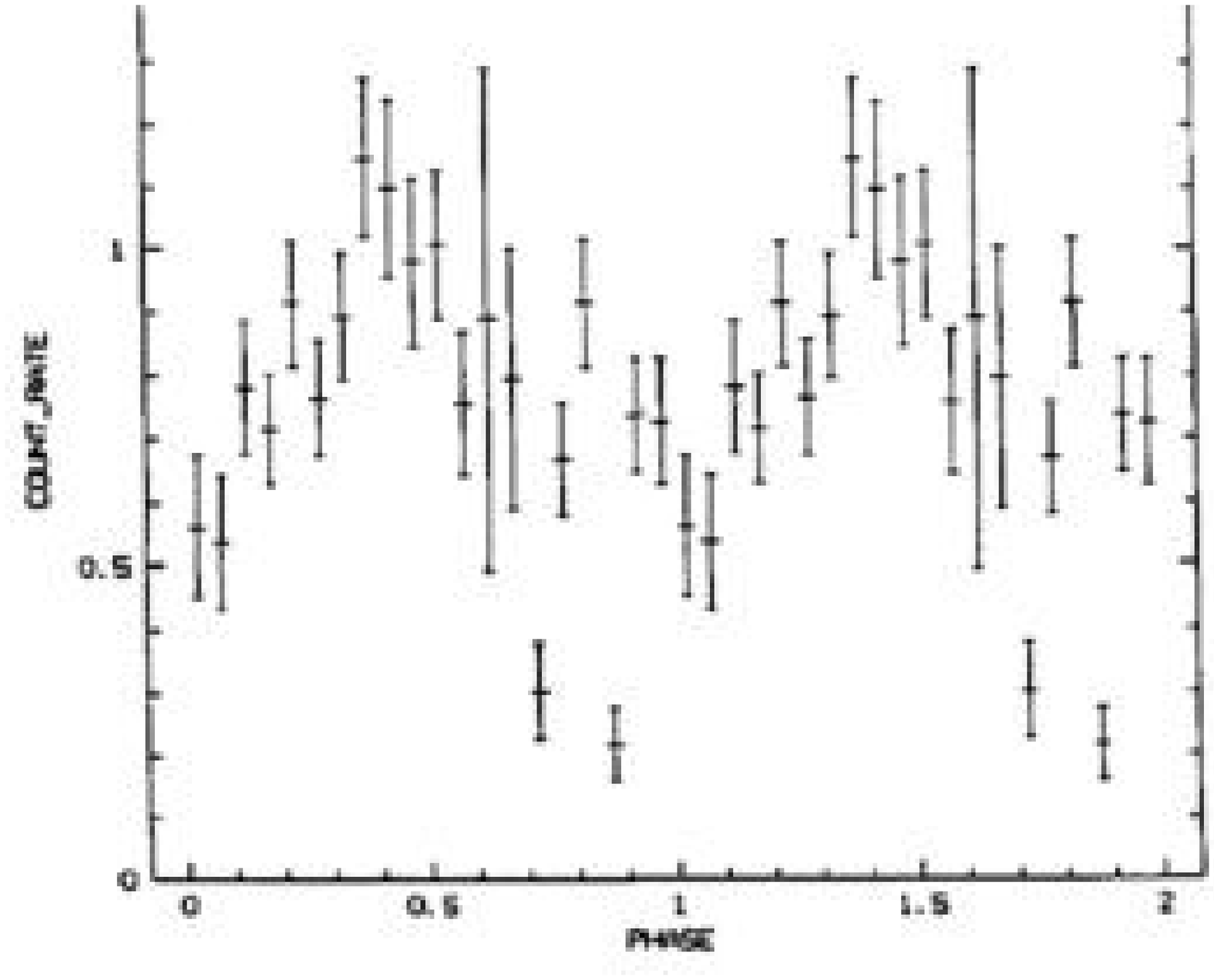}}
\hskip 0.5truecm\resizebox{0.478\textwidth}{!}{\includegraphics{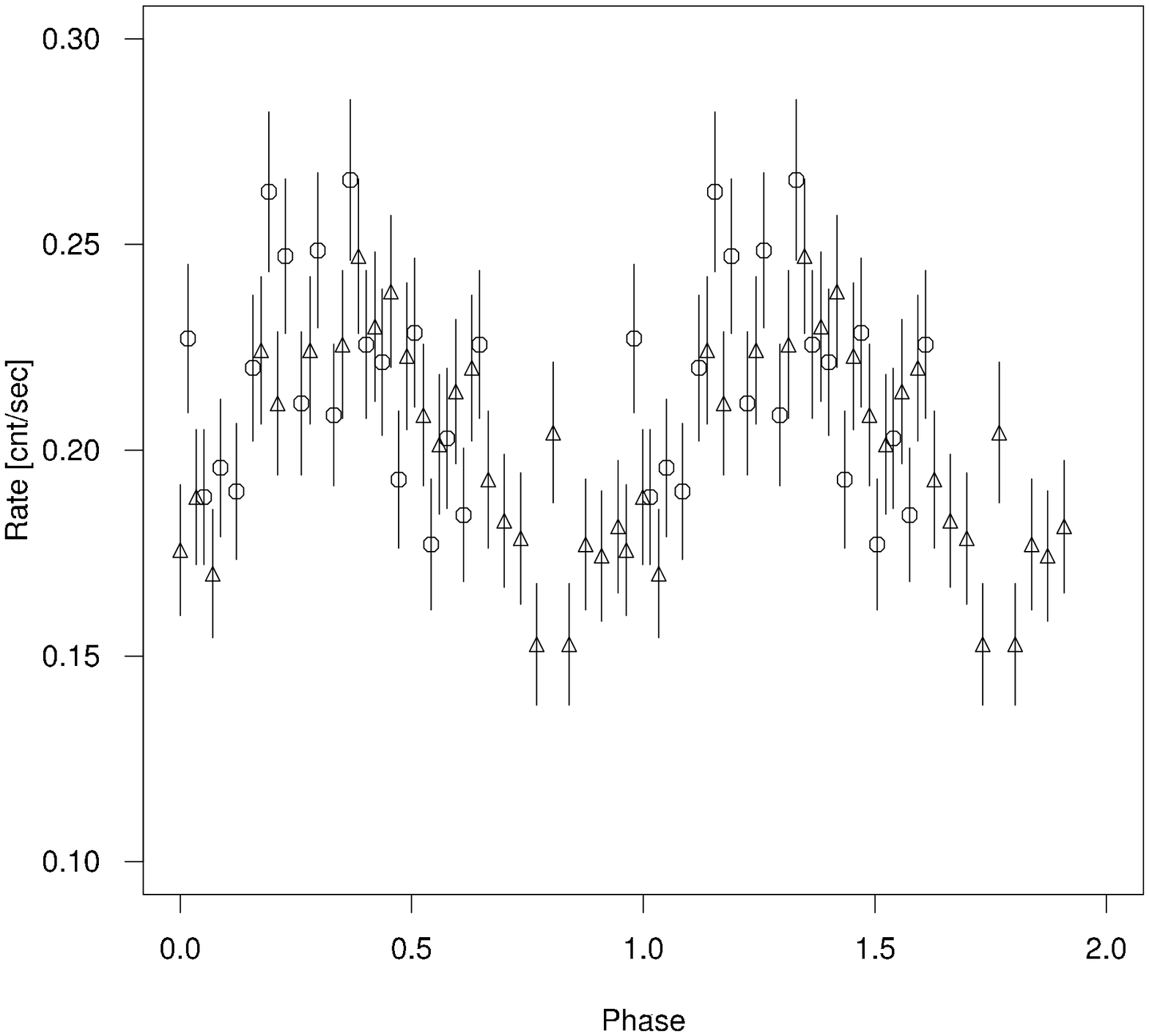}}
}
}\vskip -1truecm
\caption{\it Two examples of X-ray rotational modulation in young,
  active solar analogs: {\em a (left):} EK~Dra \citep{guedel95a}, and
  {\em b (right):} the supersaturated young solar analog VXR45
  \citep{marino03a}. Both light curves are phase-folded (reprinted with permission).}
\label{rotmodstars}
\end{figure}}

Because the X-ray luminosity in very rapidly rotating,
``supersaturated'' stars (Section~\ref{section:coronal_Xray}) is below
the empirical maximum, rotational modulation would give important
structural information on the state of such coronae. A deep modulation
in the young solar analog VXR45 (Figure~\ref{rotmodstars}b) suggests
that extreme activity in these stars is again {\it not} due to
complete coverage of the surface with active regions
(\citealt{marino03a}).

\subsubsection{Inferences from Eclipses}
\label{section:eclipses}

Most binary systems that produce coronal eclipses are close binaries
and the components are therefore in no ways similar to the young
Sun. A remarkable exception is the relatively wide binary
$\alpha$~CrB. Its X-ray active, young (age of few 100\,Myr) solar
analog of spectral type G5~V is totally eclipsed once every 17~days by
the optical primary, an A0~V star that is perfectly X-ray dark. Other
parameters are ideal as well, such as the non-central eclipse, the
eclipse time-scale of a few hours, and the relatively slow rotation
period of the secondary. Eclipse observations obtained by {\it ROSAT}
\citep{schmitt93} and by {\it XMM-Newton} \citep{guedel03b} were used
to reconstruct projected 2-D images of the X-ray structure. They
consistently reveal patches of active regions across the face of the
G~star; not much material is found significantly beyond its limb
(Figure~\ref{eclipsefig2}). The structures tend to be of modest size
($\approx 5 \times 10^9 \unit{cm}$), with large, X-ray faint areas in
between, although the star's luminosity exceeds that of the active Sun
by a factor of $\approx 30$. These observations imply moderately high
densities in the emitting active regions, and both studies mentioned
above yielded average electron densities in the brightest active
regions of a few $10^{10} \unit{cm^{-3}}$. The picture of active
coronae consisting of features similar to solar active regions thus
seems to hold also for intermediately active, young solar analogs.

\epubtkImage{}{%
\begin{figure}[t!] 
\centerline{
\includegraphics[width=0.55\textwidth]{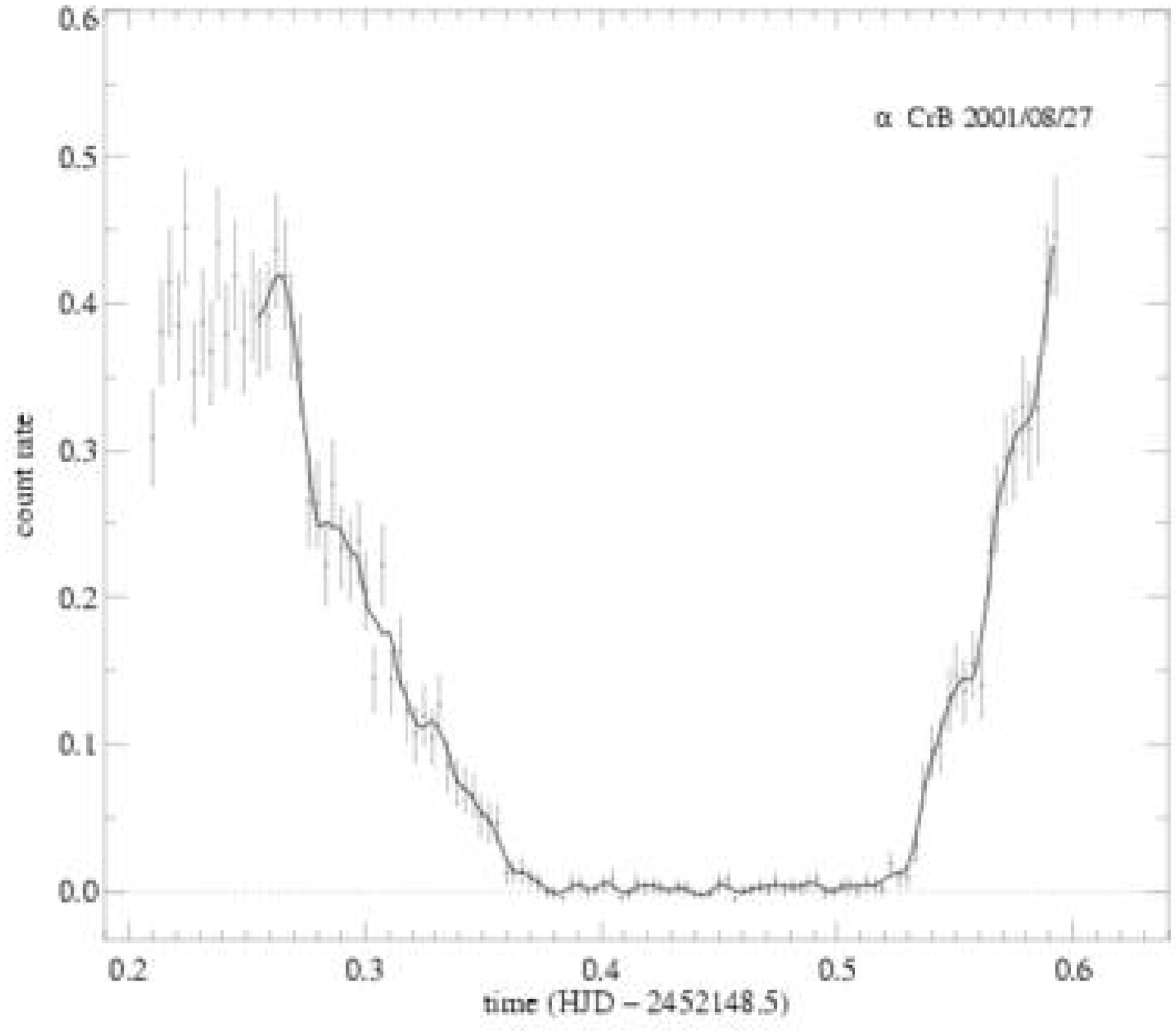}
\includegraphics[width=0.45\textwidth]{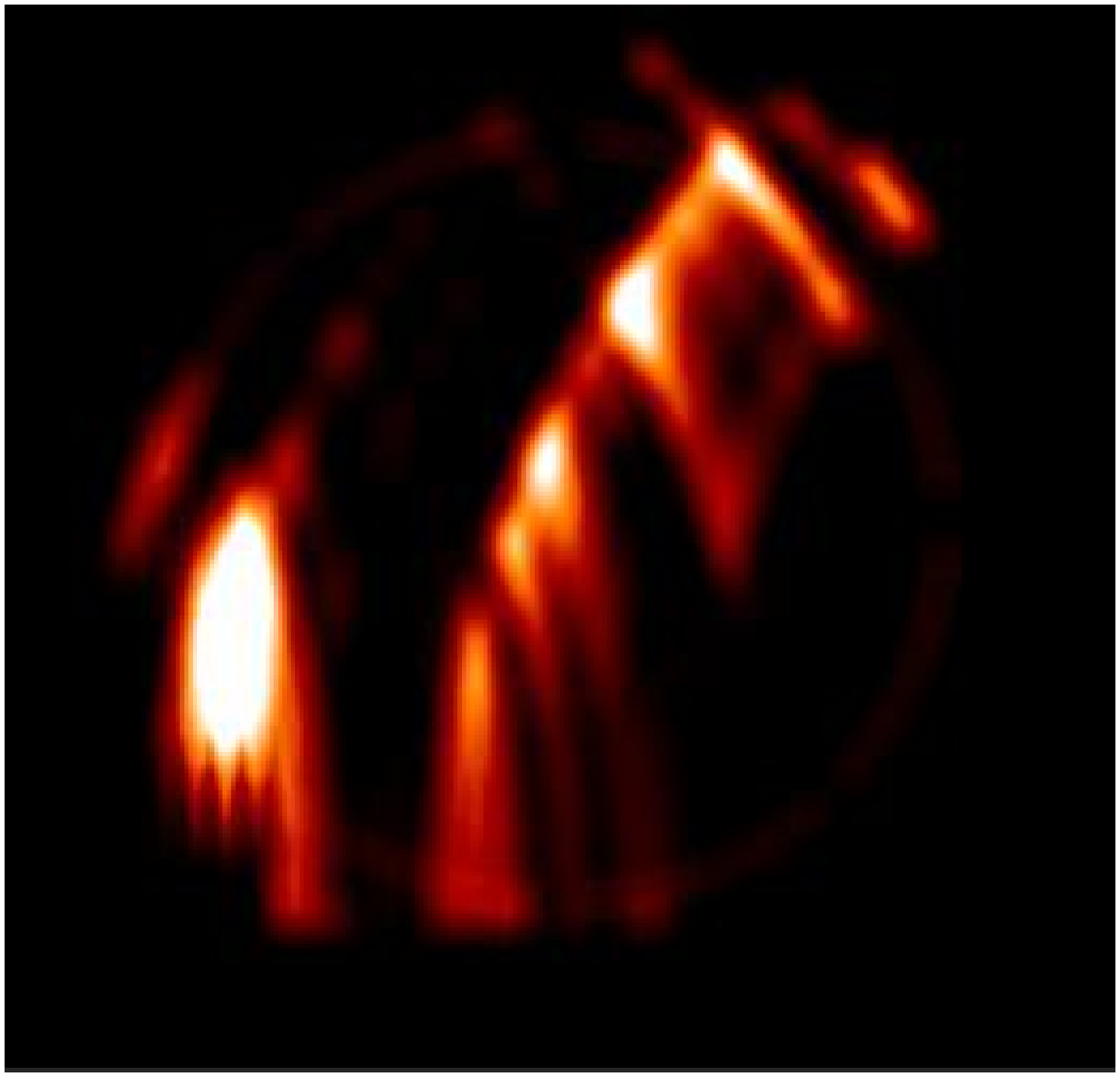}
}
\caption{\it Light curve and image reconstruction of the A+G binary
  $\alpha$~CrB. The left panel shows the light curve from observations
  with {\it XMM-Newton}, the right panel illustrates the reconstructed
  X-ray brightness distribution on the G star.  The axes are such
    that the larger, eclipsing A-type star moves from left to right
  parallel to the x-axis, i.e., the polar axis of the orbit is
  parallel to the y axis. The diameter of the star (outlined by a
  faint circle) is 0.9~solar diameters or $1.25 \times 10^{11}
  \unit{cm}$, corresponding to $\approx 365 \unit{\mu\,arcsec}$ at a
  distance of 22.9\,pc \citep[after][copyright Springer Verlag, reprinted with 
  permission]{guedel03b}.}
\label{eclipsefig2}
\end{figure}}

\subsubsection{Photospheric-Field Extrapolation to the Corona}
\label{section:fieldextrapolation}

Information on coronal structure can also be derived indirectly from
surface Zeeman-Doppler images as developed for and applied to the
stellar case by \cite{jardine02a}, \cite{jardine02b},
\cite{hussain02}, and \cite{hussain07} and further references
therein. \cite{jardine02a} and \cite{jardine02b} explored potential
field extrapolation, while \cite{hussain02} extended the models to
include some form of currents in force-free fields.

Such models also require specification of the base thermal pressure of
the plasma with respect to the local magnetic pressure, and some
cutoff of the corona at locations where the thermal pressure might
open up the coronal field lines. They can successfully recover, at
least qualitatively, the total X-ray emission measure, the average
electron density, and the low level of rotational modulation observed
on very active, young stars such as AB~Dor. The X-ray rotational
modulation is to a large extent suppressed by the highly complex
coronal structure, involving both very large magnetic features and
more compact loops anchored predominantly at polar latitudes.

\cite{schrijver02} used the model of surface magnetic-field
development devised by \cite{schrijver01a} and \cite{schrijver01b}
(Section~\ref{section:photospheric} above) to extrapolate surface
magnetic fields into the corona, assuming a potential-field
approach. The resulting magnetic structure is illustrated in
Figure~\ref{figure:schrijver_corona} for a solar-activity star, and two
examples with a 10fold and a 30fold magnetic injection rate (the
latter corresponding to a solar analog with an age of a few
100\,Myrs). Near cycle minimum, the coronae of the active stars are
dominated by a large-scale dipolar field although the low-activity
example also shows compact active regions, more so than the active
stars. At activity maximum, the active stars show magnetic-field
arcades between the polar rings of opposite polarity, weakening the
contribution from the global dipole components. Not all of these loops
will be X-ray bright, however; the loop brightness depends on its
heating rate, which in turn depends on the mode of coronal heating of
a given magnetic loop (see \citealt{schrijver02} for further
details).

\epubtkImage{}{%
\begin{figure}[htbp]
\centerline{\includegraphics[width=0.80\textwidth]{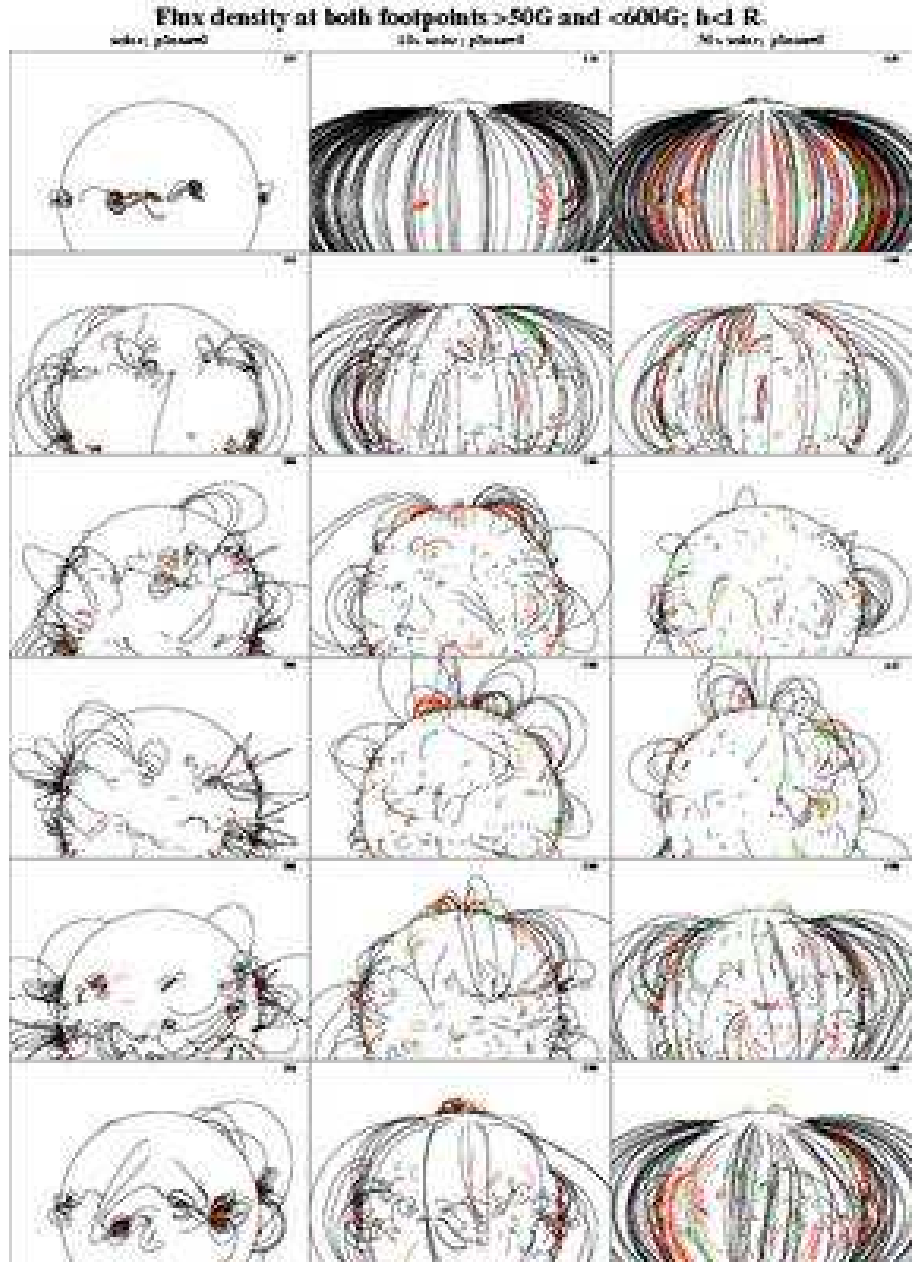}} 
\caption{\it Magnetic-field geometry of a corona of a solar analog. From
  left to right: present solar activity, 10fold higher activity,
  30fold higher activity. The panels from top to bottom show
  configurations at different phases of the activity cycle (0.00 =
  minimum, 0.16, 0.33, 0.49 $\approx$ maximum, 0.65, and 0.82,
  respectively). Only field lines with chromospheric footpoint field
  strengths between 50\,G and 600\,G are shown. Red and green curves
  show loops for which the expansion between  10,000 and 30,000\,km is
  less than a factor of 4 and 2, respectively
  \citep[from][reproduced by permission of AAS]{schrijver02}.}
\label{figure:schrijver_corona}
\end{figure}}

\subsubsection{Summary on Coronal Structure}
\label{section:structure_summary}

Despite numerous, complementary approaches to the study of coronal
structure, the results appear to be inconclusive. Compact active
regions are inferred from (and required by) the presence of rotational
modulation, and field extrapolation and numerical models suggest
compact regions as well. Extended, global fields are more difficult to
infer, and this is the consequence of two observational biases rather
than implying the absence of such structures. First, most
observational methods (e.g., X-ray eclipse and rotational modulation)
are insensitive to large structures. Second, the pressure scale
height of typical coronal plasma is less than one stellar radius
for solar analogs. The $n_\mathrm{e}^2$ dependence of X-ray emission
will therefore bias X-ray loop detection toward low heights. There is,
nevertheless, evidence for large-scale magnetic fields on young,
active stars, from two directions: surface field extrapolation based
on observed spot features or based on numerical models of flux
transport (Section~\ref{section:fieldextrapolation}) imply the presence
of structures resembling global dipole components; and spatially
resolved radio observations of active stars indeed do show very
extended magnetic structures (predominantly above the polar regions;
\citealt{benz98, mutel98}), although such evidence has not yet been
demonstrated for young solar analogs. In summary, then, observational
evidence and model simulations point to the presence of both compact
active regions and extended magnetic structures in the young Sun. The
relative proportions are likely to relate to the distribution of
magnetic field on the stellar surface, as in the models presented by
\cite{schrijver02}.

There is clear evidence that the surface magnetic filling factor
increases with increasing activity  (and decreasing rotation
  period), both from observations \citep[e.g.,][]{montesinos93} and
  theoretical and modeling studies \citep{montesinos93, fawzy02}; in
  contrast, the photospheric magnetic field strength is thought to be
  primarily restricted by pressure equilibrium with respect to the
  ambient gas pressure, i.e., the field strength is primarily
  dependent on spectral type, although a weak activity dependence is
  also present \citep[e.g.,][]{montesinos93}. The higher filling
  factors lead to less expansion of photospheric/chromospheric flux
  tubes because the tubes merge with adjacent tubes \citep{cuntz99}.
Therefore, toward more active stars, {\it coronal} magnetic fields
interact progressively more frequently due to their denser packing. A
higher rate of large flares is a consequence. Since flares enhance
the electron density along with the temperature, stars with a higher
activity level should reveal a predominance of hotter structures
\citep{guedel97a}. Such a trend is observationally well supported;
as further described in Section~\ref{section:coronal_temperature} below,
coronae at higher activity levels are systematically hotter.

\subsection{Activity Cycles in the Young Sun}
\label{section:cycles}

Records of sunspot numbers back over several hundred years show a
near-cyclic modulation that has turned out to be a central challenge
for dynamo theories. The activity period between two successive spot
maxima is approximately 11~years; because the magnetic polarity
reverses after one period, the full magnetic cycle amounts to
22\,yrs. Activity cycles corresponding to the 11\,yr solar cycle have
been found on many cool stars, mostly as a result of the Mount Wilson
HK~project \citep{baliunas95} that has collected a continuous data
stream of the chromospheric Ca\,{\sc ii} H \& K line flux diagnostic
for many stars over several decades. A subset of stars appear to lack
such cycles, however, and very active stars tend to exhibit an
irregular rather than a cyclic mode of variability
\citep{hempelmann96}. An alternative method for finding activity
cycles on magnetically active stars is the identification of cycles in
the starspot coverage.

\subsubsection{Starspot Cycles of Solar Analogs}
\label{section:starspotcycles}

\cite{messina02} specifically studied starspot cycles of stars from
the ``Sun in Time'' program. Activity cycles are found in all of them,
with periods ranging from 2.1\,yr to 13.1\,yr. A comparison with the
more comprehensive survey and the theoretical interpretations
presented by \cite{saar99} confirms the presence of two or three
branches: i) inactive solar analogs show cycles about 100 times longer
than the rotation period, \P; ii) active stars reveal cycles
200\,--\,600 times longer than \P; iii) and ``super-active'' stars
show cycles about 4 order of magnitude longer than \P
(Figure~\ref{figure:cycles}a). Among G-type stars, only EK~Dra appears
to be compatible with the third class (\citealt{messina02}).

\epubtkImage{}{%
\begin{figure}[htbp] 
\centerline{
\hbox{
\resizebox{0.56\textwidth}{!}{\includegraphics{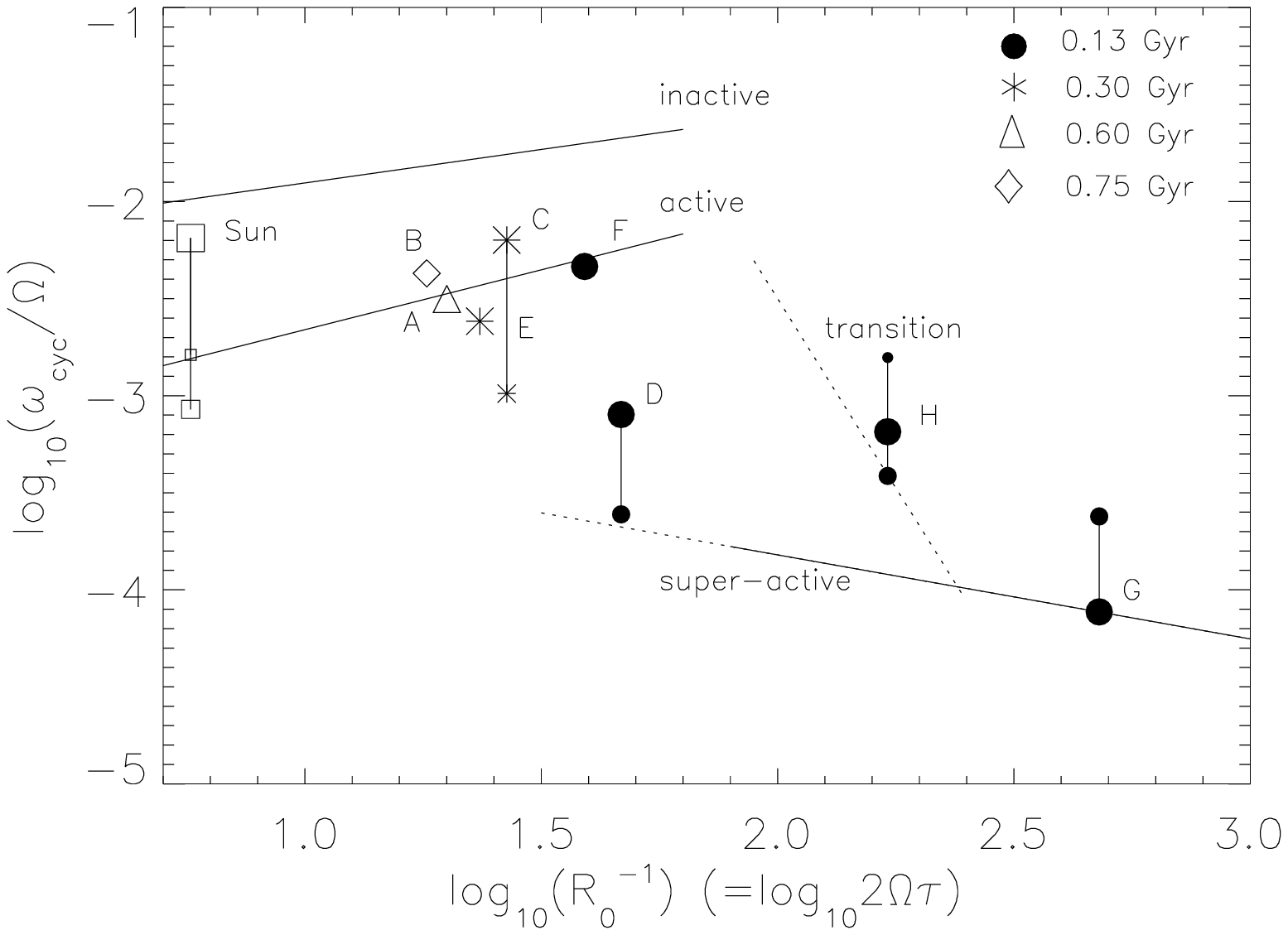}}
\hskip -0.6truecm
\resizebox{0.56\textwidth}{!}{\includegraphics{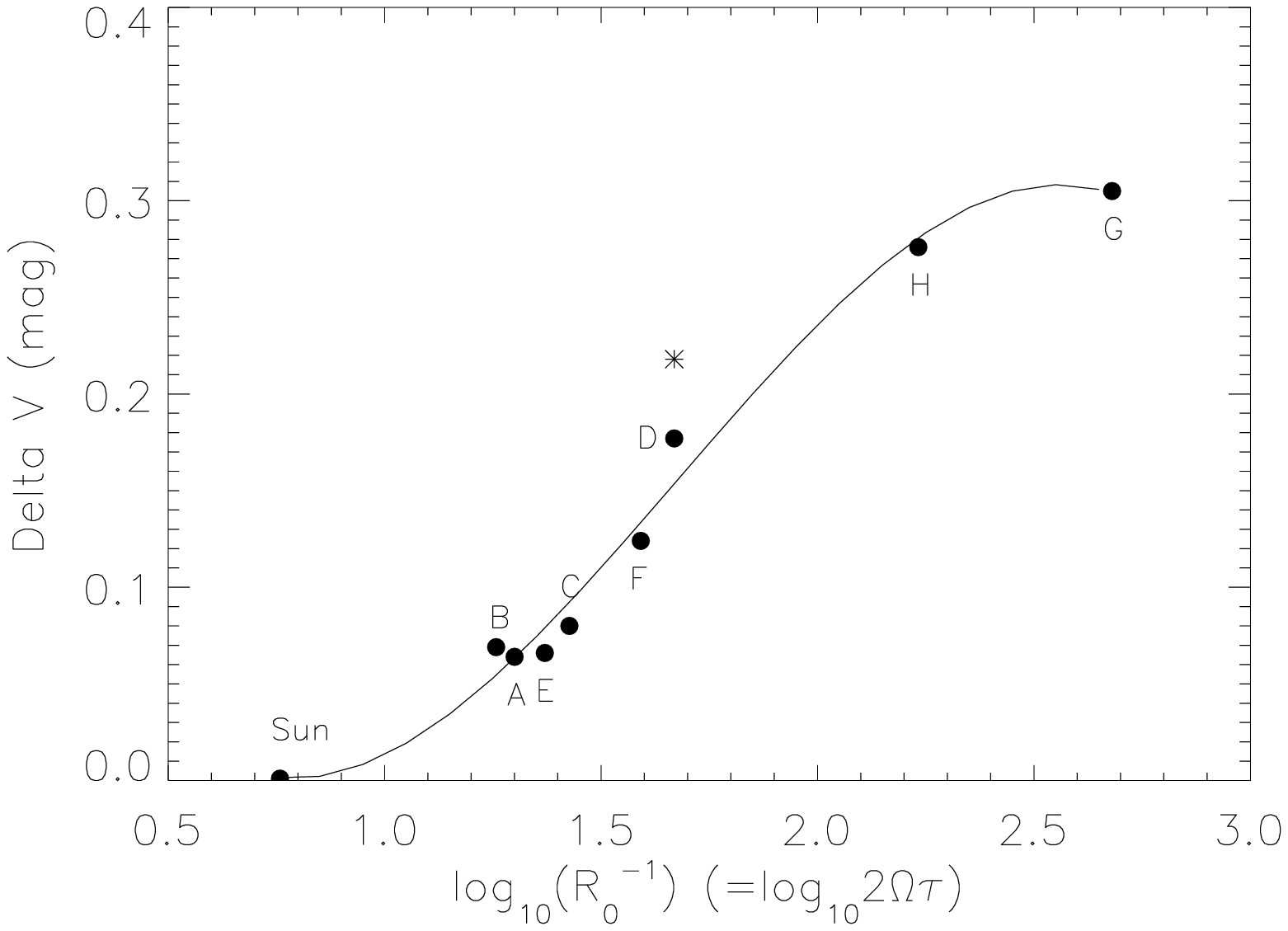}}
}
}
\caption{\it Left (a): Ratio between activity-cycle frequency and rotation
  frequency plotted vs. the inverse Rossby number, mostly for solar
  analogs. Three theoretical branches (``inactive'', ``active'', and
  ``super-active'') are shown by solid lines. Key to the labels: A =
  BE~Cet, B = $\kappa^1$~Cet, C = $\pi^1$~UMa, D = EK~Dra, E = HN~Peg,
  F = DX~Leo (K0~V), G = AB~Dor (K0~V), H = LQ~Hya (K2~V) (from
  \citealt{messina02}). -- Right (b): Amplitude of V-band variability
  as a function of the inverse Rossby number. The labels are as for
  the left figure (from \citealt{messina02}, reprinted with permission).}
\label{figure:cycles}
\end{figure}}

The brightness amplitude increases with increasing inverse Rossby
number (i.e., increasing rotation rate for constant turnover time),
indicating that spots produce progressively more modulation toward
higher activity levels (Figure~\ref{figure:cycles}b). A plateau is
suggested for the most active stars, indicating a saturation effect
when spots cover a large fraction of the stellar surface
\citep{messina02}.

\epubtkImage{}{%
\begin{figure}[b] 
\centerline{
\hbox{
\resizebox{0.58\textwidth}{!}{\rotatebox{90}{\includegraphics{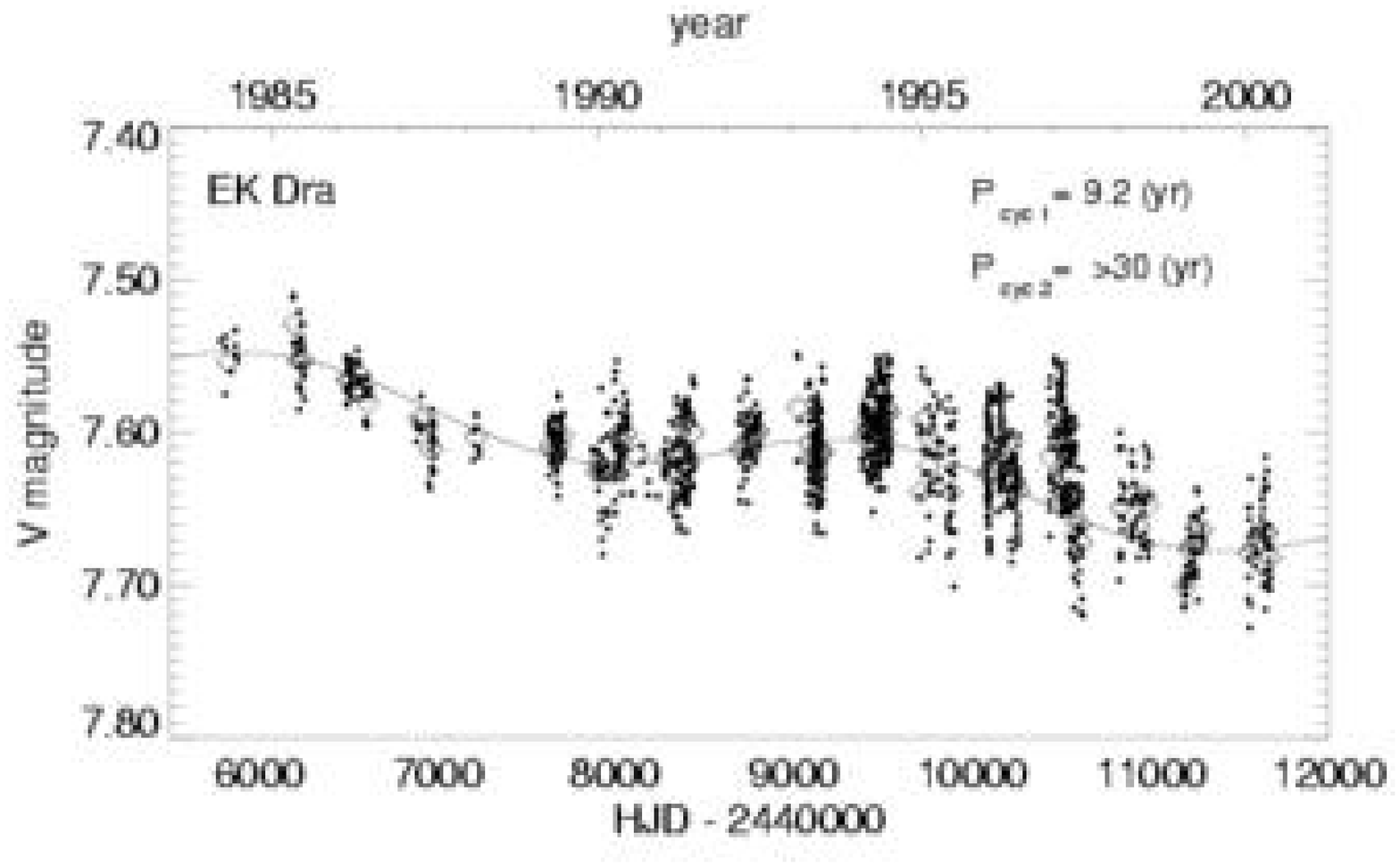}}}
\hskip -1.2truecm\resizebox{0.58\textwidth}{!}{\rotatebox{90}{\includegraphics{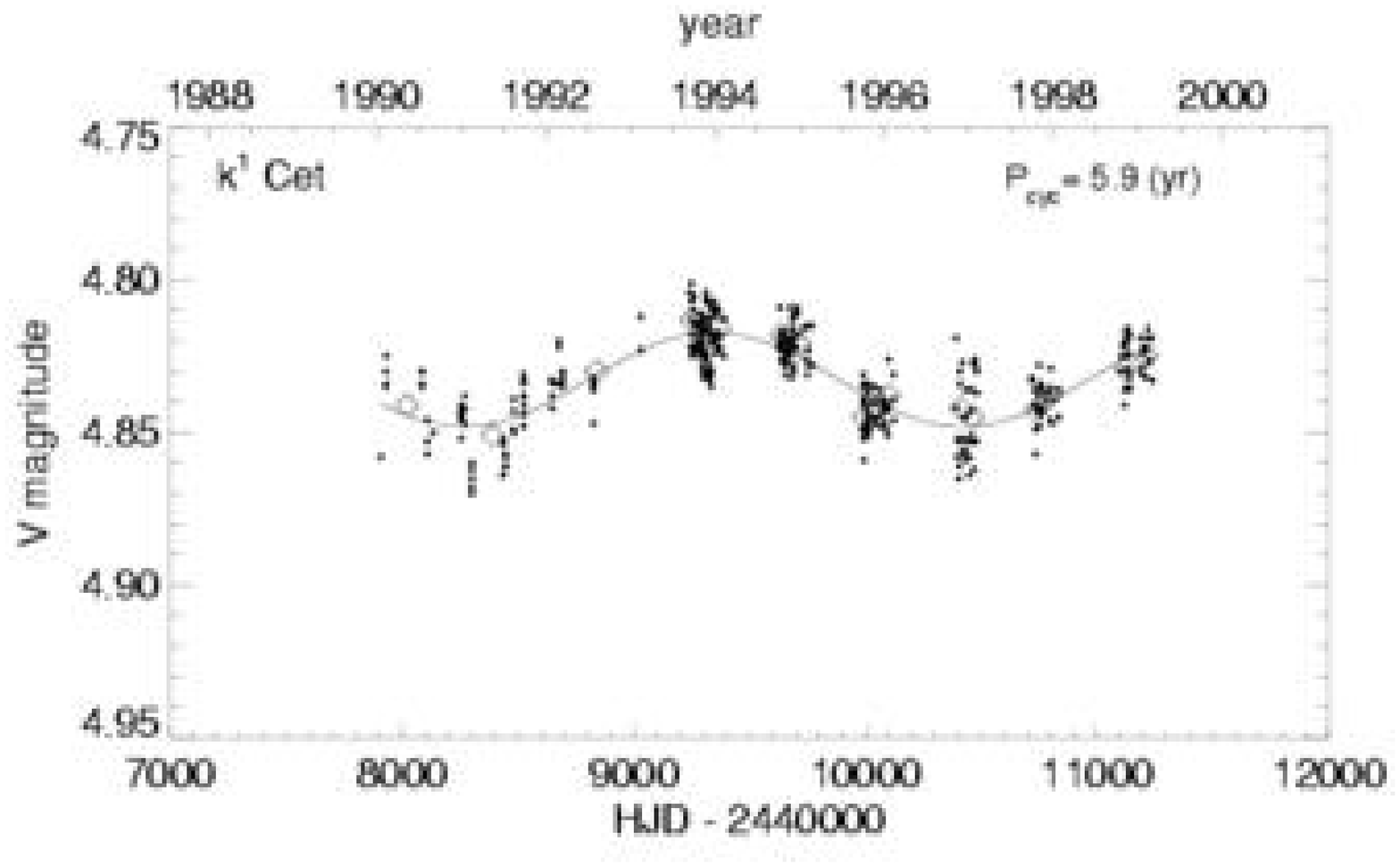}}}
}
}\vskip -1.8truecm
\caption{\it V band photometric time series and sinusoidal (plus
  long-term trend) fits for EK~Dra (left, a) and $\kappa^1$~Cet
  (right, b) \citep[from][reprinted with permission]{messina02}.}
\label{figure:spotcycles}
\end{figure}}

The standard near-ZAMS solar analog, EK~Dra, has been an important
target for cycle studies. \cite{dorren94} and \cite{dorren94c} showed
that its long-term photometric variations by $\approx 0.07 \unit{mag}$
are consistent with a period of about 12\,yrs. Variations compatible
with this time scale are also found in Ca\,{\sc ii} HK, Mg\,{\sc ii} h
and k, and the ultraviolet C\,{\sc iv}, C\,{\sc ii}, and He\,{\sc ii}
fluxes \citep{dorren94}, and in X-rays (see
Section~\ref{section:xraycycles} below). The spot periodicity has been
confirmed by \cite{messina02} although their best estimate for the
period is $9.2 \pm 0.4 \unit{yrs}$ (Figure~\ref{figure:spotcycles}). The
star is optically faintest when chromospheric and transition-region
activity is highest. On top of this cyclic behavior, there is a
long-term trend in the optical light, namely a decline of the blue
light by $0.0017 \pm 0.0004 \unit{mag\ yr^{-1}}$ for an investigated
  time span of 35\,yrs (\citealt{froehlich02}; see also
  \citealt{messina03} and \citealt{jarvinen05}). Further, two active
  longitudes shift in phase in concert with the activity cycles
  \citep{jarvinen05}. The dominant spot concentration switches between
  the two preferred longitudes with a ``flip-flop'' cycle of about
  4\,--\,4.5 yrs. These features suggest the coexistence of
  axisymmetric and non-axisymmetric dynamo modes \citep{berdyugina02}.

\cite{messina03} have studied photometric periods from spot
modulations, arguing that - as in the solar case - the varying
dominant latitudes of the spots should induce a periodic variation of
the rotation period in phase with the activity cycle, because of
differential rotation. These correlated period variations are indeed
present in solar analogs \citep{messina03}, although two patterns are
seen: either, the period decreases as the cycle proceeds (solar
behavior), or it increases (anti-solar behavior). The former effect is
due to spot migration toward the equator where the surface rotation is
faster. The second effect could be due to pole-ward acceleration of
rotation at higher latitudes where active stars predominantly show
spots, while the individual spots may still migrate toward the
equator. \cite{messina03}, however, suggest that high-latitude spots
migrate toward the (slower-rotating) pole, which induces an anti-solar
behavior in particular for stars with small inclination angles. This
model is preferred because correlations involving the fractional
variation in the rotation period show the same behavior for the two
subclasses. Support for the model comes from simulations of
magnetic-field migration toward the poles in very active stars, as
performed by \cite{schrijver01b}
(Section~\ref{section:polarspots}). Specifically, \cite{messina03} have
found a power-law relation between the rotation-period variation,
$\Delta P$, and the average rotation period, \P, of the form (see
Figure~\ref{figure:diffrot}a)

\begin{equation}
  \Delta P \propto P^{1.42 \pm 0.5}.
\end{equation}

Further, a tight correlation is found between differential rotation,
parameterized by $\Delta\Omega/\Omega$, and the activity cycle
frequency, $\omega_\mathrm{cycl}$,

\begin{equation}
  \omega_\mathrm{cycl} \propto e^{(-0.055 \pm 0.004)\Delta\Omega/\Omega}
\end{equation}

suggesting that $\Delta\Omega/\Omega$ is a key parameter controlling
the duration of the activity cycle (see Figure~\ref{figure:diffrot}b).

Differential rotation has also been measured on other very
  active stars (e.g., \citealt{donati03b}), among them a solar-like
  post-T~Tauri star \citep{donati00}, in particular based on Doppler
  imaging techniques (Section~\ref{section:doppler}). Differential
  rotation was found to be solar-like in these examples, although
  time-variable, which may hint at dynamo processes that periodically
  convert magnetic into kinetic energy and vice versa
  \citep{donati03b}.

\epubtkImage{}{%
\begin{figure}[htbp]
\centerline{
\includegraphics[width=0.53\textwidth]{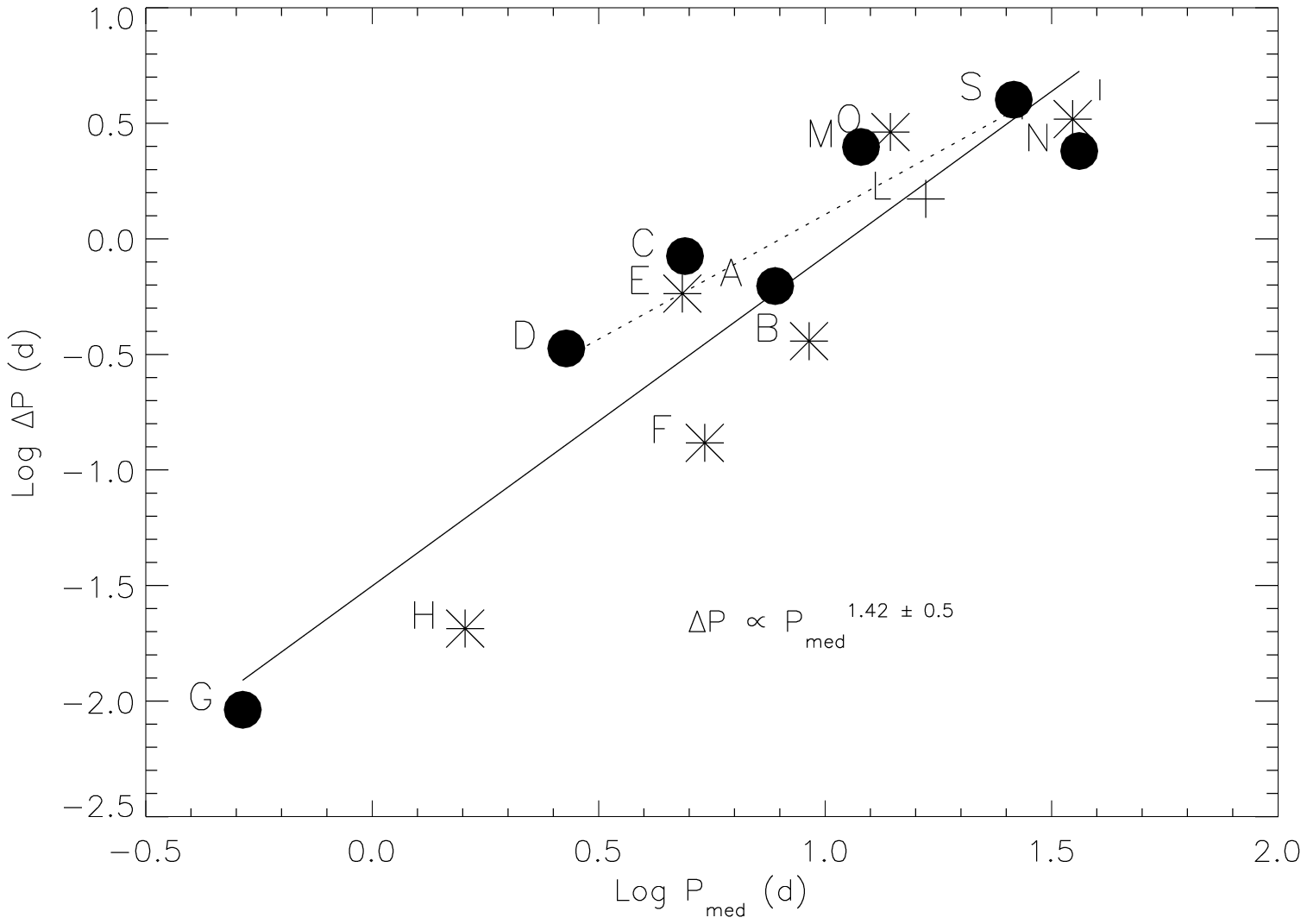}
\includegraphics[width=0.53\textwidth]{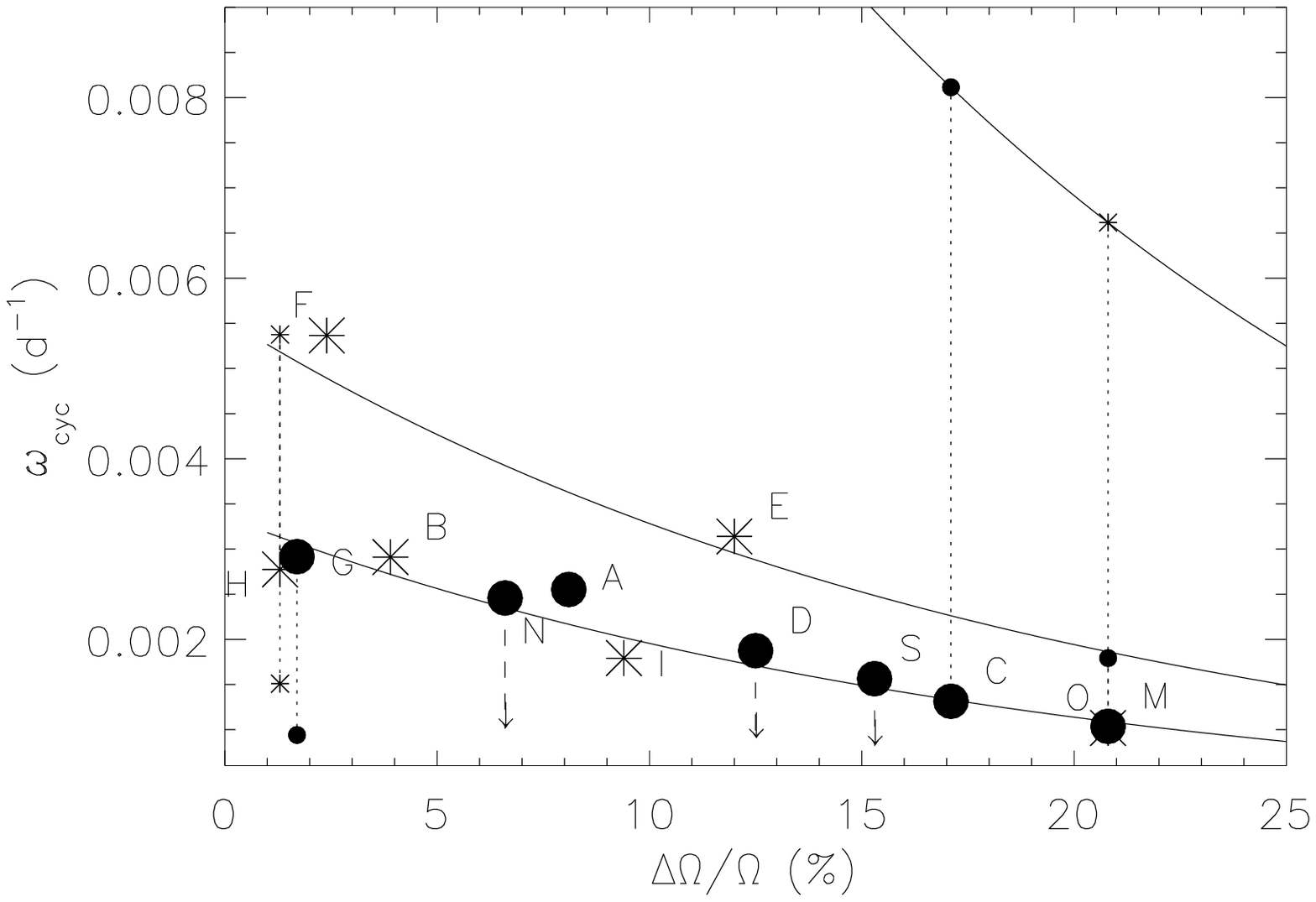}
}
\caption{\it Left: Rotation period variation as a function of the mean
  rotation period, for a sample of young solar analogs. The solid line
  is a power-law fit to the entire sample also containing early
  K~stars, while the dotted line is a fit to the G-star sample
  only. Key to the labels: A = BE~Cet, B = $\kappa^1$~Cet, C =
  $\pi^1$~UMa, D = EK~Dra, E = HN~Peg, F = DX~Leo (K0~V), G = AB~Dor
  (K0~V), H = LQ~Hya (K2~V), I = 107~Psc (K1~V), L = 61~UMa (G8~V), M
  = $\beta$~Com, N = HD~160346 (K3~V), O = 15~Sge, S = Sun
  \citep[from][]{messina03}. -- Right: The activity cycle frequency is
  shown as a function of the relative surface differential rotation
  amplitude, $\Delta\Omega/\Omega$. Vertical dotted lines connect
  multiple cycles. Three different branches are indicated by the solid
  curves \citep[from][reprinted with permission]{messina03}.}
\label{figure:diffrot}
\end{figure}}

\subsubsection{X-Ray Cycles of Solar Analogs}
\label{section:xraycycles}

Given the much stronger variability in the outer, coronal layers of a
stellar atmosphere, an activity cycle may be more easily identified
in the X-ray or radio domains. However, such cycles have eluded
detection until recently because no appropriate program had been
carried out for sufficiently long periods. A few notable examples have
now been reported from X-ray monitoring.

First tentative evidence for an X-ray cycle came from the young solar
analog EK~Dra that was monitored between 1990 and 2000 using {\it
  ROSAT}, {\it ASCA}, and {\it XMM-Newton}. Initial results were
presented in \cite{dorren95}, a more complete time series has been
published by \cite{guedel04}. There is a suggestive anti-correlation
between X-ray flux and photospheric brightness (the star is optically
brightest at its activity minimum), although the total X-ray
luminosity varies by no more than a factor of $\approx
2\mbox{\,--\,}3$.

Other reports refer to inactive solar analogs and K-type
stars. \cite{hempelmann03} and \cite{hempelmann06} have reported a
correlation between X-ray luminosity and the Ca~H \& K S index for the
two K-type stars 61~Cyg~A and B. Both show chromospheric modulations
on time scales of about 10~years, one being regular and the other
irregular. A gradual X-ray modulation was also seen during a time span
2.5~years in the G2~V star HD~81809, although there seems to be a
phase shift by about 1~year with respect to the Ca~cycle
\citep{favata04}.

Additional information on potential coronal activity cycles has been
collected from multiple observations of young open clusters and
star-forming regions. Generally, such samples indicate that
magnetically active, solar-like stars mostly lack well-expressed X-ray
cycles unless their cycle-induced variability is no more than a factor
of $\approx 2$ \citep{gagne94, gagne95a, gagne95b, stern94, stern95,
  micela96, sciortino98, grosso00, marino02, marino03b}.

\newpage


\section{Solar Radiation and Wind in Time}
\label{section:solarflux}

The Sun's magnetic activity has steadily declined throughout its MS
lifetime. This is a direct consequence of the declining dynamo as a
star spins down by losing angular momentum through its magnetized
wind. The distinguishing property of magnetic activity in the context
of stellar radiation is excess emission beyond the photospheric
thermal spectrum. Because magnetic activity expresses itself by
releasing energy, the most relevant radiation signatures of magnetic
activity are at wavelengths shorter than the dominant optical light,
i.e., at UV wavelengths (emission from active regions in the
chromosphere and the transition region), at far-ultraviolet
wavelengths (from the transition region), and the extreme-ultraviolet
and X-ray ranges (from coronal active regions). Apart from
electromagnetic radiation, particles are accelerated as a consequence
of magnetic energy release. These particles are either measured
in-situ in the solar system, or can be indirectly inferred from
non-thermal radio emission. The solar wind of course is another
particle stream that requires acceleration related to open magnetic
fields.

The present section summarizes our knowledge of these various photon
and particle losses from the Sun in Time, contrasting the young Sun's
behavior with the contemporaneous Sun by describing the long-term {\it
  evolution} of the radiation and wind signatures. High-energy
emission may have been much more important in heating and ionizing
planetary atmospheres or, at still earlier stages, circumstellar
accretion disks. To understand the young Sun's influence on its
environment, the spectral output in lines and the continuum must be
observed for solar analogs with different rotation periods
(corresponding to different magnetic activity levels) across the
ultraviolet-to-X-ray range. This ``Sun in Time'' program, introduced
in Section~\ref{section:sunintime}, has been conducted during the past
decade by various groups (e.g., \citealt{dorren94, dorren95, guedel94,
  guedel95c, guedel97a, ayres97, gaidos00, guinan03, ribas05,
  telleschi05}).

I will start with the solar wind that is responsible for the
declining solar spin rate, which in turn controls the dynamo; the
latter leads to surface activity, and hence to radiation that is again
changing on evolutionary time scales. I will discuss consequences of
the elevated high-energy output of the young Sun in the subsequent
chapters.

\subsection{The Solar Wind in Time}
\label{section:solarwind}

Stellar magnetic winds are a crucial consequence of ``stellar
activity'' but their detection in solar analogs is very
difficult. Although the wind formation and acceleration mechanisms are
still not conclusively understood in the Sun, it is clear that
magnetic fields play a major role, be it for acceleration of the
wind, for its heating, or for guiding the wind at least out to the
Alfv\'en radius. A close relation between wind and magnetic corona is
obvious \citep{parker58}, the former being related to open magnetic
field lines and the latter predominantly to closed magnetic
structures.

The best -- albeit indirect -- proof of the presence of magnetized
winds is the spin-down of convective stars on the main sequence as
such a wind carries away angular momentum from the star. I will
discuss this in Section~\ref{section:rotation}.

Direct measurement of ionized winds from solar-like stars has not yet
succeeded; potential detection methods include the measurement of
thermal radio emission from the winds \citep{lim96, gaidos00}, and
signatures of charge exchange in X-ray spectra
\citep{wargelin01}. \cite{lim96} and \cite{vdoord97} gave upper limits
to the mass-loss rates of $\dot{M}_\mathrm{w} \approx
10^{-12}\,M_{\odot} \unit{yr^{-1}}$ for ``solar-like'' winds with $T
\approx 1 \unit{MK}$ emanating from M-type dwarfs. \cite{gaidos00}
derived upper limits to $\dot{M}$ for three young solar analogs
($\pi^1$~UMa, $\kappa^1$~Cet, and $\beta$~Com), finding
$\dot{M}_\mathrm{w} \la (4-5)\times 10^{-11}\,M_{\odot}
\unit{yr^{-1}}$.

The most promising approach to date is an indirect method making use
of Ly$\alpha$ absorption in so-called ``astrospheres''; the latter are
suggested to be a consequence of interactions between stellar winds
and the interstellar medium (ISM). This subject has been extensively
reviewed in the {\it Living Reviews in Solar Physics} article by
\cite{wood04}; I will therefore only briefly summarize the essential
results.

Solar/stellar winds collide with the interstellar medium, forming,
with increasing distance from the star, a termination shock (where the
wind is shocked to subsonic speeds), a heliospause (separating the
plasma flows from the star and the ISM), and the bow shock (where the
ISM is shocked to subsonic speeds). The heliosphere is permeated by
interstellar H\,{\sc i} with $T \approx (2\mbox{\,--\,}4)\times 10^4
\unit{K}$ \citep{wood02}. Much of this gas is piled up between the
heliospause and the bow shock, forming the so-called ``hydrogen
wall'' that can be detected as an absorption signature in the
Ly$\alpha$ line. The excess absorption from the Sun's own hydrogen
wall is, due to the deceleration of the ISM relative to the star,
redshifted, while that of other astrospheres is blueshifted.

The measurable absorption depths are compared with results from
hydrodynamic model calculations \citep{wood02, wood05}. The important
point is that the amount of astrospheric absorption should scale with
the wind ram pressure, $P_\mathrm{w} \propto
\dot{M}_\mathrm{w}V_\mathrm{w}$, where $V_\mathrm{w}$ is the (unknown)
wind velocity \citep{wood98}. The latter is usually assumed to be the
same as the solar wind speed. From this, $\dot{M}_\mathrm{w}$ is
derived.

The Sun's hydrogen wall was detected in ultraviolet spectra by
\cite{linsky96}, and an equivalent astrosphere around $\alpha$~Cen A
and B was interpreted by \cite{gayley97}. Further important wind mass
loss measurements based on this method have been presented by
\cite{wood02} (and references therein) and \cite{wood05}. A systematic
study of all derived mass-loss rates shows that
$\dot{\cal{M}}_\mathrm{w}$ per unit stellar surface correlates with
the stellar X-ray surface flux,

\begin{equation}
  \dot{\cal{}M}_\mathrm{w} \propto F_\mathrm{X}^{1.34\pm 0.18}
\end{equation}

(an equivalent relation therefore holds between $\dot{M}_\mathrm{w}$
and $L_\mathrm{X}$); using the activity-age relation
(Section~\ref{section:coronal_Xray}), one finds

\begin{equation}
  \dot{\cal{M}}_\mathrm{w} \propto t^{-2.33\pm 0.55}
\end{equation}

\citep{wood05}. These two laws indicate that stellar-wind mass loss is
-- in principle -- a genuine activity indicator, the mass-loss being
higher in young, magnetically active stars. Extrapolating the above
law up to the X-ray saturation limit ($F_\mathrm{X} \approx 2\times
10^7 \unit{erg\ cm^{-2}\ s^{-1}}$) would suggest
$\dot{\cal{M}}_\mathrm{w}$ (or $\dot{M}_\mathrm{w}$) of the youngest
solar analogs to be about a thousand times higher than the present-day
solar mass loss ($\dot{M}_{\odot} \approx 2\times 10^{-14}\,M_{\odot}
\unit{yr^{-1}}$; e.g., \citealt{feldman77}). However, this power-law
relation breaks down for the most active stars with $F_\mathrm{X} \ga
8 \times 10^5 \unit{erg\ cm^{-2}\ s^{-1}}$ (\citealt{wood05},
Figure~\ref{figure:wind}). Stars at this limit show $\dot{M}_\mathrm{w}$
about 100~times the present solar value, while $\dot{M}_\mathrm{w}$
drops toward higher activity levels to about 10~times the solar
value. The reason for this breakdown between X-ray activity and
wind-mass loss may be related to the appearance of high-latitude
active regions (spots) in the most active stars
(Section~\ref{section:polarspots}); if the magnetic field becomes more
akin to a global dipole, then wind escape may be inhibited in such
stars \citep{wood05}.

\epubtkImage{}{%
\begin{figure}[htbp]
  \rotatebox{0}{\includegraphics[scale=1]{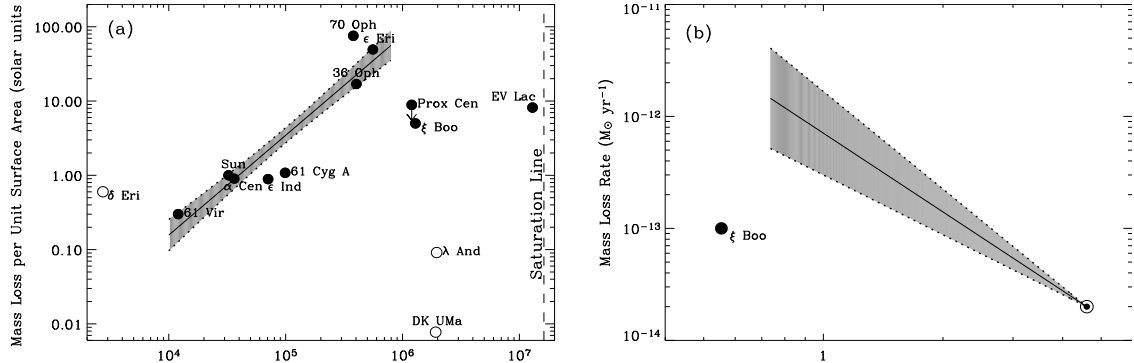}}
  \caption{\it Left (a): Mass-loss rates per unit surface area vs.\
  stellar X-ray surface fluxes. MS stars are shown by filled
  circles. The trend for inactive stars (shaded area) is not followed
  by more active stars. -- Right (b): Inferred mass-loss history of
  the Sun. Again, the trend shown for inactive stars (shaded area)
  breaks down for the most active stars \citep[from][reprinted with permission 
  of AAS]{wood05}.}
  \label{figure:wind}
\end{figure}}


\subsection{The Solar Spin in Time}
\label{section:rotation}

Stellar magnetic activity is fundamentally determined by the stellar
rotation period, as I will further discuss in, e.g.,
Section~\ref{section:coronal_Xray}. Because the magnetized stellar winds
transport angular momentum away from the star, a MS solar analog
spins down with age. The rate of change of angular momentum is related
to the mass loss rate, the spin rate, and the Alfv\'en radius,
$R_\mathrm{A}$. One finds

\begin{equation}
  {\dot{\Omega}\over \Omega} \propto {\dot{M}\over M}
  \left({R_\mathrm{A}\over R_{\odot}}\right)^m
\label{omegaM}
\end{equation}

where $m$ is between 0 and 1 depending on the magnetic field geometry
\citep{weber67, mestel84, stepien88}. One further has to couple the
average surface magnetic field strength, $B_0$, with $\Omega$. This
relation is essentially determined by the magnetic dynamo but can
reasonably be parameterized as $B_0 \propto \Omega^p$ with $p$
probably being 1 or 2 \citep{mestel84}. The approximately observed
$\Omega \propto t^{-1/2}$ law (``Skumanich law'',
\citealt{skumanich72}, see below) is recovered for a thermal wind with
$p = 1$ \citep{mestel84}, i.e., a linear dependence between average
surface magnetic field strength and rotation rate. For another
detailed study of this problem, see \cite{stepien88}.

Empirically, for solar analogs one finds,

\begin{equation}
  P = 0.21 t_6^{0.57} \unit{[d]}
\label{rotationevolution}
\end{equation}

\citep{dorren94b}, where $t_6$ is the age in Myr after arrival on the
ZAMS. Equivalently, for the rotational (equatorial) velocity and for
the rotation rate (for constant $R$),

\begin{equation}
  \Omega \propto v \propto t_6^{-0.6\pm 0.1} \unit{[d]}
\label{vrotage}
\end{equation}

\citep{ayres97} -- see Figure~\ref{figure:rotationage}. These equations
imply a decrease in rotation period from ZAMS age to the end of the MS
life of a solar analog by a factor of about 20.

\epubtkImage{}{%
\begin{figure}[t!]
  \centerline{\includegraphics[scale=0.5]{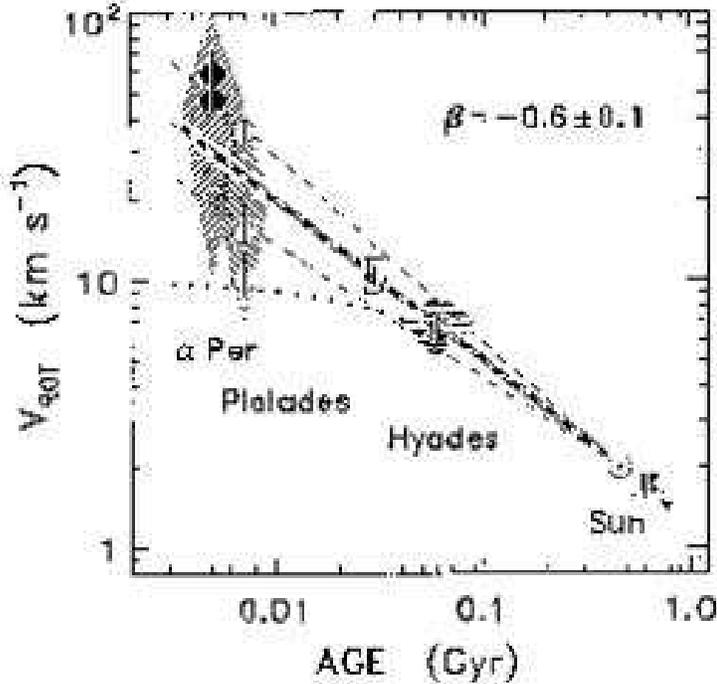}}
  \caption{\it Relation between rotational velocity, v, and age for
  solar analogs. The diamond-shaped areas show the large scatter of
  v in young clusters, before rotational convergence has been
  attained \citep[from][reprinted with permission]{ayres97}.}
  \label{figure:rotationage}
\end{figure}}

At ages of approximately 100\,Myr or younger, the stellar rotation
period is not a function of age but of the PMS history such as the
history of circumstellar-disk dispersal (e.g., \citealt{soderblom93},
see Figure~\ref{figure:rotationage}). Once the inner disk disappears,
the lack of magnetic locking via star-disk magnetic fields and the
contraction of the star toward the MS will spin-up the star and thus
determine the initial rotation period on the ZAMS. For example, the
rotation periods of G-type stars in the Pleiades and the $\alpha$ Per
clusters still scatter considerably, ranging from less than a day (the
so-called ultra-fast rotators) to many days \citep{soderblom93,
  stauffer94, randich96}, while they (and therefore the stellar X-ray
luminosities, see Section~\ref{section:coronal_evolution}) have
converged to a nearly unique value at the age of the UMa Moving Group
(300\,Myr) or the Hyades (600\,--\,700 Myr; see \citealt{soderblom93,
  stern95}).

\subsection{The Ultraviolet Sun in Time}
\label{section:uvflux}

Ultraviolet excess emission from late-type stars originates in
magnetic chromospheric and transi\-tion-zone regions that have been
heated to temperatures of order $10^4\mbox{\,--\,}10^5
\unit{K}$. Magnetic activity thus makes the UV spectra rich in
diagnostics (see Figure~\ref{figure:ribas05_5}; from
\citealt{ribas05}). \cite{dorren94} and \cite{ayres97} have studied
the evolution of ultraviolet (UV) line fluxes in detail, based on
spectral measurements obtained by {\it IUE}. \cite{ribas05} extended
these investigations to the ``Sun in Time'' sample and included
spectral information from {\it HST}. The bulk of the UV flux in the
region shortward of 1700\,\AA\ is in emission lines while the
continuum is negligible. Emission lines include those of 
O\,{\sc i}~$\lambda 1304$,  
C\,{\sc ii}~$\lambda 1335$,  
Si\,{\sc iv}~$\lambda 1400$,  
C\,{\sc iv}~$\lambda 1550$,  
He\,{\sc ii}~$\lambda 1640$, and 
C\,{\sc i}~$\lambda 1657$.

\epubtkImage{}{%
\begin{figure}[htbp]
  \centerline{\includegraphics[scale=0.52]{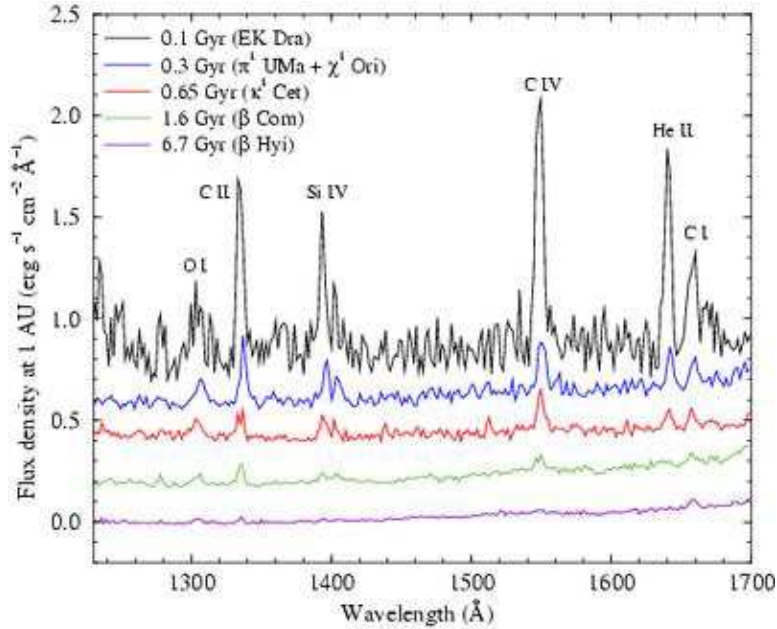}}
  \caption{\it Extracts of UV spectra of solar analogs with different
  ages. All spectral fluxes have been transformed to irradiances at
  1\,AU from the star. The spectra have been shifted along the
  ordinate, by multiples of $0.2 \unit{erg\ s^{-1}\ cm^{-2}\
  \AA^{-1}}$ \citep[from][reproduced by permission of AAS]{ribas05}.}
  \label{figure:ribas05_5}
\end{figure}}

\cite{dorren94b} studied the activity-rotation relationship for
Mg\,{\sc ii} and C\,{\sc iv} for solar analogs in the ``Sun in Time''
sample. They found (I also include their X-ray result for comparison,
to be discussed further in Section~\ref{section:coronal_Xray} below)

\begin{eqnarray}
{L_\mathrm{Mg~II}} &=&  (9.1\pm 0.2)\times 10^{29}~P^{-0.76\pm 0.096}\unit{[erg\ s^{-1}]}\\
{L_\mathrm{C~IV}}  &=&  (1.2\pm 0.4)\times 10^{29}~P^{-1.6\pm 0.15}\unit{[erg\ s^{-1}]}\\
{L_\mathrm{X}}     &=&  (9.1\pm 4.7)\times 10^{30}~P^{-2.5\pm 0.23}\unit{[erg\ s^{-1}]}
\label{activityrotation}
\end{eqnarray}

Emission from hotter regions is more strongly dependent on
rotation. Because \P decreases with age, harder emission decays more
rapidly than softer emission. I will return to this point in more
detail in Section~\ref{section:spectralevolution}.

\subsection{The Far-Ultraviolet Sun in Time}
\label{section:faruvflux}
 
\cite{guinan03} have used the {\it FUSE} satellite to probe the
transition regions of the ``Sun in Time'' stars by measuring fluxes of
key lines in the 920\,--\,1180\,\AA\ region. An example is shown in
Figure~\ref{figure:ribas05_3}. The individual flux measurements have
been given by \cite{ribas05} and are listed below in
Table~\ref{table:linefluxes} as irradiances for a distance of
1\,AU.  Measurements pertaining to the intermediately active Sun
  are also given \citep[see][for references]{ribas05}.

These line fluxes have been used to construct surface flux-rotation
relations for solar analogs \citep{guinan03}. For essentially the
entire rotation range of MS solar analogs, the line fluxes follow
power-law relations, $F \propto P^{-\alpha}$, with $\alpha \approx
1.8$. This power-law decay is steeper than for chromospheric lines
(Section~\ref{section:uvflux}), but shallower than for coronal fluxes or
luminosities (recalling that the stellar radii are all close to
$1\,R_{\odot}$ and hence the luminosities are proportional to the
surface fluxes).

A very important FUV contribution comes from the H\,{\sc i}~$\lambda
1216$~Ly$\alpha$ line \citep{ribas05}. This holds true throughout the
entire MS lifetime of a solar analog although the relative
contribution of this line with respect to higher-energy emission
increases with age (see Table~\ref{table:linefluxes}).

\epubtkImage{}{%
\begin{figure}[t]
  \centerline{\includegraphics[scale=0.42]{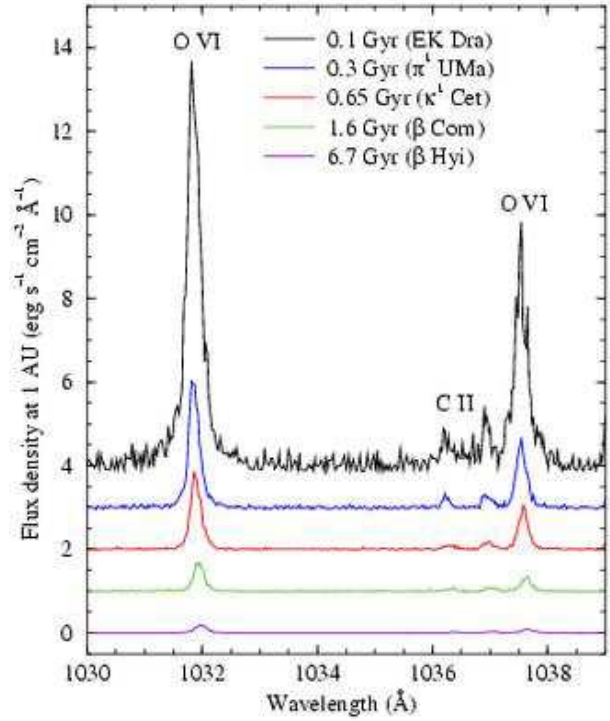}}
  \caption{Extracts of FUV spectra of solar analogs with different
  ages. The region of the {\rm O\,{\sc vi}} doublet is shown. All
  spectral fluxes have been transformed to irradiances at 1\,AU from
  the star. The spectra have been shifted along the ordinate, by
  multiples of $1 \unit{erg\ s^{-1}\ cm^{-2}\ \AA^{-1}}$
  \citep[from][reproduced by permission of AAS]{ribas05}.}
  \label{figure:ribas05_3}
\end{figure}}

\subsection{The Extreme-Ultraviolet and X-Ray Sun in Time}
\label{section:coronal_evolution}

The present-day Sun's corona can be characterized by its total
radiative output (mostly in the soft-X-ray regime) of a few times
$10^{26} \unit{erg\ s^{-1}}$ at minimum activity level to a few times
$10^{27} \unit{erg\ s^{-1}}$ at maximum activity level outside very
strong flares; its electron temperature of about 1\,--\,5\,MK,
depending on the magnetic coronal structure under consideration; and
its filling factor. The latter is difficult to define as a large
variety of (partly expanding) structures confined by closed or open
magnetic fields define the magnetic corona. Suffice it to say that the
strongest magnetic active regions are confined the volumes closely
attached to photospheric sunspot complexes, and the latter cover less
than 1\% of the solar surface even at activity maximum. Clearly, the
solar corona is far from being filled by luminous active regions.

\subsubsection{The Solar X-Ray Corona in Time}
\label{section:coronal_Xray}

Although coronae radiate across the electromagnetic spectrum, the
dominant losses occur in the soft-X-ray range; the radio regime
provides complementary diagnostics on non-thermal processes. I
concentrate on these two aspects (see
Section~\ref{section:coronal_radio} for details on radio emission).

The total X-ray output of a stellar corona depends on the available
magnetic energy and is therefore a consequence of the dynamo
operation. Younger and more rapidly rotating stars are more X-ray
luminous; as is the case for UV and FUV radiation, the X-ray output
decreases as the star ages and its rotation period increases. For
solar analogs, the decay law is,

\begin{equation}
  L_\mathrm{X} \approx (3\pm 1)\times 10^{28} t_9^{-1.5\pm 0.3} \quad
  \mathrm{[erg\ s^{-1}]}
\label{tempage}
\end{equation}

(where $t_9$ is the stellar age in Gyr) as derived from small but
well-characterized samples \citep{maggio87, guedel97a}. This decay law
holds for MS stars back in time as long as

\begin{itemize}
  \item Equation~\ref{tempage} predicts a luminosity at or below saturation
  for a solar analog, i.e., $L_\mathrm{X}/L_\mathrm{bol} \la 10^{-3}$,
  or $L_\mathrm{X} \la 4\times 10^{30} \unit{erg\ s^{-1}}$;

  \item the rotation period is a function of age for a star of given
  mass.
\end{itemize}

The second condition is fulfilled only for ages higher than (at least)
100\,Myr (\citealt{soderblom93}, see
Section~\ref{section:rotation}). Consequently, the X-ray luminosity of a
solar analog is nearly only a function of rotation period for stars
older than $\approx 100\mbox{\,--\,}200 \unit{Myr}$, and this
dependence is given by

\begin{equation}
  L_\mathrm{X}  =  10^{31.05 \pm 0.12} P^{-2.64 \pm 0.12} \unit{[erg\ s^{-1}]}
\label{activityrotationX}
\end{equation}

as derived from a sample of nearby solar analogs (\citealt{guedel97a};
see Figure~\ref{figure:rotlx}; a very similar decay law is found from
measurements conducted with broadband spectrometers on board {\it
  XMM-Newton}: $L_\mathrm{X} \propto 4.04 \times 10^{30} P^{-2.03 \pm
  0.35} \unit{[erg\ s^{-1}]}$, see \citealt{telleschi05}). The
rotation-activity law (Equation~\ref{activityrotationX}) is confirmed by
broader studies, also with regard to Rossby number $Ro$ that is
defined as the ratio between the two time scales of rotation and
convection driving the dynamo ($Ro = P/\tau_c$, where $\tau_c$ is the
convective turnover time; \citealt{noyes84, mangeney84}). All studies
find roughly $L_\mathrm{X}/L_\mathrm{bol} \propto P^{-2}$ for
late-type stars (e.g., \citealt{randich00}). Also, other activity
indicators such as the surface X-ray flux follow an equivalent law.

\epubtkImage{}{%
\begin{figure}[htbp]
\centerline{
  \includegraphics[scale=0.39]{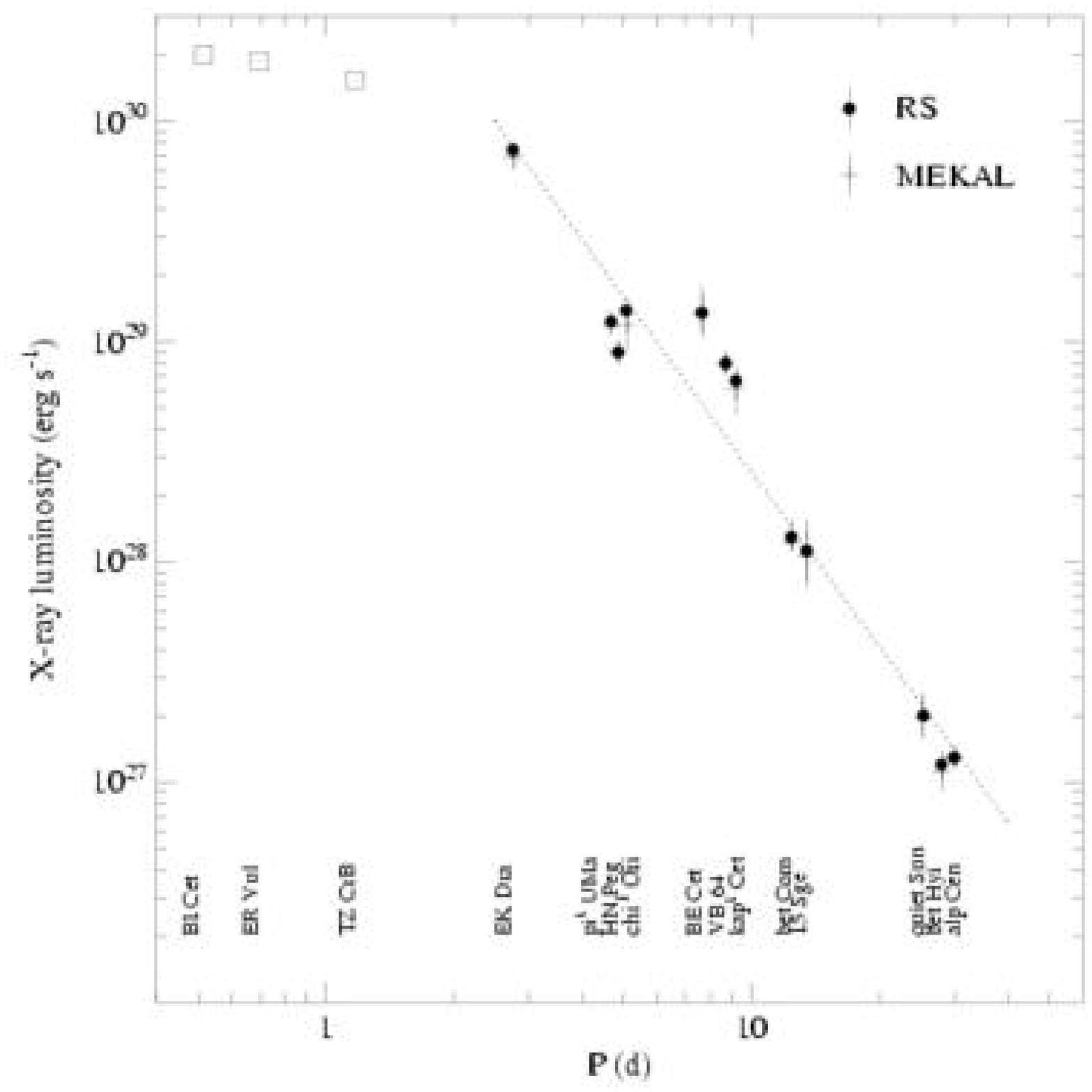}
  \includegraphics[scale=0.37]{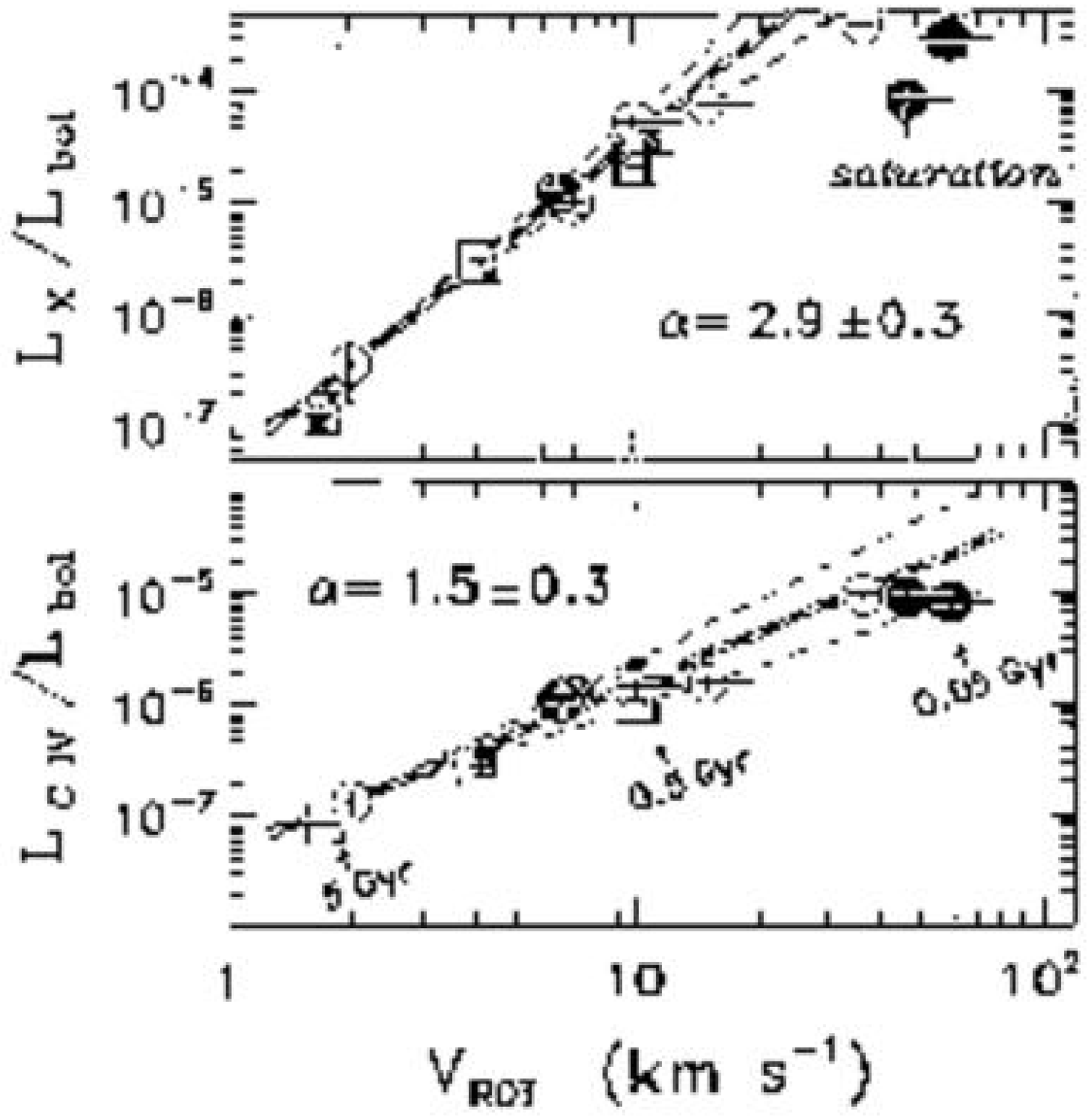}
  }
  \caption{\it Left: X-ray luminosity of solar analogs (from the ``Sun
  in Time'' sample), and power-law fit (slope = -2.6). The three
  objects marked with open squares are close binaries consisting of
  two early-G solar analogs that rotate synchronously with their orbit
  motion. These latter stars are in the saturation regime (``RS'' and
  ``MEKAL'' refer to two different atomic line codes used for the
  spectral interpretation; from \citealt{guedel97a}, reproduced by permission of AAS). -- Right:
  Normalized X-ray and {\rm C\,{\sc iv}} luminosities as a function of
  rotational velocity for solar analogs. Luminosities of the fastest
  rotators do not follow the regression laws because of saturation
  \citep[from][reprined with permission]{ayres97}.}
  \label{figure:rotlx}
\end{figure}}

Alternatively, the X-ray output roughly scales with the square of the
equatorial velocity,

\begin{equation}
  L_X \approx 10^{27}(v{\rm sin}i)^2 \quad \mathrm{[erg\ s^{-1}]}
\label{lxprot}
\end{equation}

\citep{pallavicini81, ayres80, maggio87, wood94}. {\it The X-ray
  output is an excellent indicator of dynamo activity. The coronal
  output is strongly determined by parameters that control the
  magnetic dynamo.}

For the most active stars, $L_\mathrm{X}$ is somewhat suppressed, for
reasons that are not well understood. The stars are in a
``super-saturated'' regime (\citealt{randich96}, see discussion in
\citealt{guedel04}).

The overall rotation and activity history of a solar analog star from
ZAMS to the terminal stages of its MS life thus proceeds roughly as
follows:

\begin{itemize}

  \item A star enters on the MS as a fast rotator although the
    rotation period is not unique, depending on the previous star-disk
    interaction history. The star thus typically begins its MS life in
    the saturation regime where $L_\mathrm{X} \approx
    (2\mbox{\,--\,}4) \times 10^{33} \unit{erg\ s^{-1}}$. Although it
    spins down as a consequence of its magnetic-wind mass loss, there
    is little evolution of the X-ray (and UV, FUV, EUV) output.

  \item At an age of around 20\,--\,200\,Myr, the rotation period has
  decreased sufficiently so that the star's corona drops out of the
  saturation regime (e.g., \citealt{patten96, james97}, and
  Sect.~17.5.2 in \citealt{guedel04}). For a G~star, saturation holds
  as long as $P < 1.5\mbox{\,--\,}2 \unit{d}$.

  \item Convergence of rotation periods has been achieved at ages of a
    few 100\,Myr, and by that time a solar analog shows $P \approx
    4\mbox{\,--\,}5 \unit{d}$ and $L_X \approx 10^{29} \unit{erg\
    s^{-1}}$, i.e., less than 10\% of the saturation value.

  \item From then on, the luminosity decay law holds, because the
  rotation period is a function of age and increases
  monotonically. This decay law has been observationally tested until
  the terminal stages of the MS life of a solar analog.

\end{itemize}

\subsubsection{The Coronal Temperature in Time}
\label{section:coronal_temperature}

The non-flaring corona of the contemporary Sun shows temperatures of a few MK.
Active regions tend to be hotter than the quiet corona, and consequently,
the average coronal temperature during the Sun's activity maximum is higher
than at minimum. \cite{peres00} have studied the full-disk solar coronal
emission measure distribution to find peak temperatures of $\approx 1
\unit{MK}$ and $\approx 2 \unit{MK}$ during minimum and maximum,
respectively. The corresponding X-ray luminosities amount to
$L_\mathrm{X} \approx 3\times 10^{26} \unit{erg\ s^{-1}}$ and
$L_\mathrm{X} \approx 5\times 10^{27} \unit{erg\ s^{-1}}$,
respectively. These solar observations suggest that regions of higher
magnetic activity, but also episodes of higher overall magnetic
activity, not only produce higher-luminosity plasma but also higher
temperatures. Does this reflect in coronae of stars with widely
varying activity levels?

The spectral evolution of the X-ray and EUV (coronal) Sun in Time is
illustrated in Figs.~\ref{figure:ribas05_2}, \ref{figure:rgsspec}, and
\ref{figure:epicspec}. These figures of course support the trend of
decreasing flux with increasing age as described in
Section~\ref{section:coronal_Xray}, but they reveal other important
features:

\epubtkImage{}{%
\begin{figure}[t!]
  \centerline{\includegraphics[scale=0.48]{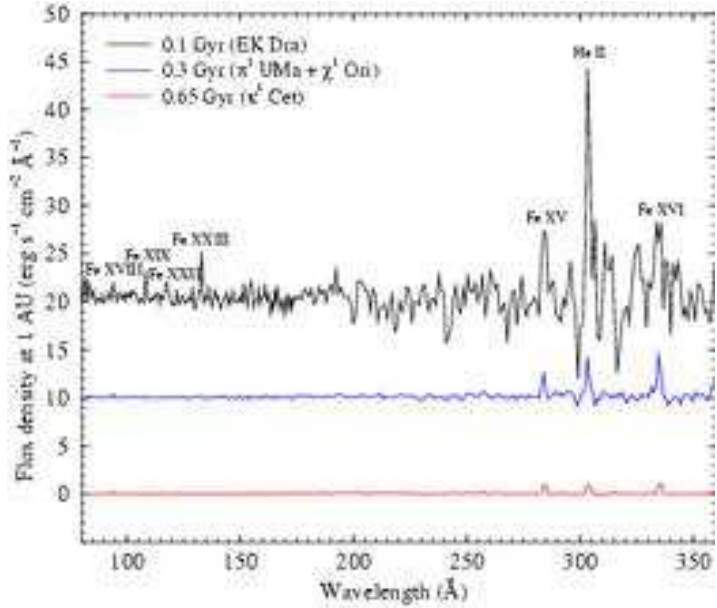}}
  \caption{\it Extracts of the EUV spectra of solar analogs with
  different ages. All spectral fluxes have been transformed to
  irradiances at 1\,AU from the star. The spectra have been shifted
  along the ordinate, by multiples of $10 \unit{erg\ cm^{-2}\ s^{-1}\
  \AA^{-1}}$ \citep[from][reproduced by permission of AAS]{ribas05}.}
  \label{figure:ribas05_2}
\end{figure}}

\epubtkImage{}{%
\begin{figure}[htbp]
  \centerline{\includegraphics[scale=0.75]{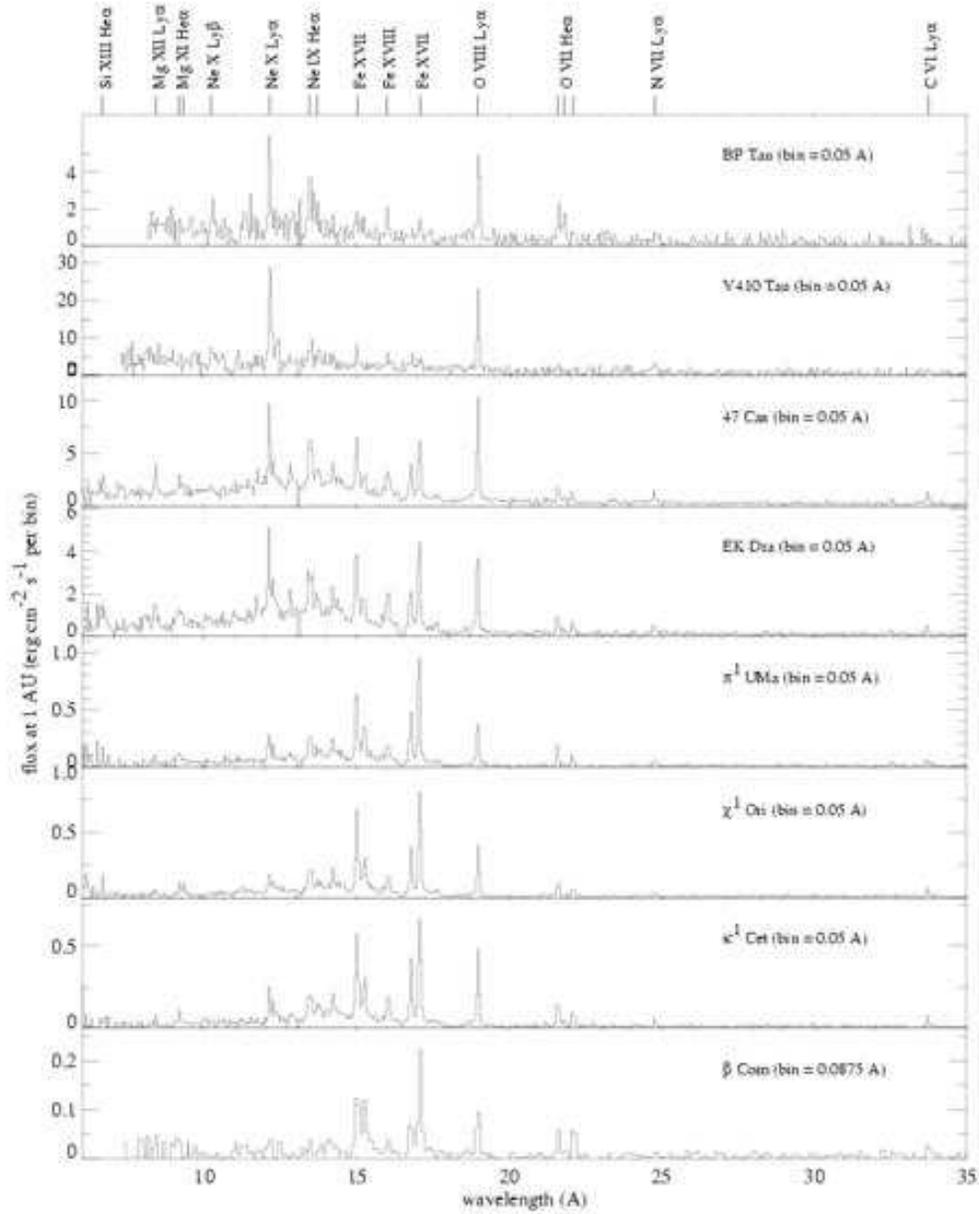}}
  \caption{\it Fluxed high-resolution X-ray spectra of solar analogs
  with different ages, obtained with the RGS instruments on board {\em
  XMM-Newton}. The topmost two spectra are from K-type T~Tauri stars
  (BP~Tau is a classical T~Tauri star, V410~Tau a weak-line T~Tauri
  star), while the other spectra are from MS solar analogs. Fluxes are
  given at a distance of 1\,AU from the star, in $\mathrm{erg\
  cm^{-2}\ s^{-1}}$ per bin, where the bin width is given in
  parentheses after the star's name in each panel (usually 0.05\,\AA\
  except for $\beta$~Com where a bin width of 0.0875\,\AA\ has been
  used). Note the anomalously strong {\rm O\,{\sc vii}} (and also {\rm
  Ne\,{\sc ix}}) lines in the CTTS BP~Tau, indicating an excessive amount of
  cool plasma (the ``soft excess''; adapted from \citealt{telleschi05}
  and \citealt{telleschi07a}).}
  \label{figure:rgsspec}
\end{figure}}

\epubtkImage{}{%
\begin{figure}[t!]
  \centerline{\includegraphics[scale=0.52]{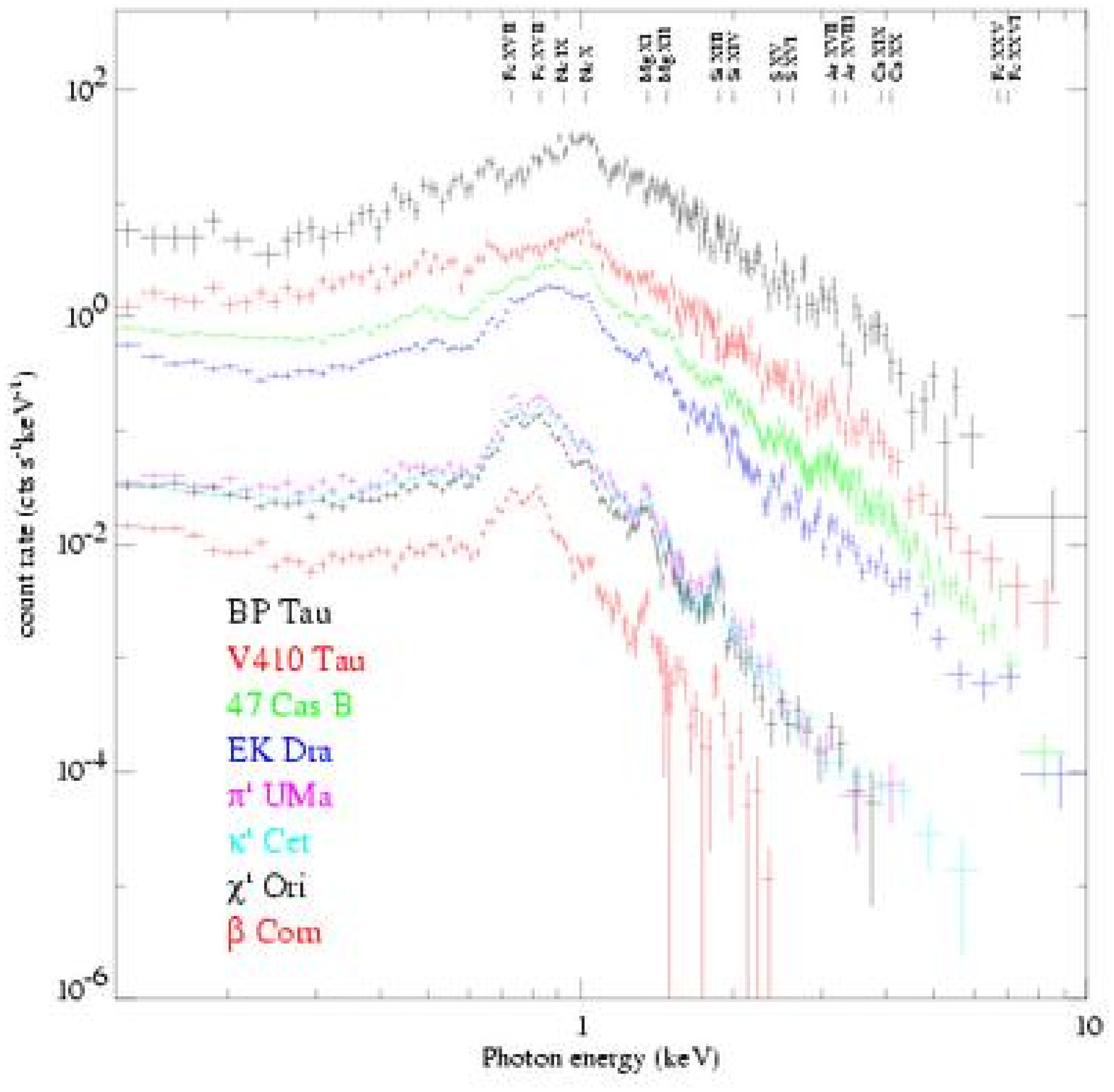}}
  \caption{\it Low-resolution CCD X-ray spectra of solar analogs with
  different ages, obtained with the EPIC MOS instrument on board {\em
  XMM-Newton}. The topmost two spectra are from K-type T~Tauri stars
  (BP~Tau is a CTTS, V410~Tau a WTTS), while the other spectra are
  from MS solar analogs. The sequence of the spectra from top to
  bottom is reflected in the lower left panel, and also by the colors
  of the spectra used for the stellar names. Count rate spectra have
  been normalized to the distance of 47~Cas (33.56\,pc); the spectrum
  of BP~Tau has been shifted upward by an additional factor of 30 for
  clarity (adapted from \citealt{telleschi05} and
  \citealt{telleschi07a}).}
  \label{figure:epicspec}
\end{figure}}

\begin{itemize}

  \item In Figure~\ref{figure:ribas05_2}, the EUV spectrum shows
  prominent lines of highly ionized Fe (Fe\,{\sc xxii, xxiii}) in the
  young EK~Dra but these lines rapidly disappear in older stars, where
  lines of Fe\,{\sc xv} and Fe\,{\sc xvi} predominate. This is a
  signature of hotter overall temperatures in the younger, more active
  solar analogs.

  \item In Figure~\ref{figure:rgsspec}, the X-ray spectra of the
  youngest, most active solar analogs (47~Cas, EK~Dra), again reveal
  strong lines from highly ionized Fe and also a prominent Ne\,{\sc x}
  line (at 12.1\,\AA); these lines are formed by plasma with
  temperatures of order 10\,MK. In contrast, in the least active star
  shown ($\beta$~Com), lines of Fe\,{\sc xvii}, Ne\,{\sc ix}, and
  O\,{\sc vii} dominate. These lines are formed at temperatures of
  2\,--\,5\,MK.

  \item Both Figure~\ref{figure:rgsspec} and ~\ref{figure:epicspec} show
  a more prominent continuum in more active stars. Continuum is
  predominantly formed by bremsstrahlung and is therefore a sensitive
  temperature indicator.

\end{itemize}

These spectral features clearly suggest higher temperatures in more
active, younger solar analogs.

From full spectral interpretation, a tight correlation has indeed
been found between the characteristic coronal temperature and the
normalized coronal luminosity $L_X/L_\mathrm{bol}$: {\it Stars at
  higher activity levels support hotter coronae} \citep{vaiana83,
  schrijver84, stern86, schmitt90, dempsey93, maggio94, gagne95a,
  schmitt95, huensch96, guedel97a, preibisch97, schmitt97, singh99}.

Figure~\ref{figure:templx} shows the correlation between average
(emission-measure weigh\-ted) coronal temperature and the total X-ray
luminosity for solar analogs, including solar maximum and minimum
values \citep[the latter from][]{peres00}. Numerically, the
correlation for the stellar sample reads

\begin{equation}
  L_\mathrm{X}  = 1.61\times 10^{26} T_\mathrm{av}^{4.05\pm 0.25} \unit{[erg\ s^{-1}]}
\label{TLx}
\end{equation}

where $T_\mathrm{av}$ is in MK, and $L_\mathrm{X}$ has been determined
in the 0.1\,--\,10\,keV band (\citealt{telleschi05}, and similar
results in \citealt{guedel97a} for the {\it ROSAT} energy band). Such
relations continue to hold into the PMS domain where exceedingly hot
coronae with temperatures up to $\approx 100 \unit{MK}$ are found
\citep{imanishi01}.

The trend seen in coronal X-ray emission is similar to the trends
described previously for the UV, FUV, and X-ray bands
(Section~\ref{section:uvflux}, \ref{section:faruvflux}, and
\ref{section:coronal_Xray}): {\it The emission from stars at higher
  activity levels is harder, implying that harder emission decays more
  rapidly in the course of stellar evolution.}

Three classes of models have been proposed to explain this
correlation:

\epubtkImage{}{%
\begin{figure}[t!]
  \centerline{\includegraphics[scale=0.6]{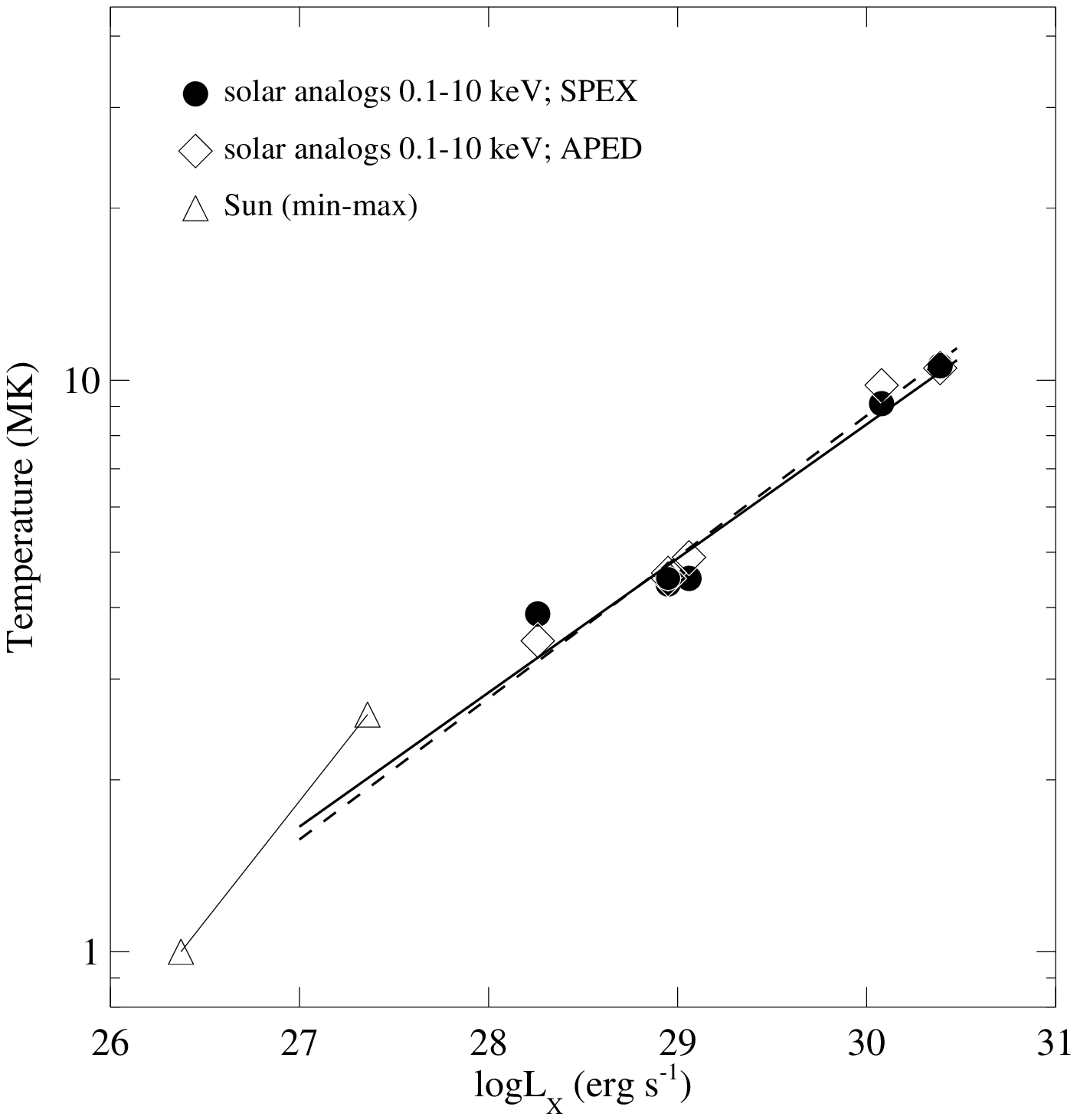}}
  \caption{\it Mean coronal temperature of solar analogs as a function
  of the X-ray luminosity. The dashed and solid lines are the
  regression fits to results based on different atomic data. The
  triangles mark the range of the solar corona between activity
  minimum and maximum \citep[figure from][reproduced by permission of AAS]{telleschi05}.}
  \label{figure:templx}
\end{figure}}

\begin{itemize}

  \item As stellar activity increases, the corona becomes
    progressively more dominated by hotter and denser features, for
    example active regions as opposed to quiet areas or coronal
    holes. Consequently, the average stellar X-ray spectrum indicates
    more hot plasma \citep{schrijver84, maggio94, guedel97a,
    preibisch97, orlando00, peres00}.

    \item Increased magnetic activity leads to more numerous
      interactions between adjacent magnetic field structures at
	the {\it chromospheric} level \citep{cuntz99} and therefore at
	the coronal level, owing to higher magnetic filling factors in
	the photosphere (Section~\ref{section:structure_summary}). The
      heating efficiency thus increases. Specifically, a higher rate
      of large flares is expected. Flares produce hot, dense plasma
      and therefore increase both the X-ray luminosity and the average
      coronal temperature of a star \citep{guedel97a}.

    \item \cite{jordan87} and \cite{jordan91} described an emission
    measure (EM)-$T$ relation based on arguments of a minimum energy
    loss configuration of the corona, assuming a fixed ratio between
    radiative losses and the coronal conductive loss. A relation
    including the stellar gravity $g$ was suggested, of the form

\begin{equation}
  EM \propto T^3g
\label{jordan}
\end{equation}

which fits to a sample of observations with $T$ taken from single-$T$
fits to stellar coronal spectra. Equation~(\ref{jordan}) holds because
coronal heating directly relates to the production rate of magnetic
fields, and the magnetic pressure is assumed to scale with the thermal
coronal pressure.

\end{itemize}

\subsection{Putting it all Together: The XUV Sun in Time}
\label{section:spectralevolution}

The above subsections provide the input for a comprehensive model of
the spectral evolution of the ``Sun in Time'' in the wavelength band
that is relevant for ionization of and chemical reactions in planetary
atmospheres and circumstellar disks, namely the 1\,--\,1700~\AA\
($\approx$0.007-10\,keV) FUV/EUV/X-ray (``XUV'') range. The results are
summarized in Tables~\ref{table:linefluxes} and \ref{table:decay}
compiled using data from \cite{ribas05} (for the UV-EUV range) and
\cite{telleschi05} (for the X-ray range).  The table also contains
  data referring to the classical T~Tauri star TW~Hya, to be discussed
  in Section~\ref{section:premainsequence}, and solar data (see
  \citealt{ribas05} for references).\epubtkFootnote{The solar line fluxes of
    O\,{\sc viii}, O\,{\sc vii}, N\,{\sc vii}, and C\,{\sc vi} were
    determined from a model spectrum synthesized in the XSPEC software
    \citep{arnaud96} using the vapec model \citep{smith01}. The model
    is based on an isothermal plasma with a temperature of 2\,MK
    normalized such that the 0.1\,--\,10\,keV luminosity is $2\times
    10^{27} \unit{erg\ s^{-1}}$. The low-FIP element abundances were set to
    values four times higher than standard photospheric abundances,
    while the high-FIP element abundances were photospheric; the S
    abundance was set to an intermediate value of 2$\times$ the
    photospheric value.} The line fluxes given in
Table~\ref{table:linefluxes} are normalized to a distance of 1\,AU and
have also been normalized to the radius our Sun had at the age of the
respective star.  Note that the Ly$\alpha$ line fluxes were
  corrected for interstellar H\,{\sc i} and D\,{\sc i} absorption,
  i.e., they represent the pure stellar contribution.

\cite{ribas05} constructed band-integrated irradiances for the
spectral ranges 1\,--\,20\,\AA\ (X-rays), 20\,--\,100\,\AA\ (soft
X-rays and EUV), 100\,--\,360\,\AA\ (EUV), and 920\,--\,1180\,\AA\
(FUV). For the wavelength range of 1180\,--\,1700\,\AA, only line
fluxes are provided because of increasing contributions from the
photospheric continuum.

All integrated irradiances correlate tightly with the stellar rotation
period or age, suggesting a rapid decay of activity at all atmospheric
levels in concert. The relations are excellently represented by power
laws, as illustrated in Figure~\ref{figure:integraldecay}a. The
power-law fits to the fluxes of the form

\begin{equation}
  F = \alpha t_9^{\beta}
\label{decaylaw}
\end{equation}

($\alpha$ and $\beta$ being constants) are given in
Table~\ref{table:decay}. Note that the inaccessible spectral range of
360\,--\,920\,\AA\ (strongly absorbed by interstellar gas) has been
interpolated between adjacent spectral ranges, assuming a decay law
with $\beta = -1$.

\begin{table}
\vbox{\vskip 4truecm
\rotatebox{90}{\vbox{\vskip -3truecm
\begin{minipage}{1.0\textheight}
\footnotesize
\caption{Integrated fluxes (in units of erg~cm$^{-2}$~s$^{-1}$) of strong
emission features normalized to a distance of 1~AU and  the radius of a one
solar mass star. UV and EUV fluxes of $\pi^1$ UMa and $\chi^1$ Ori have been averaged. Data for solar analogs are from \citet{telleschi05} and \citet{ribas05},
 and for TW Hya from \citet{herczeg02}, \citet{herczeg04}, \citet{kastner02}, and \citet{stelzer04}; the radius of TW Hya is 1$R_{\odot}$ and its distance 
is 56~pc, see \citet{herczeg04}. Fluxes of TW Hya have not been corrected for (small) photoabsorption and extinction. \label{table:linefluxes}}
\begin{tabular}{rlrrrrrrrrrr}  
\hline
\hline
$\lambda$& Line& $\log T_{\rm max}$& 0.01~Gyr & \multicolumn{2}{c}{\hrulefill\ 0.1~Gyr\ \hrulefill}    & \multicolumn{2}{c}{\hrulefill\ 0.3~Gyr\ \hrulefill}      &     0.65~Gyr          &  1.6~Gyr      &  4.56~Gyr        & 6.7~Gyr \\
(\AA)    &     &                   & TW Hya   &47 Cas       &  EK Dra         & $\pi^1$~UMa  & $\chi^1$~Ori      & $\kappa^1$~Cet        & $\beta$ Com   & Sun              & $\beta$~Hyi \\
\hline
1.85	 &Fe\,{\sc xxv}  & 7.84 & --\phb 	 &   2.57$\pm$0.63 &  1.29$\pm$0.45  &     --\pha   &	--\pha  	 &     --\pha	       &    --\pha	&    --       &  --\\
4.72	 &S\,{\sc xvi}   & 7.41 & --\phb  	 &   1.87$\pm$0.29 &	 --\ph       &   --\pha     &	--\pha  	 &     --\pha	       &    --\pha	&    --       &  --\\
10.62	 &Fe\,{\sc xxiv} & 7.27 & $< 1.4$\phb	 &   2.52$\pm$2.10 &  1.70$\pm$1.06  &   --\pha     &	--\pha  	 &     --\pha	       &    --\pha	&    --       &  --\\
6.18	 &Si\,{\sc xiv}  & 7.21 & 1.2$\pm 0.8$   &   4.62$\pm$0.87 &  2.00$\pm$0.38  &   --\pha     &	--\pha  	 &     --\pha	       &    --\pha	&    --       &  --\\
5.04	 &S\,{\sc xv}	 & 7.18 & --\ph 	 &   3.72$\pm$0.67 &  1.08$\pm$0.20  &   --\pha     &	--\pha  	 &     --\pha	       &    --\pha	&    --       &  --\\
11.74	 &Fe\,{\sc xxiii}& 7.18 & $< 1.7$\phb	 &   5.50$\pm$0.72 &  3.48$\pm$0.45  &   --\pha     &	--\pha  	 &     --\pha	       &    --\pha	&    --       &  --\\
12.29	 &Fe\,{\sc xxi}  & 7.04 & --\phb 	 &   6.72$\pm$1.43 &  4.20$\pm$0.94  & 0.39$\pm$0.12 & 0.21$\pm$0.07	 & 0.33$\pm$0.05       &    --\pha	&    --       &  --\\
6.65	 &Si\,{\sc xiii} & 7.01 & 2.3$\pm 1.1$   &   7.33$\pm$2.65 &  3.32$\pm$0.44  & 0.21$\pm$0.04 & 0.14$\pm$0.02	 & 0.19$\pm$0.03       & 0.02$\pm$0.00  &      --     &  --\\
8.42	 &Mg\,{\sc xii}  & 7.01 & 0.9$\pm 0.6$   &   8.70$\pm$0.48 &  3.46$\pm$0.51  & 0.11$\pm$0.03 & 0.07$\pm$0.02	 & 0.19$\pm$0.03       & 0.01$\pm$0.01  &      --     &  --\\
12.83	 &Fe\,{\sc xx}   & 6.98 & $< 5.1$\phb	 &   8.73$\pm$0.03 &  4.59$\pm$0.96  & 0.27$\pm$0.06 & 0.14$\pm$0.06	 & 0.24$\pm$0.05       &     --\pha	&      --     &  --\\
13.52	 &Fe\,{\sc xix}  & 6.92 & --\phb  	 &   8.32$\pm$1.41 &  3.64$\pm$1.08  & 0.46$\pm$0.07 & 0.18$\pm$0.09	 & 0.22$\pm$0.07       & 0.02$\pm$0.02  &      --     &  --\\
14.20	 &Fe\,{\sc xviii}& 6.84 & 1.6$\pm 1.3$   &  10.52$\pm$0.67 &  6.57$\pm$0.50  & 0.65$\pm$0.09 & 0.54$\pm$0.04	 & 0.57$\pm$0.04       & 0.04$\pm$0.02  &      --     &  --\\
9.17	 &Mg\,{\sc xi}   & 6.81 & --\phb  	 &  11.30$\pm$0.40 &  5.32$\pm$0.43  & 0.41$\pm$0.04 & 0.30$\pm$0.02	 & 0.45$\pm$0.03       & 0.04$\pm$0.01  &      --     &  --\\
12.13	 &Ne\,{\sc x}	 & 6.76 & 16.1$\pm 1.7$  &  17.69$\pm$1.25 &  5.30$\pm$0.69  &     --\pha    &      --\pha	 &     --\pha	       &     --\pha	&      --     &  --\\
15.01	 &Fe\,{\sc xvii} & 6.72 & $< 31.6$\phb	 &  16.89$\pm$0.69 & 12.02$\pm$0.52  & 1.96$\pm$0.09 & 1.62$\pm$0.05	 & 1.50$\pm$0.06       & 0.21$\pm$0.02  &      --     &  --\\
16.78	 &Fe\,{\sc xvii} & 6.71 &  2.4$\pm 1.6$  &   7.86$\pm$0.66 &  4.78$\pm$0.48  & 0.99$\pm$0.09 & 0.78$\pm$0.06	 & 0.86$\pm$0.05       & 0.12$\pm$0.02  &      --     &  --\\
13.45	 &Ne\,{\sc ix}   & 6.59 & 24.1$\pm 2.3$  &  10.08$\pm$1.51 &  2.78$\pm$1.15  & 0.25$\pm$0.13 & 0.32$\pm$0.08	 & 0.36$\pm$0.10       &     --\pha	&      --     &  --\\
13.70	 &Ne\,{\sc ix}   & 6.59 &  6.8$\pm 1.7$  &	  --\ph    &	  --\ph      & 0.16$\pm$0.06 & 0.19$\pm$0.04	 & 0.24$\pm$0.04       &     --\pha	&      --     &  --\\
18.97	 &O\,{\sc viii}  & 6.48 & 42.4$\pm 3.9$  &  23.03$\pm$0.63 &  9.25$\pm$0.36  & 1.05$\pm$0.05 & 0.89$\pm$0.03	 & 1.02$\pm$0.29       & 0.15$\pm$0.01  &  0.029      &  --\\
335	 &Fe\,{\sc xvi}  & 6.35 & --\phb  	 &	  --\ph    & 36.6\ph	     & \multicolumn{2}{c}{9.7\phantom{0}}& 2.6\ph	       &  --\pha	&	--    &  --   \\
361	 &Fe\,{\sc xvi}  & 6.35 & --\phb  	 &	  --\ph    & 15.7\ph	     & \multicolumn{2}{c}{6.6\phantom{0}}& 1.6\ph	       &  --\pha	& 0.016       &  --   \\
21.60	 &O\,{\sc vii}   & 6.33 & 21.3$\pm 2.6$  &   3.53$\pm$0.74 &  1.44$\pm$0.42  & 0.38$\pm$0.05 & 0.28$\pm$0.04	 & 0.37$\pm$0.05       & 0.07$\pm$0.02  & 0.063       &  --\\
22.10	 &O\,{\sc vii}   & 6.32 &  0.5$\pm 0.7$  &   1.55$\pm$0.37 &  1.25$\pm$0.37  & 0.27$\pm$0.05 & 0.20$\pm$0.04	 & 0.22$\pm$0.04       & 0.08$\pm$0.02  & 0.041       &  --\\
24.77	 &N\,{\sc vii}   & 6.32 & 12.6$\pm 2.4$  &   2.45$\pm$0.33 &  1.03$\pm$0.17  & 0.07$\pm$0.03 & 0.05$\pm$0.01	 & 0.09$\pm$0.02       & 0.01$\pm$0.01  & 0.009       &  --\\
284	 &Fe\,{\sc xv}   & 6.30 & --\phb  	 &	--\ph	   & 22.0\ph	     & \multicolumn{2}{c}{5.0\phantom{0}}& 2.4\ph	       &  --\pha	& 0.025       &  --   \\
284	 &Fe\,{\sc xv}   & 6.30 & --\phb  	 &	--\ph	   & 22.0\ph	     & \multicolumn{2}{c}{5.0\phantom{0}}& 2.4\ph	       &  --\pha	& 0.025       &  --   \\
33.73	 &C\,{\sc vi}	 & 6.13 &  6.1$\pm 1.1$  &   2.01$\pm$0.27 &  0.96$\pm$0.16  & 0.08$\pm$0.03 & 0.09$\pm$0.02	 & 0.12$\pm$0.02       & 0.03$\pm$0.01  & 0.016       &  --\\
610\&625 &Mg\,{\sc x}	 & 6.08& --\phb  	 &	  --\ph     & --\ph	 & \multicolumn{2}{c}{ --	   } &  --\pha  	   &  --\pha	   & 0.028	      &  --   \\
1032	 &O\,{\sc vi}	 & 5.42&  41.3\phb 	 &	  --\ph     & 3.1\ph	 & \multicolumn{2}{c}{0.75	   } & 0.43\pha 	   & 0.16\pha	   & 0.050	      & 0.048 \\
1038	 &O\,{\sc vi}	 & 5.42&  20.0\phb 	 &	  --\ph     & 1.5\ph	 & \multicolumn{2}{c}{0.38	   } & 0.21\pha 	   & 0.074\phb     & 0.025	      & 0.022 \\
630	 &O\,{\sc v}	 & 5.26& --\phb  	 &	  --\ph     & --\ph	 & \multicolumn{2}{c}{ --	   } &  --\pha  	   &  --\pha	   & 0.037	      &  --   \\
789	 &O\,{\sc iv}	 & 5.05& --\phb  	 &	  --\ph     & --\ph	 & \multicolumn{2}{c}{ --	   } &  --\pha  	   &  --\pha	   & 0.017	      &  --   \\
1550	 &C\,{\sc iv}	 & 5.00& 378.0\phb 	 &	  --\ph     & 9.1\ph	 & \multicolumn{2}{c}{2.21	   } & 1.02\pha 	   & 0.40\pha	   & 0.146	      & 0.082 \\
834	 &O\,{\sc ii}	 & 4.80& --\phb  	 &	  --\ph     & --\ph	 & \multicolumn{2}{c}{ --	   } &  --\pha  	   &  --\pha	   & 0.015	      &  --   \\
304	 &He\,{\sc ii}   & 4.75& --\phb  	 &	  --\ph     & 44.3\ph	 & \multicolumn{2}{c}{8.3\phantom{0}} & 2.3\ph  	    &  --\pha	   & 0.260	      &  --   \\
1640	 &He\,{\sc ii}   & 4.75& 180.0\phb 	 &	  --\ph     & 6.0\ph	 & \multicolumn{2}{c}{0.99	   } & 0.56\pha 	   &  --\pha	   & 0.040	      &  --   \\
1400	 &Si\,{\sc iv}   & 4.75&  14.6\phb 	 &	  --\ph     & 4.3\ph	 & \multicolumn{2}{c}{1.59	   } & 0.77\pha 	   & 0.28\pha	   & 0.083	      & 0.097 \\
977	 &C\,{\sc iii}   & 4.68&  58.6\phb 	 &	  --\ph     & 5.0\ph	 & \multicolumn{2}{c}{1.22	   } & 0.59\pha 	   & 0.30\pha	   & 0.150	      & 0.124 \\
1176	 &C\,{\sc iii}   & 4.68&  55.9\phb 	 &	  --\ph     & 3.4\ph	 & \multicolumn{2}{c}{0.73	   } & 0.37\pha 	   & 0.15\pha	   & 0.053	      & 0.046 \\
1206	 &Si\,{\sc iii}  & 4.40&  --\phb 	 &	  --\ph     & --\ph	 & \multicolumn{2}{c}{1.12	   } & 0.75\pha 	   &  --\pha	   & 0.095	      &  --   \\
584	 &He\,{\sc i}	 & 4.25&  --\phb 	 &	  --\ph     & --\ph	 & \multicolumn{2}{c}{ --	   } &  --\pha  	   &  --\pha	   & 0.032	      &  --   \\
1335	 &C\,{\sc ii}	 & 4.25&  45.2\phb 	 &	  --\ph     & 4.7\ph	 & \multicolumn{2}{c}{1.52	   } & 0.95\pha 	   & 0.36\pha	   & 0.188	      & 0.155 \\
1304	 &O\,{\sc i}	 & 3.85& 126.4\phb 	 &	  --\ph     & 4.3\ph	 & \multicolumn{2}{c}{1.18	   } & 0.60\pha 	   & 0.45\pha	   & 0.143	      & 0.163 \\
1657	 &C\,{\sc i}	 & 3.85&  29.2\phb 	 &	  --\ph     & 4.1\ph	 & \multicolumn{2}{c}{0.97	   } & 0.78\pha 	   & 0.47\pha	   & 0.202	      & 0.210 \\
1026	 &H\,{\sc i}	 & 3.84& --\phb  	 &	  --\ph     & --\ph	 & \multicolumn{2}{c}{3.1\phantom{5}} & 0.80\pha	   &  --\pha	   & 0.098	      &  --   \\
1216	 &H\,{\sc i}	 & 3.84&$\approx$16,000\phb  &	  --\ph     & --\ph	 & \multicolumn{2}{c}{\llap{4}2.2\phantom{0}} &\llap{2}9.3\ph &  --\pha & 6.19\phantom{0}  &  --   \\
\hline
\end{tabular}
\end{minipage}
}}
}
\end{table}

\begin{table}[htbp]
\begin{center}
\caption{\it Parameters of the power-law fits to the measured
  integrated fluxes and individual line fluxes from MS solar analogs
  $^a$ (data from \citealt{telleschi05} and \citealt{ribas05}).}
\label{table:decay}
\vskip 4mm

\begin{tabular}{llrrr}
\hline\noalign{\smallskip}
\multicolumn{3}{l}{$\lambda$ (interval) (\AA)}                 & $\alpha$    &  $\beta$          \\
\noalign{\smallskip}\hline\noalign{\smallskip}
\multicolumn{3}{l}{$\left[1-20\right]  		    $}         & 2.4\rlap{0} & \phantom{(}$-$1.92\\
\multicolumn{3}{l}{$\left[20-100\right]		    $}         & 4.4\rlap{5} & \phantom{(}$-$1.27\\
\multicolumn{3}{l}{$\left[100-360\right]		    $} & 13.5	     & \phantom{(}$-$1.20\\
\multicolumn{3}{l}{$\left[360-920\right]		    $} & 4.5\rlap{6} &  	 ($-$1.0)\\
\multicolumn{3}{l}{$\left[920-1180\right]		    $} & 2.5\rlap{3} & \phantom{(}$-$0.85\\
\multicolumn{3}{l}{$\left[1-360\right]+\left[920-1180\right]$} & 24.8        & \phantom{(}$-$1.27\\
\multicolumn{3}{l}{$\left[1-1180\right]  		    $} & 29.7	     & \phantom{(}$-$1.23\\
\noalign{\smallskip}\hline\noalign{\smallskip}
$\lambda$  (\AA)& Line         & $T_\mathrm{max}$              &            &  $\beta$           \\
\noalign{\smallskip}\hline\noalign{\smallskip}
12.29     	&Fe {\sc xxi}  & 7.04			      &     	    & \phantom{(}$-$1.77\\
 6.65     	&Si {\sc xiii} & 7.01			      &     	    & \phantom{(}$-$1.91\\
 8.42     	&Mg {\sc xii}  & 7.01			      &     	    & \phantom{(}$-$2.14\\
12.83     	&Fe {\sc xx}   & 6.98			      &     	    & \phantom{(}$-$2.09\\
13.52     	&Fe {\sc xix}  & 6.92			      &     	    & \phantom{(}$-$1.94\\
14.20     	&Fe {\sc xviii}& 6.84			      &     	    & \phantom{(}$-$1.80\\
 9.17     	&Mg {\sc xi}   & 6.81			      &     	    & \phantom{(}$-$1.80\\
15.01     	&Fe {\sc xvii} & 6.72			      &     	    & \phantom{(}$-$1.44\\
16.78     	&Fe {\sc xvii} & 6.71			      &     	    & \phantom{(}$-$1.33\\
13.45     	&Ne {\sc ix}   & 6.59			      &     	    & \phantom{(}$-$1.73\\
18.97     	&O {\sc viii}  & 6.48			      &     	    & \phantom{(}$-$1.58\\
361       	&Fe {\sc xvi}  & 6.35			      &     	    & \phantom{(}$-$1.86\\
21.60     	&O {\sc vii}   & 6.33			      &     	    & \phantom{(}$-$1.18\\
22.10     	&O {\sc vii}   & 6.32			      &     	    & \phantom{(}$-$1.01\\
24.77     	&N {\sc vii}   & 6.32			      &     	    & \phantom{(}$-$1.74\\
284       	&Fe {\sc xv}   & 6.30			      &     	    & \phantom{(}$-$1.79\\
33.73     	&C {\sc vi}    & 6.13			      &     	    & \phantom{(}$-$1.35\\
1032      	&O {\sc vi}    & 5.42			      &     	    & \phantom{(}$-$1.00\\
1038      	&O {\sc vi}    & 5.42			      &     	    & \phantom{(}$-$1.02\\
1550      	&C {\sc iv}    & 5.00			      &     	    & \phantom{(}$-$1.08\\
304       	&He {\sc ii}   & 4.75			      &     	    & \phantom{(}$-$1.34\\
1640      	&He {\sc ii}   & 4.75			      &     	    & \phantom{(}$-$1.28\\
1400      	&Si {\sc iv}   & 4.75			      &     	    & \phantom{(}$-$0.97\\
977       	&C {\sc iii}   & 4.68			      &     	    & \phantom{(}$-$0.85\\
1176      	&C {\sc iii}   & 4.68			      &     	    & \phantom{(}$-$1.02\\
1206      	&Si {\sc iii}  & 4.40			      &     	    & \phantom{(}$-$0.94\\
1335      	&C {\sc ii}    & 4.25			      &     	    & \phantom{(}$-$0.78\\
1304      	&O {\sc i}     & 3.85			      &     	    & \phantom{(}$-$0.78\\
1657      	&C {\sc i}     & 3.85			      &     	    & \phantom{(}$-$0.68\\
1026      	&H {\sc i}     & 3.84			      &     	    & \phantom{(}$-$1.24\\
1216      	&H {\sc i}     & 3.84			      &     	    & \phantom{(}$-$0.72\\
\noalign{\smallskip}\hline\noalign{\smallskip}
\multicolumn{5}{l}{$^{a}$ {Relationship of the form: $\mbox{Flux}=\alpha t_9^{\beta}$.}} \\
\end{tabular}
\end{center}
\end{table}

\epubtkImage{}{%
\begin{figure}[htbp]
  \centerline{
      \includegraphics[scale=0.37]{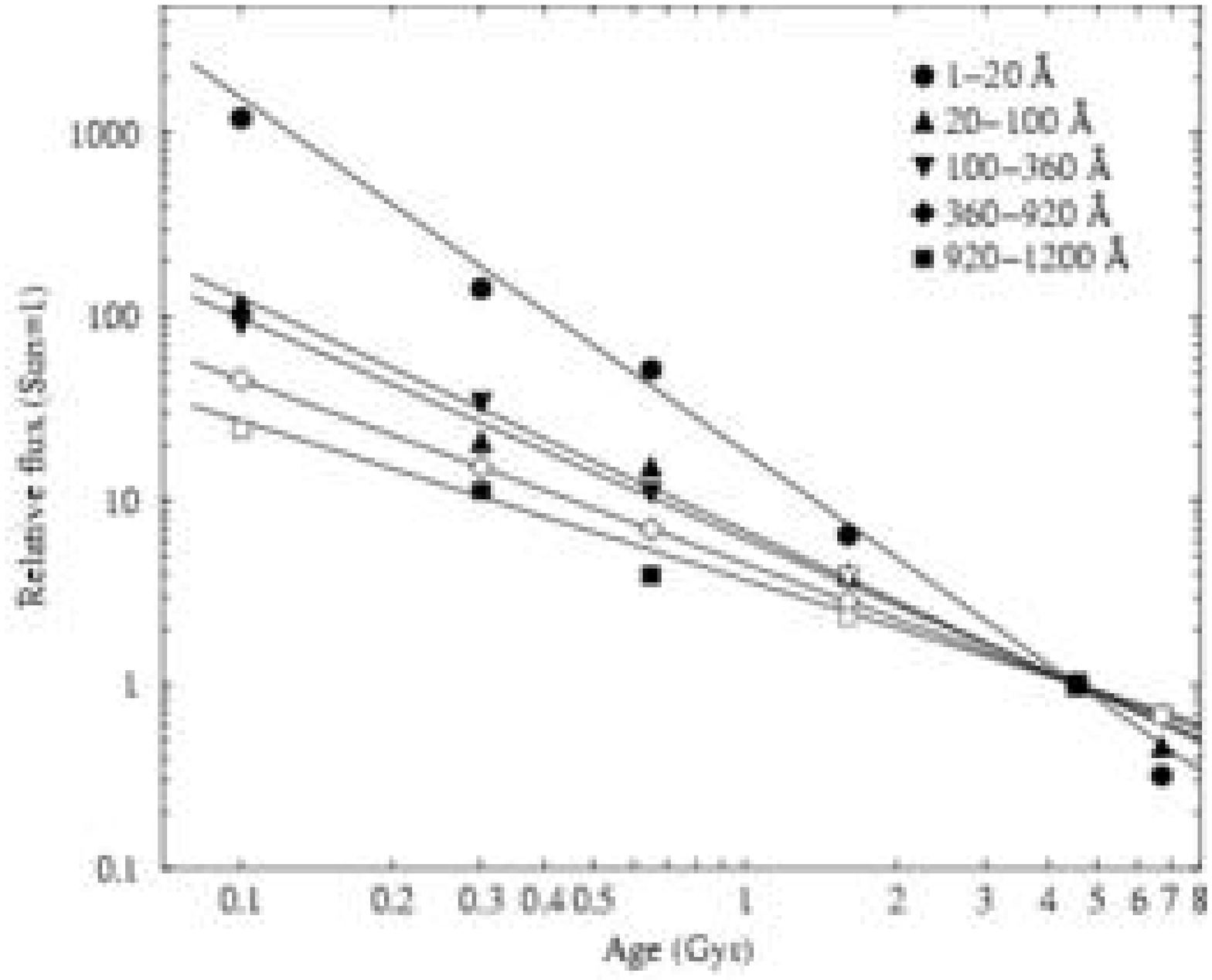}
      \includegraphics[scale=0.40]{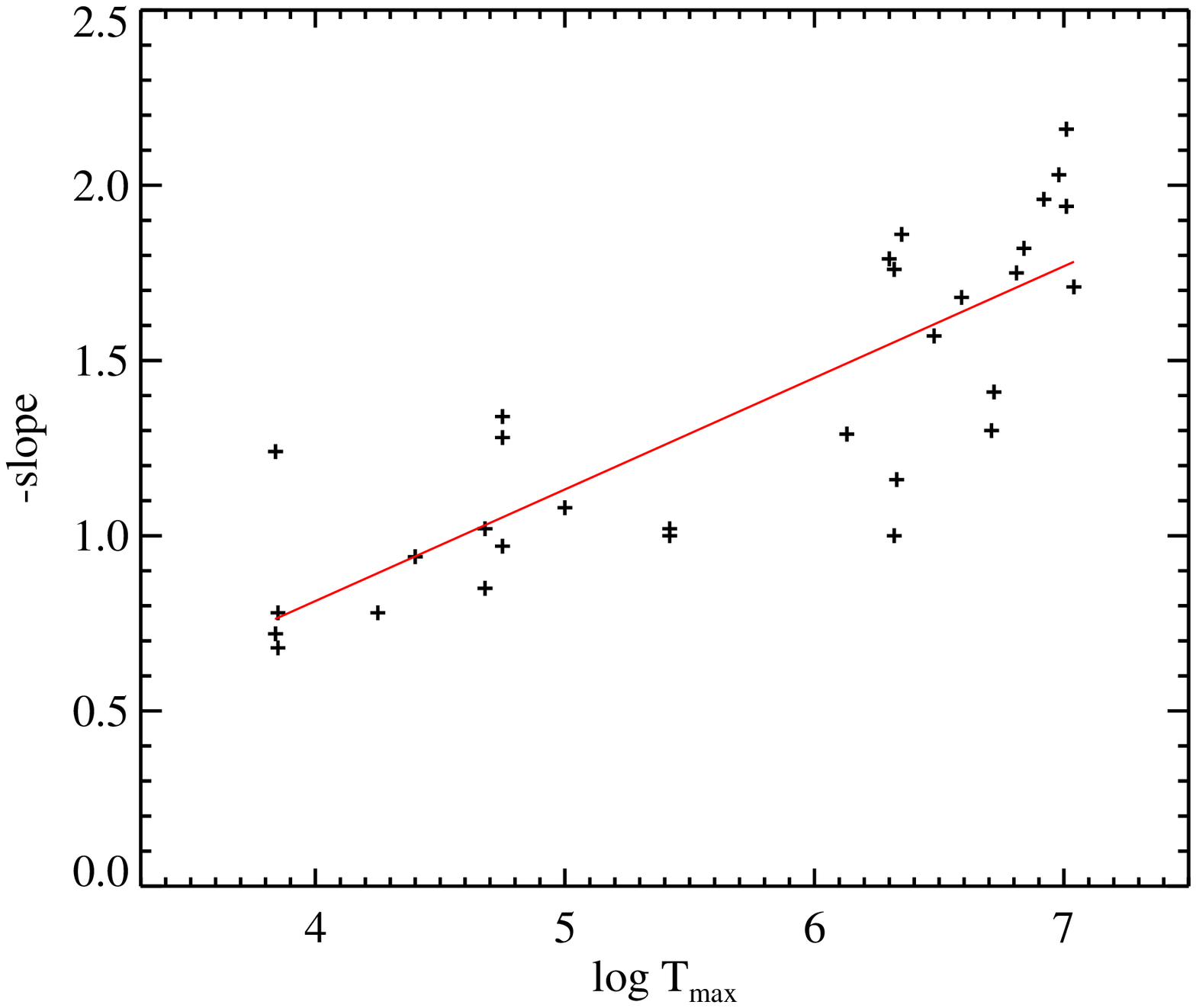}
  }
  \caption{\it Left (a): Power-law decays in time for various spectral
  ranges, normalized to the present-day solar flux. Note that the
  hardest emission decays fastest \citep[from][reproduced by permission of AAS]{ribas05}. -- Right
  (b): Flux decay slope (given as its absolute value) as a function of
  formation temperature of the respective line; each cross marks one
  emission line reported in Table~\ref{table:decay}.}
  \label{figure:integraldecay}
\end{figure}}

The power laws steepen monotonically with decreasing wavelength, i.e.,
with increasing temperature (or roughly, height) of the atmospheric
layer. The steepest decay is seen in the luminosity of the hot corona,
with $F \propto t_9^{-1.92}$, while for the ultraviolet
transition-region spectrum, $F \propto t_9^{-0.85}$. These power-laws
are compared in Figure~\ref{figure:integraldecay}a after normalization
to the present-day solar fluxes. For individual lines, the steepness
of the decay, $\beta$ in Equation~\ref{decaylaw}, is compared in
Figure~\ref{figure:integraldecay}b for a wide range of line formation
temperatures. A fit to $\beta$ as a function of formation temperature
gives

\begin{equation}
  -\beta = -0.46 + 0.32\log T_\mathrm{max}.
\end{equation} 

Although the photospheric bolometric luminosity of the Sun has
steadily increased during its MS life, starting at a level
approximately 30\% lower than today, the magnetically induced
radiation from the outer atmosphere has steeply decayed during the
same time, by factor of $\approx \mbox{1500\,--\,}2000$ for coronal
X-ray emission and a factor of $\approx 25$ for the FUV spectral
region. The total XUV emission decayed by a factor of $\approx
100$. A summary of the enhancement factors at various stages of the
solar past compared to the contemporaneous Sun is given in
Table~\ref{table:enhancement}. Given the ionizing power of some of
this radiation and the importance of UV radiation for line excitation,
this emission must have had profound consequences for the environment
of the young Sun, most notably circumstellar disks and planetary
atmospheres. This will be discussed in Section~\ref{section:environment}
and \ref{section:planets}. As an example, \citep[see][]{ribas05},
for EK~Dra, the luminosity of the C\,{\sc iii}~$\lambda 977$ line
alone exceeds the entire integrated luminosity of the present-day Sun
below 1200~$\lambda$.

\begin{table}[htbp]
\begin{center}
\caption{\it Enhancement factors of X-ray/EUV/XUV/FUV fluxes in solar
  history$^a$}
\label{table:enhancement}
\vskip 4mm

\begin{tabular}{lrrrr}
\hline\noalign{\smallskip}
Solar age  & Time before        & \multicolumn{3}{c}{\hrulefill\ Enhancement in\ \hrulefill} \\
(Gyr)      & present (Gyr)      & X-Rays (1\,--\,20\,\AA) & Soft-X (20\,--\,100\,\AA) &  FUV (920\,--\,1180\,\AA) \\
           &                    &                   & EUV (100\,--\,360\,\AA)         &  \\
           &                    &                   & XUV (1\,--\,1180\,\AA)          &  \\ 
\noalign{\smallskip}\hline\noalign{\smallskip}
0.1 	   & 4.5\phc		& 1600$^b$	  \phc    & 100\phantom{.5}\phc       &  25\phantom{.5}\phc \\
0.2 	   & 4.4\phc		&  400\phantom{$^b$}\phc  &  50\phantom{.5}\phc       &  14\phantom{.5}\phc \\
0.7 	   & 3.9\phc		&   40\phantom{$^b$}\phc  &  10\phantom{.5}\phc       &   5\phantom{.5}\phc \\
1.1 	   & 3.5\phc		&   15\phantom{$^b$}\phc  &   6\phantom{.5}\phc       &   3\phantom{.5}\phc \\
1.9 	   & 2.7\phc		&    5\phantom{$^b$}\phc  &   3\phantom{.5}\phc       &   2\phantom{.5}\phc \\
2.6 	   & 2.0\phc		&    3\phantom{$^b$}\phc  &   2\phantom{.5}\phc       &   1.6\phc \\
3.2 	   & 1.4\phc		&    2\phantom{$^b$}\phc  & 1.5 \phc                  &   1.4\phc \\
4.6 	   & 0\phantom{.5}\phc  &    1\phantom{$^b$}\phc  &   1\phantom{.5}\phc       &   1\phantom{.5}\phc \\
\noalign{\smallskip}\hline			     
\end{tabular}
\begin{list}{}{}
\item[$^{\mathrm{a}}$]{normalized to ZAMS age of 4.6\,Gyr before present}
\item[$^{\mathrm{b}}$]{large scatter possible due to unknown initial rotation period of Sun}
\end{list}
\end{center}
\end{table}

This ``softening'' of the spectral irradiance in time is illustrated
in Figure~\ref{figure:specirrad} where the entire spectral irradiance
from 1\,--\,3500\,\AA\ (except the strongly absorbed EUV range) is
shown for various solar analogs. While the UV flux varies moderately
and mostly due to emission lines (Figure~\ref{figure:uvirrad}), the EUV
and X-ray level drops dramatically along the MS evolution.

An alternative illustration is provided by the luminosity-luminosity
diagram in Figure~\ref{figure:fluxflux} in which the relation between
normalized coronal X-ray and transition-region UV emission is shown
for solar-like stars. The two emissions follow a power-law with a
slope of 1.9, indicating that toward higher activity levels,  X-rays
increase more rapidly than UV fluxes \citep{ayres97}.

\epubtkImage{}{%
\begin{figure}[htbp]
  \centerline{\includegraphics[scale=0.57]{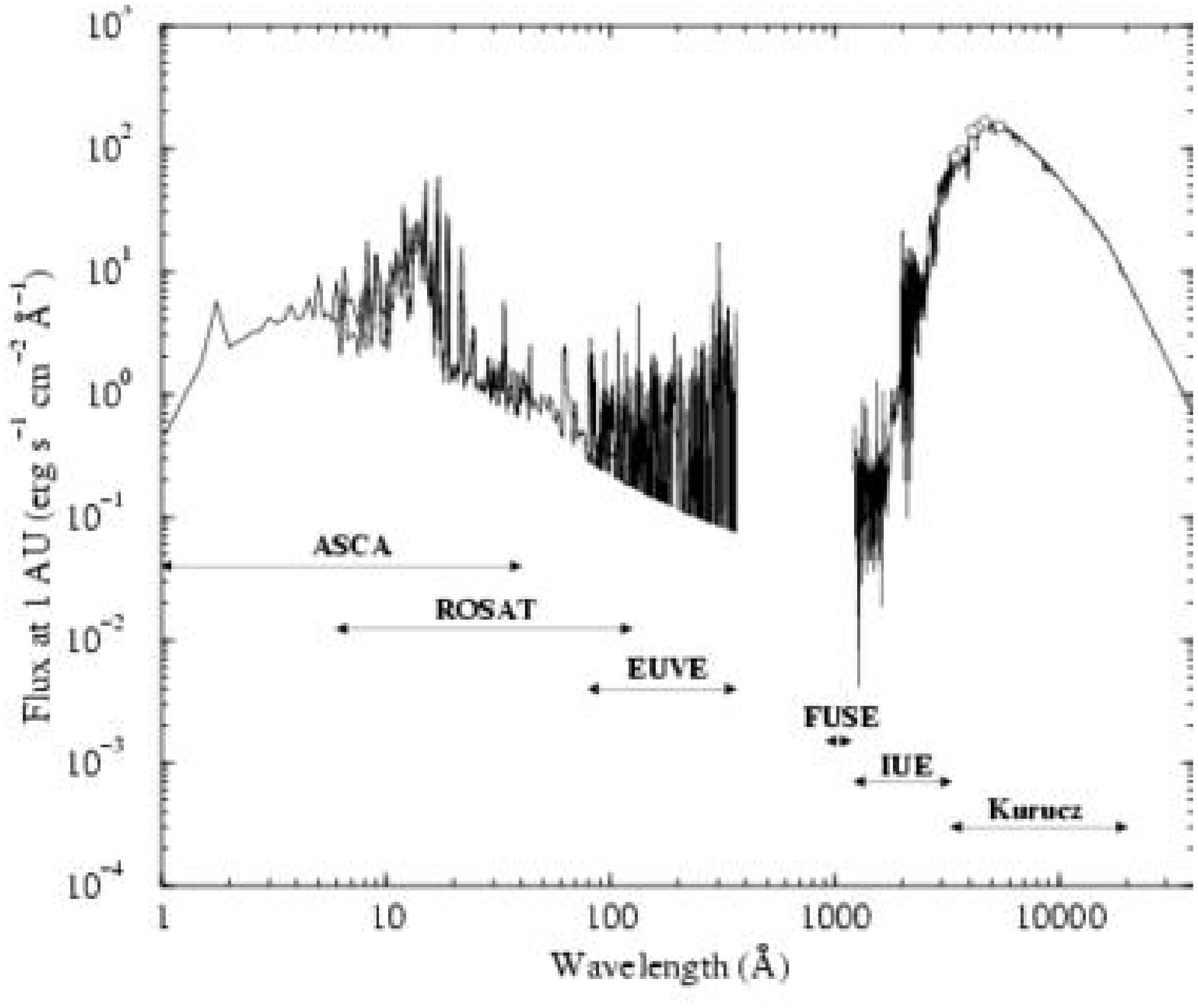}}
  \centerline{\includegraphics[scale=0.57]{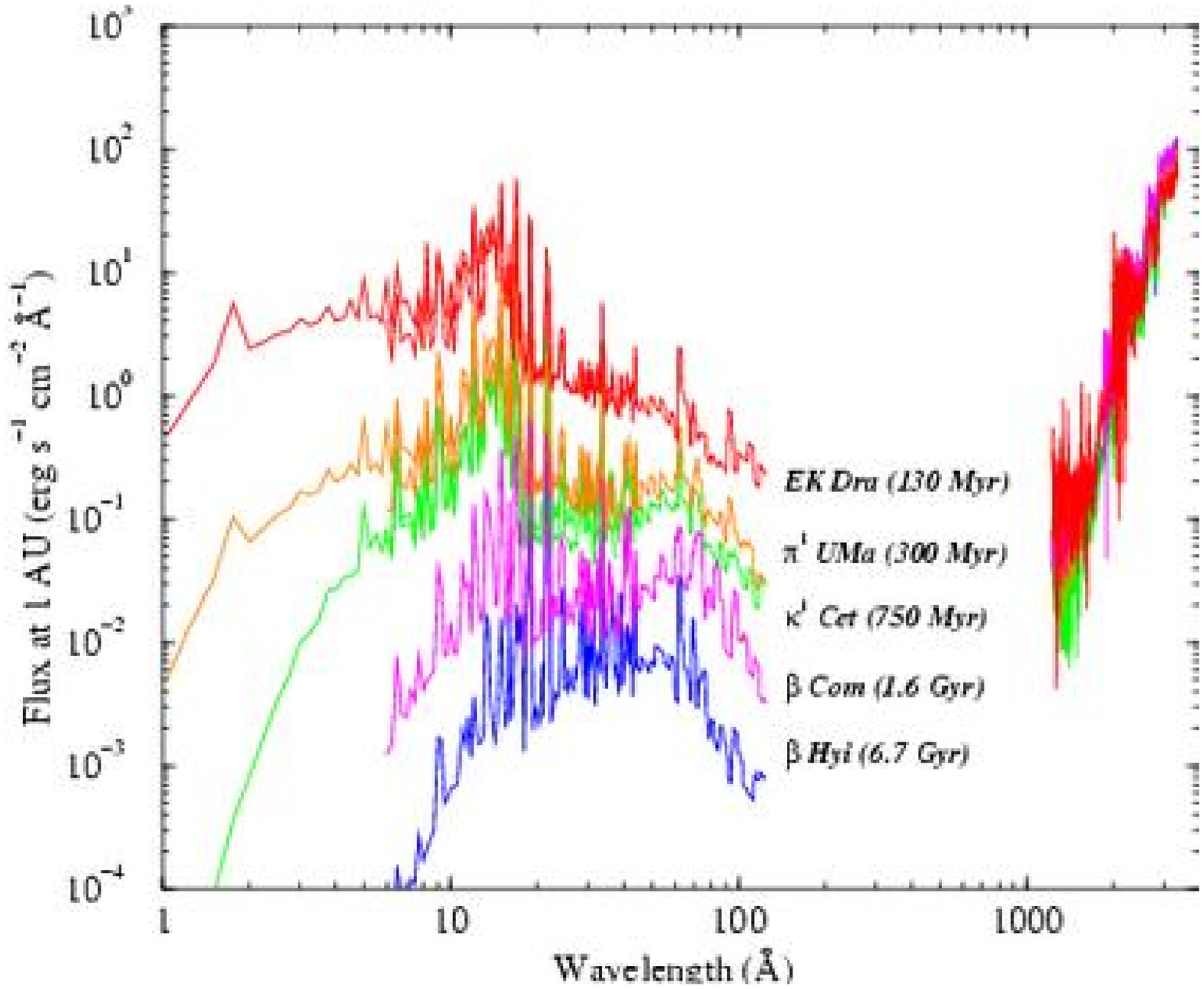}}
  \caption{\it Top: Spectral irradiance of EK~Dra for a distance of
  1\,AU. The various instruments used for the reconstruction are
  marked. Bottom: Irradiances at 1\,AU from solar analogs with
  different ages \citep[from][reprinted with permission of ASP]{guinan02a}.}
  \label{figure:specirrad}
\end{figure}}

\epubtkImage{}{%
\begin{figure}[htbp]
  \centerline{\includegraphics[scale=0.5]{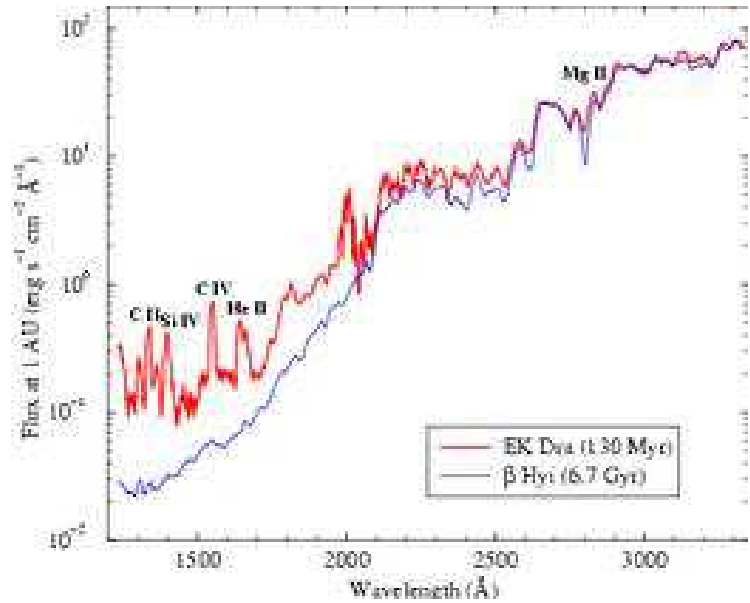}}
  \caption{\it UV irradiances of EK~Dra (upper spectrum) and the very
  old solar analog, $\beta$~Hyi (lower spectrum). Note progressively
  higher flux levels in EK~Dra toward shorter wavelengths
  \citep[from][reprinted with permission of ASP]{guinan02a}.}
  \label{figure:uvirrad}
\end{figure}}

\epubtkImage{}{%
\begin{figure}[htbp]
  \centerline{\includegraphics[scale=0.35]{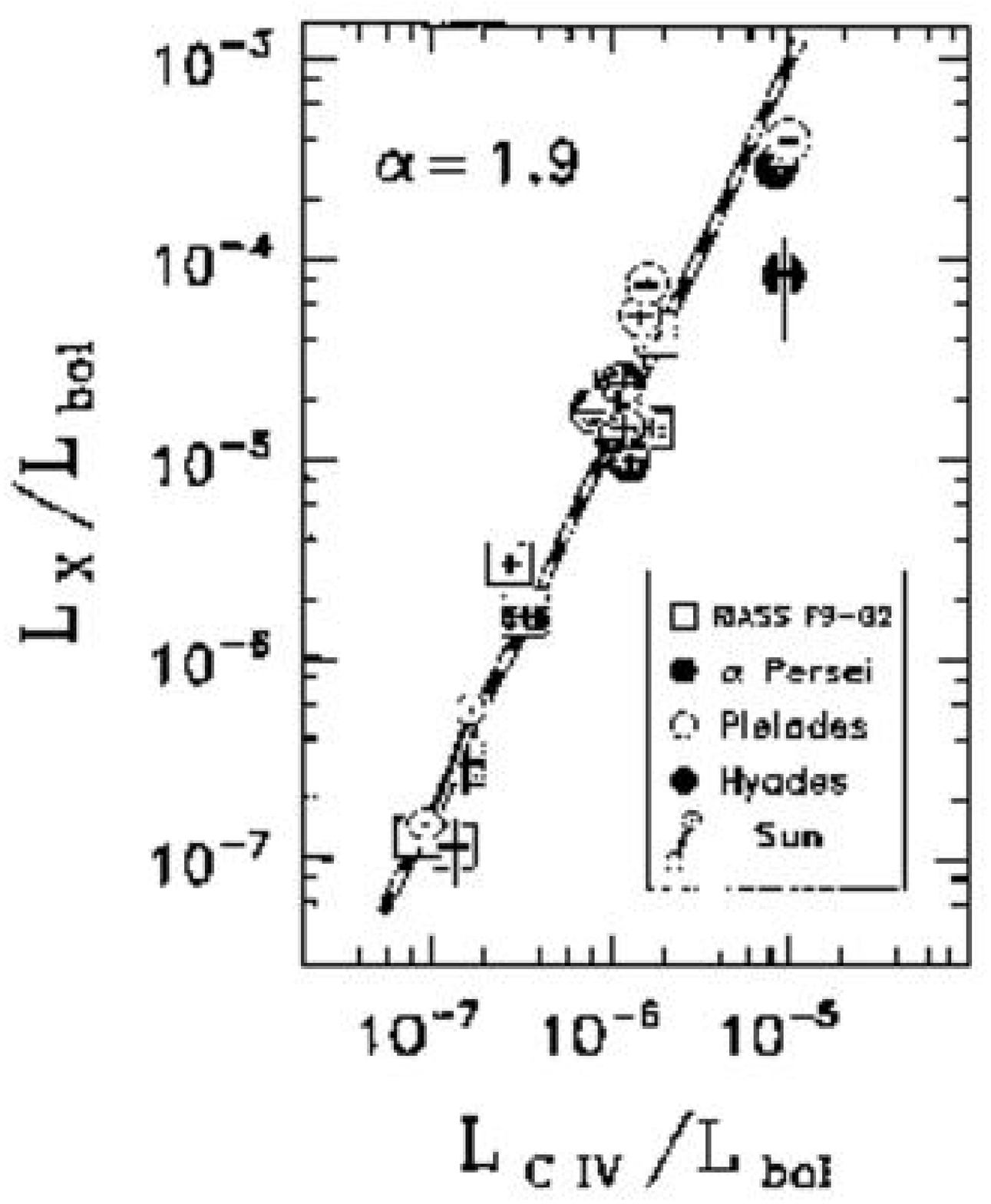}}
  \caption{\it Correlation between normalized coronal X-ray and and
  transition-region ({\rm C\,{\sc iv}}) UV emission for solar-like field and
  cluster stars (from the {\em ROSAT/IUE All Sky Survey} [RIASS], see
  \citealt{ayres97}, reprinted with permission).}
  \label{figure:fluxflux}
\end{figure}}

\subsection{The Radio Sun in Time}
\label{section:coronal_radio}

Magnetically active stars are sources of vigorous radio emission,
which is mostly due to gyroemission from accelerated electrons that
are trapped in coronal magnetic fields. Radio emission thus provides
diagnostics for coronal magnetic fields and at the same time for
high-energy electron populations.

\subsubsection{Overview}

Solar radio emission is a mixture of various kinds of radiation, some
of which are thermal while others are of non-thermal origin. The total
solar radio emission is strongly variable and is dominated by
different radiation types at different times.

{\it Thermal} radiation dominates as long as no strong flares are
occurring. Three principal types of thermal radiation are observed
\citep[e.g.,][]{gary94}:

\begin{description}

  \item [1.] Thermal bremsstrahlung from chromospheric layers is seen
  across the solar disk. The emission is dominated by contributions
  from the optically thick layer. The brightness temperature is of
  order $10^4 \unit{K}$, depending somewhat on wavelength.

  \item [2.] Above active regions with strong magnetic fields,
  gyroresonance (cyclotron) emission at low harmonics of the
  gyrofrequency is often observed. The dominant contributions come
  from the highest harmonic that is optically thick. This radiation is
  patchy and covers only a small fraction of the solar
  disk. Gyroresonance emission may also be accompanied by
  bremsstrahlung that originates from optically thick layers with
  coronal temperatures.

  \item [3.] Some faint, optically thin coronal bremsstrahlung can
  also be seen from coronal loops that are filled with hot plasma.

\end{description}

The third contribution is usually rather weak; the non-flaring solar
radio emission is dominated by the first and second
contributions. Although variable, the average radio luminosity at
3.6\,cm wavelength from these contributions is $\log L_\mathrm{R}
\approx 10.80\mbox{\,--\,}11.15 \unit{[erg\ s^{-1}\ Hz^{-1}]}$
\citep{drake93}. The picture changes entirely during solar flares,
when two new contributions dominate the solar radio output:

\begin{description}

  \item [4.] Various forms of non-thermal emission from coherent
  radiation processes such as electron cyclotron masers or plasma
  radiation mechanisms are signatures of unstable, non-thermal
  electron populations. The emission is often (but not exclusively)
  narrow-band and occurs in short bursts (from milliseconds to a few
  seconds). More persistent coherent emission is occasionally seen.

  \item [5.] Gyrosynchrotron radiation originates from non-thermal
  electron populations that are trapped in magnetic fields. This
  emission shows a broad spectrum with optically thick and thin
  components, with a turnover from the former to the latter in the
  1\,--\,10\,GHz range. Gyrosynchrotron emission is a direct tracer of
  accelerated particle populations.

\end{description}

\subsubsection{Observational Results}

The discovery of solar analogs in radio waves \citep{guedel94} was
motivated by a correlation between radio and X-ray luminosities that
holds for M-type MS stars, but also extends to RS CVn-type binaries
and seems to hold similarly for solar flares \citep{guedel93,
  benz94}. The following systematics apply:

\begin{itemize}

  \item The radio emission of detected solar analogs follows the same
  radio vs. X-ray luminosity relation as found for active M dwarfs.

  \item Only stars with ``hard'' X-ray spectra, i.e., X-ray emission
  from very hot coronal plasma (10\,MK and more), are non-thermal
  radio sources.

  \item The detected solar analogs that are not components of spun-up
  binaries are therefore young ($\la$ a few 100\,Myr), magnetically
  active stars.

\end{itemize}

Figure~\ref{figure:radiodecay} shows a summary of normalized radio
luminosities of solar analogs with different rotation periods. Like
X-ray emission, the radio output decays with increasing rotation
period (i.e., decreasing magnetic activity), although the decay is
much steeper. The radio decay is reminiscent of the very steep decay
of the harder portion of soft X-ray emission which is also an
excellent tracer of magnetic activity. The steep decline of
non-thermal radio emission therefore extends the trend observed above
(Section~\ref{section:spectralevolution}): Radio emission originates
from the most energetic electrons in the stellar atmosphere.

Figure~\ref{figure:tempradio} shows the radio luminosity as a function
of the average coronal temperature of solar analogs; only radio upper
limits are available for the intermediately active coronae with
temperatures $T_\mathrm{av} \approx 3\mbox{\,--\,}5 \unit{MK}$; a
regression fit to the points including the upper limits therefore
provides a lower limit to the slope:

\begin{equation}
  L_\mathrm{R}  \approx 1.69\times 10^9 T_\mathrm{av}^{5.29 \pm 0.74}
  \quad \mathrm{[erg\ s^{-1}\ Hz^{-1}]}
\label{tempradio}
\end{equation}

(based on the APEC emission line code; \citealt{telleschi05}). This
empirical relation shows again, and explicitly, that strong
non-thermal radio emission develops only in very active stars with hot
coronae. Because non-thermal electrons need to be replenished on
short time scales, presumably in flare-like processes, a model in
which the very hot plasma component is formed by the same flares is
suggestive. Alternatively, non-thermal electron distributions may
  be accelerated continuously out of a Maxwellian distribution into a
  runaway tail by electric fields \citep{holman86}, a concept also
  proposed to operate in solar flares \citep{holman92}. Such
  mechanisms of course work best if very hot plasmas are present.

\epubtkImage{}{%
\begin{figure}[htbp]
  \centerline{\includegraphics[scale=0.5]{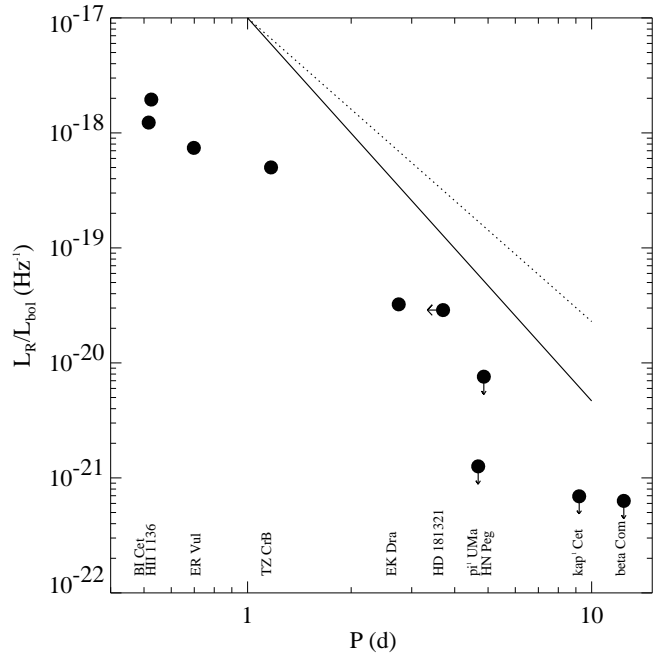}}
  \caption{\it Relation between $L_\mathrm{R}/L_\mathrm{bol}$ and the
  stellar rotation period. Note upper limits for slow rotators. The
  dotted line illustrates the slope of the X-ray decay law as a
  function of P (Equation~\ref{activityrotationX}), while the solid line
  shows the slope of the most rapidly decaying, ``hot'' X-ray lines
  (slopes from Table~\ref{table:decay} and using
  Equation~\ref{rotationevolution}; adapted from \citealt{guedel01b}).}
  \label{figure:radiodecay}
\end{figure}}

\epubtkImage{}{%
\begin{figure}[htbp]
  \centerline{\includegraphics[scale=0.45]{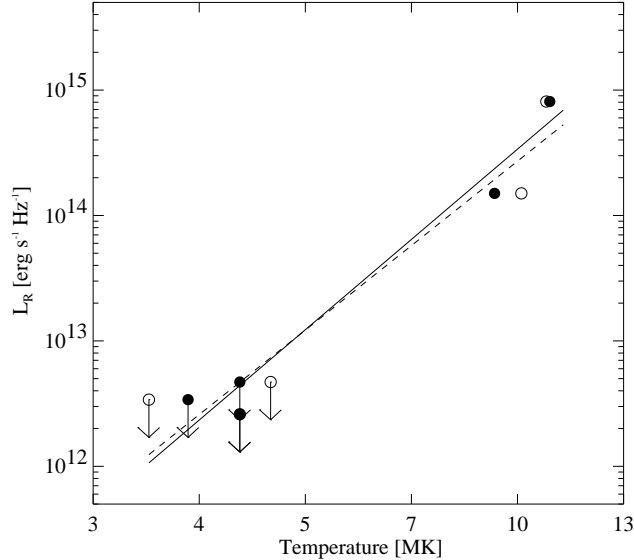}}
  \caption{\it Relation between non-thermal radio emission and average
  coronal temperature \citep[from][reproduced by permission of AAS]{telleschi05}. Open and filled
  circles refer to different atomic emission line codes used for the
  temperature determination.}
  \label{figure:tempradio}
\end{figure}}

Other types of radio emission are too faint to be detected from solar
analogs. Signatures for thermal gyroresonance emission have been found
in an M~dwarf \citep{guedel89}, and \cite{drake93} reported evidence
for non-flaring radio emission probably related to chromospheric
bremsstrahlung and some active-region gyroresonance emission in the
nearby F-type subgiant Procyon; more systematic observations of solar
analogs during all phases of their MS life will require more sensitive
radio telescopes than hitherto available.

\subsection{Coronal Flares in Time}
\label{section:flares}

\subsubsection{Flare Energy Distributions and Coronal Heating}
\label{section:energydistribution}

Magnetic flares in the solar and stellar atmospheres heat plasma
explosively to $\ga 10 \unit{MK}$ on time scales of minutes. Large
amounts of emission measure are produced, presumably by bringing up
heated material from the chromospheric layers, thus increasing the
density in coronal regions of magnetic loops. It has been suggested
that much, if not all, of the hot coronal plasma has been heated (and
evaporated) by flares that may be too small and too frequent to be
detected individually. This hypothesis is known as the  ``microflare''
or ``nanoflare'' hypothesis in solar physics \citep{parker88}. There
is observational evidence that numerous small-scale flare events
occur in the solar corona at any time
\citep[e.g.,][]{lin84}. Hard-X-ray studies have shown that they are
distributed in energy according to a power law,

\begin{equation}
  \frac{dN}{dE} = k E^{-\alpha}
\label{powerlaw}
\end{equation}

where $dN$ is the number of flares per unit time with a total energy
in the interval [$E,E+dE$], and $k$ is a constant. If $\alpha\ge 2$,
then the energy integration (for a given time interval,
$\int_{E_\mathrm{min}}^{E_\mathrm{max}} E [dN/dE]dE$) diverges for
$E_\mathrm{min} \rightarrow 0$, that is, if the power law is
extrapolated to small flare energies, a lower cut-off is required for
the power-law distribution; arbitrary energy release powers are
possible depending on the value of $E_\mathrm{min}$. From solar
studies, $\alpha \approx 1.6\mbox{\,--\,}1.8$ for ordinary solar flares
\citep{crosby93}, but some recent studies of low-level flaring suggest
$\alpha = 2.0 - 2.6$ \citep{krucker98, parnell00}.

\subsubsection{Phenomenological Evidence}
\label{section:phenomenology}
 
There is abundant evidence that flare-like processes are important in
magnetically active stars:

\begin{itemize}

  \item A linear correlation between the time-averaged power from
  optical flares and the low-level, ``quiescent'' X-ray luminosity
  suggests that flares are important contributors to the overall
  stellar coronal heating \citep{doyle85, skumanich85, whitehouse85}.

  \item Over the entire solar magnetic cycle, the monthly average soft
  X-ray luminosity scales in detail and linearly with the rate of
  detected H$\alpha$ flares \citep{pearce92}.

  \item In analogy, the stellar ``quiescent'' X-ray luminosity
  correlates approximately linearly with the rate of X-ray flares
  (above a given lower energy threshold; \citealt{audard00}).

  \item The ``quiescent'' X-ray luminosity of magnetically active
  stars correlates linearly with the contemporaneous (non-flaring)
  radio gyrosynchrotron emission. The latter requires accelerated
  electrons, for which flares are an obvious source \citep{guedel93}.

  \item This latter correlation holds similarly for many solar and
  stellar flares \citep{benz94}.

  \item The ratio between energy losses in coronal X-rays and in the
  chromospheric Mg\,{\sc ii} lines is the same in flares and in
  quiescence \citep{haisch90}.

  \item The ratio between energy losses in coronal X-rays and in the
  chromospheric UV emission measured in broad filter bands is the same
  in flares and in quiescence \citep{mitra05}.

  \item A tight correlation exists between $L_\mathrm{X}$ and
  H$\gamma$ luminosity that it the same for flares and for quiescence
  \citep{mathioudakis90}.

  \item Many observations have shown continuous low-level variability
  in the stellar X-ray output, with time scales reminiscent of coronal
  flares (\citealt{audard03b}; see Figure~\ref{figure:lightekdra} for
  the X-ray light curve of EK~Dra). In particular, a strong temporal
  correlation between H$\gamma$ flare flux and simultaneous low-level
  X-ray flux in dMe stars suggests that a large number of flare-like
  events are always present \citep{butler86}. Up to 50\% of the X-ray
  output in PMS stars may be due to relatively strong flares
  \citep{montmerle83}, which has become particularly evident in recent
  observations of the Orion region \citep{feigelson02a, wolk05}.

\end{itemize}

\epubtkImage{}{%
\begin{figure}[htbp]
  \centerline{\includegraphics[scale=0.5]{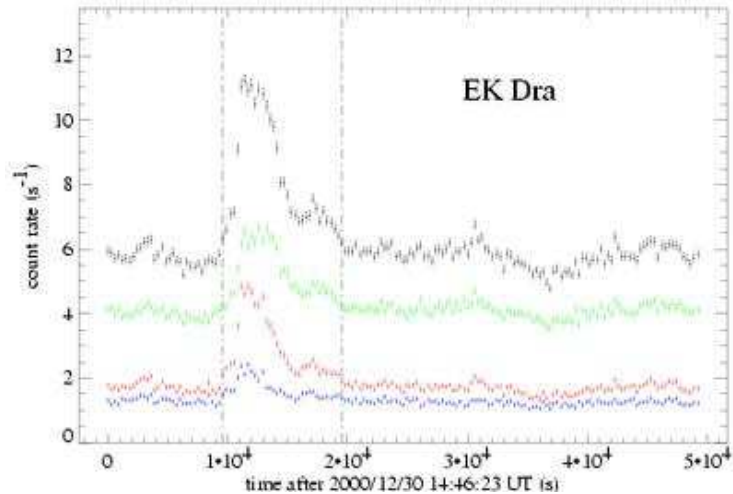}}
  \caption{\it X-ray light curve of the young (ZAMS) solar analog
  EK~Dra, obtained with {\em XMM-Newton}. The curves refer to the
  total 0.2\,--\,10\,keV photon energy range (black), the soft band
  (0.2\,--\,1\,keV; green), the hard band ($>$1\,keV; red), and the
  ratio of hard/soft (blue; from \citealt{telleschi05}, reproduced by permission of AAS).}
  \label{figure:lightekdra}
\end{figure}}

\subsubsection{Stellar Flare Energy Distributions}
\label{section:energydistributions}

Motivated by the above phenomenology, several studies have addressed
flare distributions in stellar coronae. Earlier work by
\cite{collura88} and \cite{pallavicini90} indicated power laws with
$\alpha < 2$ although the sensitivities used in those observations
were quite limited, and flares from different stars at different
distances were lumped together. This also holds for a study by
\cite{osten99} for RS~CVn-type binaries.

To avoid bias with regard to detection limits at lower energies, new
methods have been devised by \cite{audard99}, \cite{audard00},
\cite{kashyap02}, \cite{guedel03a}, and \cite{arzner04}, the latter
three using Monte Carlo forward methods and analytical inversion. Work
by \cite{wolk05}, \cite{arzner07}, and \cite{stelzer07} has extended
flare statistics into the PMS domain.
 
Most of these recent studies converge to $\alpha \approx
2.0\mbox{\,--\,}2.5$ (Figure~\ref{figure:flarerate}), indicating a
potentially crucial role of flares in coronal heating if the power-law
flare energy distribution extends by about 1\,--\,2~orders of
magnitude below the actual detection limit in the light curves. The
X-ray coronae of active stars would be an entirely hydrodynamic
phenomenon rather than an ensemble of hydrostatic loops.

\epubtkImage{}{%
\begin{figure}[t!]
  \centerline{\includegraphics[scale=0.47]{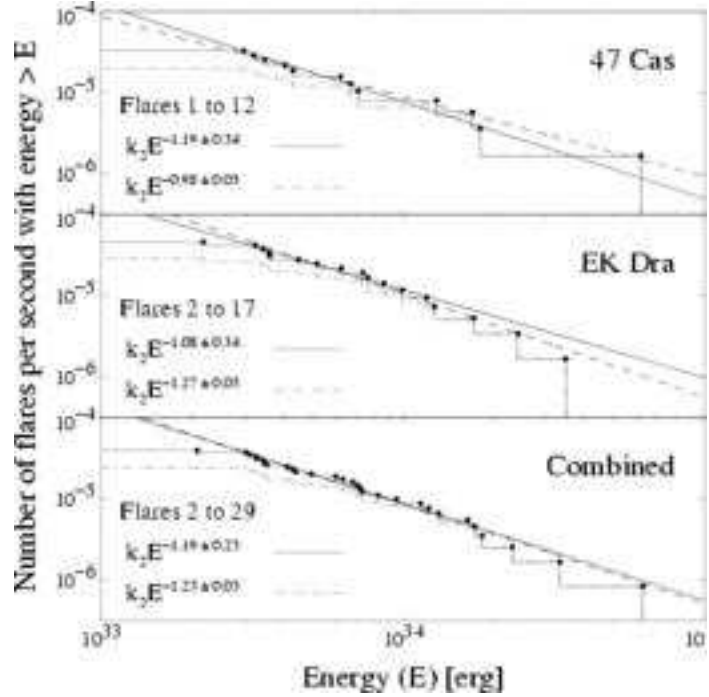}}
  \caption{\it Cumulative flare rate distributions in energy for
  47~Cas, EK~Dra, and both combined.  The indices of power-law
  fits to the cumulative distributions are smaller than the those of
  the differential distributions by one, $\alpha -1$. The
  distributions plotted as solid histograms have been corrected for
  flare overlap. Also shown are power-law fits to the corrected
  distributions (solid lines; see \citealt{audard99} for more
  details; reproduced by permission of AAS).}
  \label{figure:flarerate}
\end{figure}}

\subsubsection{Stochastic Flares and Coronal Observations}
\label{section:stochasticflares}

This mechanism has been studied specifically for young solar analogs
by \cite{guedel97a}, \cite{guedel97c}, \cite{audard99}, and
\cite{telleschi05}, in particular with regard to synthetic light
curves and emission measure distributions.

\cite{guedel97a} found that the {\it time-averaged} emission measure
distribution of solar flares adds a characteristic, separate emission
measure component at high temperatures ($> 10 \unit{MK}$); numerical
simulations support this picture \citep{guedel97c}, and corresponding
hot emission measure components are indeed also identified in active
solar analogs from X-ray spectroscopic observations. If a full
distribution of flares contributes, including small flares with lower
temperature, then the entire observed X-ray emission measure could be
formed by the continually heating and cooling plasma in flares. The
predicted steep low-$T$ slope (up to $\approx 4$) of an emission
measure distribution induced by stochastic flaring compares very
favorably with observations of active stars \citep{guedel03a}. This
holds true specifically for solar analogs for which $\alpha =
2.2\mbox\,{\,-\,}2.8$ is implied from the emission measure slopes
\citep{telleschi05}.
 
\cite{audard99} and \cite{audard00} derived the flare energy
 distribution of young solar analogs (Figure~\ref{figure:flarerate},
 where the cumulative distributions are plotted, with a power-law
 index that is shallower by one unit) and studied consequences for
 the thermal structure of the active coronae. They speculated that the
 higher temperatures in active coronae are due to a larger rate of
 (larger) flares, although they left undecided whether there is a
 larger rate of reheating events of the same coronal structures or
 whether a higher number of active regions produce the higher flare
 rate.

\subsubsection{Summary: The Importance of Stochastic Flares}
\label{section:flareimportance}

The above overview suggests that the young solar corona may have been
strongly driven by flares erupting as a consequence of magnetic
instabilities. There is no conclusive proof that stochastic flares
dominate coronal heating although there is clear evidence that active
stellar coronae are continuously variable on time scales similar to
flare events, and there is ample circumstantial albeit indirect
  evidence supporting this model, as described above. There are
important implications from flare-heating coronal models:

\begin{itemize}

  \item The corona can no longer be treated as an ensemble of static
  magnetic loops. Scaling laws often used to assess ``coronal
  structure'' are meaningless.

  \item Solar flares are often accompanied by coronal mass ejections,
  contributing to the overall mass loss rate. A high rate of coronal
  flaring may contribute significantly to an enhanced mass loss in the
  young Sun (see Section~\ref{section:solarwind} and
  Section~\ref{section:brightsun}).

  \item Solar flares accelerate electrons and ions to high energies,
  both within closed magnetic structures (evident by synchrotron
  emission from mildly relativistic electrons or non-thermal hard
  X-rays and gamma-rays from collisions in chromospheric gas) and
  along open field lines toward interplanetary space (measured in situ
  as ``solar energetic particles''). Copious coronal flares in young
  solar analogs will eject large numbers of energetic particles that
  are important for non-thermal interactions with planetary
  atmospheres (Section~\ref{section:interactions}) and for the
  irradiation of circumstellar disks and meteoritic material,
  potentially producing (some of the) well-known isotopic anomalies.

\end{itemize}

\subsection{The Solar Coronal Composition in Time}
\label{section:abundances}

From a naive point of view, the coronal and solar-wind material is
expected to show the same overall composition as that of the
underlying star, and in particular its photosphere from where the
material has been brought up. However, measurements of the composition
of the solar corona and the solar wind have revealed what is commonly
known as the ``First Ionization Potential (FIP) Effect''; this
abundance anomaly results in elements with a FIP below $\approx 10
\unit{eV}$ (e.g., Si, Mg, Ca, Fe) being enriched by factor of a few
with respect to hydrogen when compared to the photospheric mixture,
while elements with a higher FIP (e.g., C, N, O, Ne, Ar) show the
photospheric composition. The solar measurements have been summarized
comprehensively in the extensive work by \cite{meyer85a, meyer85b} and
\cite{feldman92}. The same FIP effect is in fact also present in solar
energetic particles and -- unexpectedly -- in cosmic
rays.\epubtkFootnote{Meyer normalized the abundances such that the
  low-FIP element abundances were photospheric and the high-FIP
  abundances depleted, whereas the present-day view is that high-FIP
  elements are at photospheric levels, see \cite{feldman92}.}

A discussion of the solar FIP effect is not within the scope of this
review, nor are the various physical models that have been proposed
for more than two decades. I refer the interested reader to the
presentations by \cite{henoux95}, \cite{jordan98}, \cite{drake95}, and
\cite{laming95},  and also the series of papers in {\it Space
    Science Reviews} volume 85 (p. 283--418, 1998). In short, a
fractionation process, probably involving electric and/or magnetic
fields or pressure gradients, occurs at chromospheric levels where
low-FIP elements are predominantly ionized and high-FIP elements are
predominantly neutral. Because ions and neutrals are affected
differently by electric and magnetic fields, acceleration or drift of
elements with different FIP may occur at different speed or with
different efficiency. The FIP effect varies considerably in strength
between various types of coronal structure; see, e.g.,
\cite{jordan98}. Although it is also expressed in full-disk solar
spectral data, it is confined to coronal heights where $T \ga 1
\unit{MK}$, while material at lower temperatures shows the
photospheric composition \citep{laming95}.

Why are abundance anomalies interesting for the study of the young
Sun? The study of solar analogs at different ages has, in fact, shown
systematics in the coronal composition that a successful, future model
for the generation and heating of stellar coronae must explain. I
briefly review the various findings.

\subsubsection{Abundances in Stellar Coronae}

Although detailed studies of coronal composition require
high-resolution X-ray spectra, anomalies in the coronal composition of
magnetically active stars have been recognized already in the early
days of X-ray astronomy. Most notably, a significant deficiency of
coronal Fe, by factors of a few when compared to the solar
composition, has been noted (e.g., \citealt{swank81, white94}). There
are two caveats with regard to early abundance measurements: i)
low-resolution X-ray spectroscopy does not isolate individual spectral
lines of any element with the exception of Fe, which produces a line
system at 6.7\,keV that can be measured by low-resolution (e.g.,
CCD-type) detectors. This line, however, forms at very high
temperatures (formation temperature $\approx 60\mbox{\,--\,}70
\unit{MK}$), and knowledge of the thermal structure at those high
temperatures is required if the element abundances are to be trusted;
ii) more severely, for most stars no reliable photospheric abundances
are known (again, the exception is mostly Fe). Because the
photospheric composition of stars may significantly differ from the
solar photosphere, any coronal abundance anomaly in stars should
really be defined with respect to the underlying photosphere. In any
case, before the advent of spectroscopy with {\it XMM-Newton} and {\it
  Chandra}, a sufficient body of data clearly showed that the most
active stars reveal strong depletion of most heavy elements, with
little FIP-related systematics \citep{singh99}.

Early observations of the extremely active subgiant HR~1099
\citep{brinkman01} and the MS~K-type ZAMS star AB~Dor
\citep{guedel01a} with the {\it XMM-Newton} Reflection Grating
Spectrometer uncovered a new, systematic FIP-related bias in
magnetically active stars: in contrast to the solar case, {\it low-FIP
  abundances are systematically depleted with respect to high-FIP
  elements}, a trend now known as the ``inverse FIP effect''
(IFIP). The ratio between the abundances of Ne (highest FIP) and Fe
(low FIP) is therefore unusually large, of order 10 in the most
extreme cases (when compared to the solar photospheric mixture). This
pattern has been confirmed for many further active stars (e.g.,
\citealt{drake01, huenemoerder01, huenemoerder03} etc). Only highly
active stars show the IFIP pattern. Toward intermediately active
stars, the effect weakens, until eventually flat abundance
distributions are recovered \citep{audard03a}.

\subsubsection{The Composition of the Young Solar Corona}

Solar analogs provide the best basis for a systematic study of
abundance trends because they directly link to the solar
case. Fortuitously, many well-studied solar analogs with different
ages and activity levels are available for study in the solar
vicinity. Again, the ``Sun in Time'' sample is outstanding in that the
photospheric mix has been well documented, often for several elements
apart from Fe. The principal result is that all of them show
photospheric abundances close to solar (see summary table in
\citealt{telleschi05}). The only independent parameter for this sample
with regard to coronal abundance anomalies is therefore again the {\it
  activity level} or, equivalently, {\it age}.

The sample shows a clear systematic development from an inverse FIP
effect in the youngest examples (with underabundances of Fe and Si) to
a marked FIP effect in objects at ages of about 300\,Myr and older
(see Figure~\ref{figure:abundances}). As long as the IFIP pattern is
present, all abundances appear to be sub-solar, but the Fe/H
abundance ratio gradually rises with decreasing coronal activity. The
abundance pattern reverts to a normal, solar-type FIP anomaly for
stars at activity levels of $\log L_X/L_\mathrm{bol} \la -4$
\citep{telleschi05}. Unexpectedly, this transition also seems to
coincide with i) the transition from coronae with a prominent hot ($T
\ga 10 \unit{MK}$) component to cooler coronae, and ii) with the
transition from prominent non-thermal radio emission to the absence
thereof \citep{telleschi05}.

\epubtkImage{}{%
\begin{figure}[t!]
  \centerline{\includegraphics[scale=0.45]{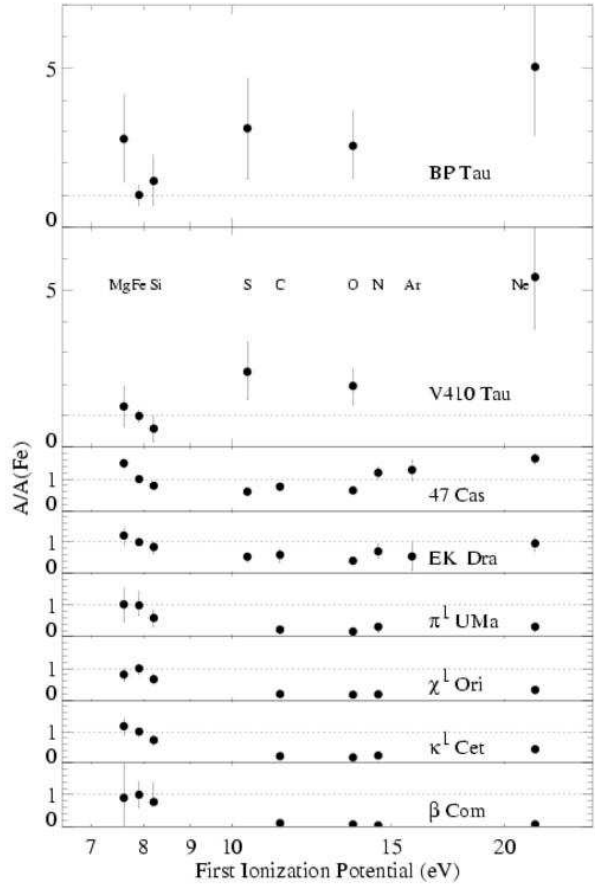}}
  \caption{\it Coronal element abundances normalized to the Fe
  abundance, with respect to the solar photospheric mixture as a
  function of FIP; the total X-ray luminosities are, from top to
  bottom: $\log L_\mathrm{X}$ = 30.06 (BP~Tau), 30.66 (V410~Tau),
  30.39 (47~Cas~B), 30.08 (EK~Dra), 29.06 ($\pi^1$~UMa), 28.95
  ($\chi^1$~Ori), 28.95 ($\kappa^1$~Cet), 28.26 ($\beta$~Com)
  \citep[after][]{telleschi05, telleschi07a}.}
  \label{figure:abundances}
\end{figure}}

\subsubsection{The Ne/O Abundance Ratio: Subject to Evolution?}

Figure~\ref{figure:abundanceratios} shows an anomaly for the O/Ne ratio
which is found at values of 0.3\,--\,0.7~times the solar ratio, apparently
{\it regardless of the stellar activity level}. Because both O and Ne
are high-FIP elements, their abundance ratio {\it could} reflect the
photospheric ratio. But then, the Sun's composition would be
anomalous.

The tabulations of several solar element abundances have recently been
revised \citep{asplund05a}, resulting in significant discrepancies
between solar models and helioseismological results
\citep[see][]{antia05} and references therein) {\it unless} further
solar abundances were also different. A Ne abundance higher by a
factor of at least 2.5 than hitherto assumed would be
needed. Therefore, \cite{telleschi05} were the first to point out that
the systematically non-solar O/Ne abundance ratio (by a similar
factor) calls for a revision of the solar Ne abundance tabulation
which at the same time would solve the solar helioseismology
problem. This was further elaborated by \cite{drake05a} who suggested
a factor of 2.7 upward revision of the adopted solar Ne abundance.

\epubtkImage{}{%
\begin{figure}[htbp]
  \centerline{
  \includegraphics[scale=0.38]{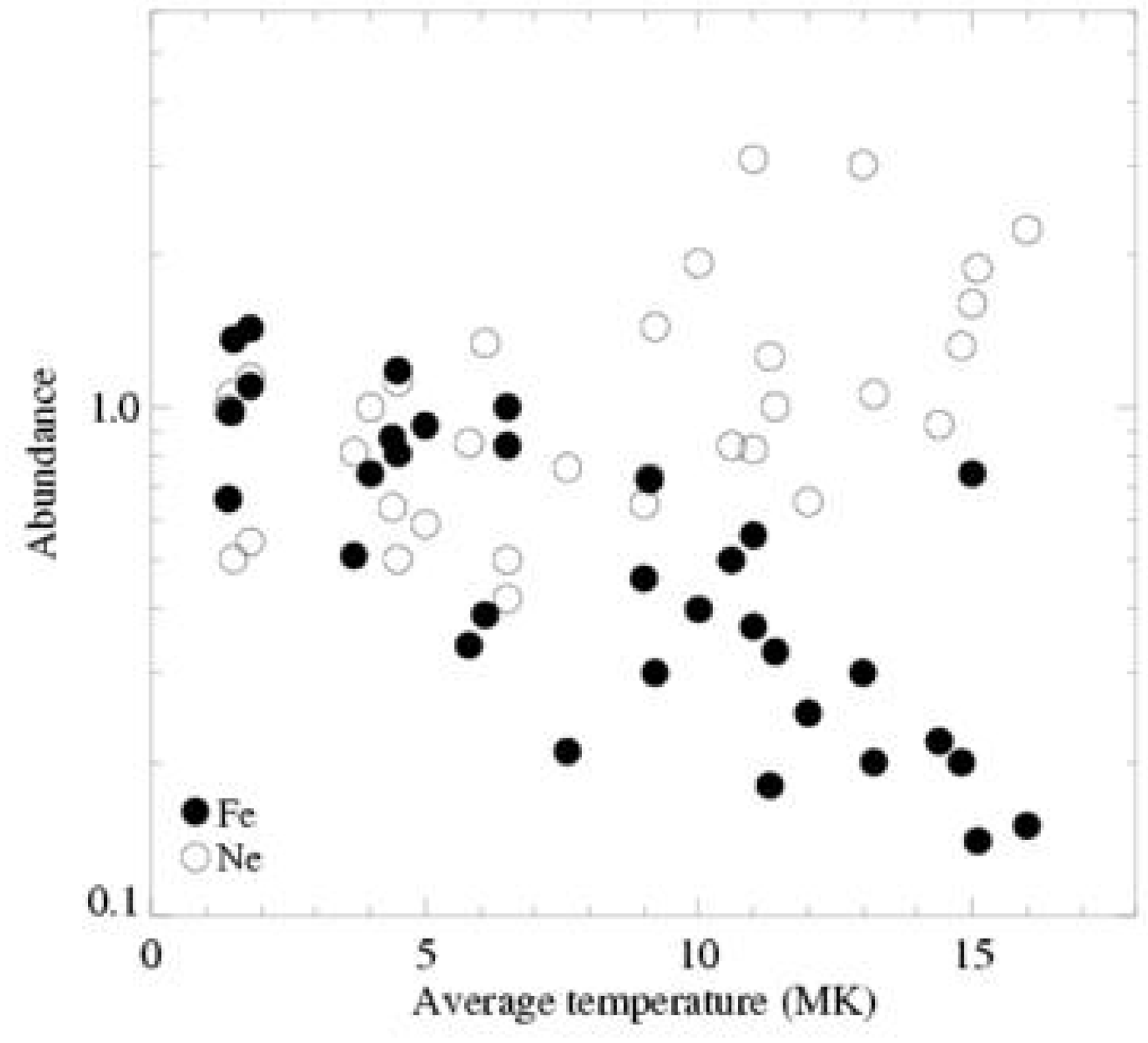}
  \includegraphics[scale=0.38]{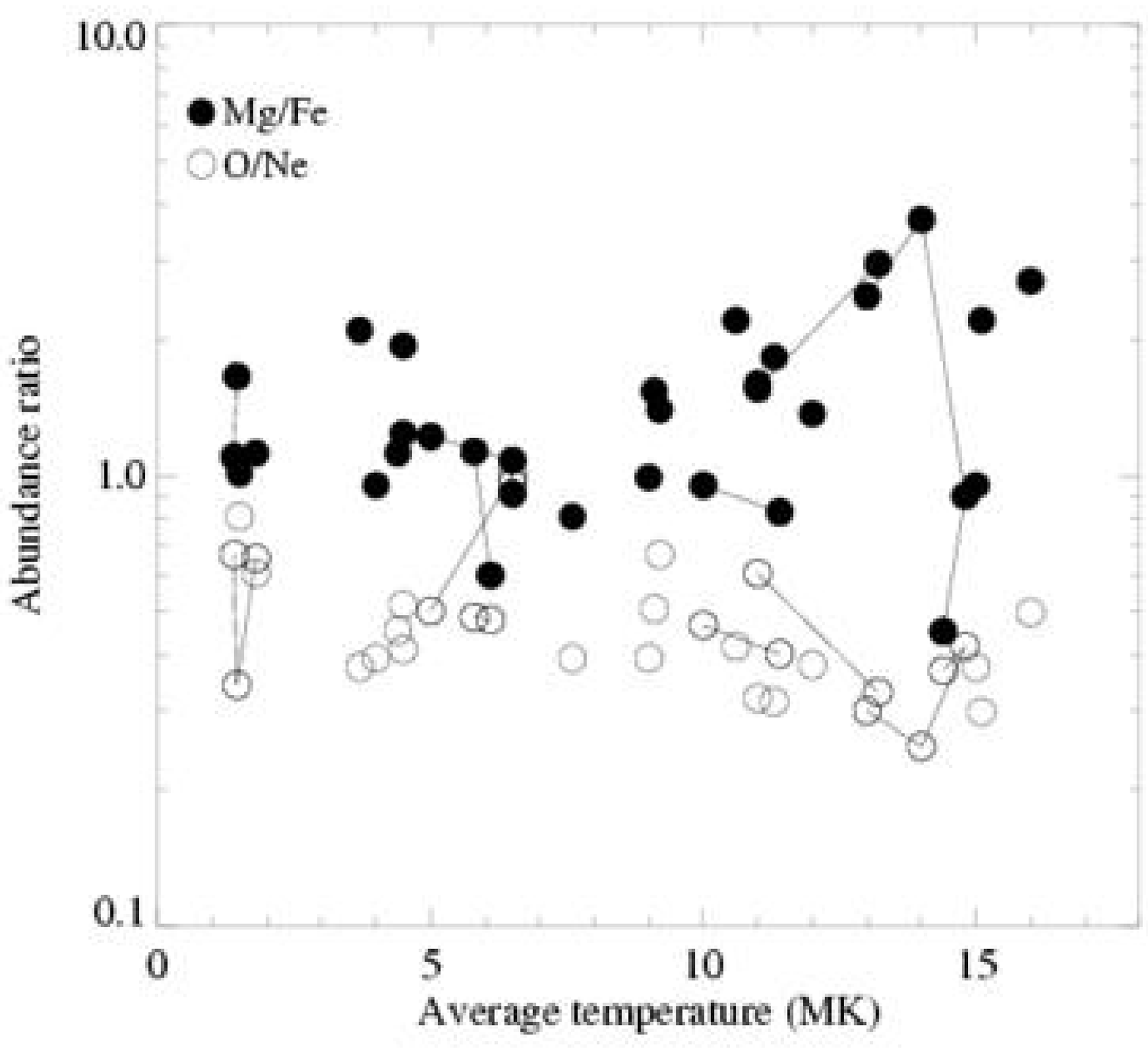}
  }
  \caption{\it  Coronal abundances of {\rm Fe} (low FIP) and Ne (high
  FIP -- left plot), and ratios of {\rm Mg/Fe} (low FIP) and {\rm
  O/Ne} (high FIP -- right plot) for various stars, shown as a
  function of the average coronal temperature
  \citep[from][copyright Springer Verlag, reprinted with permission]{guedel04}.}
  \label{figure:abundanceratios}
\end{figure}}

Two directions should be taken to verify this suggestion. First,
further coronal abundances of low-activity solar-analog stars should
be derived. So far, agreement with solar values has in fact been
reported for $\alpha$~Cen~A \citep{raassen03}, Procyon
\citep{raassen02}, and for $\beta$ Com, the latter with large error
bars, however \citep{telleschi05}. There seems to be a trend toward
higher O/Ne ratios in Figure~\ref{figure:abundanceratios} indeed. New
results for $\alpha$~Cen have been reported by \cite{liefke06},
placing the Ne/O abundance ratio significantly lower than values for
active stars, albeit not as low as the presently accepted solar
photospheric value. Second, verification of the solar Ne/O ratio is
needed. Although Ne cannot be measured directly in the photosphere,
the Ne/O ratio can be derived from coronal measurements the same way
as done for stellar observations. Recent analysis of solar active
region X-ray spectra and of EUV spectra from transition-region levels
\citep{schmelz05,young05} both report the standard Ne/O abundance
ratios, rejecting an upward correction of Ne. The issue should
therefore be considered to be open for the time being. A FIP-related,
i.e., ``evolutionary'' enrichment of Ne in active corona remains a
viable possibility \citep{asplund05b}.

\subsection{Summary: The Young, Active Sun}

Although the MS Sun started as a ZAMS star with a bolometric
luminosity that was significantly lower than today -- by about 30\% --
a very different picture must be drawn from high-energy observations
of solar analogs, including UV, FUV, EUV, X-ray, and non-thermal radio
emissions. All signatures of magnetic activity, among them dark
starspots, chromospheric and transition-region line emission, coronal
X-rays, and radio gyrosynchrotron emission in young solar analogs
point to levels of magnetic activity orders of magnitude above the
present solar level. The present-day Sun in fact resides at the lowest
levels a solar-like star can attain -- near the end of its MS life, as
a spun-down star.

Short-wavelength radiation and non-thermal radio emission show
power-law decays during the solar evolution that steepen toward higher
energies of the underlying electron population. In parallel, the
rotation rate of the Sun also decayed, again in a power-law
fashion. Rotation is indeed thought to be the ultimate driver of these
decay laws, as it is one of the two fundamental ingredients of the
magnetic dynamo, the other being convection which does not strongly
change during the Sun's main-sequence life.

The solar wind is another expression of magnetic activity. The mass
loss induced by the wind also decays according to a power-law although
this does not seem to hold for the youngest and most active examples
which show a suppressed wind.

The high level of magnetic activity in the young Sun must have been a
pivotal ingredient in shaping the solar environment, in particular
planetary atmospheres. The most extreme expressions of magnetic
activity, magnetic flares that appear to be dominating contributors to
high-energy photon and particle production in young solar analogs,
need to be accounted for in planetary atmospheric modeling. I will
discuss some of these aspects in Section~\ref{solarsystem}.


\newpage

\section{Further Back in Time: The Pre-Main Sequence Sun}
\label{section:premainsequence}

\subsection{Where was the Cradle of the Sun?}

The present-day galactic environment is in no ways related anymore to
the place where the Sun was born. The contemporaneous Sun is obviously
not a member of a bound cluster. There has consequently been some
speculation regarding the environment in which the Sun formed. The
large-scale solar environment controls interactions with the
outflowing solar wind and is therefore relevant for the influx of
cosmic rays into the inner solar system.

During the past $10^6$~years, the Sun has moved from the very hot
($\approx 10^6 \unit{K}$) low-density ``local bubble'' environment
into the cooler and denser ($\approx 6900 \unit{K}$, $n(\mathrm{H}^0 +
\mathrm{H}^+) \approx 0.3 \unit{cm^{-3}}$) ``local interstellar
cloud'' \citep{frisch00, frisch05, frisch06}. The pressure
characteristics of these two environments are similar, so that the
radii of the heliosphere would have been similar (120\,--\,130\,AU;
\citealt{frisch00}). However, the heliospheric structure would shrink
to ten percent of its present size if the Sun encountered molecular
clouds, and such encounters are likely to have occurred several times
in the past \citep{frisch00}.

Moving back in time, the evidence becomes indirect. The best
``memory'' of the earliest episodes of the Sun's life is contained in
meteorites (Section~\ref{section:meteoritics}). The presence of live
$^{60}\mathrm{Fe}$ in the early solar system (see
Section~\ref{section:meteoritics}) inferred from meteoritic trace
elements \citep{tachibana03} cannot be explained by local processes
(such as stellar flares, see Section~\ref{section:meteoritics}) but is
thought to require supernova explosions \citep{meyer00}. The local
environment of the forming solar system was therefore likely
reminiscent of a high-mass star-forming region like the Orion region;
the young Sun and its circumstellar disk may have resided in a H\,{\sc
  ii} region for a considerable amount of time \citep{hester04}; the
intense ultraviolet radiation field from massive stars might have
contributed to the evaporation of the molecular environment of the Sun
(so-called proplyds in the Orion Nebula, \citealt{odell01}, or
``evaporating gaseous globules'' [EGGs], \citealt{hester96}). These
structures contain protostars that can be detected in X-rays
\citep{kastner05, linsky07}, i.e., deterioration of the environment is
due both to the larger-scale ``interstellar'' environment and the
stellar magnetic activity itself.

\subsection{New Features in the Pre-Main Sequence Sun}

\subsubsection{Evolutionary stages: Overview}

Modern theory of star formation together with results from
comprehensive observing programs have converged to a picture in which
a forming low-mass star evolves through various stages with
progressive clearing of a contracting circumstellar envelope. In its
``class 0'' stage (according to the mm/infrared classification
scheme), the majority of the future mass of the star still resides in
the contracting molecular envelope. ``Class~I'' protostars have
essentially accreted their final mass while still being deeply
embedded in a dust and gas envelope and surrounded by a thick
circumstellar disk. Jets and outflows may be driven by these optically
invisible ``infrared stars''. Once the envelope is dispersed, the
stars enter their ``Classical T~Tauri'' stage (CTTS, usually
belonging to infrared class~II with an IR excess)
with excess H$\alpha$ line emission if they are still surrounded by a
massive circumstellar disk; the latter results in an infrared
excess. ``Weak-line T~Tauri'' stars (also ``naked T~Tauri stars'',
\citealt{walter86}; usually with class~III characteristics, i.e.,
essentially showing a photospheric spectrum) have lost most of their
disk and are dominated by photospheric light \citep{walter88}.

\subsubsection{New Features: Accretion, Disks, and Jets}
\label{section:newfeatures}

Moving back in time from the MS into the PMS era of solar evolution,
we encounter changes in the Sun's internal structure and in
fundamental stellar parameters such as radius and
$T_\mathrm{eff}$. The typical ``T~Tauri'' Sun at an age of
0.5\,--\,3\,Myr was bolometrically 1\,--\,4~times more luminous and
1.7\,--\,3.6~times larger in radius, while its surface $T_\mathrm{eff}
\approx 4260 \unit{K}$, corresponding to a K5 star
\citep[after][]{siess00}. The interior of T~Tauri stars evolving on
the Hayashi track is entirely convective. The operation of an
$\alpha\Omega$ dynamo should not be possible, yet very high levels of
magnetic activity are clearly observed on T~Tauri stars. Alternative
dynamos such as convective dynamos may be in operation. ``Solar
analogy'' no longer holds collectively for all stellar parameters,
except (roughly) for stellar mass. Also, there is a complex stellar
environment, including an accretion disk, outflows and jets, and
probably a large-scale stellar magnetosphere that interacts with
these structures (\citealt{camenzind90, koenigl91, collier93, shu94};
Figure~\ref{figure:ttaustar}). Magnetic interactions between the star
and its environment are important in the following contexts:

\begin{itemize}

  \item Star-disk magnetic fields may form large magnetospheres that
  lead to star-disk locking and disk-controlled spin rates of the
  star. Generally, disk-surrounded CTTS rotate relatively slowly,
  probably due to disk locking, with \P = 5\,--\,10\,d, compared to
  only a few days for diskless, non-accreting WTTS \citep{bouvier93};
  if a rotation-induced magnetic dynamo is at work in CTTS, then it
  may be dampened compared to dynamos in freely (and rapidly) rotating
  diskless WTTS.

  \item Non-synchronized rotation may be possible in particular in
  very young systems in spite of magnetic fields connecting the star
  with the inner border of the disk. The different rotation rates
  produce shear in the star-disk magnetic fields, eventually leading
  to field lines winding up around the star and reconnecting,
  releasing energy and plasmoids that stream out along large-scale
  magnetic fields \citep{hayashi96, montmerle00}. This may be a viable
  model for the production of protostellar jets.

  \item Star-disk magnetic fields are thought to define the basic
    route of material accreting from the disk to the star
    \citep{uchida84, camenzind90, koenigl91}. This star-disk
    magnetospheric complex (Figure~\ref{figure:ttaustar}) is in the
    center of present-day studies of PMS magnetic ``activity'' as it
    links the accretion process to stellar magnetic fields and
    probably to outflows and jets. Although the details of the
    magnetic interface between the star and the circumstellar disk are
    far from being understood, the physics of mass-loading of the
    magnetic field lines is being addressed in detail based on
    theoretical \citep[e.g.,][]{shu94, ferreira07} and numerical
    investigations \citep[e.g.,][]{romanova06}. Magnetically funneled
    material forms  ``hot spots'' on the stellar surface that can be
    observed as an ultraviolet excess \citep[e.g.,][]{calvet98}.

\end{itemize}

\epubtkImage{}{%
\begin{figure}[htbp]
  \centerline{\includegraphics[scale=0.48]{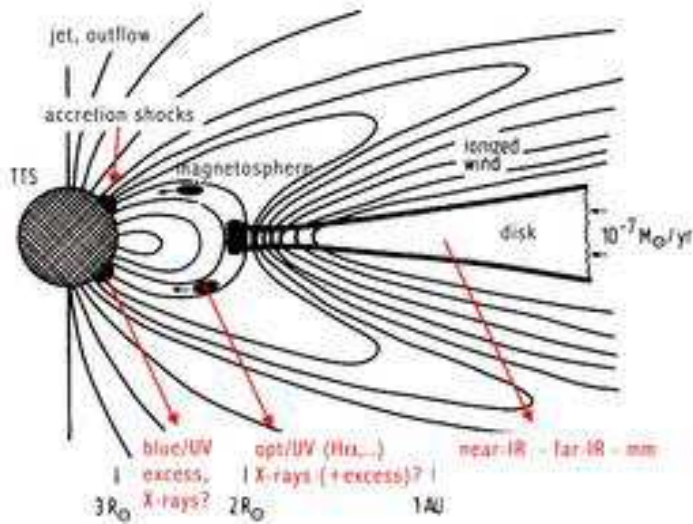}}
  \caption{\it Sketch of the environment of a classical T~Tauri
  star or a protostar, showing the star, the circumstellar disk,
  magnetic field lines, and closed star-disk magnetic field structures
  that funnel material from the disk to the star \citep[adapted
  from][Copyright Wiley-VCH Verlag GmbH \& Co. KGaA. Reproduced
with permission]{camenzind90}.}
  \label{figure:ttaustar}
\end{figure}}

\subsubsection{New Emission Properties: Solar-Like or Not?}

Given the similarity between WTTS and active, ZAMS stars, ``activity''
in WTTS as seen in spots, chromospheric and transition-region lines,
and X-ray emission has conventionally been attributed to solar-like
magnetic activity. Non-thermal gyrosynchrotron radio emission in WTTS
suggests processes similar to those observed in active MS stars and
subgiant binaries, processes that are thought to be related to
electron acceleration in coronal flares \citep{white92a, white92b,
  guedel02}. As for CTTS, strong optical and ultraviolet emission
lines were initially assumed to give evidence for extremely active
chromospheres and transition regions (\citealt{joy45, herbig80}, see
review by \citealt{bertout89}). Flares and ``flashes''
\citep[e.g.,][]{ambartsumian82} also suggest analogy with magnetic
flaring in more evolved stars and the Sun. Arguments in favor of
solar-like magnetic {\it coronal} activity in {\it all} TTS include i)
dominant, typical electron temperatures of order $10^7 \unit{K}$
that require magnetic confinement, and ii) the presence of flares
\citep{feigelson81a, feigelson81b, feigelson89, walter84, walter88}.

However, the solar analogy seems to break down in the optical/UV range
where strong flux excesses are recorded; these excesses are
uncorrelated with X-rays \citep{bouvier90} but seem to relate to the
accretion process (Section~\ref{section:uvtts}). A similar excess
recently discovered in soft X-rays may be related to both coronal
magnetic fields and accretion (\citealt{guedel07c}, see
Section~\ref{section:excess}). Also, thermal radio emission observed in
CTTS, probably due to optically thick winds but also due to bipolar
jets, reaches beyond the solar analogy \citep{cohen82,
  bieging84}. Clearly, emission across the electromagnetic spectrum
needs to be understood in the context of the new features around
accreting stars, and in particular the related magnetic fields,
summarized in Section~\ref{section:newfeatures}.

A summary of the present view of activity and other properties of PMS
stars from the earliest stages to the arrival on the MS is given in
Figure~\ref{figure:ysotable} \citep[from][]{feigelson99}.

\epubtkImage{}{%
\begin{figure}[t!]
  \centerline{\includegraphics[scale=0.76]{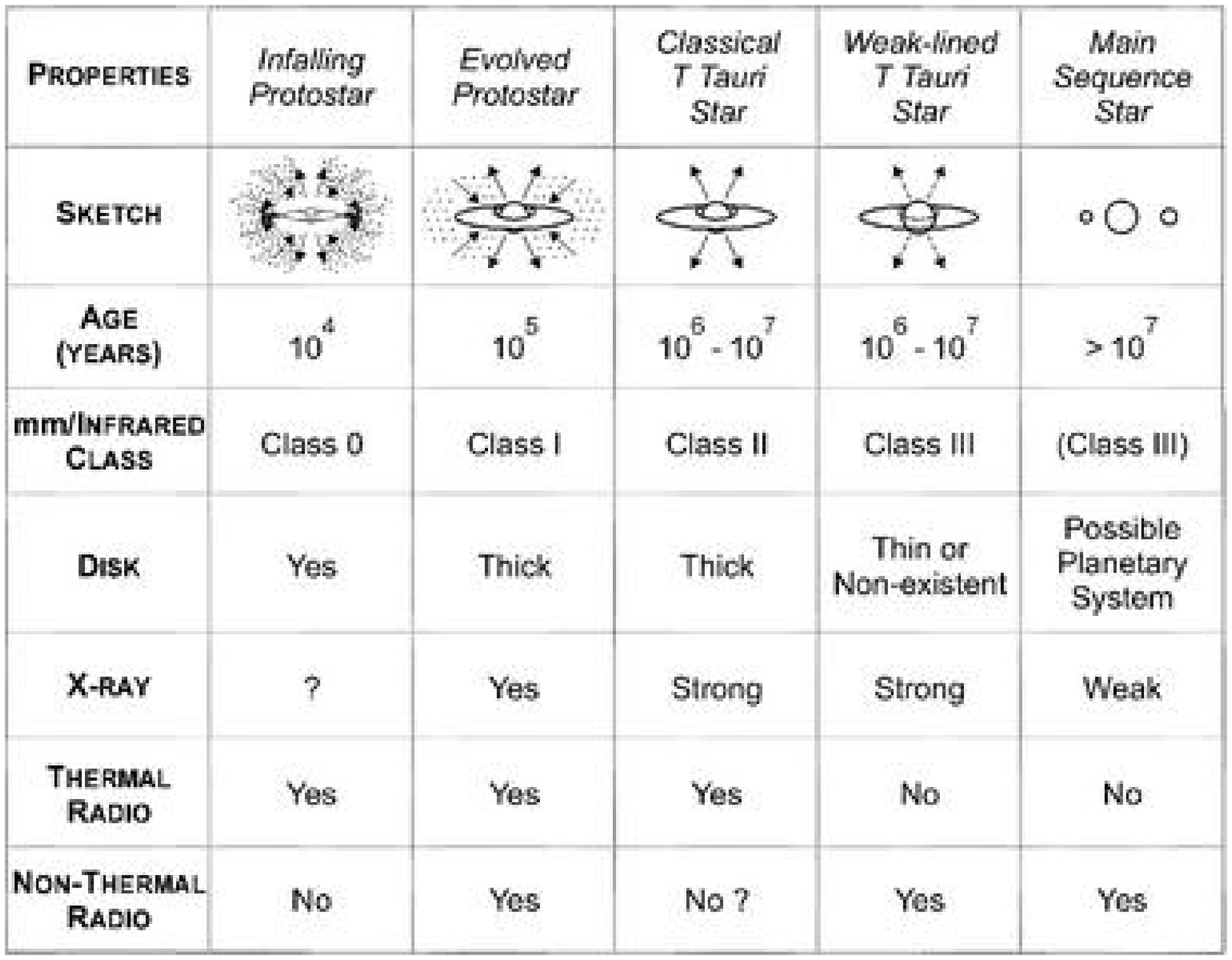}}
  \caption{\it  Summary of properties of PMS objects, in comparison
  with MS stars \citep[from][reprinted, with permission, from the Annual 
  Review of Astronomy and Astrophysics, Volume 37 \copyright\ 1999 by Annual Reviews,
  www.annualreviews.org]{feigelson99}.}
  \label{figure:ysotable}
\end{figure}}

\subsection{The T~Tauri Sun}

\subsubsection{The Magnetic Field of the T~Tauri Sun}

Surface magnetic fields have been successfully measured on
``solar-analog'' T~Tauri stars, using Zeeman broadening
\citep{krull99, valenti04, krull07}. For example, BP~Tau (a
$0.65\,M_{\odot}$ CTTS) maintains a mean magnetic field strength
($\sum Bf$) of $2.6 \pm 0.3 \unit{kG}$, i.e., the equivalent of
sunspots for $f = 1$. Such field strengths exceed the equipartition
value of T~Tauri photospheres ($\approx 1 \unit{kG}$,
\citealt{krull99, krull07}), which may be a consequence of near-total
surface filling \citep{solanki94, krull07}. T~Tauri photospheres are
therefore dominated by magnetic pressure rather than thermal pressure,
in contrast to MS stars and the Sun but similar to stellar coronae in
general. The average surface magnetic fields of CTTS also exceed
  a prediction from X-rays, based on a correlation between X-ray
  luminosity and photospheric magnetic flux valid for MS stars and the
  Sun \citep{pevtsov03}. Upper limits of the net polarization of
photospheric lines suggest that the photospheric magnetic fields form
predominantly in small-scale structures, although a dipole may
dominate at large distances \citep{krull99, valenti04, krull07}. Large
dipole components are also suggested from observations of spots
concentrated at the poles of T~Tauri stars, similar to active MS solar
analogs \citep[e.g.,][]{joncour94, hatzes95, rice96}. In any case, the
observed fields should be strong enough to truncate circumstellar
disks indeed \citep{krull99, krull07},  even if rather complex
  (non-dipolar) surface magnetic field distributions are assumed
  \citep{gregory06}.

Zeeman-Doppler Imaging techniques have successfully been applied to
extremely active solar analogs in the PMS phase. A particularly clear
case was presented by \cite{donati00}, finding solar-like differential
rotation on a post-T~Tauri star (see also Section~\ref{section:starspotcycles}).
More recently, \cite{donati07} have reconstructed a rather complex, large-scale
magnetic topology on the CTTS V2129 Oph; they found a relatively 
weak dipole but stronger octupolar fields that are tilted against the rotation
axis, with strong near-polar spots. The accretion footpoints are also found 
to be located at high latitudes. An attempt was made at extrapolating the fields
to the inner rim of the disk, showing that some field lines should successfully
accrete toward the observed hot spots.

At coronal levels, X-ray rotational modulation provides information on
the large-scale distribution of stellar magnetic fields. Rotational
modulation is widespread among extremely active T~Tauri stars
\citep{flaccomio05}. Some 10\% of the studied stars in the Orion X-ray
sample show such evidence, suggesting that i) the X-ray emitting
active regions are not homogeneously distributed on the surface, i.e.,
despite the X-ray saturation level reached by these stars, the surface
cannot be filled with X-ray-bright magnetic loops; ii) the X-ray
emitting regions responsible for the rotational modulation are
directly associated with the surface and cannot extend much beyond
$R_*$ \citep{flaccomio05}. A comparison of the modulation depth with
the Sun's modulation in fact shows that the longitudinal
inhomogeneities are similar \citep{flaccomio05}.

\subsubsection{The Ultraviolet T~Tauri Sun}
\label{section:uvtts}

A defining property of (accreting) classical T~Tauri stars is their
strong line emission of, e.g., H$\alpha$ or Ca\,{\sc ii} H \& K. These
strong lines were initially thought to be evidence of massive
chromospheres similar to those seen on the Sun or in cool stars (see
review by \citealt{bertout89}), and the discovery of strong UV lines
such as those of Si\,{\sc ii}, Si\,{\sc iv}, and C\,{\sc iv} --
equivalent to ``transition region'' lines in the Sun formed above
$10^4 \unit{K}$ -- supported this picture.

However, when compared with MS stars, including chromospherically very
active examples, UV line and continuum emission is up to $10^2-10^4$ times
stronger in CTTS (see example of TW~Hya in
Table~\ref{table:linefluxes}; \citealt{canuto82, canuto83, bouvier90,
  valenti00}), regardless of the photospheric effective temperature or
the stellar rotation period but correlated with the mass accretion
rates derived from optical continuum data (\citealt{bouvier90,
  johnskrull00}; Figure~\ref{figure:softexcess}a below). Further, UV or
H$\alpha$ line surface fluxes of CTTS show, in contrast to more
evolved stars, no correlation with coronal X-rays, the latter being in
the range of RS CVn-type active binary systems or very active MS stars
but the UV/H$\alpha$ lines showing a wide range of excess flux
\citep{bouvier90}.

Coronal and ``chromospheric/transition region'' fluxes are thus not
correlated in CTTS, contrasting strongly with MS and subgiant stars
for which a sharp correlation is taken as evidence for a common
physical heating mechanism (operating in related magnetic fields;
Section~\ref{section:spectralevolution}, Figure~\ref{figure:fluxflux}). An
additional mechanism must be responsible for the optical/UV line flux
excess. Apart from the line excess fluxes, there is also a strong blue
continuum excess that leads to ``veiling'' in the optical spectrum,
i.e., a filling-in of absorption lines by continuum emission; this
emission is also not compatible with chromospheric radiation. The most
obvious property common only to CTTS among the stars considered above
is accretion; downfalling material could provide the energy to
generate the optical/UV excess \citep{bertout88, basri89}.

Nearly free-falling gas can be heated to maximum temperatures

\begin{equation}
  T_s = 8.6 \times 10^5 \unit{K}\left({M\over 0.5\,M_{\odot}}\right)
  \left({R\over 2\,R_{\odot}}\right)^{-1}
\end{equation}

in shocks forming at the bottom of magnetic accretion funnels
\citep{calvet98}. UV and optical line emission could thus provide
diagnostics for the accretion velocity, the mass accretion rate, and
possibly the surface filling factor of accretion funnels.

The present consensus, based on such concepts as well as line profile
properties and correlations with the mass accretion rate, is that the
UV excess emission originates from material heated in accretion shocks
\citep[e.g.,][]{calvet98, gullbring98}. Some of the emission lines
(e.g., H$\alpha$, Ca\,{\sc ii}) may also form in the accretion funnels
themselves, or in stellar winds \citep{ardila02}.

\subsubsection{The X-ray T~Tauri Sun in Time}

\cite{feigelson02a}, \cite{wolk05}, and \cite{telleschi07a} presented
X-ray studies of near-solar-mass stars (stars in the ranges of
$0.7\,M_{\odot} \le M \le 1.4\,M_{\odot}$ and $0.9\,M_{\odot} \le M
\le 1.2\,M_{\odot}$, respectively,  in the former two studies of
  the Orion Nebula cluster, and wider in the latter study of the
  Taurus star-forming region). The sample ages typically comprise the
  $\log t = 5.5\mbox{\,--\,}7$ range and contain both disk-surrounded
and disk-less T~Tauri stars. The median X-ray luminosity in the Orion
sample is found at $\log L_\mathrm{X} = 30.25$, i.e., three orders of
magnitude above the average solar X-ray output, but there is
  evidence for a slow decay with age, $\log L_\mathrm{X} \propto
t^{-1.1}$ \citep{feigelson02a, wolk05}. A shallower decay was reported
by \cite{preibisch05b} for  the same stellar cluster, with an
exponent between $-0.2$ and $-0.5$, but when considering normalized
$L_\mathrm{X}/L_\mathrm{bol}$ or average surface X-ray flux, then both
Feigelson et al.'s and Preibisch \& Feigelson's studies indicate that
the $L_\mathrm{X}$ decay law is roughly compatible with full
saturation  (i.e., $L_\mathrm{X}/L_\mathrm{bol} \approx 10^{-3}$) as
the star descends the Hayashi track and its bolometric luminosity is
decreasing \citep{feigelson02a}. \cite{telleschi07a} used the
  Taurus sample over a wider mass range but removed the strong
  $L_\mathrm{X}$ versus mass correlation in order to normalize the
  X-ray evolutionary behavior to a solar-mass star. The slope of the
  $L_\mathrm{X}$ vs. age correlation is fully compatible with the
  Orion results, with a power-law index of $-0.36\pm 0.11$ although
  the correlation is dominated by scatter from other sources, and its
  significance is marginal.

The evolutionary $L_\mathrm{X}$ decay is thus {\it qualitatively}
different from that in MS stars: it is due to stellar contraction (and
perhaps a change in the internal dynamo while the star transforms from
a fully convective to a convective-radiative interior); in contrast,
the decay of $L_\mathrm{X}$ in MS stars is due to stellar spin-down
while the stellar structure and size remain nearly
constant. Figure~\ref{figure:lxevolution} shows the long-term
evolution of the median X-ray output from PMS stages to the end of the
MS evolution, for G-type stars with ages $>10 \unit{Myr}$ and K-type
stars with ages $<10 \unit{Myr}$ (because the predecessors of MS~G
stars are PMS K stars; data from \citealt{guedel04}). The slight trend
toward decreasing $L_\mathrm{X}$ at ages $<10 \unit{Myr}$ follows
approximately $L_\mathrm{X} \propto t^{-0.3}$, in agreement with the
individual trends for the Orion and the Taurus samples,  albeit the
scatter is large. No ``onset'' of activity can be seen back to ages
$<1 \unit{Myr}$.

\epubtkImage{}{%
\begin{figure}[htbp]
  \centerline{\includegraphics[scale=0.8]{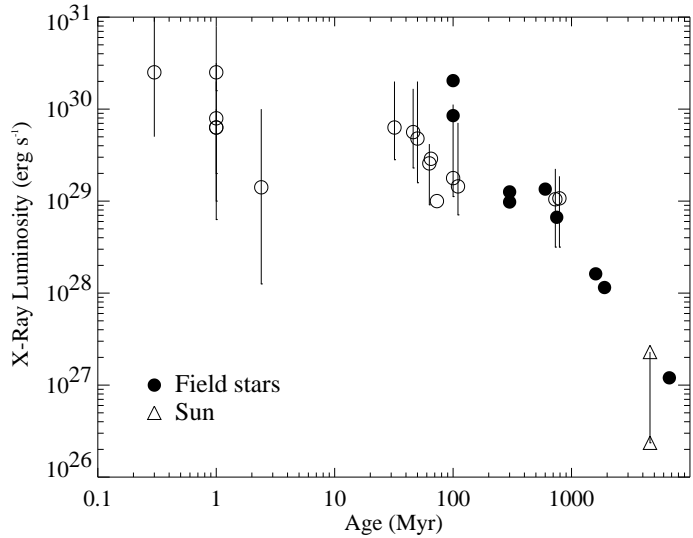}}
  \caption{\it Evolution of the X-ray luminosity of a near-solar mass
  star. A sample of field stars from the ``Sun in Time'' program is
  shown by filled circles. Open circles show the median $L_\mathrm{X}$
  for G-type stars in open clusters for ages $>$10\,Myr, and for
  K-type stars in open clusters and star-forming regions for ages
  $<$10\,Myr. The error bars show the approximate $1\sigma$ scatter in
  the samples \citep[adapted from][original data from references given
  therein; copyright Springer Verlag, reprinted with permission]{guedel04}.}
  \label{figure:lxevolution}
\end{figure}}

In summary, the age evolution of the X-ray output is modest in PMS
stars, the bulk of the X-ray output being determined by other stellar
properties. There are at least four such parameters that have been
discussed in the recent literature: bolometric luminosity, mass,
rotation, and mass accretion rate. I briefly summarize these parameter
dependencies in turn:

\begin{itemize}

  \item Similar to young MS stars, CTTS and WTTS straddle the
  empirical X-ray saturation limit, i.e., $L_\mathrm{X} \approx
  10^{-3.5} L_\mathrm{bol}$ \citep[e.g.,][]{preibisch05a,
  telleschi07a}, pointing to a dynamo process somehow related to the
  dynamo in MS stars. The long-term evolution of $L_\mathrm{bol}$ may
  then be the principal parameter for the long-term evolution of
  $L_\mathrm{X}$.

  \item In contrast to MS stars, PMS stars show a correlation between
      $L_\mathrm{X}$ and mass \citep[e.g.,][]{feigelson93,
      preibisch05a, telleschi07a}. However, if one restricts the MS
      sample to saturated, young stars, then they follow such a
      relation as well, owing to the well-known mass-$L_\mathrm{bol}$
      relation on the MS, $L_\mathrm{bol} \propto M^3$. Conversely,
      the flatter $L_\mathrm{bol}-M$ relation for a given PMS
      isochrone combined with the saturation law yields an
      $L_\mathrm{X}$ vs.\ mass correlation that is compatible with the
      observed relation \citep{telleschi07a}, i.e., the two relations
      are interdependent. \item Rotation is one of the main drivers of
      the magnetic dynamo and therefore of magnetic activity in MS
      stars. The near-saturation state of most PMS stars, however,
      suggests that rotation cannot be a key parameter. This is borne
      out by explicit correlation studies of the Orion
      \citep{preibisch05a} and the Taurus \citep{briggs07} samples of
      T~Tauri stars that show little in the way of a correlation as
      seen in MS~stars. On the contrary, some apparent trends in this
      direction are the result of population bias
      \citep{briggs07}. The absence of a decrease in the
      $L_\mathrm{X}/L_\mathrm{bol}$ ratio up to rotation periods of at
      least 10\,d (in contrast to MS stars) can be explained by the
      convective turnover time in PMS stars being much larger,
      yielding smaller Rossby numbers for a given rotation period and
      therefore saturation up to longer rotation periods
      \citep{preibisch05a, briggs07}.

      \item Accretion could influence $L_\mathrm{X}$ either by
      generating X-rays itself, or by suppressing coronal X-ray
      production. Empirically, $L_\mathrm{X}$ is correlated with the
      mass accretion rate $\dot{M}$ but this is not a physical
      correlation for the following reason. \cite{muzerolle03},
      \cite{muzerolle05}, and \cite{calvet04} described a relation
      between $\dot{M}$ and the stellar mass $M$, approximately
      reading $\dot{M} \propto M^2$. Because $L_\mathrm{X}$ correlates
      with $L_\mathrm{bol}$ and therefore, for a typical age
      isochrone, with mass, $L_\mathrm{X}$ must correlate with
      $\dot{M}$. \cite{telleschi07a} removed this trend from the
      Taurus sample. Although the two parameters still do not seem to
      be fully independent, a clear correlation cannot be
      established. Accretion does seem to influence magnetic energy
      output in more subtle ways, however, to be discussed in the
      Section~\ref{section:excess}.

\end{itemize}

\subsubsection{Coronal Excesses and Deficits Induced by Activity?}
\label{section:excess}

If the photoelectric absorption by the accreting gas is small, then
the softest X-ray range may reveal the high-temperature tail of the
shock emission measure thought to be responsible for the UV excesses
(Section~\ref{section:uvtts}). \cite{telleschi07a} and \cite{guedel07b}
identified an excess in the O\,{\sc vii}/O\,{\sc viii} Ly$\alpha$ flux
(or luminosity) ratio in CTTS when compared with WTTS or MS stars, the
so-called X-ray {\it soft excess} of CTTS
(Figure~\ref{figure:softexcess}b). In the most extreme case of the CTTS
T~Tau, the excess O\,{\sc vii} flux is such that this line triplet,
formed at only $\approx 2 \unit{MK}$, is the strongest in the soft
X-ray spectrum (see Figure~\ref{figure:ttauspec}; \citealt{guedel07c}).

\epubtkImage{}{%
 \begin{figure}[t!]
 \centerline{
    \includegraphics[scale=0.32]{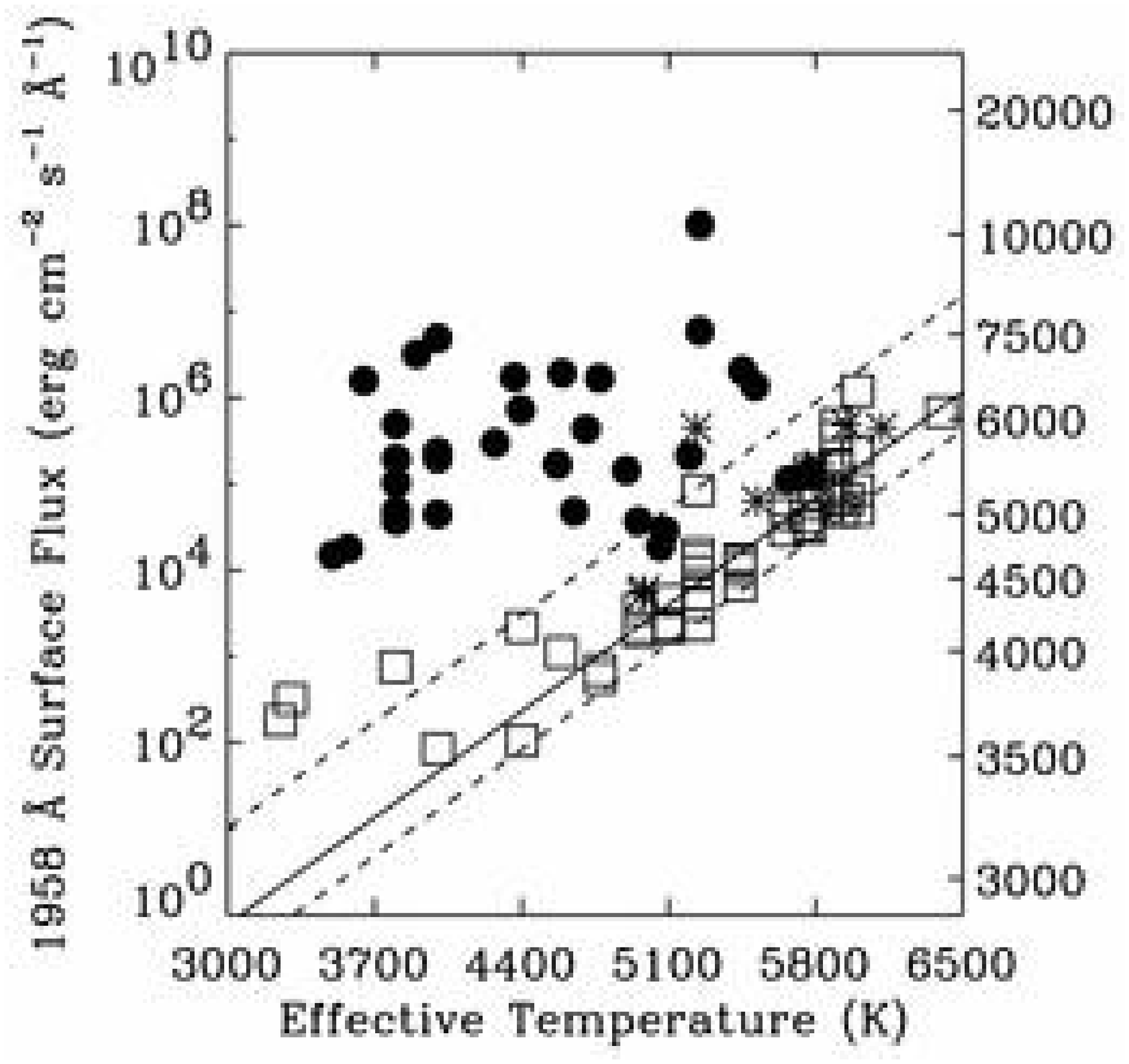}
    \includegraphics[scale=0.56]{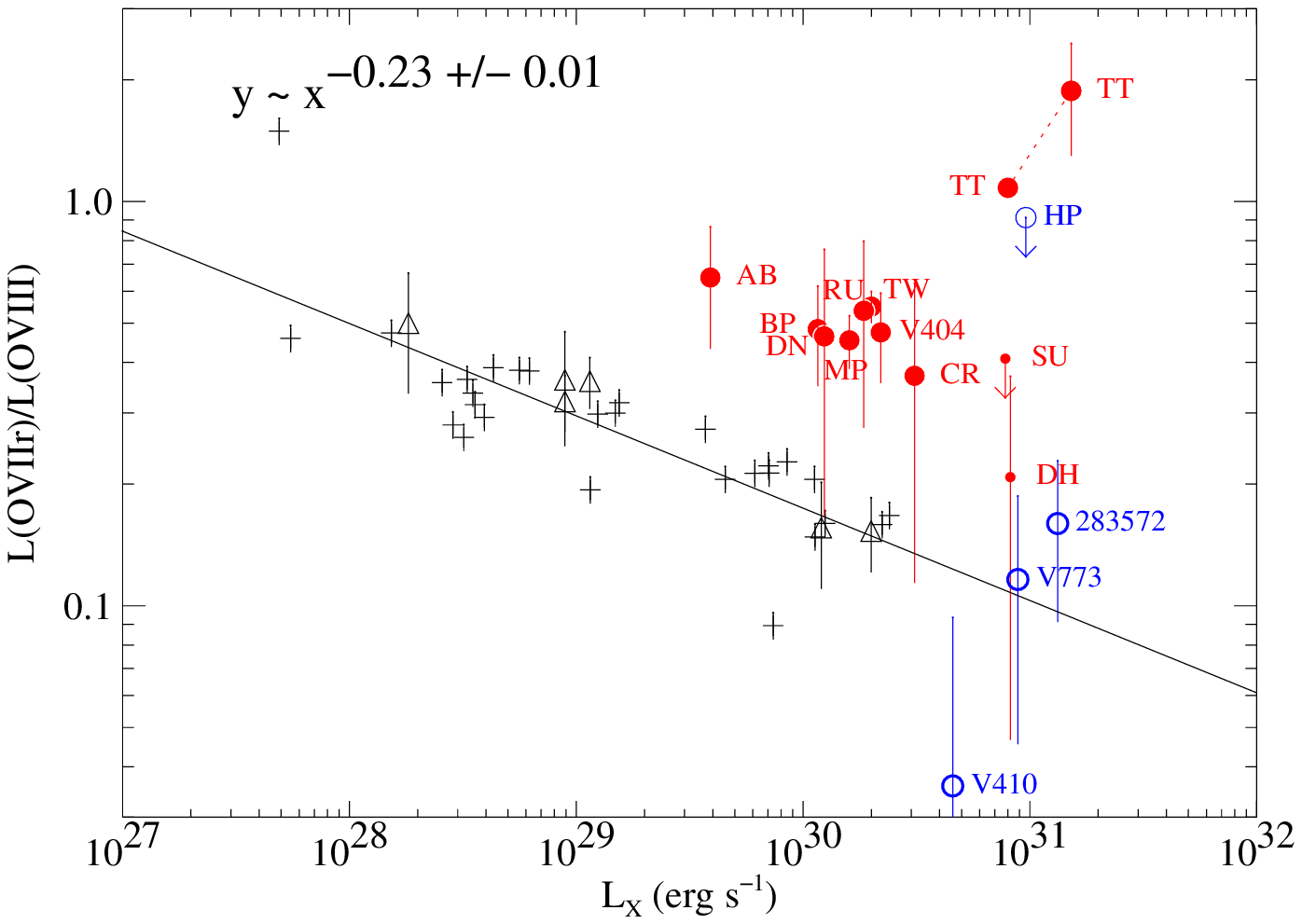}
  }
  \caption{\it  {\em Left:} The {\it ultraviolet excess} of
  CTTS. The figure shows the mean continuum surface flux at
  $\approx$1958\,\AA\ vs.\ the stellar effective temperature. CTTS are
  shown by the solid circles, naked (diskless) TTS by asterisks, and
  MS stars by squares. The solid lines define a fit to the MS stars,
  the dashed lines giving the lower and upper bounds
  \citep[from][reproduced by permission of AAS]{johnskrull00}. -- {\em Right:} The {\it X-ray soft
  excess} in CTTS. The figure shows the ratio between {\rm O\,{\sc
  vii}} r and {\rm O\,{\sc viii}} Ly$\alpha$ luminosities (each in
  units of $\mathrm{erg\ s^{-1}}$) vs.\ $L_\mathrm{X}$. Labels give
  initial letters of stellar names. Crosses mark MS stars
  \citep[from][]{ness04}, triangles solar analogs
  \citep[from][]{telleschi05}, filled (red) circles CTTS, and open
  (blue) circles WTTS. Two solutions (high- and low-absorption) are
  given for T~Tau, connected by a dotted line. The solid line is a
  power-law fit to the MS~stars with $L_\mathrm{X} > 10^{27}
  \unit{erg\ s^{-1}}$ \citep[from][reprined with permission]{guedel07c}.}
  \label{figure:softexcess}
\end{figure}}

\epubtkImage{}{%
\begin{figure}[t!]
  \centerline{\includegraphics[scale=0.55]{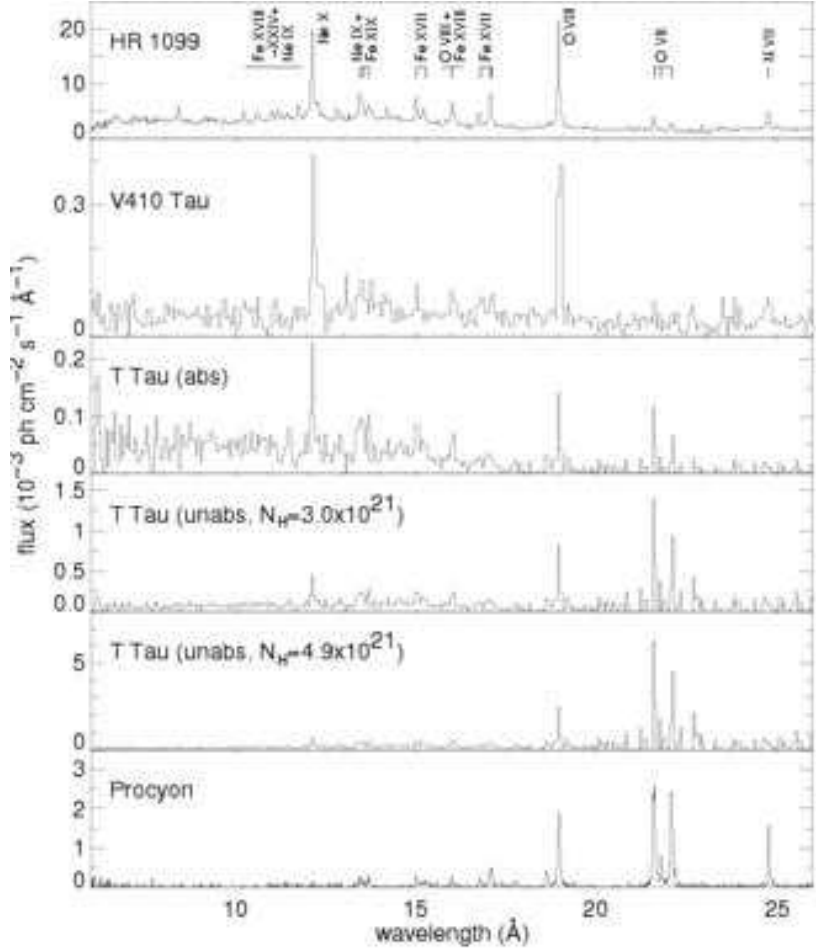}}
  \caption{\it Comparison of fluxed X-ray photon spectra of (from
	 top to bottom, all from {\it XMM-Newton RGS}) the active
	 binary HR~1099, the WTTS V410~Tau, the CTTS T~Tau, the T~Tau
	 spectrum modeled after removal of absorption (two versions,
	 using $N_\mathrm{H} = 3 \times 10^{21} \unit{cm^{-2}}$ and
	 $4.9 \times 10^{21} \unit{cm^{-2}}$, respectively), and the
	 inactive MS star Procyon.
	 The bins are equidistant in wavelength (from top to bottom,
	 the bin widths are, respectively, 0.025\,\AA, 0.070\,\AA,
	 0.058\,\AA, 0.058\,\AA, 0.058\,\AA, and 0.010\,\AA; from
	 \citealt{guedel07c}, reprinted with permission).}
  \label{figure:ttauspec}
\end{figure}}

Interestingly, however, the X-ray soft excess in CTTS is comparatively
moderate, the $L$(O\,{\sc vii} $r$)/$L$(O\,{\sc viii}) line flux ratio
being enhanced by factors of typically $\approx$3\,--\,4~times that of
equivalent MS stars or WTTS
(Figure~\ref{figure:softexcess}). Furthermore, the excess X-ray line
fluxes do not seem to be correlated with the UV line excesses but are
correlated with the overall stellar {\it coronal}  activity level
traced, for example, by the O\,{\sc viii} Ly$\alpha$ line flux
\citep{guedel07c}. It appears that the X-ray soft excess depends on
the level of magnetic (``coronal'') activity although it is, at the
same time,  related to the presence of accretion. The two dependencies
may point to an interaction between accretion and magnetic activity at
``coronal'' heights.

The shock interpretation of the softest X-rays and the X-ray soft
excess is appealing, but remains controversial until a larger sample
of CTTS with various accretion properties has been interpreted. In
particular, given the appreciable accretion rates, high shock
densities of order $10^{12}\mbox{\,--\,}10^{14} \unit{cm^{-3}}$ are
expected, as first indeed reported from density-sensitive line
diagnostics of O\,{\sc vii} and Ne\,{\sc ix} in the CTTS TW~Hya,
forming at only a few MK \citep{kastner02, stelzer04}.  However, some
accreting PMS stars show much lower densities, such as AB~Aur
\citep{telleschi07a} and T~Tau \citep{guedel07b}; the same discrepancy
between expected and observed densities has also been reported from
ultraviolet density diagnostics \citep{johnskrull00}.

In stark contrast to the X-ray soft excess and the UV excess described
above, it is now well established that CTTS show a moderate {\it
  suppression} of 0.1\,--\,10\,keV soft X-ray emission, typically by a
factor of $\approx$2 when compared with WTTS of similar properties
\citep{strom94, damiani95, neuhaeuser95, stelzer01, flaccomio03,
  preibisch05a, telleschi07b}. Although selection/detection bias or
different photoelectric absorption has been quoted to be responsible
for these differences (see \citealt{guedel04} and references therein),
the luminosity deficit in CTTS is now thought to be real; it appears
that accretion suppresses coronal heating in a fraction of the coronal
volume \citep{preibisch05a}, or at least leads to larger amounts of
cooler plasma, which is perhaps the same plasma inferred from the
X-ray soft excess \citep{telleschi07a, guedel07c}. Alternatively, the
presence of a circumstellar disk could strip the outer parts of the
stellar corona, thus reducing $L_\mathrm{X}$ \citep{jardine06}.

\subsubsection{X-Ray Flaring of the T~Tauri Sun}

A high level of near-continuous flaring is found in PMS solar-mass
stars. As much as half of the emitted X-ray energy, if not more, may
be due to strong flares \citep{montmerle83}, and many TTS are nearly
continuously variable probably also owing to flares \citep{mamajek00,
  feigelson02b, preibisch02, skinner03}. Examples with extreme
luminosities and temperatures up to 100\,MK have been reported
\citep[e.g.,][]{preibisch95, skinner97, tsuboi98, tsuboi00,
  imanishi02}. The most extreme flares are found on CTTS and
protostars, a possible hint at star-disk magnetic interactions during
flares. \cite{wolk05} studied frequency and properties of flares in
the Orion Nebula cluster, concluding that the median peak luminosity
of their sample was $\log L_\mathrm{X} = 30.97$, with extremely hard
spectra at peak time. The median electron temperature was found at
7\,keV. An analogous study has been presented by \cite{stelzer07} for
T~Tauri stars in the Taurus Molecular Cloud. The extreme flaring
recorded on these PMS stars may have an important bearing on coronal
heating (see Section~\ref{section:flares}) and on the alteration of
solids in the young stellar environment (see
Section~\ref{section:meteoritics}).

\subsubsection{The Radio T~Tauri Sun in Time}

Early VLA surveys quickly reported strong radio emission from both
CTTS and WTTS. Somewhat unexpectedly, however, radio emission comes in
two principal flavors. The early, pioneering studies by
\cite{cohen82}, \cite{bieging84}, \cite{cohen86}, \cite{schwartz84},
and \cite{schwartz86} recognized thermal wind-type emission with
rising spectra and in cases large angular sizes for several CTTS. This
radio emission can then be used to estimate mass loss rates; these are
found to range up to  $\la 10^{-7}\,M_{\odot} \unit{yr^{-1}}$
\citep{andre87}. The partly enormous kinetic wind energy derived under
the assumption of a uniform spherical wind suggests anisotropic
outflows while structural changes in the radio sources indicate
variable outflows, probably along jet-like features; at shorter radio
wavelength, dust emission from the disk becomes apparent as well
\citep{cohen86, rodriguez92, rodriguez94, wilner96}. The thermal radio
emission tells us nothing about the presence or absence of stellar
magnetic fields. As described earlier, CTTS do show many signatures of
magnetic activity, but whatever the possible accompanying radio
emission, it seems to be absorbed by the circumstellar ionized wind.
 
The situation is different in WTTS in which the presence of huge
flares \citep{feigelson85, stine88, stine98}, longer-term variability,
and falling spectra clearly point to non-thermal gyrosynchrotron
emission \citep{bieging84, kutner86, bieging89, white92a, felli93,
  phillips96} analogous to radio emission observed in more evolved
active stars.  Conclusive radio evidence for the presence of
solar-like magnetic fields in WTTS came with the detection of weak
circular polarization during flares but also in quiescence
\citep{white92b, andre92, skinner93}. Extremely energetic particles
radiating synchrotron emission may be involved, giving rise to linear
polarization in flares on the WTT star HD 283447
\citep{phillips96}. VLBI observations showing large ($\sim 10\,R_*$)
magnetospheric structures with brightness temperatures up to $T_b
\approx 10^9 \unit{K}$ fully support the non-thermal picture
\citep{phillips91}.
 
As a WTT star ages, its radio emission drops rapidly on time scales
of a few million years from luminosities as high as $10^{18}
\unit{erg\ s^{-1}\ Hz^{-1}}$ to values around or below $10^{15}
\unit{erg\ s^{-1}\ Hz^{-1}}$ at ages beyond 10\,Myr. Young age of a
star is thus favorable to strong radio emission \citep{oneal90,
  white92a, chiang96}, whereas toward the subsequent ZAMS stage it is
only the very rapid rotators that keep producing radio emission at the
$10^{15} \unit{erg\ s^{-1}\ Hz^{-1}}$ level \citep{carkner97,
  magazzu99, mamajek99}.

\subsubsection{The Composition of the T~Tauri Sun's Corona}

Initial studies of a few accreting T~Tau stars, in particular the old
($\approx 10 \unit{Myr}$) TW~Hya, have shown an abundance pattern in
the X-ray source similar to the IFIP effect although the Ne/Fe
abundance ratio is unusually high, of order 10 with respect to the
solar photospheric ratio, and the N/O and N/Fe ratios are enhanced by
a factor of $\approx$3.

These anomalous abundance ratios have been suggested \citep{stelzer04,
  drake05a} to reflect depletion of Fe and O in the accretion disk
where almost all elements condense into grains except for N
\citep{savage96,charnley97} and Ne \citep{frisch03} that remain in the
gas phase which is accreted onto the star. If accretion occurs
predominantly from the gas phase in the higher layers of the disk
  while the grains grow and/or settle at the disk midplane, then the
  observed abundance anomaly may be a consequence.

Larger systematics have made this picture less clear, however. Several
CTTS {\it and} WTTS have revealed large Ne/Fe ratios ($\approx$4 or
higher), much larger than in MS active solar analogs \citep{kastner04,
  argiroffi05, argiroffi07, telleschi05, telleschi07a, guenther06} but
similar to RS~CVn binaries \citep{audard03a}. In contrast, the CTTS SU
Aur reveals a low Ne/Fe abundance ratio of order unity
\citep{robrade06, telleschi07a}, similar to some other massive CTTS
\citep{telleschi07a}.

Partial clarification of the systematics has been presented by
\cite{telleschi07a} \citep[see also][]{guedel07b} who found that

\begin{itemize}

  \item the abundance trends, and in particular the Ne/Fe abundance
  ratios, do not depend on the accretion status but seem to depend on
  spectral type or surface $T_\mathrm{eff}$, the later-type stars
  showing a stronger IFIP effect (larger Ne/Fe abundance ratios); see
  Figure~\ref{figure:fene}.

  \item the same trend is also seen in disk-less ZAMS stars.

\end{itemize}

Anomalously high Ne/O abundance ratios remain, however, for TW~Hya
\citep{stelzer04} and V4046~Sgr \citep{guenther06} when compared to
the typical level seen in magnetically active stars, including PMS
objects. The initial idea proposed by \cite{drake05b} was that the
selective removal of some elements from the accretion streams should
occur only in old accretion disks such as that of TW~Hya where
cogulation of dust to larger bodies is ongoing, whereas younger
T~Tauri stars still accrete the entire gas and dust phase of the inner
disk. However, the old CTTS MP~Mus does not show any anomaly in the
Ne/O abundance ratio \citep{argiroffi07}. Larger samples are needed
for clarification.

\epubtkImage{}{%
\begin{figure}[htbp]
  \centerline{\includegraphics[scale=0.7]{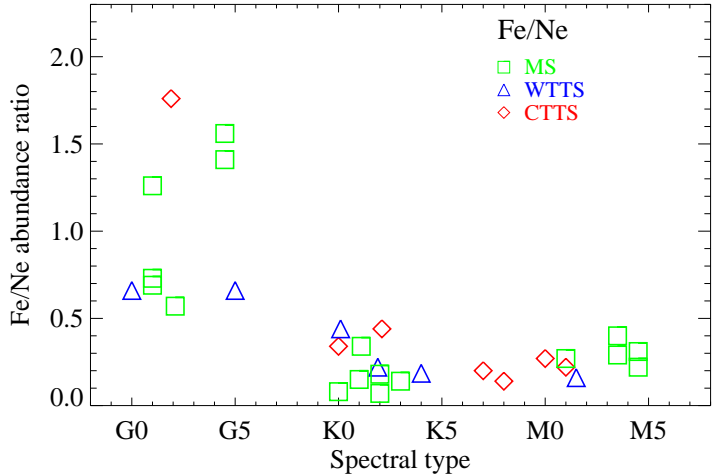}}
  \caption{\it The coronal {\rm Fe/Ne} abundance ratio (with respect
  to the solar photospheric mixture) in the X-ray sources of various
  PMS stars and very active MS stars as a function of spectral class. Symbols
  mark different types of stars: squares: active MS stars; triangles: WTTS;
  diamonds: CTTS. For details and references, see \cite{telleschi07a}
  and \cite{guedel07b}.}
  \label{figure:fene}
\end{figure}}

\subsection{The Protostellar Sun}

\subsubsection{Magnetic Activity in the Protostellar Sun}

The protostellar Sun was deeply embedded in its molecular cloud
envelope. Direct optical observation of protostars is preluded by high
extinction; the best access to these objects is through infrared and
X-ray observations, the former, however, picking up much of the
emission from dust in the disk and the envelope. Recent efforts have
succeeded in obtaining {\it photospheric} spectra of Class~I
protostars from light scattered off the bipolar cavities carved by the
outflows \citep{white04}. Somewhat perplexingly, the analysis of the
Taurus sample of embedded objects reveals very little statistical
difference between Class~I and CTTS objects in Taurus, for example
with respect to rotation rates, accretion rates, bolometric
luminosities, spectral classes, and disk masses, providing evidence
that this sample is co-eval with the CTTS sample; the embedded objects
seem to be those with the longest envelope dispersal time, while they
are all past their main accretion phase \citep{white04}.

Magnetic field measurements are commensurately
challenging. \cite{krull07b} recently succeeded in obtaining
near-infrared diagnostics to measure the photospheric magnetic field
on a class~I protostar. He reports a field strength of 3.6\,kG, making
this the highest mean surface field so far detected on any young
stellar object.

Direct evidence for magnetic activity is seen in X-rays. Strong X-ray
activity is found in considerable numbers of ``Class~I'' protostars
thanks to {\it Chandra}'s and {\it XMM-Newton}'s hard-band sensitivity
\citep[see, e.g.,][]{imanishi01, preibisch01, preibisch02,
  preibisch03b, getman02, guedel07a}. Their measured characteristic
temperatures are very high, of order 20\,--\,40\,MK
\citep{tsujimoto02, imanishi01}. Some of these values may, however, be
biased by strongly absorbed (``missing'') softer components in
particular in spectra with limited signal-to-noise ratios. It is
correspondingly difficult to characterize the $L_\mathrm{X}$ values in
traditional soft X-ray bands for comparison with more evolved stars.

\subsubsection{Magnetic Flaring of the Protostellar Sun}

Strong X-ray flaring is a characteristic of protostellar solar
analogs. Many of these events are extremely large, with total soft
X-ray energies of up to $\approx 10^{37} \unit{erg}$ \citep{koyama96,
  kamata97, grosso97, ozawa99, imanishi01, preibisch03a,
  imanishi03}. Such flares realistically require large volumes, in
fact to an extent that star-disk magnetic fields become a possibility
for the flaring region (\citealt{grosso97} for YLW~15 in $\rho$~Oph),
with important consequences for the irradiation of the stellar
environment by high-energy photons and particles (see
Section~\ref{section:environment}).

\subsubsection{Radio Emission from the Protostellar Sun}

At radio wavelengths, genuine, embedded class I protostars have most
often been detected as thermal sources, and this emission is
predominantly due to collimated thermal winds or jets. These jets are
probably ionized by neutral winds that collide with the ambient medium
at distances of around 10\,AU and that are aligned with molecular
outflows (e.g., \citealt{bieging85, snell85, brown87, curiel89,
  rodriguez89, rodriguez03, anglada95, anglada98}). Ionized
circumstellar material and winds easily become optically thick and
therefore occult any non-thermal, magnetic emission from close to the
star. However, the discovery of polarization in T~Tau(S)
\citep{phillips93, smith03}, in IRS~5 \citep{feigelson98}, in
protostellar jet sources \citep{yusef90} and the jet outflows
themselves \citep{curiel93, hughes97, ray97}, as well as variability
and negative spectral indices in T~Tau(S) \citep{skinner94} provided
definitive evidence for magnetic fields and particle acceleration
around these class~I objects.

\subsection{The Pre-Main Sequence Sun's Environment in Time}
\label{section:environment}

The molecular and dust environment of the very young (PMS) Sun was
affected in several ways by ultraviolet radiation, high-energy
radiation and high-energy particle streams emitted by the Sun. These
often neglected effects have recently attracted considerable
attention. Although they apply to the environment of any forming star
that is magnetically active, they have been considered in particular
for solar-like stars and specifically for the young Sun itself because
of various traces that may be observable in the solid bodies of the
present solar system. I give a brief overview of the themes in so far
as they may relate to the past solar activity.

\subsubsection{Circumstellar Disk Ionization}

In terms of physical processes, ionization of the disk is important
for the operation of the magnetocentrifugal instability (MRI;
\citealt{balbus91}) thought to be the main driver of accretion in
young stellar objects. Although cosmic rays have long been suspected
to be an effective disk ionization source \citep{gammie96}, the
high-level, hard coronal emission and frequent stellar flares may be
more effective ionizing sources \citep{glassgold97,
  feigelson02a}. This is even more so as young solar analogs in the
T~Tau stage drive very strong winds that are very likely magnetized;
such winds effectively shield the inner disk from cosmic rays, as does
the present-day solar wind, at least for cosmic rays with energies
$<100 \unit{MeV}$.

The distance to which stellar X-ray ionization dominates over that
produced by galactic cosmic rays can be estimated to be
\citep{glassgold97, montmerle01}

\begin{equation}
  D \approx 0.02 \unit{pc} 
  \left(
       {\left[{E \over 1 \unit{keV}} \right]^{-2.485}} {L_\mathrm{X}
        \over 10^{29} \unit{erg\ s^{-1}}} \left[{\zeta \over 10^{-17}
        \unit{s^{-1}}} \right]^{-1}J_0
  \right)^{1/2}
\end{equation}

where $E$ is the photon energy in the range of 1\,--\,20\,keV (for
other energies, the first term in the parentheses must be generalized
to $\sigma(kT_\mathrm{X})/\sigma(1 \unit{keV})$), $\zeta$ is the
cosmic ray ionization rate, and $J_0$ is an attenuation factor ($J_0
\approx 0.13$ for optical depth of unity for a 1\,keV photon). The
values given in the parentheses are characteristic for our situation,
with the cosmic-ray ionization rate referring to a UV shielded
molecular core. Stellar X-ray ionization therefore dominates cosmic
ray ionization out to about 1000\,AU, i.e., across most of the radius
of a typical circumstellar disk. Taking into account frequent, strong
flares, significant portions of molecular cores may predominantly be
ionized by the central star rather than by cosmic rays.

\cite{glassgold97} and \cite{igea99} modeled ionization and heating
of circumstellar disks by stellar coronal X-ray sources. The incoming
X-ray photons are subject to Compton scattering and photoelectric
absorption as they propagate through the disk. X-ray photons may
interact with molecules or atoms by ejecting a fast (primary)
photoelectron. This photoelectron collisionally produces on average
27~secondary electrons and ions (for a photon energy of
1\,keV). Harder photons on average penetrate deeper and thus ionize
layers of the disk closer to the equatorial plane, while softer X-rays
ionize closer to the disk surface. The disk ionization fraction is
then determined when an equilibrium between ionization and
recombination has been reached. Electron fractions of
$10^{-15}\mbox{\,--\,}10^{-10}$ are obtained at vertical disk column
densities of $N_\mathrm{H} = 10^{27}-10^{21} \unit{cm^{-2}}$ (as
measured from infinity) for distances of 0.1\,--\,10\,AU from the
central star. The precise results depend somewhat on the hardness of
the X-ray spectrum (a modest $L_\mathrm{X} = 10^{29} \unit{erg\
  s^{-1}}$ has been assumed).

\epubtkImage{}{%
\begin{figure}[t!]
  \centerline{\includegraphics[scale=0.64]{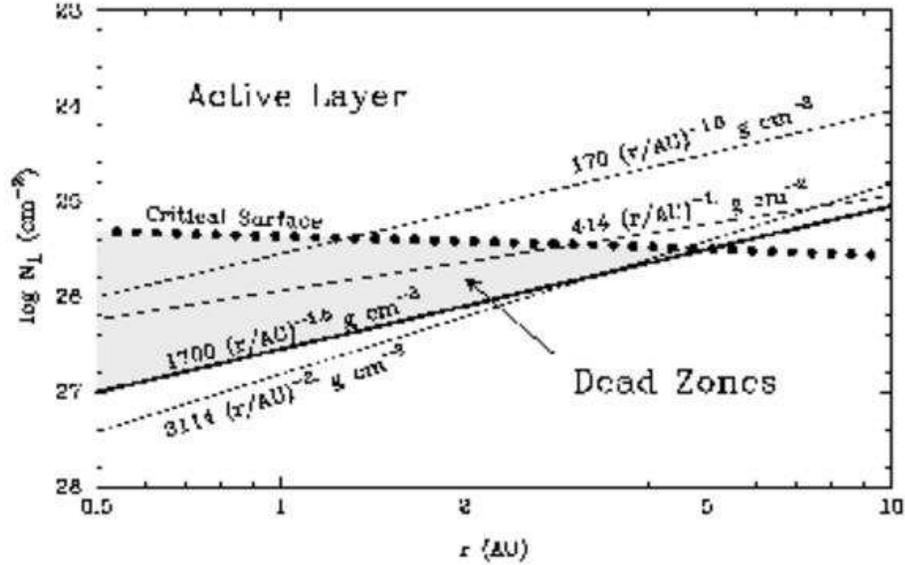}}
  \caption{\it Consequences of X-ray irradiation of a circumstellar
  disk, shown in the r vs.\ $N_{\perp}$ (vertical gas column density)
  plane. The critical surface (filled circles) separates regions that
  are coupled to the magnetic field based on a sufficiently high
  ionization fraction (above) from deeper regions that are uncoupled
  (below). The various solid lines define the disk midplanes for
  specific disk models, the bold line characterizing the minimum solar
  nebula \citep[from][reproduced by permission of AAS]{igea99}.}
  \label{figure:diskirradiation}
\end{figure}}

The important points here are, 1) that the ionization fraction at the
top of the disk is orders of magnitude higher than the ionization
fraction that would result from standard cosmic-ray irradiation; and
2)  that at vertical column densities of $N_\mathrm{H} =
10^{24}\mbox{\,--\,}10^{25} \unit{cm^{-2}}$ and less, the disk is
sufficiently ionized to become unstable against MRI. The disk surface
will thus couple to the magnetic field and accrete to the star. In
contrast, the deeper layers remain decoupled and therefore
``quiescent'', at least within 5\,AU
(Figure~\ref{figure:diskirradiation}, \citealt{igea99}). These are the
likely sites of planet formation \citep{glassgold97}. Modifications of
these calculations by introducing trace heavy metals and diffusion
have been discussed by \cite{fromang02} and \cite{ilgner06}.

Disk irradiation and photoionization by stellar UV photons is further
responsible for photoevaporation of gaseous disks \citep{hollenbach94,
  clarke01,  alexander06a, alexander06b}, and therefore the long-term
accretion history of the star-disk system. Additional X-ray
irradiation is, however, of secondary importance only
\citep{alexander04}.

\subsubsection{Circumstellar Disk Heating}

 Apart from disk ionization, X-ray irradiation also leads to disk
 heating \citep{igea99, glassgold04}. While dust disks are heated by
 the central star's optical and UV light to a few 100\,K at distances
 up to a few AU, the gas component may thermally decouple in
 particular in the upper layers where the density is small. A model
 calculation based on accretion viscosity heating combined with X-ray
 heating due to the central star shows that the upper layers of the
 gaseous disk ($N_\mathrm{H} \la 10^{21} \unit{cm^{-2}}$) can be
 heated up to $\approx 5000 \unit{K}$
 (Figure~\ref{figure:diskheating}). This holds even for low viscous
 heating efficiency where the X-ray heating contribution entirely
 dominates \citep{glassgold04}. At the same time, the strong
 temperature gradients in the temperature inversion region lead to the
 production of large amounts of ``warm'' CO. Similar calculations
 by \cite{gorti04} support the above picture of X-rays dominating gas
 heating at the disk surface.

\epubtkImage{}{%
\begin{figure}[t!]
  \centerline{\includegraphics[scale=0.6]{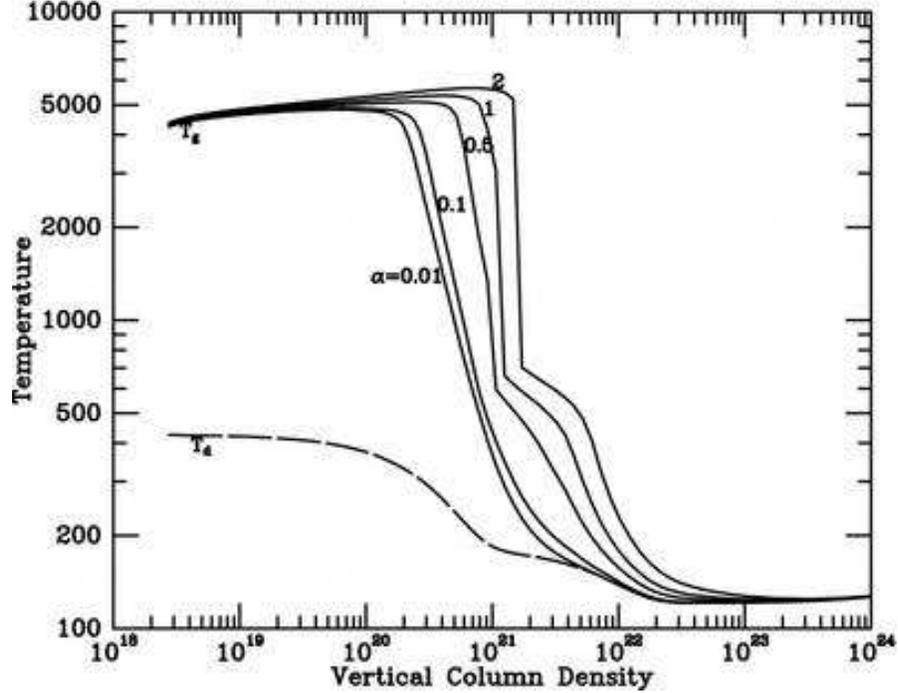}}
  \caption{\it Gas and dust temperatures of a circumstellar disk as a
  function of vertical column density. Different solid lines give the
  gas temperature for various viscous-heating rates. The leftmost
  curve is predominantly due to X-ray heating, in the rightmost curve,
  viscous heating is important. The dashed line gives the dust
  temperature from a standard dust-disk model \citep[figure
  from][reproduced by permission of AAS]{glassgold04}.}
  \label{figure:diskheating}
\end{figure}}

\subsubsection{Observational Evidence of Disk Irradiation}

The elevated ultraviolet and X-ray activity level of young
low-mass stars leads to significant irradiation of circumstellar
accretion disks. Interactions between high-energy photons and disk
matter is evident from X-ray photoabsorption in star-disk systems seen
edge-on, but also  from reprocessed starlight: Spatially
  unresolved FUV fluorescence lines of H$_2$ have been detected from
  several CTTS \citep{brown81,valenti00,ardila02,herczeg02,herczeg06},
  but usually not from WTTS \citep{valenti00}; this emission is
  reprocessed stellar Ly$\alpha$  emission most intensely radiated
  from the accretion spots. At least in cases where the line radial
  velocities are coincident with stellar radial velocities, an origin
  of the fluorescence in a hydrogenic surface layer of the inner
  accretion disk at temperatures of 2000\,--\,3500\,K is likely
  \citep{herczeg02, herczeg04} although significantly blueshifted
  $\mathrm{H_2}$ emission points to outflow-related fluorescence in
  some systems \citep{brown81,ardila02,walter03,saucedo03,herczeg06}.
Fluorescent emission is also generated by reprocessing of X-ray
photons; X-ray fluorescence is seen in particular in the 6.4\,keV line
of cold iron \citep{imanishi01, tsujimoto05, favata05}.

Simple energy considerations are revealing: \cite{herczeg04} estimate
the rate of energy deposited in the environment of the CTTS TW~Hya as
a result of Ly$\alpha$ photoexcitation and subsequent far-ultraviolet
fluorescence of $\mathrm{H}_2$ to be $1.4 \times 10^{29} \unit{erg\
  s^{-1}}$. The total soft X-ray luminosity of $\approx 1.4 \times
10^{30} \unit{erg\ s^{-1}}$ will at least partially heat the disk
surface layer further \citep{igea99}.

Warm $\mathrm{H}_2$ has also been detected through infrared $2.12
\unit{\mu\,m}$ ro-vibrational emission from several T~Tauri stars
\citep{weintraub00, bary03}. This emission is thought to be excited by
collisions between $\mathrm{H}_2$ molecules and X-ray induced
non-thermal electrons, or by an UV radiation field. High temperatures,
of order 1000\,--\,2000\,K, are required, but such temperatures are
predicted from disk irradiation by X-rays out to several AU
\citep{glassgold04}, or by UV radiation from the central star out to
$\approx 10 \unit{AU}$ {\it if} the star shows an UV excess
\citep{nomura05}.

\cite{glassgold07} proposed forbidden [Ne\,{\sc ii}] and [Ne\,{\sc iii}] 
infrared line emission at $12.81 \unit{\mu\,m}$ and $15.55 \unit{\mu\,m}$,
respectively, to be indicative of X-ray irradiation. The high first
ionization potential of Ne (21.6\,eV) indeed requires Ly continuum or
X-ray photons for ionization (or cosmic rays, which are unlikely to be
abundant in the inner disk region). The transitions are collisionally
excited in warm gas, requiring temperatures of a few 1000\,K, attained
in disk surface layers out to about 20\,AU for X-ray irradiated disks
\citep{glassgold04}. The [Ne\,{\sc ii}] $12.81 \unit{\mu\,m}$
transition has indeed been detected in several CTTS \citep{pascucci07,
  lahuis07, ratzka07}.

As a further consequence, specific chemical reactions may be
induced. For example, Ly$\alpha$ itself can dissociate molecules
  like $\mathrm{H}_2$ and $\mathrm{H_{2}O}$ and can ionize Si and C
  \citep{herczeg04}. Ly$\alpha$ radiation photodissociates HCN (but
  not CN), which leads to an enhancement of CN relative to HCN
  \citep{bergin03}.

\subsection{The T~Tauri Sun's Activity and Meteoritics}
\label{section:meteoritics}
 
The presence of chondrules and isotopic anomalies in chondritic
meteorites has posed one of the most outstanding problems in our
understanding of the young solar system. {\it Chondrules} are
millimeter-sized spheres of igneous rock embedded in the meteoritical
matrix; their content and structure suggests that they were heated to
melting temperatures ($\approx 2000 \unit{K}$) of solid iron-magnesium
silicates for only an hour or less. They must have cooled in an
ambient magnetic field of $\approx 10 \unit{G}$ \citep{shu01}. {\it
  Calcium-aluminum-rich inclusions (CAIs)} are structures in
meteorites that vary in shape; they may derive from melt or partial
melts.  CAIs contain evidence for short-lived radionuclides in the
young solar system, in particular of $^{26}\mathrm{Al}$,
$^{41}\mathrm{Ca}$, $^{53}\mathrm{Mn}$, and $^{60}\mathrm{Fe}$ with
half-lives of 1.1, 0.15, 5.3, and 2.2\,Myr, respectively \citep{lee98,
  gounelle01}.

Conventionally, it has been assumed that these short-lived isotopes
were injected by external stellar nucleosynthetic events (ejecta from
AGB stars, Wolf--Rayet stars, supernova explosions) that at the same
time triggered the collapse of the parent molecular cloud, to form the
solar system. The fundamental problem with external seeds is the short
time required between the formation and injection of live
radionuclides and their incorporation into solid CAI structures; this
time span should not exceed $10^5 \unit{yrs}$ \citep{lee98}, i.e., the
trigger for the formation of the solar system must have been extremely
fast. Observations of galactic star-formation regions show
star-forming molecular cloud cores to be rarely within the immediate
environment of Wolf--Rayet wind bubbles or supernova shells
\citep{lee98};  the association of asymptotic giant branch stars
  with star-forming regions is also very small, making injection of
  $^{60}\mathrm{Fe}$ into the solar system from such a star improbable
  \citep{kastner94}. Also, the assembly of CAIs and chondrules into
planetesimals takes much longer, of order 5\,Myr \citep{lee98}.

These problems can be removed if the short-lived radionuclides were
formed locally, namely by bombardment with ``cosmic rays'' ejected by
stellar flares \citep{lee98}. While this alternative for the
production of the radionuclides is not unanimously accepted or may not
be responsible for {\it all} isotopic anomalies in meteorites
\citep[e.g.,][]{goswami01, wadhwa07}, I discuss only this hypothesis
here as it is directly related to the (undisputed) high activity level
of the young Sun in its T~Tauri stage.

There is indeed substantial evidence for an active early Sun not only
from inferences from active, young solar analogs
(Section~\ref{section:solarflux}), but also from large enrichments of
spallation-produced $^{21}\mathrm{Ne}$ and $^{38}\mathrm{Ar}$ in
``irradiated'' meteorite grains (i.e., grains that show radiation
damage trails from solar-flare Fe-group nuclei), compared to
``non-irradiated'' grains (\citealt{caffee87}; a summary of further,
earlier, albeit ambiguous evidence can be found in
\citealt{newkirk80}). Galactic cosmic-ray irradiation would require
exposure times in excess of $10^8$~yr for some of these grains,
incompatible with other features of the meteorites
\citep{caffee87}. Alternatively, energetic solar flare protons could
be responsible, but the present-day level would again be insufficient
to explain the anomaly. \cite{caffee87} concluded that an elevated
particle flux, related to the enhanced magnetic activity of the young
Sun, naturally explains the meteoritic spallation-produced $^{21}$Ne
enrichment. A flux several orders of magnitude in excess of
present-day values and a harder energy spectrum would be required.
 
Energetic protons required for the generation of radionuclides could
be generated in various places in the extended stellar
magnetosphere. In the ``$x$-wind'' model proposed by \cite{shu97,
  shu01}, magnetic reconnection flares occur at the inner border of
the accretion disk where closed stellar magnetic fields and open disk
fields converge. Flares would flash-melt protochondrules, and the
$x$-wind would eject them to larger solar distances. Radionuclides
would be synthesized by flare proton bombardment.

Alternatively, the elevated activity of the central star itself may be
sufficient to produce the required proton flux at planetary
distances. \cite{feigelson02a} estimated the proton flux at 1\,AU of a
solar analog in its T~Tauri phase, from a statistical X-ray study of
T~Tauri stars in the Orion Nebula Cluster. They found that frequent
flares on T~Tauri stars are $10^{1.5}$~times more luminous than the
largest solar flares (or $10^4$~times more than solar flares that
occur with a daily frequency). These same flares occur at a rate about
$10^{2.5}$ higher than the rate of the largest solar flares. As solar
proton fluxes scale non-linearly with the solar X-ray luminosity,
\cite{feigelson02a} estimated a proton flux about $10^5$ times higher
than at present (i.e., $10^7 \unit{protons\ cm^{-2}\ s^{-1}}$ at
1\,AU). Given the high flare rate, this flux was probably present
almost continuously.

Regardless of the location of the proton acceleration (flaring)
source, I now summarize the relevant results for various
radionuclides. For example,  $^{41}\mathrm{Ca}$ is predominantly
produced through

\begin{eqnarray}
  \mathrm{^{42}Ca(p, pn)^{41}Ca} \\
  \mathrm{^{40}Ca(\alpha, ^3He)^{41}Ca} \\
  \mathrm{^{40}Ca(\alpha, ^3He)^{41}Sc}
\end{eqnarray}

where $^{41}\mathrm{Sc}$ electron-captures to $^{41}\mathrm{Ca}$
\citep{lee98}. Note that the abundance ratio for
$^{42}\mathrm{Ca}/^{40}\mathrm{Ca}$ is $6.7\times 10^{-3}$
\citep[][and references therein]{lee98}. Summing all three production
channels, an isotopic ratio $^{41}$Ca/$^{40}$Ca as inferred from CAIs
requires a proton flux of $5 \times 10^3\mbox{\,--\,}10^4$~times the
present-day value during an irradiation time scale of $5\times
10^5-10^6$~yrs \citep{goswami01}. This approximately matches the
observational implications from T~Tauri flares by \cite{feigelson02a}.

Similar considerations for $^{26}\mathrm{Al}$ lead to an
underproduction by a factor of 20 under the same conditions
\citep{lee98}. \cite{goswami01} require a proton flux $10^5$ times as
strong as the present flux at 1\,AU to explain the inferred $^{26}$Al
abundance in the forming CAIs, and irradiation times of about 1\,Myr;
this is in excellent agreement with the observational inferences made
by \cite{feigelson02a}.  However, $^3\mathrm{He}$ bombardment of
$^{24}\mathrm{Mg}$ may efficiently produce $^{26}\mathrm{Al}$ as
well. $^3\mathrm{He}$ is preferentially accelerated in solar {\it
  impulsive} (as opposed to gradual) flares (see discussion in
\citealt{lee98} and references therein). The problem then arises that
$^{41}\mathrm{Ca}$ is overproduced by two orders of magnitude through
reactions involving $^3\mathrm{He}$. \cite{shu97, shu01} therefore
proposed that CAIs consisting of refractory, Ca-Al rich material are
surrounded by thick mantles of less refractory, Mg-rich
material. $^3$He nuclei would therefore be stopped in the outer mantle
where $^{26}\mathrm{Al}$ is produced from $^{24}\mathrm{Mg}$, while
the $^{40}\mathrm{Ca}$-rich interior remains less affected, i.e.,
$^{41}\mathrm{Ca}$ production is suppressed. Canonical isotopic ratios
can then indeed be derived for most of the species of interest
\citep{gounelle01}.

The most promising support for local irradiation by solar (or
possibly, trapped cosmic ray) protons has been the discovery of
$^{10}\mathrm{Be}$ (\citealt{mckeegan00}; half-life of 1.5\,Myr) and
possibly also the extremely short-lived $^7\mathrm{Be}$
(\citealt{chaussidon06}; half-life of 53\,d). The $^{10}\mathrm{Be}$
isotope could be entirely produced by solar protons and
$^4\mathrm{He}$ nuclei at asteroidal distances \citep{gounelle01,
  marhas04} while it is destroyed in the alternative nucleosynthetic
production sources such as massive stars or supernova explosions. If
the presence of $^7\mathrm{Be}$ in young meteorites can be confirmed,
then its short half-life precludes an origin outside the solar system
altogether and requires a local irradiation source.

Despite the successful modeling of radionuclide anomalies in early
CAIs, at least the case of $^{60}\mathrm{Fe}$ remains unsolved in this
context. It is difficult to synthesize by cosmic-ray reactions; the
production rate falls short of rates inferred from observations by two
orders of magnitude \citep{lee98, goswami01}  and requires stellar
  nucleosynthesis or, most likely, a supernova event \citep{meyer00}.
 
Although the formation of radionuclides in early meteorites is under
debate \citep{goswami01} and may require several different production
mechanisms (\citealt{wadhwa07}; for example, to explain the
  simultaneous presence of the $^{60}\mathrm{Fe}$ and the
  $^{10}\mathrm{Be}$ isotopes), the above models are at least
promising in explaining some nuclear processing of solar-system
material without external irradiation source but with sources whose
presence cannot be disputed, namely high-energy particle populations
that are a direct consequence of the magnetic activity of the young
Sun. Isotopic anomalies in meteorites have opened a window to the
violent environment of the young solar system.


\subsection{Summary: The Violent Pre-Main Sequence Sun}

Both the T~Tauri and the protostellar Sun were extremely magnetically
active, as far as we can tell from observations of contemporaneous
objects at these stages. There is no definitive evidence for a turn-on
of magnetic activity \citep[for a contrasting view, see][]{linsky07},
but the stellar environment (gas and dust disks and envelopes) makes
observations challenging. On the other hand, perhaps the most
interesting aspect of magnetic activity in this phase is indeed its
influence on the environment itself. High X-ray and UV fluxes produced
both by magnetic activity and magnetically funneled accretion flows
onto the star heat and ionize the circumstellar disk, thus controlling
mechanisms as diverse as gas-disk photoevaporation, accretion through
the magnetorotational instability, and chemical networks across the
disk. Strong flaring and observations of non-thermal gyrosynchrotron
emission in T~Tauri and protostars further indicate the presence of
strongly elevated particle fluxes in the pre-main sequence Sun's
environment. The impact on solid-state matter forming in the
circumstellar disk may be visible to the present day, in the form of
daughter products of short-lived radioactive isotopes formed by proton
impact. The study of the young Sun's environment under the aspect of
magnetic activity of the central star is still in its early stage and
rapidly developing.


\newpage

\section{The Solar System in Time: Solar Activity and Planetary Atmospheres}
\label{solarsystem}

\subsection{The Faint Young Sun Paradox: Greenhouse or Deep Freeze?}
\label{section:faintsun}

Standard solar models imply that the Sun's bolometric luminosity has
monotonically increased during the past 4.6\,Gyr. At its ZAMS stage
($t = -4.6 \unit{Gyr}$), the solar luminosity was only 70\,--\,75\%
its present-day level. In the absence of an atmosphere, the Earth's
surface equilibrium temperature would then have amounted to 355\,K
\citep{sagan97}. Assuming an albedo and an atmospheric composition
equal to values of the present-day Earth, the mean surface temperature
would have been below the freezing point of seawater until $\approx
2 \unit{Gyr}$ ago \citep{sagan72}. The increase of the surface
temperature in the presence of an atmosphere is due to the greenhouse
effect, i.e., the property of the atmospheric mixture to be
transparent to incoming optical and near-infrared light but to absorb
mid-infrared emission that has been re-emitted from the surface. The
present-day atmosphere of the Earth raises the surface temperature by
only 33\,K compared to the atmosphere-free equilibrium value
(\citealt{sagan72, kasting03}, see Figure~\ref{figure:greenhouse}).

\epubtkImage{}{%
\begin{figure}[htbp]
  \centerline{\includegraphics[scale=0.57]{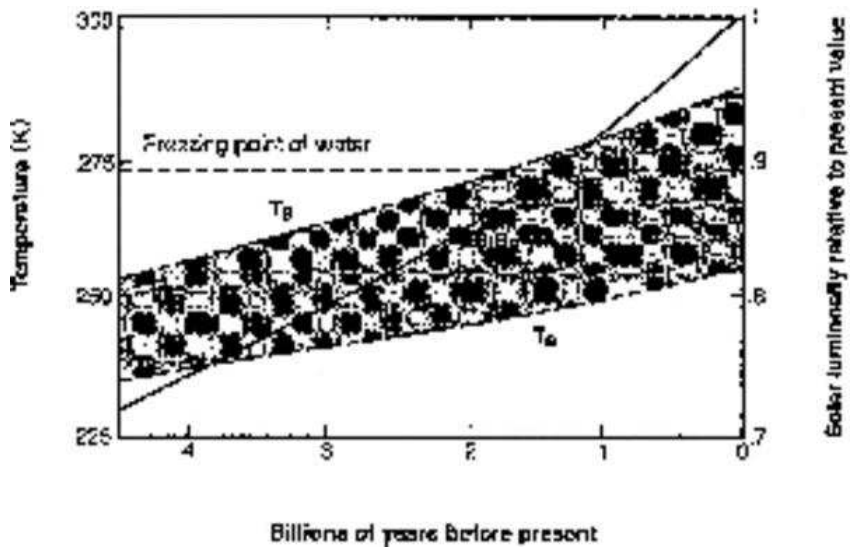}}
  \caption{\it Illustration of the greenhouse and the Faint Young Sun
  Paradox for the Earth. The solid line indicates the solar luminosity
  relative to the present value (right y axis); the lower dashed curve
  is the effective radiating temperature of the Earth (i.e., its
  surface being treated as a blackbody radiator); the upper dashed
  curve shows the calculated, mean global surface temperature affected
  by the greenhouse ($\mathrm{CO}_2$ mixing ratio and relative
  humidity have been kept fixed; from \citealt{kasting03}, reprinted, 
  with permission, from the Annual Review of Astronomy and Astrophysics, Volume 41 
  \copyright\ 2003 by Annual Reviews, www.annualreviews.org).}
  \label{figure:greenhouse}
\end{figure}}

On the other hand, there is clear geologic evidence for a warm climate
in the early Earth's history, with average temperatures perhaps even
significantly above present-day values (\citealt{kasting89a,
  kasting89b, karhu86}, see also summary of further evidence in
\citealt{sackmann03}). Sedimentary rocks \citep{bowring89} and
indirect evidence of microbial life in rock dated to 3.8\,Gyr ago
(\citealt{mojzsis96}; see also summaries by \citealt{nisbet00} and
\citealt{nisbet01}) clearly suggest the presence of liquid water
\citep[see also][]{kasting89a}.

There is similar geological evidence for a warmer young Mars, as seen
in particular in extensive channels formed by massive streams of
liquid water. But again, the mean Martian surface temperature is too
low for the presence of liquid water, and the less intense
photospheric light from the younger Sun would obviously have
aggravated this problem.

The apparent contradiction between implications from the standard
solar model and the geologic evidence for a warm early climate on the
Earth and Mars is known as the ``Faint Young Sun Paradox''
\citep{kasting89a, kasting91b}. This problem has been addressed along
two major lines of argumentation of which one (assuming a higher mass
loss rate for the young Sun) indirectly relates to the magnetic
activity of the young Sun; the other (arguing with greenhouse gases)
may be related in as of yet unknown ways to magnetic activity as well
(via atmospheric chemistry, see Section~\ref{section:chemistry}),
although other processes in no ways related to solar magnetic fields
may be relevant. After reviewing these two approaches, I will discuss
model calculations that specifically address the influence of magnetic
activity on planetary atmospheres (Section~\ref{section:planets}).

\subsubsection{The Relevance of Greenhouse Gases}
\label{section:greenhouse}

Today's modest greenhouse effect is due to atmospheric $\mathrm{CO}_2$
and $\mathrm{H_{2}O}$ \cite[see, e.g.,][for a summary]{kasting89a}. A
  stronger greenhouse  could have been effective in a different
  atmosphere of the young Earth:

1) A higher content of atmospheric gaseous $\mathrm{CO}_2$ might
increase the greenhouse effect \citep{owen79, cess80} as in the
present-day atmosphere of Venus, but a 100fold increase compared to
present-day levels would be required \citep{kasting89a}; such levels
of $\mathrm{CO}_2$ are plausible because the carbonate-silicate
geochemical cycle (which binds $\mathrm{CO}_2$ dissolved in rainwater
to silicate minerals in the soil) operates in such a way that removal
of atmospheric $\mathrm{CO}_2$ increases with increasing temperature,
thus inducing a negative feedback loop between CO$_2$ greenhouse
warming and $\mathrm{CO}_2$ removal \citep{kasting89a}. However, the
absence of siderite in old soils argues against the required high
levels of $\mathrm{CO}_2$ \citep{rye95}. For Mars, high levels of
$\mathrm{CO}_2$ in a higher-pressure atmosphere would condense in
clouds. The resulting increased global albedo would in fact lead to a
net cooling \citep{kasting91a}. Although the same clouds may also
back-scatter radiation and therefore support the greenhouse 
\citep{forget97}, experiments suggest that this mechanism is too small
to rise the temperatures above the freezing point \citep{glandorf02}.

2) Greenhouse gases such as $\mathrm{NH}_3$ \citep{sagan72} and
   $\mathrm{CH}_4$ \citep{sagan97} could have been present in
   appreciable amounts in the young atmospheres, in analogy to the
   present-day atmosphere of Titan. However, $\mathrm{NH}_3$
   dissociates rapidly due to solar UV radiation (\citealt{kuhn79};
   see also \citealt{owen79}). While $\mathrm{CH}_4$ is subject to UV
   dissociation as well, its lifetime is much longer, and biological
   activity could regenerate it at sufficiently high levels
   \citep{pavlov00}. Moreover, its photolysis may produce a
   high-altitude haze of organic solids that shields ammonia
   sufficiently from UV dissociation \citep{sagan97}.

How the changes in the atmospheres came about is not entirely clear
but may partly be related to the past solar activity (apart from,
e.g., weathering, plate tectonics, volcanism, and biological
activity).  Although enhanced levels of solar EUV and X-ray emission
will not directly alter the lower planetary atmospheres but only
affect the higher thermosphere (where this radiation is absorbed) and
the exosphere (see Section~\ref{section:planets} below), the complex
chemistry induced by photoionization,  photodissociation, and heating
through enhanced high-energy irradiation may be a key factor in
determining what greenhouse gases were available in the young
planetary atmospheres, as speculated by \cite{ribas05}. For example,
enhanced photodissociation may have influenced the abundances of
ammonia and methane. Also, photochemistry and subsequent production of
UV-shielding $\mathrm{O}_3$ \citep{canuto82, canuto83} was important
for the formation and evolution of life, and life itself eventually
altered the composition of the young terrestrial atmosphere very
significantly.

\subsubsection{A Bright Young Sun?}
\label{section:brightsun}

A more radical remedy of the Faint Young Sun Paradox would be a Sun
that was in fact not faint, i.e., did not follow the standard solar
model calculations \citep[see][for a discussion on controversies related to possible greenhouse effects -- or their need -- in the early atmospheres of Earth and Mars]{sackmann03}. 
Such would be possible if the ZAMS Sun had been more massive, having lost its mass
in a wind at rates considerably higher than the present-day solar
wind. The latter results in a mass loss of $(2\mbox{\,--\,}3) \times
10^{-14}\,M_{\odot} \unit{yr^{-1}}$ \citep[][and references therein]{wood04}, and the radiative losses of energy transformed in
thermonuclear reactions amount to about 3~times this rate. If the Sun
had been subject to these present-day losses for its entire lifetime,
its ZAMS mass would have been only 0.05\% higher than the present-day
value \citep{minton07}. This would change the young Sun's bolometric
luminosity negligibly (recall the mass-luminosity relation for MS
stars, which requires approximately $L_\mathrm{bol} \propto M^3$;
based on \citealt{siess00} ZAMS calculations for low-mass stars).

Higher wind mass-loss rates would be an interesting alternative
\citep{graedel91}. \cite{willson87} hypothesized that
intermediate-mass stars may lose appreciable amounts of mass during
their MS life, in particular in the pulsation-instability strip; early
G-type stars would then be descendants of A-type stars. \cite{hobbs89}
concluded that a wind mass loss of $0.041\,M_{\odot}$ since the Sun's
arrival on the ZAMS would suffice to explain the low Li values
observed in the present-day photosphere (because Li would be diluted
when the wind-driving surface layer is progressively mixed with
Li-free material entering from lower, hotter layers; see also
\citealt{schramm90}; note, however, that there are other, and more
important, processes that deplete Li, see \citealt{sackmann03}). A
higher mass loss rate for the young Sun is in fact supported by
meteoritic and lunar evidence, suggesting that 2.5\,--\,3.5\,Gyr ago
(solar age of 1\,--\,2\,Gyr), the wind mass loss was on average
10~times higher than at present \citep{geiss91}. This would, however,
result in a solar mass still only $\approx 0.1\%$ higher at $t = -3
\unit{Gyr}$ than now \citep{sackmann03}. To simultaneously fulfill the
requirement of liquid water on young Mars, the initial solar mass
would have to be $\ga 1.03\,M_{\odot}$ \citep{sackmann03}.

\cite{gaidos00} used radio observations of three solar analogs at
ages of a few 100\,Myr to set upper limits to their present mass-loss
rate. Because the spin rates of solar analogs reveal a power-law
decay in time, $\Omega \propto t^{-0.6}$ (Equation~\ref{vrotage}, also
\citealt{skumanich72} who gave an exponent of $-0.5$),
\cite{gaidos00} argued for a power-law decay of the mass-loss rate as
well (Equation~\ref{omegaM}), which, together with the radio upper limits,
results in a maximum cumulative mass loss of 6\% of the solar mass
during the past 4\,Gyr. This is close to the suggested mass losses to
dilute Li (\citealt{hobbs89}, but note other Li depletion processes),
is in agreement with the minimum loss of 3\% required to explain
liquid water on Mars \citep{sackmann03}, and is also slightly lower
than the upper limit of 7\% of $M_{\odot}$ to avoid runaway greenhouse
on Earth \citep{whitmire95, kasting88} (the runaway greenhouse would
evaporate the entire water ocean so that all water would be present in
the atmosphere as steam; photodissociation and rapid loss of hydrogen
by hydrodynamic escape [see Section~\ref{section:interactions} below]
would lead to a dry planet -- analogous to present-day Venus; see
\citealt{ingersoll69}).

Corresponding models have been computed by, among others,
\cite{boothroyd91}, \cite{guzik95}, and \cite{sackmann03}. The upper
limit for the ZAMS Sun allowed by the Li depletion purely due to
wind-mass loss was found to be $1.1M_{\odot}$ \citep{boothroyd91,
  guzik95}. Helioseismology constraints are compatible with model
calculations starting with ZAMS solar masses up to
$(1.07\mbox{\,--\,}1.10)\,M_{\odot}$ \citep{boothroyd91, guzik95} but
the consequent enhanced mass loss should be confined to the earliest
$\approx 200 \unit{Myr}$ of the Sun's life on the MS, implying loss
rates as high as $5 \times 10^{-10}\,M_{\odot} \unit{yr^{-1}}$
\citep{guzik95}. Somewhat depending on the precise mass-loss law, the
solar flux starts at values up to 7\% higher than the present-day
value (corresponding to mass-loss rates of $\approx
10^{-11}\mbox{\,--\,}10^{-10}\,M_{\odot} \unit{yr^{-1}}$ at ZAMS age)
to drop to a minimum no less than 80\% after 1\,--\,2\,Gyr, and to
increase again in agreement with the evolution of the standard solar
model (Figure~\ref{figure:variablewind}). The highest acceptable initial
solar mass is $1.07\,M_{\odot}$ to ensure that the young Earth would
not lose its water via a greenhouse effect, photodissociation and
subsequent loss of hydrogen into space (\citealt{sackmann03}; see also
Section~\ref{section:planets} below).

\epubtkImage{}{%
\begin{figure}[htbp]
  \centerline{\includegraphics[scale=0.55]{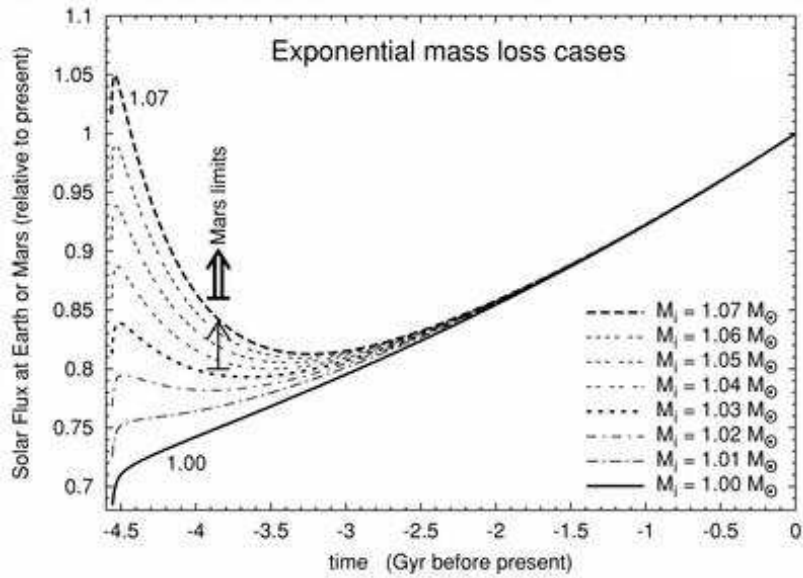}}
  \caption{\it Solar flux in time relative to present, for a Sun that
  was subject to strong mass loss in its past. Different curves show
  calculations for different initial masses and corresponding
  mass-loss rates such that the present-day values are obtained. The
  mass-loss rate declines exponentially in time; the flux increase at
  later times is due to the luminosity increase of the Sun for nearly
  constant mass. The double arrow indicates the lower limit for the
  presence of liquid water on early Mars, the thin arrow an
  (unrealistic) extreme lower limit \citep[from][reproduced by permission of AAS]{sackmann03}.}
  \label{figure:variablewind}
\end{figure}}

However, the indirect inferences for the mass-loss rates of the young
Sun derived by \cite{wood02} and \cite{wood05} (see
Section~\ref{section:solarwind}) would again not support a significantly
brighter ZAMS Sun. Using the power-law mass-loss decay relation of
\cite{wood02} back to ZAMS, a total mass-loss of about
$0.01\,M_{\odot}$ would result \citep{sackmann03}, with an uncertainty
of a factor of a few. Most of the mass loss would occur in the first
few 100\,Myr. The suppressed mass loss at early times, however
\citep{wood05}, suggests that no more than $0.003\,M_{\odot}$  could
be lost during the Sun's MS life \citep{minton07}.

In summary, the main problem with the ``bright young Sun'' model
remains the disagreement between climatic requirements for the
young-Sun mass (i.e., a ZAMS solar mass of
$[1.03\mbox{\,--\,}1.07]\,M_{\odot}$) and the indirectly measured
mass-loss rates \citep{minton07} that tend to be too small
(\citealt{wood05}, i.e., resulting in a ZAMS mass of no more than
$1.01\,M_{\odot}$), although radio upper limits \citep{gaidos00} are
still compatible with the required mass-loss rates.

\subsubsection{Cosmic Rays and a Stronger Solar Wind}
\label{section:cosmicrays}

\cite{shaviv03} suggested a link between the cosmic ray flux and
average global temperatures on Earth. Although a physical basis and an
accepted proof are still missing, there is suggestive evidence that
elevated cosmic-ray fluxes have a cooling effect on the Earth's
atmosphere. In this picture, cosmic rays ionize tropospheric layers,
and charged ion clusters lead to condensation nuclei that eventually
form clouds. Low-lying clouds have a cooling effect \citep{shaviv03}.

Given that wind of the young Sun was stronger
(Section~\ref{section:solarwind}), the cosmic-ray flux reaching the
inner solar system was suppressed compared to present-day
fluxes. Cloud formation would thus be suppressed, leading to a warmer
climate. Model calculations (also including effects due to variable
star-formation rate in the solar vicinity on the cosmic-ray
generation, a more rapid rotation of the Earth, and a smaller land
mass) suggest that about 2/3 of the temperature reduction associated
with the fainter young Sun can be compensated.

\subsection{The Sun's Activity in the Young Solar System}
\label{section:planets}

Stellar magnetic activity strongly influences planetary surfaces and
atmospheres in at least two ways:

\begin{itemize}

  \item The solar wind and high-energy particle streams interact with
  ions and neutrals in the upper planetary atmospheres, but also with
  the solid surfaces in the absence of atmospheres. Electron impact
  ionization, charge exchange, and other processes lead to important
  alterations of atmospheric or surface materials and to their erosion
  \citep[e.g.,][]{chassefiere04, kulikov07}.

  \item High-energy (XUV) radiation affects planetary atmospheres
  through ionization, heating, dissociation and recombination and
  induced chemistry \citep[e.g.,][]{canuto82}. Hard radiation also
  interacts with solid surfaces.

\end{itemize}

The evolution of planetary surfaces and atmospheres can therefore be
understood only in the context of the evolution of the enhanced
high-energy solar particle and photon fluxes. For this reason, the
past, much higher solar XUV and particle fluxes are crucial elements
in our understanding of planet evolution, in particular questions on
evaporation of their atmospheres and the loss of water
reservoirs. After a short summary of relevant physical effects, I
address key issues related to solar-system planets and Saturn's moon
Titan, and conclude with aspects related to ``hot Jupiters'' around
solar analogs.

The evolution of the Earth's atmosphere is sketched in
Figure~\ref{figure:earthatmosphere}. It indicates a large water
reservoir at the earliest times, part of which is thought to have
subsequently escaped from Earth (see also Section~\ref{section:earth}
below). Note also the rapid increase of biologically produced
$\mathrm{O}_2$ starting around 2.2\,Gyr before present.

\epubtkImage{}{%
\begin{figure}[t!]
  \centerline{\includegraphics[scale=0.6]{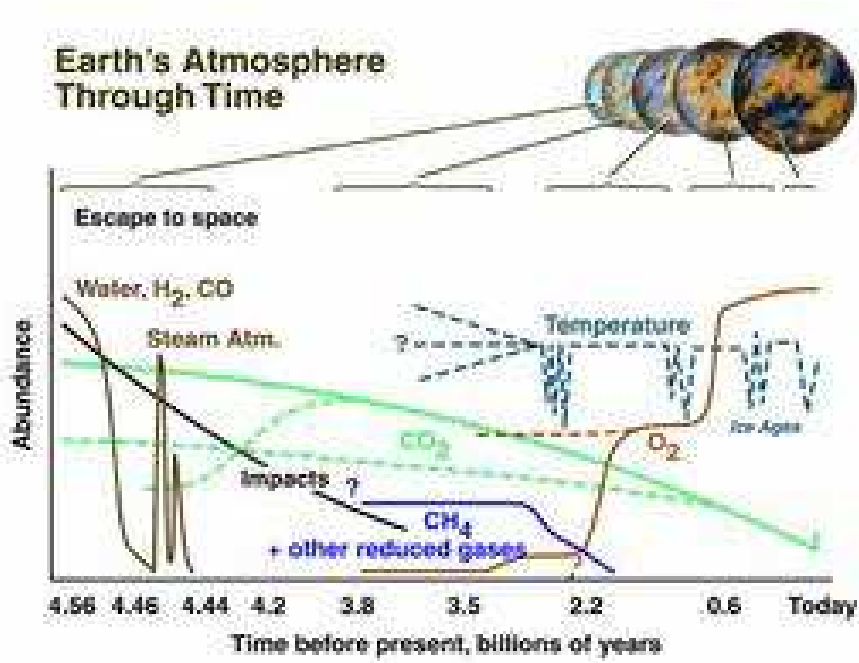}}
  \caption{\it Evolution of the constituents of the Earth's atmosphere
  in time (adapted from
  \url{http://science.nasa.gov/headlines/y2002/10jan\_exo-atmospheres.htm?list161037},
  after D.D.\ Marais and K.J.\ Zahnle).}
  \label{figure:earthatmosphere}
\end{figure}}

\subsubsection{Planetary Atmospheric Chemistry Induced by High-Energy
  Radiation}
\label{section:chemistry}

The young Sun's elevated ultraviolet flux had considerable impact on
the photochemistry of planetary atmospheres. \cite{ayres97} estimated
the photoionization rates based on the evolution of ionizing flux of
solar analogs. He concluded that photorates scale approximately as
$t^{-1}$, implying rates that were nearly 100~times stronger at a
solar age of 100\,Myr than at present (Figure~\ref{figure:photorates}).

\epubtkImage{}{%
\begin{figure}[t!]
  \centerline{\includegraphics[scale=0.45]{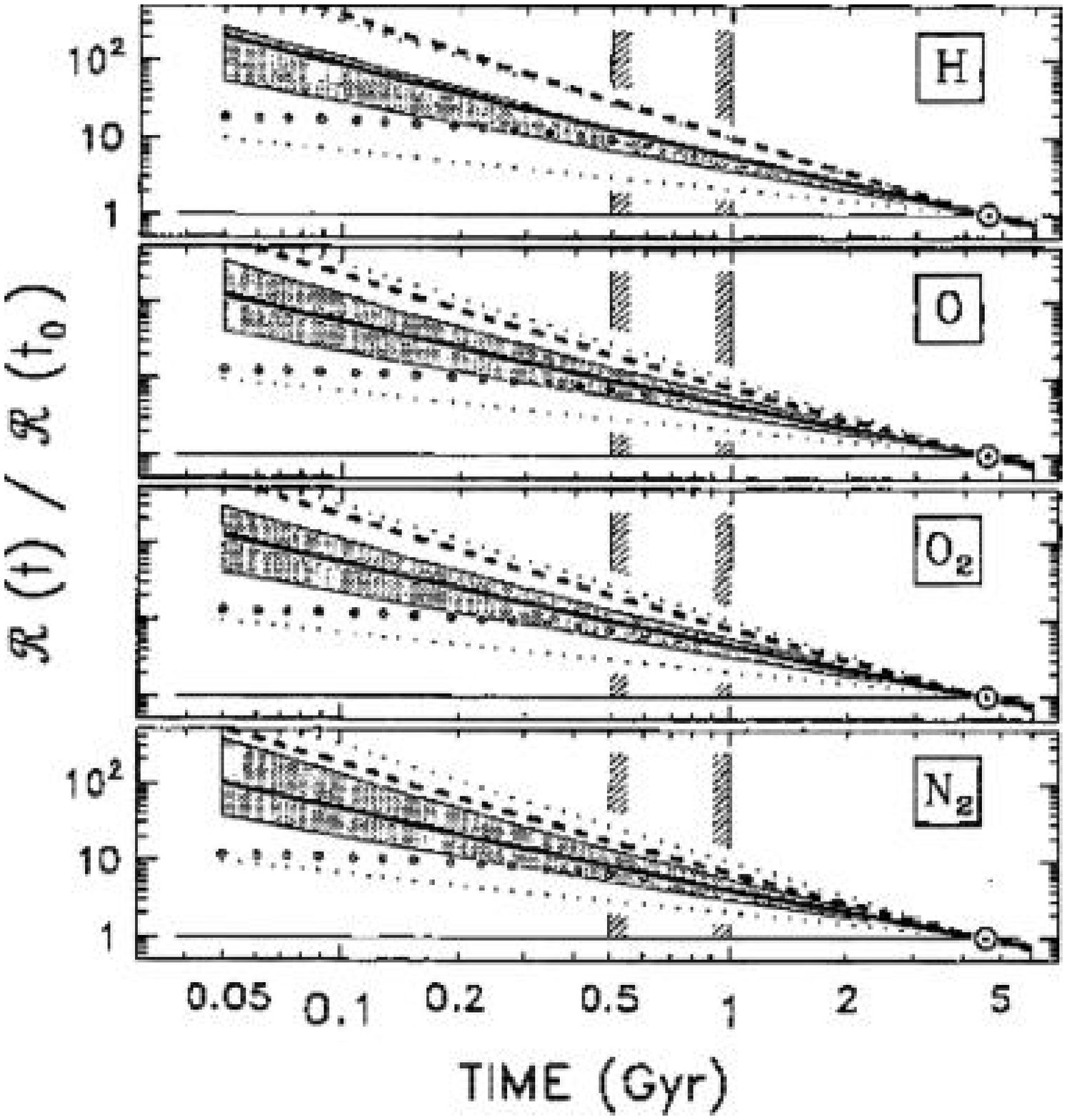}}
  \caption{\it Photoionization rates for  $\mathrm{H}$, $\mathrm{O}$,
  $\mathrm{O}_2$, and  $\mathrm{N}_2$ for different ages of the Sun,
  derived from the level of ionizing flux of solar analogs. The thick
  solid lines give the best estimates, the shaded areas give the
  allowed ranges based on uncertainties in the rotation-activity
  relationship \citep[from][reprinted with permission]{ayres97}.}
  \label{figure:photorates}
\end{figure}}

UV radiation initiated the photochemical processes leading to the
formation of  $\mathrm{O}_2$ and $\mathrm{O}_3$ in the prebiotic
terrestrial atmosphere \citep{canuto82}. Ozone absorbs strongly in the
200\,--\,300\,nm band and therefore protects life from lethal doses of
UV radiation.

UV radiation produces oxygen by photolysis of water and
$\mathrm{CO}_2$:

\begin{eqnarray}
   \mathrm{H_{2}O} + h\nu    & \rightarrow & \mathrm{OH} + \mathrm{H}; \quad \lambda \le 240 \unit{nm} \\
   \mathrm{OH} + \mathrm{OH} & \rightarrow & \mathrm{O} + \mathrm{H_{2}O} \\
   \mathrm{CO_2} + h\nu      & \rightarrow & \mathrm{CO} + \mathrm{O}; \quad \lambda \le 230 \unit{nm}
\end{eqnarray}

after which $\mathrm{O}_2$ is formed \citep{canuto82}. The atomic H
can escape into space (see Section~\ref{section:interactions}
below). Ozone is formed through

\begin{equation}
  \mathrm{O} + \mathrm{O_2} + \mathrm{M} \rightarrow \mathrm{O_3} + \mathrm{M} \\
\end{equation}

(M stands for other molecules; there are also several reactions that
destroy $\mathrm{O}_2$ and $\mathrm{O}_3$; see
\citealt{canuto82}). Atmospheric chemical models subject to enhanced
UV irradiation show that the surface mixing ratio of O$_2$ increases
by a factor of $10^4\mbox{\,--\,}10^6$ above the standard pre-biotic
value of $10^{-15}$ due to enhanced levels of solar UV radiation
(corresponding to 1\,--\,300~times the present-day Sun).

The UV irradiation of the young Earth was particularly strong during
the Sun's T~Tauri phase, up to factors of $10^3$ above the present
solar values. The potential influence on planet formation and the
subsequent generation of planetary atmospheres should therefore have
been very significant at those epochs -- if any of the planetary
atmospheres were indeed forming at such early times already. The UV
irradiation could increase the total column of $\mathrm{O}_3$ by
1\,--\,3~orders of magnitude \citep{canuto83}, thus shielding incoming
solar UV radiation efficiently and protecting early forms of life.

\subsubsection{High-Energy Radiation and the Planetary Biospheres in
  Habitable Zones}
\label{section:biospheres}

Ionizing radiation is of crucial importance also for the evolution of
life. While high doses are lethal for life, modest doses are important
drivers of genetic mutations and therefore biological
evolution. Although at present most high-energy radiation is absorbed
in the high atmosphere of the Earth, significant effects may be found
in thinner atmospheres (e.g., Mars) or in the environment of the young
Sun where the irradiation by XUV photons was hundreds of times
stronger. \cite{smith04} have computed the propagation of ionizing
radiation through model atmospheres of terrestrial-like planets.

The incident spectra are those of supernovae explosions, gamma-ray
bursts and, relevant for us, coronal X-ray flares that occurred at a
much higher rate and up to much higher total energies in the young Sun
compared to the present  \citep{telleschi05, audard00}. Monte Carlo
methods were used to calculate the transfer of incident X-ray and
gamma-ray photons via Compton scattering and photoabsorption. The
incident high-energy photons ionize molecules, thereby creating
primary electrons,  which collide further to produce secondary
photoelectrons. These excite molecules and create aurora-like
emission. A single incident photon can cause ionization of tens of
thousands of molecules. For a neutral gas, most of the
secondary-electron energy, however, goes into molecule excitation
rather than ionization (different from ionized gas;
\citealt{smith04}).

Strong re-radiation of the energy will occur in the $\mathrm{N}_2$
bands in the UV region \citep{smith04}, and at this point, the
composition of the atmosphere becomes relevant as a variety of opacity
sources and Rayleigh scattering will redistribute the
radiation. \cite{smith04} studied in particular ``habitable''
atmospheres,  requiring column densities of at least
$30\mbox{\,--\,}100 \unit {g\ cm^{-2}}$ for Earth-like planet (the
Earth atmosphere's column density is $1024 \unit{g\ cm^{-2}}$). Their
principal results can be summarized as follows:

\begin{itemize}

  \item While the direct transmission of stellar flare X-ray photons
       to a planetary surface is negligible, in a thin atmosphere
       with column density of $30\mbox{\,--\,}100 \unit{g\ cm^{-2}}$,
       pure atmospheric Rayleigh scattering (appropriate for a young
       terrestrial atmosphere) would transmit $\approx 10\%$ of the
       incident energy flux as UV radiation; about 1\,--\,3\% are
       transmitted as UV photons if only a terrestrial-like
       $\mathrm{O_3/O_2}$ molecular absorption screen is present.

       \item In denser, Earth-like atmospheres ($\ga 100 \unit{g\
       cm^{-2}}$), direct transmission of X-ray photons is essentially
       zero, but still of order $2 \times 10^{-3}$ of the incident
       flux reaches the surface as UV radiation even in the presence
       of an $\mathrm{O_3/O_2}$ shield. In a pure Rayleigh-scattering
       atmosphere (important for the young Earth), the transmission is
       of order 5\%.

\end{itemize}

Even planets with thick atmospheres that shield evolving life from
direct ionizing radiation will shower their surfaces with UV radiation
owing to stochastic flares occurring on the host star. The present-day
Sun produces about one flare every 100\,yrs with a transmitted UV
surface flux equal to or larger than the continuous solar UV flux in
the 200\,--\,320\,nm band \citep{smith04}. Whether the UV irradiation
effect is of prime importance for a planet around a young solar-like
star is not clear, but it appears to be important for planets orbiting
lower-mass M~dwarf flare stars \citep{smith04}.

\subsubsection{Planetary Atmospheric Loss and High-Energy Radiation
  and Particles}
\label{section:interactions}

Various mechanisms induced by solar magnetic activity can lead to {\it
  escape} of planetary atmospheric species, both neutrals and ions.
  The principal losses strongly depend on the magnetic activity of the
  host star. {\it Thermal escape} processes relate to a Maxwellian
  particle distribution and are consequently determined by the
  temperature of the respective atmospheric layer, and therefore by
  its heating rate. Thermal escape is predominantly controlled by
  photon {\it irradiation} from the host star. {\it Non-thermal}
  escape is due to other processes in which energization and escape
  are related to microscopic non-thermal processes (e.g., particle
  acceleration; \citealt{lundin04, lundin07}). Key factors are
  high-energy particle streams from the central star, i.e., the
  solar/stellar winds, and the magnetic environment of the planet
  itself.

The heating of planetary upper atmospheres and the generation of
ionospheres is primarily due to the solar XUV radiation. The {\it
  thermosphere} is the layer in which most of this radiation is
absorbed and is transformed to heat, leading to a positive (upward)
temperature gradient. In the {\it exosphere}, the mean free path of
atmospheric species is large, so that collisions are negligible. The
lighter species may escape from this layer into space if their thermal
velocity exceeds the gravitational escape velocity. The base of the
exosphere, the {\it exobase}, is located at the height where the mean
free path is equal to the local scale height, $H =
kT_\mathrm{exo}/mg$, where $k$ is the Boltzmann constant,
$T_\mathrm{exo}$ is the exobase temperature, $m$ is the mass of the
main atmospheric species, and $g$ is the gravitational acceleration
\citep{kulikov07}.

Let us define the thermal escape parameter

\begin{equation}
  X = {GMm\over kT_\mathrm{exo}r}
\end{equation}

as the ratio between the potential energy of a particle and the
thermal energy $kT_\mathrm{exo}$, where $M$ is the planet's mass, $m$
is the particle mass (in the following: mostly atomic hydrogen), $r$
is the planetocentric distance, and $T_\mathrm{exo}$ is the exospheric
temperature. Let us now assume a Maxwellian distribution for a given
species. If $X > 1.5$, e.g. for low temperatures, only the particles
in the tail of the Maxwellian have kinetic energies sufficient for
escape. The particle loss through a spherical surface at radius $r$ is
given by the Jeans formula (``Jeans escape'', e.g., \citealt{jeans25,
  opik63, kulikov06}),

\begin{equation}
  F = 4\pi r^2 \left({ v_0\over 2\pi^{1/2}}\right) n_\mathrm{exo} (1 + X) e^{-X}.
\end{equation}

Here, $v_0 = (2kT_\mathrm{exo}/m)^{1/2}$ is the most probable velocity
of the particle \citep[see, e.g.,][]{chassefiere04, kulikov06}.

If the mean thermal energy of the particles ($3kT_\mathrm{exo}/2$)
exceeds the gravitational energy, i.e., for $X < 1.5$ or analogously
for temperatures $T_\mathrm{exo} > 2GMm/3kr$ or $kT \ga
mv_\mathrm{esc}^2/2$, the exosphere becomes unstable and hydrostatic
equilibrium does not apply. The top of the atmosphere ``blows off'',
i.e., moves radially away from the planet with a velocity near its
thermal velocity. To produce sufficiently high temperatures, a high
level of EUV irradiation is required. The incoming EUV energy is then
no longer converted into thermal energy but directly into kinetic
energy. ``Hydrodynamic escape'' calculations are required
\citep{opik63, hunten73, watson81}.

The most important heating and cooling processes in the upper
atmosphere of the Earth are, following \cite{kulikov07} and references
therein,

\begin{itemize}

  \item heating of the thermosphere due to photoionization of
  $\mathrm{CO}_2$, $\mathrm{N}_2$, $\mathrm{O}_2$ and $\mathrm{O}$ by
  solar XUV irradiation,

  \item heating due to photodissociation of $\mathrm{O}_2$ and
  $\mathrm{O}_3$ by solar UV irradiation,

  \item chemical heating in exothermic binary and 3-body reactions,

  \item molecular heat conduction in neutral gas,

  \item cooling by infrared emission in the ro-vibrational bands of
  $\mathrm{CO}_2$, $\mathrm{NO}$, $\mathrm{O}_3$, $\mathrm{OH}$,
  $\mathrm{NO}^+$, $^{14}\mathrm{N}^{15}\mathrm{N}$, $\mathrm{CO}$,
  $\mathrm{O}_2$, and others,

  \item heating and cooling by contraction and expansion of
  thermosphere,

  \item turbulent energy dissipation and heat conduction.

\end{itemize}

Models of XUV heating of the present-day Earth's thermosphere predict
an average exospheric temperature of $\approx 1000 \unit{K}$
\citep[][and references therein]{kulikov07}, in agreement with
measurements. For Venus, $T_\mathrm{exo} = 270 \unit{K}$
\citep{kulikov06, kulikov07} where the low value is due to efficient
cooling by $\mathrm{CO}_2$. And for Mars, $T_\mathrm{exo} =
220\mbox{\,--\,}240 \unit{K}$ \citep{kulikov07}.

{\it Non-thermal processes} are also relevant in outer planetary
atmospheres. There are several main types of non-thermal escape
processes \citep{chassefiere04, lundin07}:

\begin{itemize}

  \item {\it Dissociative recombination} (photochemical escape): after
  UV photoionization, ions recombine by forming {\it energetic}
  neutrals that may escape. Dissociative recombination of ionospheric
  molecular ions and photodissociation of molecular neutrals are
  sources of suprathermal H, C, N, O, and CO in the exosphere of
  (present) Venus \citep[][and references therein]{kulikov07}. These
  atoms are eventually thermalized through collisions, or are lost if
  they reach the exobase with energies exceeding the escape energy.

  \item {\it Ion pick-up escape:} Ions are produced by
  photo-ionization, electron impact, or charge exchange and are
  dragged along by the magnetic field; a fraction of the particles
  escape in the solar wind.

  \item {\it Ionospheric outflow:} ionospheric matter is accelerated
  away by the electric field in the solar wind.

  \item {\it Ion sputtering:} A fraction of the ions produced by
  photo-ionization, electron impact, or charge exchange re-impact the
  neutral atmosphere (owing to their large gyroradii) and eject
  neutral particles. This process is particularly important in the
  absence of a strong planetary magnetic field.

\end{itemize}

I now discuss results from applications to solar-system bodies and
extrasolar planets.

\subsubsection{Mercury: Erosion of Atmosphere and Mantle?}
\label{section:mercury}

Mercury stands out among solar-system planets by showing the largest
average density, and its core is large compared to other terrestrial
planets (60\% of its radius). It is possible that strong winds and
enhanced XUV irradiation from the young Sun removed much of the early
atmosphere and the outer mantle of Mercury \citep{tehrany02,
  ribas05}. It is thus possible that Mercury started as a planet
similar in size to Earth but has lost its mantle as a consequence of
interactions with the young, active Sun.

\subsubsection{Venus: Where has the Water Gone?}
\label{section:venus}

The atmosphere of Venus is extremely dry when compared to the
terrestrial atmosphere (a total water content of $2 \times 10^{19}
\unit{g}$ or 0.0014\% of the terrestrial ocean, vs.\ $1.39 \times
10^{24} \unit{g}$ on Earth, \citealt{kasting83}). The similar masses
of Venus and Earth and their formation at similar locations in the
solar nebula would suggest that both started with similar water
reservoirs \citep{kasting83}. The Venus atmosphere also reveals a
surprisingly high deuterium-to-hydrogen (D/H) abundance ratio of $(1.6
\pm 0.2) \times 10^{-2}$ or 100~times the terrestrial value of $1.56
\times 10^{-4}$ \citep{donahue82, kasting83}. These observations have
motivated a basic model in which water was abundantly available on the
primitive Venus. Any water oceans would evaporate, inducing a
``runaway greenhouse'' in which water vapor would become the major
constituent of the atmosphere \citep{ingersoll69}. This vapor was
then photodissociated in the upper atmosphere by {\it enhanced} solar
EUV irradiation \citep{kasting83}, followed by the escape of hydrogen
(but not deuterium) into space \citep{watson81, kasting83}. Oxygen
either (partly)  followed the same route or (partly) reacted with
reduced atmospheric gases or minerals in the crust
\citep{kasting84}. Judged from the D/H abundance ratio, at least 0.3\%
of a terrestrial ocean should have been outgassed \citep{donahue82}. I
summarize some of the results and problems discussed in the literature
as far as they refer to solar activity and its much higher level in
the early Sun.

Detailed atmospheric models have been calculated by, e.g.,
\cite{watson81}, \cite{kasting83}, \cite{kasting84}, \cite{kasting88},
\cite{chassefiere96b}, and \cite{kulikov06}. These models solve the
equations of continuity and diffusion (e.g., constrained to the
vertical direction in 1-D approaches) for a number of species,
hydrostatic and heat balance equations, and equations of vibrational
kinetics for radiating molecules. They include irradiation by solar
XUV and UV and their long-term evolution \citep[see][for a
  review]{kulikov07}.

\cite{kulikov06} and \cite{kulikov07} presented results for Venus,
assuming no atmospheric (in particular, hydrogenic) loss, i.e., the
upper atmospheres are {\it not} hydrogen-rich and not humid
(Figure~\ref{figure:venus}). For present-day solar XUV irradiation, the
calculated exospheric temperature is 270\,K, in good agreement with
measurements. However, increasing the XUV flux to levels of the young
Sun, i.e., a 100fold EUV flux, thermospheric temperatures up to
$\approx 8200 \unit{K}$ are reached 4.5\,Gyrs ago, if the
$\mathrm{CO}_2$ content is the same as today (96\%). Higher
temperatures are obtained for lower $\mathrm{CO}_2$ content.

\epubtkImage{}{%
\begin{figure}[htbp]
  \centerline{\includegraphics[scale=0.53]{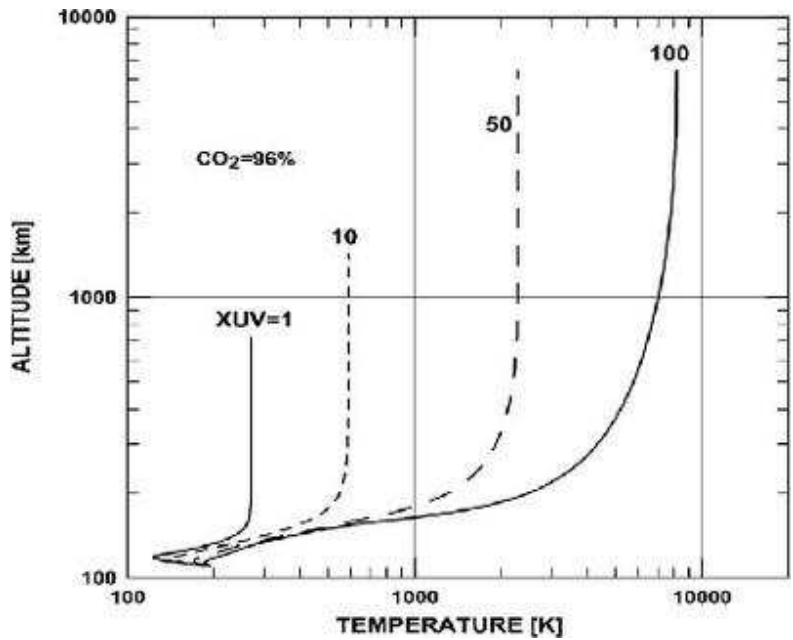}}
  \caption{\it Modeled temperature profiles for Venus for various
  solar XUV levels: 1 = present-day, 10~times (3.8\,Gyr ago), 50~times
  (4.33\,Gyr ago), and 100~times (4.5\,Gyr ago) present levels, based on
  a 96\% $\mathrm{CO}_2$ thermosphere \citep[from][]{kulikov07}. Copyright
  Springer Verlag, reprinted with permission.}
  \label{figure:venus}
\end{figure}}

The blow-off limit in the exobase is reached at 4000\,K, permitting
hydrogenic blow-off during the first 250\,Myr after the Sun's arrival
on the ZAMS. Detailed hydrodynamic computations by
\cite{chassefiere96b} show high Jeans escape flux even for present-day
solar XUV irradiation for a Venus-like planet with a hydrogen
atmosphere.

The question on water on the young Venus then concentrates on the
availability of water vapor in the young Venus' upper atmosphere, that
can photo-dissociate so that H escapes. The critical solar flux
leading to complete evaporation of a water ocean (a ``runaway
greenhouse'', \citealt{ingersoll69}) is $\approx 1.4\,S_\mathrm{sun}$
nearly independently of the amount of cooling $\mathrm{CO}_2$ in the
atmosphere \citep{kasting88}, where $S_\mathrm{sun} = 1360
\unit{Wm^{-2}}$ is the present-day solar constant. This critical flux
is just about the flux expected from the young Sun at the orbit of
Venus (and is about 70\% of the present-day value at the same solar
distance).

These model calculations suggest that the amount of water in the
present terrestrial ocean could have escaped from Venus in only
50\,Myr if the $\mathrm{H_{2}O}$  mass fraction (mass mixing
    ratio) was at least 0.46, and the XUV flux was 70\,--\,100~times
  the present values \citep{kulikov07}.

The remaining problem then is to explain the extremely low oxygen
content in the present-day atmosphere of Venus
\citep{chassefiere96a}. In a strong hydrodynamic escape process on
young Venus, heavier gases such as O may be dragged with H
\citep{zahnle86, zahnle90} although \cite{chassefiere96a} concluded
that such loss was modest. Ion pick-up by the solar wind and
ionospheric erosion by Kelvin--Helmholtz instabilities may be the
dominant $\mathrm{O}^+$ loss mechanism in the present atmosphere of
Venus \citep{lammer06, chassefiere97}. The denser, young solar wind
\citep{wood02} could in fact easily remove the expected amount of
$\mathrm{O}^+$ from a massive water ocean \citep{chassefiere97,
  kulikov06}. The equivalent of one terrestrial ocean could be
removed in only 10\,Myr if the solar wind had been
$10^3\mbox{\,--\,}10^4$~times stronger than today, i.e., the escape
would be ``instantaneous'' compared with outgassing time scales
\citep{chassefiere97} although it is doubtful whether the young solar
wind was indeed as strong as required for such massive losses
(\citealt{wood05}, see Section~\ref{section:solarwind}).

\subsubsection{Earth: A $\mathrm{CO}_2$ Atmosphere and Magnetic
  Protection}
\label{section:earth}

There is evidence for early hydrodynamic escape from the Earth's
atmosphere. Compared to Venus, the lighter noble gases Ne and Ar are
depleted in the Earth's atmosphere with respect to the heavier Kr and
Xe; the isotopic composition of the noble gases points in the same
direction \citep[][and references therein]{zahnle86, zahnle90,
  chassefiere04}.

The \cite{kulikov07} model of the terrestrial atmosphere shows that a
ten-fold increase of the XUV flux, corresponding to the Sun $>3.8
\unit{Gyr}$ ago, leads to exobase temperatures of $>10,000 \unit{K}$,
which would imply escape rates not only for H but also
for $\mathrm{H}_2$, He, N, O, or C. The hydrogenic blow-off temperature
is only about 5000\,K. As a consequence of dissociation of
$\mathrm{H_{2}O}$, large amounts of water would thus have been lost
  unless there was a significantly higher $\mathrm{CO}_2$ content in
  the early atmosphere, because $\mathrm{CO}_2$ is efficiently cooling
  the upper atmosphere by radiation in the 15$\mu m$ band
  \citep{kulikov07}. Given the high early solar XUV flux, the
  atmospheric $\mathrm{CO}_2$ content in the young Earth's atmosphere
  may have been at similar levels as in the atmospheres of the
  present-day Venus and Mars ($\approx 95\%$;
  \citealt{kulikov07}). Its reduction and consequently, the rise of
  $T_\mathrm{exo}$, due to weathering onto the surface and the
  seafloor during the first 500\,Myr was accompanied by a rapid decay
  of the solar EUV irradiation so that $T_\mathrm{exo}$ was kept below
  blow-off conditions.

An important factor of the young Earth's atmosphere is the relatively
strong (as compared to Venus or evolved Mars) terrestrial
magnetosphere that protects the atmosphere from solar-wind pickup
erosion. An early on-set of the dynamo was therefore essential to
protect the atmosphere and the large water reservoir of the Earth
\citep{kulikov07, lundin07}.

\subsubsection{Mars: Once Warmer and Wetter?}
\label{section:mars}
 
There is substantial evidence for a warmer and wetter climate on
ancient Mars \citep{carr86}. The presence of valley networks in the
highlands as well as some crater morphology require the presence of
liquid water (e.g., \citealt{craddock93}).  Using Mars Global Surveyor
images and altimeter data, \cite{carr03} estimated the depth of a
primitive Martian ocean at 150\,m (over the whole planet). Judged from
the near-absence of valleys in the younger northern areas, liquid
water disappeared suddenly 3.5\,--\,4\,Gyr ago
\citep{chassefiere04}. Temperatures above $0^{\circ} \unit{C}$ and a
massive atmosphere were required at earlier epochs, with a substantial
content of greenhouse gases \citep{chassefiere04}.

Massive atmospheric loss is relatively easy to explain, given Mars'
low surface gravity and the absence of a global magnetic field
protecting the atmosphere from the solar wind after about 700\,Myr of
the planet's formation \citep{acuna98}. Intense erosion by direct
interaction between solar wind and the extended atmosphere is thus
possible although significant suppression of non-thermal escape by the
magnetic field would be possible during the first 200\,--\,300\,Myrs
\citep{hutchins97, kulikov07}. In present-day Mars, hydrogen is
predominantly lost through thermal escape (five times more efficiently
than by non-thermal escape), while loss of oxygen to space requires
non-thermal processes \citep{lammer03a}.

Like for other planets, the thermal escape of the young atmosphere of
Mars depended strongly on the solar EUV irradiation, and non-thermal
escape processes would also have depended on the ancient solar wind
strength. Blow-off conditions are reached for an exospheric
temperature $T_\mathrm{exo} \ga 1000 \unit{K}$ \citep{lammer03a}. This
requires an EUV flux $\approx$100~times the present solar value
(\citealt{kulikov07}, see Figure~\ref{figure:mars}), i.e., blow-off may
apply to the the earliest times of the solar evolution, namely the
first few 100\,Myr after the Sun's arrival on the ZAMS (when a strong
solar wind would perhaps provide additional support,
\citealt{chassefiere96b, chassefiere97, chassefiere04}). For solar
conditions 2.5\,Gyr and 3.5\,Gyr ago, \cite{lammer03a} concluded that
the exospheric temperature was only $T_\mathrm{exo} = 350 \unit{K}$
and 600\,K, respectively (with a 3~times and 6~times higher solar EUV
flux, respectively),  insufficient for blow-off unless a significantly
lower $\mathrm{CO}_2$ mixing ratio applied \citep{kulikov07}, although
Jeans escape and non-thermal escape of H remained operational.

\epubtkImage{}{%
\begin{figure}[htbp]
  \centerline{\includegraphics[scale=0.53]{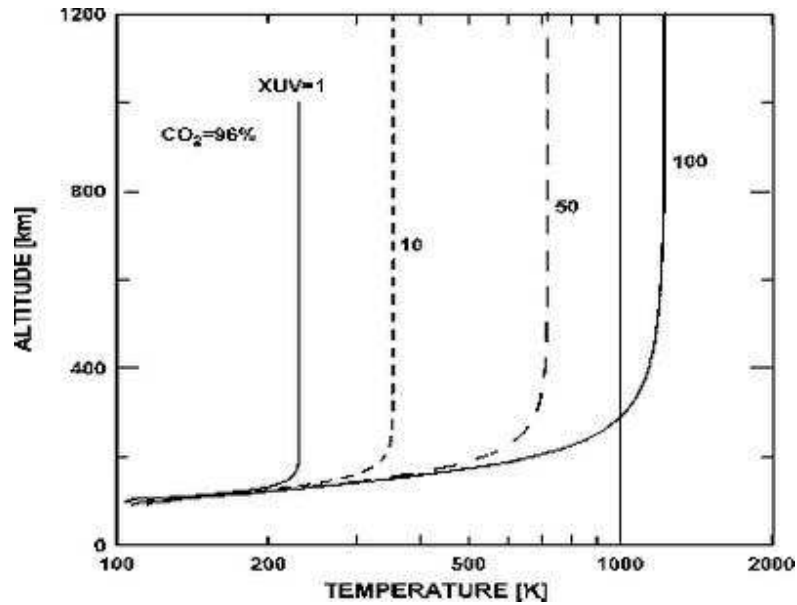}}
  \caption{\it Modeled temperature profiles for Mars for various solar
  XUV levels: 1 = present-day, 10~times (3.8\,Gyr ago), 50~times
  (4.33\,Gyr ago), and 100~times (4.5\,Gyr ago) present levels, based
  on a 96\% $\mathrm{CO}_2$ thermosphere \citep[from][]{kulikov07}. Copyright
  Springer Verlag, reprinted with permission.}
  \label{figure:mars}
\end{figure}}

Calculations show that in contrast to the situation on Venus and
Earth, a considerable fraction of O (from the dissociated
$\mathrm{H_{2}O}$ molecules)  may have been dragged off by the escaping
  H in the very young Mars (4\,--\,4.6\,Gyr ago;
  \citealt{chassefiere96a}). Non-thermal escape processes (of O) alone
  may explain the loss of a  12\,--\,15\,m Martian ocean during the
  past 3.5\,Gyrs (\citealt{lammer03a, chassefiere04}, and further
  references therein converging to similar numbers). \cite{lammer03a}
  and \cite{chassefiere04} summarized escape rates for three different
  epochs (present, 2.5\,Gyr and 3.5\,Gyr ago) to find that oxygen loss
  to space in the early epochs is predominantly due to sputtering
  (and also ion pick-up; see also \citealt{luhmann92, jakosky94}), in
  contrast to the present-day situation. During the last 2\,Gyr,
  however, O escape was not sufficient to balance H escape from
  dissociated water, implying that oxygen must have reacted
  efficiently with surface material.

\subsubsection{Venus, Earth, Mars: Similar Start, Different Results?}
\label{section:comparison}

Even if Venus, Earth, and Mars might all have started with dense,
$\mathrm{CO}_2$ rich atmospheres and a large reservoir of water, the
outcome of atmospheric evolution shows three very different planets: a
very arid Venus with its dense, greenhouse-effective hot
$\mathrm{CO}_2$ atmosphere; a water-rich Earth with a biologically
transformed, intermediate-density atmosphere;  and a cold Mars with
frozen water reservoirs and a greenhouse-ineffective $\mathrm{CO}_2$
atmosphere. Key factors that led to the variations between the
present-day planets are, among others: the close distance of Venus to
the Sun, potentially allowing for a runaway greenhouse leading to the
complete escape of water into space; the Earth's magnetic field that
protected the atmosphere from erosion by the early and stronger solar
wind, together with the biological activity unfolding on young Earth;
and the low surface gravity of Mars coupled with the absence of a
magnetic field that led, except during the initial few 100\,Myrs, to
more efficient erosion of the early atmosphere. Further factors such
as continents (allowing for a carbonate-silicate geochemical cycle and
thus the removal of $\mathrm{CO}_2$, see
Section~\ref{section:greenhouse}),  plate tectonics (allowing for
release of $\mathrm{CO}_2$, see \citealt{kasting89a}), or volcanic
activity may matter as well. Nevertheless, in all cases, the level of
XUV emission and the strength of the solar wind in the early solar
system were of prime importance. Figure~\ref{figure:comparison} shows a
sketch of the different evolutionary paths of the atmospheres of
Venus, Earth, and Mars considering the long-term evolution of the
Sun's XUV radiation and the solar-wind strength (from
\citealt{kulikov07}; see also \citealt{lundin07}).

\epubtkImage{}{%
\begin{figure}[htbp]
  \centerline{\includegraphics[scale=0.7]{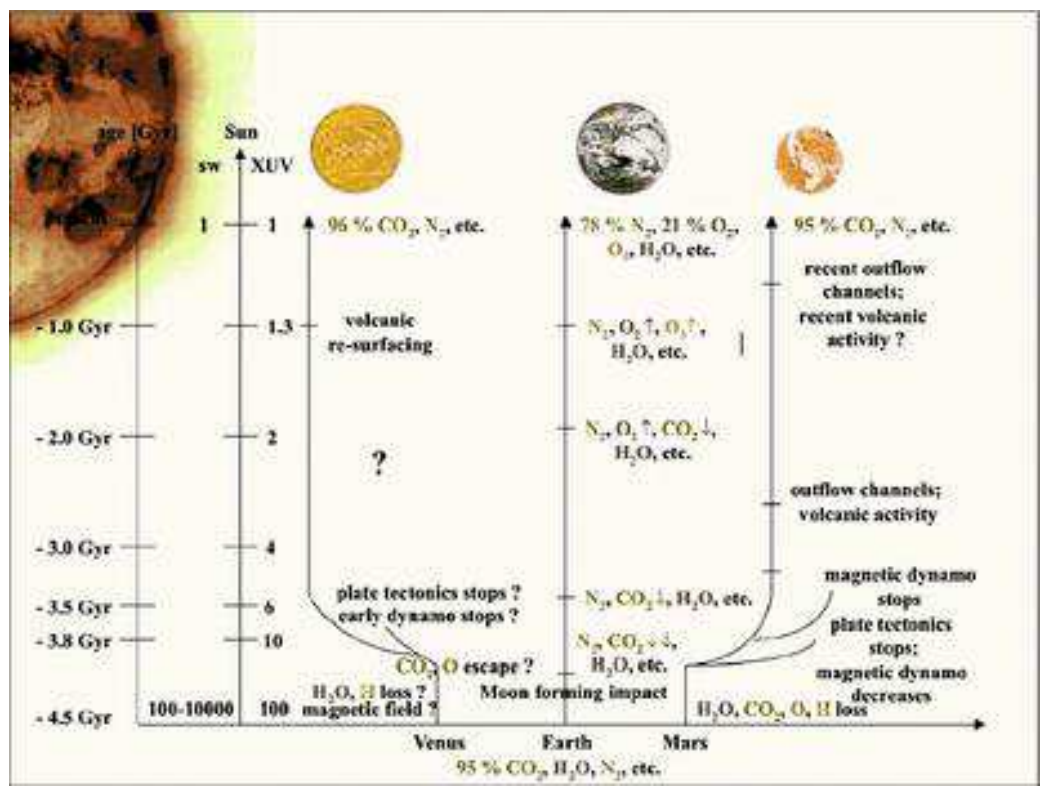}}
  \caption{\it Comparison of evolutionary paths of Venus, Earth, and
  Mars, accounting for the influence of solar XUV and particle
  radiation \citep[adapted from][]{kulikov07}. Copyright
  Springer Verlag, reprinted with permission.}
  \label{figure:comparison}
\end{figure}}

\subsubsection{Titan: Early Atmospheric Blow-Off?}
\label{section:titan}
 
The same atmospheric methodology as described for Venus and Mars can
be applied to other solar-system atmospheres. There is a particularly
interesting isotopic anomaly in the atmosphere of Titan, namely a
$^{15}$N enrichment with respect to  $^{14}$N by a factor of
3.9\,--\,4.5 \citep[][and references therein]{penz05}. Non-thermal
escape processes alone cannot account for the enrichment unless an
unrealistically strong solar wind is assumed for the first 500\,Myrs
after the origin of the solar system \citep{lammer00, penz05}.

It is important to note that present-day Titan's orbit resides
entirely within Saturn's magnetosphere, i.e., its atmosphere is
protected from the solar wind. Because the latter has decayed during
the Sun's evolution (Section~\ref{section:solarwind}), Titan spent most
of its time outside Saturn's magnetosphere at ages of about 100\,Myr,
and still more than half of the time at an age of 1\,Gyr; its
atmosphere was therefore subject to direct solar-wind
interactions. Nevertheless, sputtering and ion pick-up are not
sufficient to explain the anomalous $^{15}\mathrm{N}/^{14}\mathrm{N}$
ratio \citep{penz05}.

Thermal escape is an alternative. The present exospheric temperature
is in the range of $T_\mathrm{exo} \approx 150\mbox{\,--\,}180
\unit{K}$, while for blow-off, $T_\mathrm{exo} > 8040 \unit{K}$ is
required \citep[see][and references therein]{penz05}. Scaling the
present $T_\mathrm{exo}$ by using the XUV flux decay law derived for
solar analogs (Section~\ref{section:spectralevolution}), one finds
blow-off conditions for Titan during the first 200\,Myr. Only 50\,Myr
would in fact suffice to remove 30 Titan atmospheres, needed to
explain the anomalous $^{15}\mathrm{N}/^{14}\mathrm{N}$ isotopic ratio
(\citealt{penz05}; see also \citealt{lunine99}).

\subsubsection{Hot Jupiters: Mass Evolution by Evaporation?}
\label{section:hotjupiter}

The discovery of extrasolar, Jupiter-class planets very close to their
parent stars \citep{mayor95} immediately brought up the question on
the stability of such objects. Gaseous planets close to a solar-like
star will efficiently evaporate. However, as discussed above, it is
not the radiative effective temperature, $T_\mathrm{eff}$, that is
relevant for evaporation, but rather the exospheric temperature,
$T_\mathrm{exo}$, which is determined by the stellar XUV flux, and is
therefore a consequence of stellar activity.

A ``Hot Jupiter''  orbiting a star like the present-day Sun at a
distance of 0.04\,AU is, with regard to solar XUV irradiation, an
analog to a Jupiter-like planet in orbit around the ZAMS Sun at a
distance of 1.3\,AU. The analogy does not hold for the stellar wind,
and there may also be a substantial effect due to synchronous (i.e.,
slow) rotation of hot Jupiters \citep{griessmeier04}.

\cite{lammer03b} derived the escape rate of a hydrogen atmosphere of a
Jupiter-like exoplanet based on exospheric XUV absorption and
heating. They used the time-dependent XUV fluxes from the ``Sun in
Time'' program (Section~\ref{section:spectralevolution}). Although for
Jupiter in its 5.2\,AU orbit around the present-day Sun, XUV heating
leads to $T_\mathrm{exo} \approx 140 \unit{K}$ only, a close-in
hydrogen-rich Jupiter at distances $<0.3 \unit{AU}$ may suffer
blow-off temperatures $T_\mathrm{exo} > 2 \times 10^4 \unit{K}$. The
XUV-absorbing atmospheric layer (the ``exobase'') expands to several
planetary radii, which has been directly confirmed by a planetary
transit observation \citep{vidal03}. The initial calculations by
\cite{lammer03b} led to very high mass-loss rates, initially
criticized by \cite{yelle04} who derived 20 times smaller escape rates
by also including atmospheric chemistry in their calculations; in
particular, they considered thermospheric cooling by $\mathrm{H}_3^+$
and the lowering of the solar energy deposition rate by ionization of
$\mathrm{H}$ to $\mathrm{H}^+$, which either escapes, or recombines by
emitting a photon, which escapes as well. In this case, the escape
rate will not be large enough to substantially affect the planetary
evolution. This was substantially confirmed by new calculations
performed by the previous group \citep[see][]{penz07} who considered
Roche-lobe effects to the hydrodynamic loss, and used realistic
heating efficiencies (rather than an energy-limited, 100\% efficiency
assumption) and IR cooling terms for HD~209458b, a Jupiter-sized
planet in orbit around an old, inactive solar analog. The temperatures
at $1.5\,R_\mathrm{planet}$  then reach 6000\,--\,10,000\,K
(Figure~\ref{figure:exoplanets}) which, however, leads to blow-off
thanks to gravitational effects by the Roche lobe. Loss rates of $7
\times 10^{10} \unit{g\ s^{-1}}$ are obtained, in agreement with
calculations by \cite{yelle04} and \cite{yelle06}. Extrapolated to
young MS ages of the host star (applying elevated XUV irradiation,
i.e., a factor of 100 at an age of 0.1\,Gyr, see
Table~\ref{table:enhancement}), the loss rate may exceed $10^{12}
\unit{g\ s^{-1}}$, leading to an integrated mass loss of 1\,--\,3.4\%
of the present mass over the entire MS life time of the host star, a
fraction that is not leading to substantial alterations in the
planet. However, higher evaporation fractions are possible for planets
orbiting closer to the host star.

\epubtkImage{}{%
\begin{figure}[htbp]
  \centerline{\includegraphics[scale=0.55]{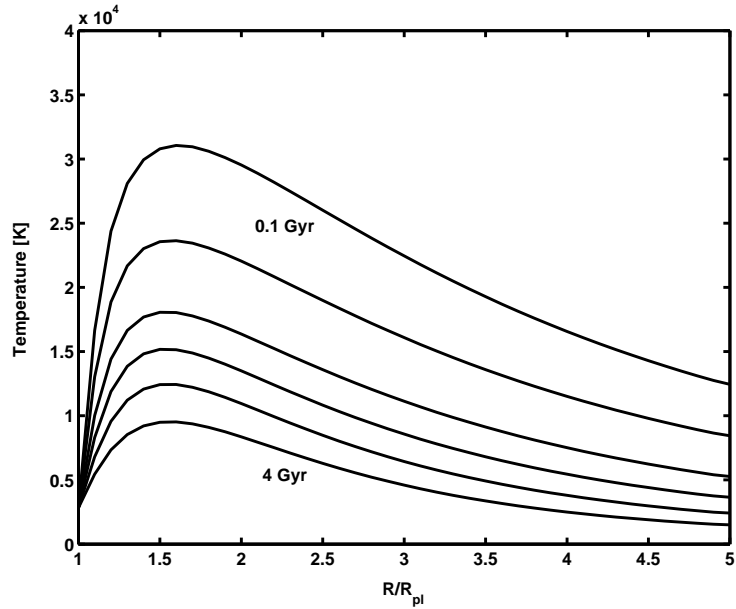}}
  \caption{\it Temperature profile through the atmosphere of the ``hot
  Jupiter'' HD~209458b in a 0.046\,AU orbit around its solar-analog
  host star, for different XUV radiation levels corresponding to ages
  of  4, 2, 1, 0.5, 0.2, and 0.1\,Gyr \citep[from][reprinted with permission]{penz07}.}
  \label{figure:exoplanets}
\end{figure}}

In summary, hot Jupiters around young solar analogs should be in a
state of extensive atmospheric loss for four reasons. The first three,
close distance, high XUV flux, and strong wind of the host star have
been discussed above. The fourth is the lack of a strong planetary
dynamo owing to the slow, synchronous rotation of the planet. This
prevents the planet from building up an extensive magnetosphere that
protects the upper atmospheric layers from erosion by the solar
wind. In fact, during the first 500\,Myr, the magnetosphere may not
protect the atmospheres of hot Jupiters even if tidal locking were
ignored \citep{griessmeier04}. The ``hot Jupiter'' would therefore
interact with the solar/stellar wind in a similar way as present-day
Venus. Such interactions are even further enhanced by collisions with
coronal mass ejections (CMEs). Given that these were probably much
more frequent in the young Sun, significant effects are expected for
planets at larger distances. \cite{khodachenko07} found that in order
to prevent atmospheric erosion down to the planetary core, a
sufficiently strong planetary magnetic field must be present to
deflect CME plasma at stand-off distances of at least
$1.3\,R_\mathrm{planet}$ for hot Jupiters like HD~209458b. The very
existence of numerous Jupiter-mass planets very close to their host
stars may indicate that the planets are able to retain substantial
magnetic moments \citep{khodachenko07}.

\subsection{Summary: The High-Energy Young Solar System}

The young, ZAMS Sun was fainter than the present-day Sun in optical
light by about 30\%, and the consequent lowered irradiance of
solar-system planets poses serious problems as the Earth should have
been in deep freeze if its atmosphere was not fundamentally different
from today's; however, for both the Earth and Mars, there is evidence
for a mild climate in the early days of the Sun's main-sequence
life. But the optical and infrared radiation is not the only, and
perhaps not even the key player. Ultraviolet radiation affects the
chemistry of the upper atmospheres, and EUV and X-ray radiation is
absorbed in the upper layers, ionizing and heating the atmospheric
gas. Subsequent thermal or non-thermal processes, notably also
interactions with the solar wind, led to escape of atmospheric
constituents. Irradiation by UV, EUV, and X-ray photons as well as the
solar wind were tens to hundreds of times stronger in the early solar
system and should have played a key role in affecting and modifying
young planetary atmospheres. Most notably, the loss of water, for
example on Venus, is an inevitable product of photodissociation and
various loss mechanisms, in particular for the light hydrogen. It may
be the Earth's strong magnetic field that protected it sufficiently
from suffering a similar fate as either Venus or Mars.

\newpage


\section{Summary and Conclusions}
\label{section:summary}

Only a few decades ago, the past of our Sun since its arrival on the
ZAMS was gauged essentially by its modest bolometric
luminosity. Stellar evolution calculations have consistently shown
that the young (ZAMS) Sun was in fact {\it fainter} than the
present-day Sun, albeit by only about 30\% or so. Small as this
difference is, it was at least sufficient to recognize a fundamental
problem that might have arisen in the young solar system
\citep{sagan72}: Both Earth and Mars should have been in deep freeze,
perhaps being covered by glaciers all over, not permitting water to
flow and shape the Earth's and Mars' surface and to eventually serve
as the medium in which life formed.

About 3 decades ago, satellites such as {\it IUE} and {\it Einstein}
opened the window to stellar ultraviolet and X-ray sources, entirely
changing our view of stellar evolution in particular in the context of
magnetic fields and the ``activity'' they induce. Young stars seen in
open clusters and in star-forming regions are not feeble UV and X-ray
sources only growing up to the mature solar level. On the contrary,
almost every young (ZAMS, PMS) solar analog dwarfs the present-day
magnetically induced radiative output of the Sun {\it by orders of
  magnitude.} A common X-ray luminosity level of a ZAMS Sun is
1000~times the present solar level. The same holds true for the
production of high-energy particles as seen by their radio
emission. The contemporaneous Sun does not get close to a young star's
radio output even during the strongest flares. While observationally
more challenging, evidence for winds has been found for solar analogs,
and the picture is the same: The present-day Sun occupies a place at
the low end of the range of stellar winds, as it does with respect to
X-ray, ultraviolet,  and radio emission.

The total energy output in the magnetically induced radiation (mostly
in coronal soft X-rays and in chromospheric/transition region
ultraviolet) from the most active, young solar analogs is still
relatively small, amounting to about 1\% of the bolometric energy
loss of the Sun \citep{ribas05}. So, why should we worry?

The real power of magnetically induced radiation is not in its total
luminosity but in the ability of individual photons (and high-energy
particles) to affect the stellar environment. Ultraviolet and X-ray
photons are, in contrast to optical photons, strongly absorbed at
large heights in planetary atmospheres. They ionize and heat these
layers, leading to complex chains of reactions including chemical
networks producing, for example, $\mathrm{O}_2$ and $\mathrm{O}_3$
that shield the surface from any lethal dose of ultraviolet light; and
they heat the thermosphere and exosphere, leading to escape of light
elements in possibly significant numbers.

The impact may have been enormous. It is quite possible that Venus,
the Earth, and Mars started as similar bodies in the young solar
system, but the present-day differences between them could hardly be
larger: the extremely arid, dense, and greenhouse-heated atmosphere of
Venus,  the biologically active surface of the Earth, and the cold
but water-bearing Mars may owe much of their present state to the
action of an extremely magnetically active young Sun - through its
magnetically induced high-energy emission rather than the optical and
infrared emission. Similar things may apply to other solar-system
bodies and also to extrasolar planets. Once high levels of
short-wavelength radiation from young stars had been reported from
observations, its potential role in transforming young planetary
atmospheres was rapidly recognized (e.g., \citealt{sekiya80, zahnle82,
  canuto82, canuto83, kasting83}).

The magnetic, solar wind may have been responsible for the fate of
solar-system bodies as well. It is an important factor in non-thermal
escape processes occurring in upper planetary atmospheres, but it
might even have had some indirect influence on the lower atmospheres
by removing mass from the young Sun. If the young Sun were indeed
slightly more massive and therefore more luminous, the ``Faint Young
Sun Paradox'' may disappear. This solution is currently not preferred
given estimates for young solar winds that would fall short of the
required mass-loss rates by orders of magnitude. However, the picture
of the past solar wind is still not very complete. A direct detection
of any wind of a solar analog (e.g., through its radio emission) has
yet to be reported.

Magnetic activity was perhaps even more fundamentally important in the
PMS Sun when the latter was surrounded by a circumstellar disk in
which planets formed. Not only are magnetic fields crucial for guiding
accretion flows onto the star and perhaps lock the stellar rotation to
the orbit period of the disk; the magnetically induced radiation is
also important in ionizing and partly heating the upper layers of the
accretion disk, thus driving accretion toward the stars and chemical
networks across the disk. Further, high-energy particles produced by
gigantic stellar flares may have left their traces in meteoritic
material, providing a direct window to past conditions in the young,
forming solar system.
 
The big picture of the young Sun's activity and its influence on the
solar-system environment is, in many ways, rather incomplete and at
places controversial. What could be classified as solid knowledge, and
where is more research investment needed?

While even the answer to this question may not be unanimous among
various researchers, the following is probably safe to say, as far as
``established facts'' and ``issues open to speculation''  are
concerned:

\begin{itemize}

  \item Judged from all observational evidence, the young Sun was much
  more magnetically active, due to its higher rotation rate. We are
  not precisely sure what the maximum activity level (and
  correspondingly, the shortest rotation period) of the young Sun was,
  but a rotation period about ten times shorter than at present and
  consequently elevated high-energy (X-ray) losses of order 100~times
  the present level are likely, at least as long as we consider the
  Sun to follow typical trends rather than being a maverick. Higher
  activity levels are possible but difficult to prove from statistical
  stellar studies.

  \item Direct interactions between contemporaneous ``solar activity''
  and upper planetary atmospheres are well established and can be
  studied in-situ in present-day atmospheres
  \citep[e.g.,][]{lammer03a}. The detailed mechanisms having occurred
  in the young solar system are still subject to some speculation as
  several ingredients are difficult to know: what was the composition
  of the initial atmospheres, was there plate tectonics, were there
  continents, how much water was present on the early planets, how
  much $\mathrm{CO}_2$ was in the atmospheres? In what ways did early
  life alter the chemical composition of the Earth's atmosphere? Was
  there a magnetic field strong enough to shield the atmospheres from
  the strong solar wind? Regardless of these partly open questions,
  sophisticated models are available that make predictions that can
  be tested. In particular, these models include the higher activity
  level of the young Sun and can successfully explain the loss of
  water from Venus and the loss of a dense atmosphere from Mars.

  \item The magnetic activity of the young, PMS Sun was undoubtedly
  affecting the complex,  molecular environment of the forming solar
  system through ionization, heating, nuclear reactions, and induced
  chemistry. Such mechanisms are beginning to be studied and modeled
  in detail in other stellar systems. A precise assessment is
  difficult for the young Sun as we have little information on its
  actual UV, X-ray and particle output at those phases, but again,
  indirect evidence may shed light on these issues. The presence of
  chondrules, CAIs with their isotopic anomalies as well as traces of
  irradiation damage in meteoritical material (by nuclei ejected from
  the Sun) may at least partly be due to elevated activity of the
  young Sun, although alternative theories need to be tested.
      
\end{itemize}
    
The young solar system was a place permeated by short-wavelength
radiation, high-energy particles, and an intense wind. Interactions
with solar-system bodies and the initial circumsolar disk were of
fundamental importance much more so than in the relatively quiescent
present-day solar system. The basic driver of all these mechanisms was
the enhanced solar magnetic field. Solar magnetic activity is by no
means a phenomenon localized in the immediate surrounding of the Sun
itself; it has been of relevance in shaping the entire solar system
and the individual bodies back to the earliest times of its
evolution.

\newpage


\section{Acknowledgements}
\label{section:acknowledgements}

It is a pleasure to thank Sami Solanki and the Editorial Board of the
{\it Living Reviews of Solar Physics} for inviting me to write the
present review article, and the editorial staff of the {\it Living
  Reviews} for their professional help and their patience when
unforeseen events took over. I am much indebted to my two referees,
Jeffrey L.\ Linsky and Thierry Montmerle, whose thorough and critical
reviews of this paper have improved both content and
presentation. Special thanks go to Helmut Lammer, Yuri Kulikov, and
Thomas Penz for providing me their new manuscripts before
publication.  Discussions with Tom Ayres and Helmut Lammer have
stimulated inclusion of issues that I would otherwise have failed to
mention. I thank the following colleagues for giving me permission to
reproduce their original figures: Marc Audard, Tom Ayres, Max
Camenzind, Alfred Glassgold, Christopher Johns-Krull, Eric Feigelson,
James Kasting, Helmut Lammer, Antonietta Marino, Stephen Marsden,
Sergio Messina, Thomas Penz, Ignasi Ribas, I.-Juliana Sackmann,
Carolus Schrijver, Klaus Strassmeier, Alessandra Telleschi, and Brian
Wood.

\newpage



\bibliography{refs}

\end{document}